\documentclass[a4paper, 10pt, oneside]{article}
\usepackage[a4paper,inner=2cm,outer=2cm,top=3cm,bottom=2.5cm]{geometry}

\usepackage{amsmath}
\usepackage{amssymb}
\usepackage{amsthm}
\usepackage{dsfont}
\usepackage{slashed}
\usepackage{mathrsfs}
\usepackage{mathtools}
\usepackage{color}

\usepackage{bibgerm}

\usepackage[english]{babel}

\usepackage[utf8x]{inputenc}

\usepackage{graphicx,epsfig}

\usepackage{pstricks,pstricks-add,pst-pdf}

\usepackage{setspace}

\usepackage{booktabs}

\usepackage{float}

\usepackage{nextpage}

\usepackage{newclude}

\usepackage[small,bf]{caption}
\usepackage{subcaption}

\usepackage{url}

\usepackage[colorlinks=true,
	linkcolor=black,
	citecolor=black,
	urlcolor=black,
	filecolor=black
]{hyperref}

\usepackage[T1]{fontenc}

\usepackage{cancel}

\usepackage{pifont}

\usepackage{etex}
\usepackage{bchart}

\usepackage{appendix}

\usepackage[absolute]{textpos}

\newcommand{\ChPT}{$\chi$PT}

\renewcommand{\O}{\mathcal{O}}
\newcommand{\p}{\partial}
\newcommand{\pvint}{\hspace{0.75mm}\mathcal{P}\hspace{-3.5mm}\int}

\renewcommand{\Re}{\mathrm{Re}}
\renewcommand{\Im}{\mathrm{Im}}

\newcommand{\<}{\langle}
\renewcommand{\>}{\rangle}

\newcommand{\dprime}{{\prime\prime}}
\newcommand{\tprime}{{\prime\prime\prime}}

\newcommand{\remark}[1]{}

\newcommand{\m}{\hphantom{-}}

\newcommand{\cmark}{\ding{51}}
\newcommand{\xmark}{\ding{55}}

\makeatletter
\renewcommand{\maketag@@@}[1]{\hbox{\m@th\normalsize\normalfont#1}}%
\makeatother

\makeatletter
\renewcommand\paragraph{\@startsection{paragraph}{4}{\z@}%
  {-3.25ex \@plus -1ex \@minus -0.2ex}%
  {0.01pt}%
  {\bfseries}%
}
\makeatother

\usepackage{abstract}

\begin{document}

\mbox{}

\bigskip

\begin{center}
{\LARGE{\bf A Dispersive Treatment of $K_{\ell4}$ Decays}}

\vspace{0.5cm}

Gilberto Colangelo${}^a$, Emilie Passemar${}^{b,c,d}$, Peter Stoffer${}^{a,e}$

\vspace{1em}

\begin{center}
\it
${}^a$Albert Einstein Center for Fundamental Physics, Institute for Theoretical Physics, \\
University of Bern, Sidlerstrasse~5, CH-3012 Bern, Switzerland \\
\mbox{} \\
${}^b$Department of Physics, Indiana University, Bloomington, IN 47405, USA \\
\mbox{} \\
${}^c$Center for Exploration of Energy and Matter, Indiana University, Bloomington, IN 47403, USA \\
\mbox{} \\
${}^d$Theory Center, Thomas Jefferson National Accelerator Facility, Newport News, VA 23606, USA \\
\mbox{} \\
${}^e$Helmholtz-Institut für Strahlen- und Kernphysik (Theory) and Bethe Center for Theoretical Physics, University of Bonn, D-53115 Bonn, Germany
\end{center} 

\end{center}

\vspace{1em}

\hrule

\begin{abstract}
$K_{\ell4}$ decays offer several reasons of interest: they allow an
accurate measurement of $\pi\pi$-scattering lengths; they provide the best
source for the determination of some low-energy constants of \ChPT{};
one form factor is directly related to the chiral anomaly, which can be
measured here. 
We present a dispersive treatment of $K_{\ell4}$ decays that provides a
resummation of $\pi\pi$- and $K\pi$-rescattering effects. The free
parameters of the dispersion relation are fitted to the data of the
high-statistics experiments E865 and NA48/2. The matching to \ChPT{} at NLO
and NNLO enables us to determine the LECs $L_1^r$, $L_2^r$ and $L_3^r$.
With recently published data from NA48/2, the LEC $L_9^r$ can be
determined as well. In contrast to a pure chiral treatment, the dispersion
relation describes the observed curvature of one of the form factors, which
we understand as a rescattering effect beyond NNLO.
\end{abstract}

\setlength{\TPHorizModule}{1cm}
\setlength{\TPVertModule}{1cm}
\textblockorigin{\paperwidth}{0cm}
\begin{textblock}{4}(-6,2)
\raggedleft
JLAB-THY-15-1998
\end{textblock}

\hrule

\vspace{1em}

\setcounter{tocdepth}{3}
\tableofcontents

\clearpage


\section{Introduction}

$K_{\ell4}$ denotes the semileptonic decay of a kaon into two pions and a
lepton pair. Its amplitude has a similar structure to that of $K\pi$
scattering, with the difference that in $K_{\ell4}$ decays one of the axial
currents couples to an external field, the $W$ boson, which decays into the
lepton pair -- the $q^2$ of this axial current is therefore variable rather
than being stuck at $M_K^2$ as in $K\pi$ scattering. This difference has
the important consequence that in $K_{\ell4}$ decays the allowed
kinematical region reaches down to lower energies, $E \leq M_K$, whereas in
$K \pi$ scattering $E \geq M_K+ M_\pi$.  From the point of view of chiral
perturbation theory (\ChPT{}) \cite{Weinberg1968, GasserLeutwyler1984,
  GasserLeutwyler1985}, the low-energy effective theory of QCD, $K_{\ell4}$
decays offer similar information as $K\pi$ scattering, but in a kinematical
region where the chiral expansion is more reliable.

Due to its two-pion final state, $K_{\ell4}$ is also one of the cleanest
sources of information on $\pi\pi$ interaction
\cite{Shabalin1963,Cabibbo1965,Batley2010}.

The latest high-statistics $K_{\ell4}$ experiments E865 at BNL
\cite{Pislak2001, Pislak2003} and NA48/2 at CERN \cite{Batley2010,
  Batley2012} have achieved an impressive accuracy. The statistical errors
of the $S$-wave of one form factor reach in both experiments the
sub-percent level. Matching this precision requires a theoretical treatment
beyond one-loop order in the chiral expansion. A first treatment beyond one
loop, based on dispersion relations, was already done twenty years ago
\cite{Bijnens1994}. The full two-loop calculation became available in 2000
\cite{Amoros2000}. However, as we will show below, even at two loops
\ChPT{} is not able to predict the curvature of one of the form factors.

Here, we present a new dispersive treatment of $K_{\ell4}$ decays. The form
of the dispersion relation we solve is not exact, but relies on an
assumption (absence of $D$- and higher wave contributions to
discontinuities) that is violated only starting at $\O(p^8)$ in the chiral
expansion. It resums two-particle rescattering effects, which we expect to
be the most important contribution beyond two loops. Indeed, we observe
that the dispersive description is able to reproduce the curvature of the
form factor. 

The dispersion relation is parametrised by subtraction constants, which are
not constrained by unitarity. These have to be determined by theoretical
input or by a fit to data. It turns out that the available data does not
constrain all the subtraction constants to a sufficient precision.
Therefore, we use the soft-pion theorem, a low-energy theorem for
$K_{\ell4}$ that receives only $SU(2)$ chiral corrections, as well as some
chiral input to constrain the parameters that are not well determined from
data alone.

The present treatment of $K_{\ell4}$ decays represents an extension and a
major improvement of our previous dispersive framework \cite{Stoffer2010,
  Colangelo2012, Stoffer2013}. The modifications and improvements concern
the following aspects:
\begin{itemize}
	\item Instead of a single linear combination of form factors, now
          we describe the two form factors $F$ and $G$ simultaneously. This
          allows us to include more experimental data in the fits. 
	\item The new framework is valid also for non-vanishing invariant
          energies of the lepton pair. In the previous treatment, we
          neglected the dependence on this kinematic variable. This
          approximation is no longer used and the observed dependence on
          the lepton invariant energy can be taken into account. 
	\item We apply corrections for isospin-breaking effects in the
          fitted data that have not been taken into account in the
          experimental analysis. 
	\item We perform the matching to \ChPT{} directly on the level of
          the subtraction constants, which avoids the mixing with the
          treatment of rescattering effects. 
	\item Besides a matching to one-loop \ChPT{}, we also study the
          matching at two-loop level. 
\end{itemize}
The first two points required a substantial modification and extension of
the dispersive framework from the very start, but rendered it much more
powerful. The old treatment can be understood as a limiting case of the new
framework. 

The outline is as follows: in section~\ref{sec:Kl4DispersionRelation}, we
derive the dispersion relation for the $K_{\ell4}$ form factors, which has
the form of a set of coupled integral equations. In
section~\ref{sec:Kl4NumericalSolutionDR}, we describe the numerical
procedure that is used to solve this
system. Section~\ref{sec:Kl4SubtractionConstantsDetermination} is devoted
to the determination of the free parameters of the dispersion relation and
the derivation of matching equations to \ChPT{}. In
section~\ref{sec:Kl4Results}, we present the results of the fit to data and
the values of the low-energy constants $L_1^r$, $L_2^r$ and $L_3^r$
obtained in the matching to \ChPT{}. 
Section~\ref{sec:ConclusionOutlook} concludes the main text.
The appendices contain several details on the
kinematics, the derivation of the dispersion relation and explicit
expressions for the matching equations. Further details that are omitted
here can be found in \cite{Stoffer2014a}.


\section{Dispersion Relation for $K_{\ell4}$}

\label{sec:Kl4DispersionRelation}

\subsection{Decay Amplitude and Form Factors}

$K_{\ell4}$ are semileptonic decays of a kaon into two pions and a lepton-neutrino pair:
\begin{align}
	K^+(k) \rightarrow \pi^+(p_1) \pi^-(p_2) \ell^+(p_\ell) \nu_\ell(p_\nu) ,
\end{align}
where $\ell\in\{e,\mu\}$ is either an electron or a muon. There exist other
decay modes involving neutral mesons. Their amplitudes are related to the
above decay by isospin symmetry -- in our dispersive treatment of
$K_{\ell4}$, we will work in the isospin limit and could therefore describe
as well the neutral mode. In the present analysis, however, we only consider the 
charged mode because it is the one which has been measured more accurately.

In the standard model, semileptonic decays are mediated by $W$
bosons. After integrating out the $W$ boson from the standard model
Lagrangian, we end up with a Fermi type effective current-current
interaction. The matrix element of $K_{\ell4}$ then splits up into a
leptonic times a hadronic part. The leptonic matrix element can be treated
in a standard way. The hadronic matrix element exhibits the usual $V-A$
structure of weak interaction: 
\begin{align}
	{}_\mathrm{out}\<\pi^+(p_1)\pi^-(p_2) \ell^+(p_\ell)\nu_\ell(p_\nu) \big| K^+(k)\>_\mathrm{in} = i (2\pi)^4 \delta^{(4)}(k-p_1-p_2-p_\ell-p_\nu) \mathcal{T} , \\
	\mathcal{T} = \frac{G_F}{\sqrt{2}} V_{us}^* \bar u(p_\nu) \gamma^\mu(1-\gamma_5)v(p_\ell) \< \pi^+(p_1) \pi^-(p_2) \big| V_\mu(0)-A_\mu(0) \big| K^+(k) \> ,
\end{align}
where $V_\mu = \bar s \gamma_\mu u$ and $A_\mu = \bar s \gamma_\mu \gamma_5
u$. Note that although we drop the corresponding labels, the meson states
are still in- and out-states with respect to the strong interaction. 

The Lorentz structure of the currents allows us to write the two hadronic
matrix elements as 
\begin{align}
	\mathcal{V}_\mu^{+-} := \big\< \pi^+(p_1) \pi^-(p_2) \big| V_\mu(0) \big| K^+(k)\big\> &= -\frac{H}{M_K^3} \epsilon_{\mu\nu\rho\sigma} L^\nu P^\rho Q^\sigma , \\
	\mathcal{A}_\mu^{+-} := \big\< \pi^+(p_1) \pi^-(p_2) \big| A_\mu(0) \big| K^+(k) \big\> &= -i \frac{1}{M_K} \left( P_\mu F + Q_\mu G + L_\mu R \right) ,
\end{align}
where $P = p_1 + p_2$, $Q = p_1 - p_2$, $L = k - p_1 - p_2$. The form
factors $F$, $G$, $R$ and $H$ are dimensionless scalar functions of the
Mandelstam variables: 
\begin{align}
	\begin{split}
		s &= (p_1 + p_2)^2 = (k - L)^2 , \\
		t &= (k - p_1)^2 = (p_2 + L)^2 , \\
		u &= (k - p_2)^2 = (p_1 + L)^2 .
	\end{split}
\end{align}

We further define the invariant squared energy of the lepton pair $s_\ell =
L^2$. For the hadronic matrix element, we regard $s_\ell$ as a fixed
external quantity. 

\subsection{Analytic Structure}

\label{sec:AnalyticStructure}

\begin{figure}[H]
	\centering
	\psset{unit=0.248cm}
	\begin{pspicture*}(-30,-30)(30,30)
		\pspolygon[linestyle=none, fillstyle=solid, fillcolor=lightgray](-17.7014,4)(17.7014,4)(0,-26.6597)
		\pspolygon[linewidth=2pt, fillstyle=solid, fillcolor=gray](0.,12.5112)(-0.483109,11.4744)(-0.8654,10.6123)(-1.1701,9.88454)(-1.41382,9.26242)(-1.6087,8.72486)(-1.76389,8.25607)(-1.88635,7.84396)(-1.98151,7.47914)(-2.05365,7.15419)(-2.10619,6.86318)(-2.14191,6.60133)(-2.16306,6.3647)(-2.17151,6.15006)(-2.16882,5.95472)(-2.15629,5.77641)(-2.13503,5.61324)(-2.10597,5.46357)(-2.06991,5.32602)(-2.02755,5.19941)(-1.97946,5.08269)(-1.92618,4.97498)(-1.86814,4.87551)(-1.80573,4.78359)(-1.73931,4.69865)(-1.66916,4.62016)(-1.59554,4.54766)(-1.51869,4.48077)(-1.43881,4.41912)(-1.35607,4.36243)(-1.27062,4.31043)(-1.1826,4.26289)(-1.09212,4.21961)(-0.999265,4.18044)(-0.904118,4.14524)(-0.806736,4.11391)(-0.707165,4.08637)(-0.605428,4.06259)(-0.501534,4.04253)(-0.395473,4.02624)(-0.28721,4.01376)(-0.176688,4.00518)(-0.063822,4.00067)(0.0510095,4.00043)(0.164146,4.00447)(0.274925,4.0126)(0.383438,4.02465)(0.489745,4.04052)(0.593881,4.06015)(0.695859,4.08352)(0.795675,4.11063)(0.893304,4.14153)(0.988704,4.1763)(1.08182,4.21502)(1.17257,4.25782)(1.26088,4.30488)(1.34662,4.35637)(1.42967,4.41252)(1.50988,4.47359)(1.58708,4.53988)(1.66107,4.61173)(1.73162,4.68953)(1.79848,4.77373)(1.86135,4.86483)(1.9199,4.96343)(1.97374,5.07017)(2.02243,5.18584)(2.06546,5.3113)(2.10226,5.44756)(2.13214,5.5958)(2.15432,5.75739)(2.16788,5.9339)(2.17173,6.12723)(2.1646,6.33958)(2.14497,6.57359)(2.11099,6.83243)(2.06047,7.11995)(1.99068,7.44082)(1.8983,7.80083)(1.77915,8.20721)(1.62795,8.6691)(1.43792,9.19823)(1.20022,9.80993)(0.90309,10.5246)(0.530513,11.3699)(0.0601512,12.3846)
		\psline[linewidth=2pt, fillstyle=solid, fillcolor=gray](-10.2708,31.5486)(-10.1568,31.346)(-10.0428,31.1435)(-9.92882,30.9409)(-9.81485,30.7383)(-9.70091,30.5357)(-9.58699,30.333)(-9.4731,30.1303)(-9.35922,29.9275)(-9.24537,29.7247)(-9.13154,29.5218)(-9.01774,29.319)(-8.90396,29.116)(-8.7902,28.9131)(-8.67647,28.7101)(-8.56277,28.507)(-8.44909,28.3039)(-8.33544,28.1007)(-8.22182,27.8975)(-8.10822,27.6943)(-7.99466,27.491)(-7.88112,27.2876)(-7.76762,27.0842)(-7.65415,26.8808)(-7.54071,26.6772)(-7.4273,26.4737)(-7.31393,26.27)(-7.20059,26.0663)(-7.08729,25.8626)(-6.97402,25.6588)(-6.8608,25.4549)(-6.74761,25.2509)(-6.63446,25.0469)(-6.52135,24.8428)(-6.40828,24.6387)(-6.29525,24.4344)(-6.18227,24.2301)(-6.06934,24.0257)(-5.95645,23.8213)(-5.84361,23.6167)(-5.73081,23.4121)(-5.61807,23.2074)(-5.50538,23.0025)(-5.39275,22.7976)(-5.28017,22.5926)(-5.16764,22.3875)(-5.05518,22.1823)(-4.94277,21.977)(-4.83043,21.7716)(-4.71815,21.5661)(-4.60594,21.3604)(-4.4938,21.1547)(-4.38173,20.9488)(-4.26973,20.7428)(-4.15781,20.5366)(-4.04596,20.3303)(-3.9342,20.1239)(-3.82252,19.9173)(-3.71093,19.7106)(-3.59943,19.5038)(-3.48802,19.2967)(-3.37671,19.0895)(-3.26549,18.8821)(-3.15439,18.6746)(-3.04339,18.4668)(-2.9325,18.2589)(-2.82173,18.0508)(-2.71108,17.8424)(-2.60056,17.6338)(-2.49017,17.425)(-2.37992,17.216)(-2.26981,17.0067)(-2.15986,16.7972)(-2.05005,16.5873)(-1.94042,16.3772)(-1.83095,16.1668)(-1.72167,15.9561)(-1.61257,15.7451)(-1.50367,15.5337)(-1.39497,15.322)(-1.2865,15.1099)(-1.17825,14.8973)(-1.07025,14.6844)(-0.962501,14.471)(-0.855021,14.2572)(-0.747825,14.0429)(-0.640931,13.828)(-0.534356,13.6126)(-0.42812,13.3966)(-0.322245,13.18)(-0.216753,12.9627)(-0.111671,12.7447)(-0.00702684,12.526)(0.,12.5112)(0.105138,12.7311)(0.21072,12.9502)(0.316717,13.1686)(0.423103,13.3864)(0.529855,13.6035)(0.636949,13.82)(0.744367,14.0359)(0.85209,14.2514)(0.9601,14.4663)(1.06838,14.6807)(1.17692,14.8947)(1.28571,15.1083)(1.39472,15.3215)(1.50396,15.5343)(1.61341,15.7467)(1.72305,15.9588)(1.83289,16.1706)(1.9429,16.382)(2.05309,16.5932)(2.16345,16.804)(2.27396,17.0146)(2.38462,17.2249)(2.49543,17.435)(2.60638,17.6448)(2.71746,17.8544)(2.82867,18.0638)(2.94,18.273)(3.05145,18.4819)(3.16301,18.6907)(3.27469,18.8993)(3.38646,19.1077)(3.49834,19.3159)(3.61032,19.524)(3.72238,19.7319)(3.83454,19.9396)(3.94679,20.1472)(4.05912,20.3546)(4.17154,20.5619)(4.28403,20.7691)(4.39659,20.9761)(4.50924,21.183)(4.62195,21.3898)(4.73473,21.5964)(4.84758,21.803)(4.96049,22.0094)(5.07347,22.2157)(5.18651,22.4219)(5.2996,22.628)(5.41276,22.834)(5.52597,23.04)(5.63923,23.2458)(5.75254,23.4515)(5.86591,23.6572)(5.97933,23.8627)(6.09279,24.0682)(6.2063,24.2736)(6.31985,24.4789)(6.43345,24.6841)(6.5471,24.8893)(6.66078,25.0944)(6.77451,25.2994)(6.88827,25.5044)(7.00208,25.7093)(7.11592,25.9141)(7.2298,26.1188)(7.34371,26.3235)(7.45766,26.5282)(7.57164,26.7328)(7.68566,26.9373)(7.7997,27.1417)(7.91378,27.3461)(8.02789,27.5505)(8.14204,27.7548)(8.25621,27.9591)(8.3704,28.1633)(8.48463,28.3674)(8.59889,28.5715)(8.71317,28.7756)(8.82747,28.9796)(8.94181,29.1836)(9.05616,29.3875)(9.17055,29.5914)(9.28495,29.7952)(9.39938,29.999)(9.51383,30.2028)(9.62831,30.4065)(9.74281,30.6102)(9.85732,30.8138)(9.97186,31.0174)(10.0864,31.221)(10.201,31.4246)
		\psline[linewidth=2pt, fillstyle=solid, fillcolor=gray](-26.3958,-30.4736)(-26.2831,-30.2736)(-26.1704,-30.0736)(-26.0577,-29.8736)(-25.9451,-29.6736)(-25.8325,-29.4736)(-25.7199,-29.2736)(-25.6074,-29.0736)(-25.4949,-28.8736)(-25.3824,-28.6736)(-25.27,-28.4736)(-25.1577,-28.2736)(-25.0454,-28.0736)(-24.9331,-27.8736)(-24.8209,-27.6736)(-24.7087,-27.4736)(-24.5965,-27.2736)(-24.4845,-27.0736)(-24.3724,-26.8736)(-24.2604,-26.6736)(-24.1485,-26.4736)(-24.0366,-26.2736)(-23.9248,-26.0736)(-23.813,-25.8736)(-23.7013,-25.6736)(-23.5896,-25.4736)(-23.478,-25.2736)(-23.3664,-25.0736)(-23.2549,-24.8736)(-23.1435,-24.6736)(-23.0322,-24.4736)(-22.9209,-24.2736)(-22.8096,-24.0736)(-22.6985,-23.8736)(-22.5874,-23.6736)(-22.4764,-23.4736)(-22.3654,-23.2736)(-22.2546,-23.0736)(-22.1438,-22.8736)(-22.0331,-22.6736)(-21.9224,-22.4736)(-21.8119,-22.2736)(-21.7014,-22.0736)(-21.5911,-21.8736)(-21.4808,-21.6736)(-21.3706,-21.4736)(-21.2605,-21.2736)(-21.1505,-21.0736)(-21.0406,-20.8736)(-20.9308,-20.6736)(-20.8212,-20.4736)(-20.7116,-20.2736)(-20.6021,-20.0736)(-20.4928,-19.8736)(-20.3836,-19.6736)(-20.2745,-19.4736)(-20.1655,-19.2736)(-20.0567,-19.0736)(-19.948,-18.8736)(-19.8394,-18.6736)(-19.731,-18.4736)(-19.6227,-18.2736)(-19.5146,-18.0736)(-19.4066,-17.8736)(-19.2988,-17.6736)(-19.1912,-17.4736)(-19.0837,-17.2736)(-18.9764,-17.0736)(-18.8693,-16.8736)(-18.7624,-16.6736)(-18.6557,-16.4736)(-18.5492,-16.2736)(-18.4429,-16.0736)(-18.3369,-15.8736)(-18.231,-15.6736)(-18.1254,-15.4736)(-18.0201,-15.2736)(-17.915,-15.0736)(-17.8102,-14.8736)(-17.7056,-14.6736)(-17.6013,-14.4736)(-17.4974,-14.2736)(-17.3937,-14.0736)(-17.2904,-13.8736)(-17.1874,-13.6736)(-17.0847,-13.4736)(-16.9824,-13.2736)(-16.8805,-13.0736)(-16.779,-12.8736)(-16.6779,-12.6736)(-16.5773,-12.4736)(-16.4771,-12.2736)(-16.3774,-12.0736)(-16.2781,-11.8736)(-16.1794,-11.6736)(-16.0813,-11.4736)(-15.9837,-11.2736)(-15.8868,-11.0736)(-15.7904,-10.8736)(-15.6948,-10.6736)(-15.5998,-10.4736)(-15.5056,-10.2736)(-15.4122,-10.0736)(-15.3197,-9.87356)(-15.228,-9.67356)(-15.1372,-9.47356)(-15.0475,-9.27356)(-14.9587,-9.07356)(-14.8711,-8.87356)(-14.7847,-8.67356)(-14.6996,-8.47356)(-14.6158,-8.27356)(-14.5335,-8.07356)(-14.4527,-7.87356)(-14.3736,-7.67356)(-14.2963,-7.47356)(-14.2209,-7.27356)(-14.1476,-7.07356)(-14.0765,-6.87356)(-14.0079,-6.67356)(-13.9419,-6.47356)(-13.8787,-6.27356)(-13.8187,-6.07356)(-13.7621,-5.87356)(-13.7092,-5.67356)(-13.6605,-5.47356)(-13.6163,-5.27356)(-13.5772,-5.07356)(-13.5436,-4.87356)(-13.5163,-4.67356)(-13.496,-4.47356)(-13.4836,-4.27356)(-13.4801,-4.07356)(-13.4868,-3.87356)(-13.5052,-3.67356)(-13.537,-3.47356)(-13.5844,-3.27356)(-13.6502,-3.07356)(-13.7376,-2.87356)(-13.851,-2.67356)(-13.9959,-2.47356)(-14.1794,-2.27356)(-14.4113,-2.07356)(-14.7048,-1.87356)(-15.0789,-1.67356)(-15.5613,-1.47356)(-16.195,-1.27356)(-17.0504,-1.07356)(-18.2527,-0.873558)(-20.0516,-0.673558)(-23.0417,-0.473558)(-29.1745,-0.273558)(-54.0702,-0.0735576)(-60,-30)
		\psline[linewidth=2pt, fillstyle=solid, fillcolor=gray](26.3958,-30.4736)(26.2831,-30.2736)(26.1704,-30.0736)(26.0577,-29.8736)(25.9451,-29.6736)(25.8325,-29.4736)(25.7199,-29.2736)(25.6074,-29.0736)(25.4949,-28.8736)(25.3824,-28.6736)(25.27,-28.4736)(25.1577,-28.2736)(25.0454,-28.0736)(24.9331,-27.8736)(24.8209,-27.6736)(24.7087,-27.4736)(24.5965,-27.2736)(24.4845,-27.0736)(24.3724,-26.8736)(24.2604,-26.6736)(24.1485,-26.4736)(24.0366,-26.2736)(23.9248,-26.0736)(23.813,-25.8736)(23.7013,-25.6736)(23.5896,-25.4736)(23.478,-25.2736)(23.3664,-25.0736)(23.2549,-24.8736)(23.1435,-24.6736)(23.0322,-24.4736)(22.9209,-24.2736)(22.8096,-24.0736)(22.6985,-23.8736)(22.5874,-23.6736)(22.4764,-23.4736)(22.3654,-23.2736)(22.2546,-23.0736)(22.1438,-22.8736)(22.0331,-22.6736)(21.9224,-22.4736)(21.8119,-22.2736)(21.7014,-22.0736)(21.5911,-21.8736)(21.4808,-21.6736)(21.3706,-21.4736)(21.2605,-21.2736)(21.1505,-21.0736)(21.0406,-20.8736)(20.9308,-20.6736)(20.8212,-20.4736)(20.7116,-20.2736)(20.6021,-20.0736)(20.4928,-19.8736)(20.3836,-19.6736)(20.2745,-19.4736)(20.1655,-19.2736)(20.0567,-19.0736)(19.948,-18.8736)(19.8394,-18.6736)(19.731,-18.4736)(19.6227,-18.2736)(19.5146,-18.0736)(19.4066,-17.8736)(19.2988,-17.6736)(19.1912,-17.4736)(19.0837,-17.2736)(18.9764,-17.0736)(18.8693,-16.8736)(18.7624,-16.6736)(18.6557,-16.4736)(18.5492,-16.2736)(18.4429,-16.0736)(18.3369,-15.8736)(18.231,-15.6736)(18.1254,-15.4736)(18.0201,-15.2736)(17.915,-15.0736)(17.8102,-14.8736)(17.7056,-14.6736)(17.6013,-14.4736)(17.4974,-14.2736)(17.3937,-14.0736)(17.2904,-13.8736)(17.1874,-13.6736)(17.0847,-13.4736)(16.9824,-13.2736)(16.8805,-13.0736)(16.779,-12.8736)(16.6779,-12.6736)(16.5773,-12.4736)(16.4771,-12.2736)(16.3774,-12.0736)(16.2781,-11.8736)(16.1794,-11.6736)(16.0813,-11.4736)(15.9837,-11.2736)(15.8868,-11.0736)(15.7904,-10.8736)(15.6948,-10.6736)(15.5998,-10.4736)(15.5056,-10.2736)(15.4122,-10.0736)(15.3197,-9.87356)(15.228,-9.67356)(15.1372,-9.47356)(15.0475,-9.27356)(14.9587,-9.07356)(14.8711,-8.87356)(14.7847,-8.67356)(14.6996,-8.47356)(14.6158,-8.27356)(14.5335,-8.07356)(14.4527,-7.87356)(14.3736,-7.67356)(14.2963,-7.47356)(14.2209,-7.27356)(14.1476,-7.07356)(14.0765,-6.87356)(14.0079,-6.67356)(13.9419,-6.47356)(13.8787,-6.27356)(13.8187,-6.07356)(13.7621,-5.87356)(13.7092,-5.67356)(13.6605,-5.47356)(13.6163,-5.27356)(13.5772,-5.07356)(13.5436,-4.87356)(13.5163,-4.67356)(13.496,-4.47356)(13.4836,-4.27356)(13.4801,-4.07356)(13.4868,-3.87356)(13.5052,-3.67356)(13.537,-3.47356)(13.5844,-3.27356)(13.6502,-3.07356)(13.7376,-2.87356)(13.851,-2.67356)(13.9959,-2.47356)(14.1794,-2.27356)(14.4113,-2.07356)(14.7048,-1.87356)(15.0789,-1.67356)(15.5613,-1.47356)(16.195,-1.27356)(17.0504,-1.07356)(18.2527,-0.873558)(20.0516,-0.673558)(23.0417,-0.473558)(29.1745,-0.273558)(54.0702,-0.0735576)(60,-30)
		\psline(66.1131, -100)(-49.357, 100)
		\psline(57.735, 114.511)(-57.735, -85.4888)
		\psline(-123.848, 0)(107.092, 0)
		\psline[linestyle=dotted](116.625, 12.5112)(-114.315, 12.5112)
		\psline[linestyle=dotted](121.539, 4)(-109.401, 4)
		\psline[linestyle=dotted](116.625, 12.5112)(-114.315, 12.5112)
		\psline[linestyle=dotted](-69.62, 93.9258)(45.85, -106.074)
		\psline[linestyle=dotted](-42.343, -100)(73.127, 100)
		\psline[linestyle=dotted](-61.4514, 108.074)(54.0186, -91.9258)
		\psline[linestyle=dotted](-58.3124, 113.511)(57.1577, -86.4888)
		\psline[linestyle=dotted](-58.6803, -100)(56.7898, 100)
		\psline[linestyle=dotted](-64.9584, -100)(50.5117, 100)
		\put(-29.5,0.5){$s=0$}
		\put(-29.5,4.75){$s=4M_\pi^2$}
		\put(-29.5,13.25){$s=M_K^2$}
		\rput{-60}(16,-15){$t=0$}
		\rput{-60}(11,-20){$t=(M_K-M_\pi)^2$}
		\rput{-60}(-2.5,-25){$t=(M_K+M_\pi)^2$}
		\rput{60}(4,23){$u=0$}
		\rput{60}(9.5,21){$u=(M_K-M_\pi)^2$}
		\rput{60}(21,12){$u=(M_K+M_\pi)^2$}
		\put(-4,6){decay region}
		\put(-3.5,27){$s$-channel}
		\put(-27,-10){$t$-channel}
		\put(20,-10){$u$-channel}
		\put(-5,-10){real amplitude}
	\end{pspicture*}
	\caption{Mandelstam diagram for $K_{\ell4}$ for the case $s_\ell = 0$}
	\label{img:MandelstamDiagram1}
\end{figure}
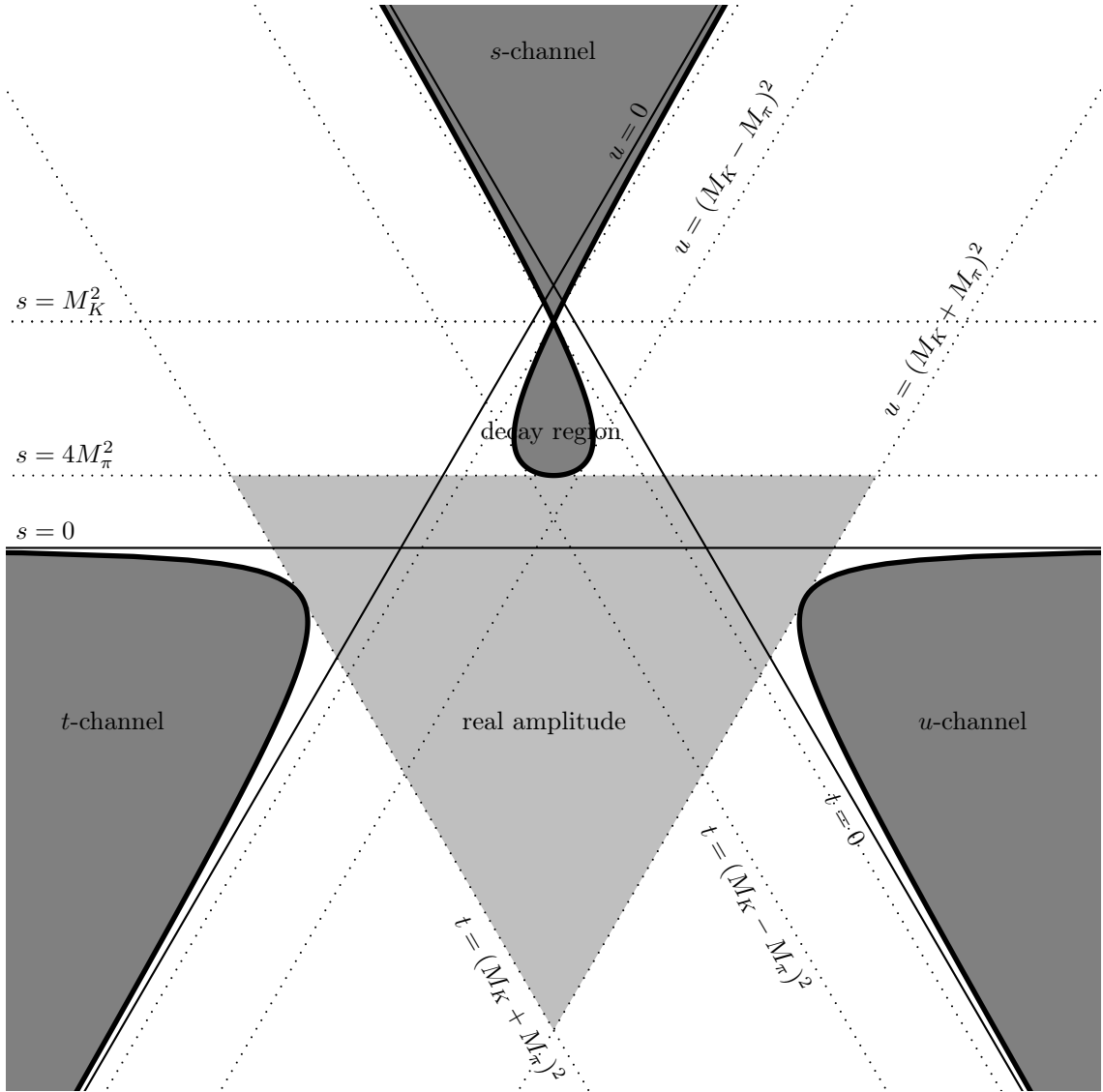

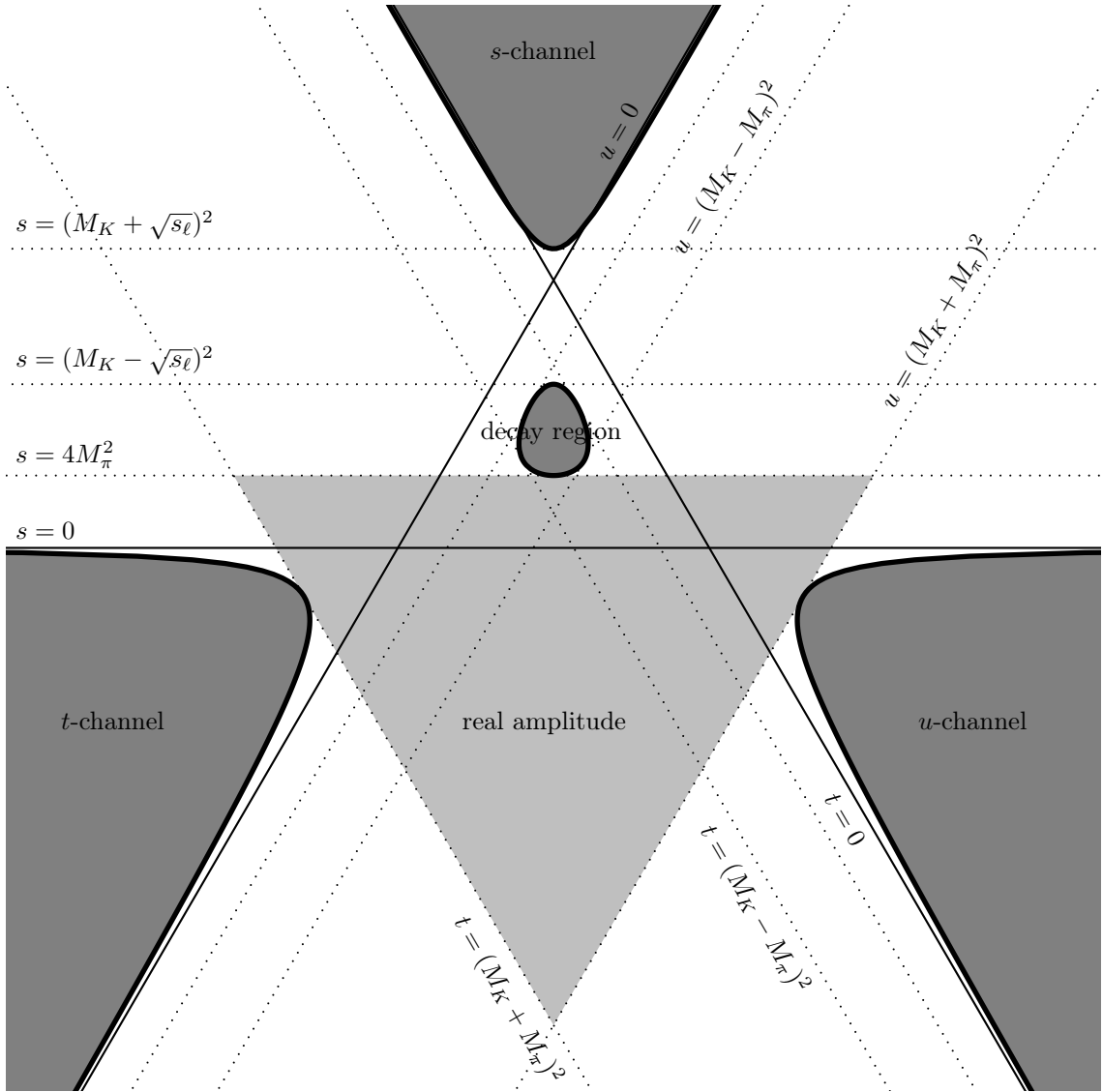
\begin{figure}[H]
	\centering
	\psset{unit=0.248cm}
	\begin{pspicture*}(-30,-30)(30,30)
		\pspolygon[linestyle=none, fillstyle=solid, fillcolor=lightgray](-17.5397,4)(17.5397,4)(0,-26.3797)
		\pspolygon[linewidth=2pt, fillstyle=solid, fillcolor=gray](-1.79221,5.07298)(-1.87635,5.41871)(-1.89766,5.65562)(-1.89763,5.85557)(-1.88566,6.03484)(-1.86574,6.20033)(-1.84,6.35575)(-1.80973,6.50332)(-1.77578,6.64452)(-1.73874,6.78037)(-1.69903,6.91159)(-1.65696,7.03872)(-1.61277,7.16218)(-1.56663,7.28226)(-1.51868,7.39921)(-1.46901,7.51319)(-1.4177,7.62432)(-1.3648,7.73268)(-1.31032,7.83832)(-1.25428,7.94125)(-1.19665,8.04144)(-1.13741,8.13884)(-1.07651,8.23335)(-1.01386,8.32485)(-0.949383,8.41316)(-0.882938,8.49808)(-0.816672,8.5767)(-0.556377,8.82755)(-0.370319,8.94981)(-0.21788,9.01384)(-0.0858075,9.04259)(0.0322829,9.04713)(0.140056,9.0338)(0.239845,9.00664)(0.333238,8.9684)(0.42137,8.92105)(0.505081,8.86604)(0.585009,8.80448)(0.661648,8.73723)(0.735389,8.66495)(0.806542,8.58819)(0.875355,8.50738)(0.942029,8.42286)(1.00672,8.33492)(1.06957,8.24377)(1.13067,8.14959)(1.19009,8.05251)(1.24789,7.95263)(1.30411,7.85001)(1.35876,7.74467)(1.41185,7.63661)(1.46334,7.52579)(1.51319,7.41214)(1.56134,7.29553)(1.57232,7.26788)(1.73916,6.7789)(1.82563,6.42913)(1.87303,6.14704)(1.89519,5.90865)(1.89901,5.70204)(1.88858,5.52011)(1.86661,5.35815)(1.83504,5.21284)(1.79528,5.0817)(1.74843,4.96284)(1.69537,4.85476)(1.63678,4.75624)(1.57324,4.66629)(1.50522,4.5841)(1.43312,4.50899)(1.35726,4.44038)(1.27793,4.37778)(1.19536,4.3208)(1.10974,4.26909)(1.02125,4.22236)(0.930015,4.18038)(0.836147,4.14297)(0.739728,4.10997)(0.640816,4.08129)(0.539446,4.05687)(0.435625,4.03669)(0.329333,4.0208)(0.220519,4.00927)(0.109095,4.00226)(-0.00507317,4.)(-0.122177,4.00284)(-0.242487,4.01122)(-0.36638,4.02581)(-0.494394,4.04753)(-0.627314,4.07776)(-0.766342,4.11856)(-0.913462,4.17338)(-1.07234,4.24856)(-1.25134,4.3586)(-1.4829,4.55967)(-1.50522,4.5841)(-1.57324,4.66629)(-1.63678,4.75624)(-1.69537,4.85476)(-1.74843,4.96284)(-1.79528,5.0817)
		\psline[linewidth=2pt, fillstyle=solid, fillcolor=gray](-10.3668,32.3344)(-10.2517,32.1337)(-10.1366,31.933)(-10.0216,31.7323)(-9.90648,31.5316)(-9.7914,31.3309)(-9.67632,31.1303)(-9.56124,30.9296)(-9.44615,30.7289)(-9.33105,30.5283)(-9.21595,30.3276)(-9.10084,30.127)(-8.98573,29.9264)(-8.87061,29.7258)(-8.75548,29.5252)(-8.64034,29.3246)(-8.52519,29.1241)(-8.41003,28.9235)(-8.29486,28.723)(-8.17967,28.5225)(-8.06447,28.3221)(-7.94926,28.1216)(-7.83404,27.9212)(-7.71879,27.7208)(-7.60353,27.5204)(-7.48825,27.3201)(-7.37295,27.1198)(-7.25763,26.9196)(-7.14228,26.7193)(-7.02691,26.5192)(-6.91151,26.3191)(-6.79607,26.119)(-6.68061,25.919)(-6.56511,25.719)(-6.44958,25.5191)(-6.334,25.3193)(-6.21838,25.1196)(-6.10272,24.9199)(-5.987,24.7203)(-5.87123,24.5209)(-5.75541,24.3215)(-5.63951,24.1222)(-5.52355,23.9231)(-5.40752,23.724)(-5.29141,23.5252)(-5.17521,23.3264)(-5.05891,23.1279)(-4.94251,22.9295)(-4.82601,22.7313)(-4.70938,22.5333)(-4.59262,22.3355)(-4.47572,22.138)(-4.35866,21.9407)(-4.24143,21.7438)(-4.12402,21.5471)(-4.0064,21.3509)(-3.88856,21.155)(-3.77047,20.9595)(-3.6521,20.7645)(-3.53344,20.57)(-3.41445,20.3761)(-3.29509,20.1829)(-3.17532,19.9903)(-3.0551,19.7986)(-2.93436,19.6077)(-2.81305,19.4178)(-2.6911,19.229)(-2.56843,19.0415)(-2.44493,18.8554)(-2.32049,18.6709)(-2.19498,18.4883)(-2.06824,18.3078)(-1.94008,18.1298)(-1.81025,17.9547)(-1.67848,17.7829)(-1.5444,17.6152)(-1.40759,17.4521)(-1.26749,17.2948)(-1.12336,17.1444)(-0.974271,17.0027)(-0.818964,16.8717)(-0.65571,16.7544)(-0.482072,16.6552)(-0.294486,16.5801)(-0.0874935,16.5386)(0.147764,16.5461)(0.426863,16.6295)(0.780557,16.8421)(1.28196,17.3106)(2.18255,18.4704)(2.21315,18.5145)(2.3385,18.6974)(2.46279,18.8821)(2.58617,19.0684)(2.70873,19.2562)(2.83058,19.4451)(2.9518,19.6351)(3.07246,19.8262)(3.19262,20.018)(3.31233,20.2107)(3.43163,20.4041)(3.55057,20.598)(3.66919,20.7926)(3.78751,20.9877)(3.90556,21.1832)(4.02337,21.3791)(4.14096,21.5755)(4.25835,21.7721)(4.37555,21.9691)(4.49258,22.1664)(4.60946,22.364)(4.7262,22.5618)(4.84281,22.7598)(4.9593,22.9581)(5.07569,23.1565)(5.19197,23.3551)(5.30815,23.5538)(5.42426,23.7527)(5.54028,23.9518)(5.65623,24.1509)(5.77211,24.3502)(5.88793,24.5496)(6.00369,24.7491)(6.1194,24.9487)(6.23506,25.1484)(6.35067,25.3481)(6.46624,25.548)(6.58177,25.7479)(6.69726,25.9478)(6.81272,26.1478)(6.92815,26.3479)(7.04354,26.548)(7.15891,26.7482)(7.27426,26.9484)(7.38958,27.1487)(7.50488,27.349)(7.62015,27.5493)(7.73541,27.7497)(7.85065,27.9501)(7.96588,28.1505)(8.08109,28.351)(8.19628,28.5514)(8.31146,28.7519)(8.42663,28.9525)(8.54179,29.153)(8.65694,29.3536)(8.77208,29.5541)(8.88721,29.7547)(9.00233,29.9553)(9.11744,30.1559)(9.23255,30.3566)(9.34765,30.5572)(9.46274,30.7579)(9.57783,30.9585)(9.69292,31.1592)(9.808,31.3599)(9.92307,31.5606)(10.0381,31.7612)(10.1532,31.9619)(10.2683,32.1626)
		\psline[linewidth=2pt, fillstyle=solid, fillcolor=gray](-26.8011,-31.0616)(-26.6878,-30.8616)(-26.5746,-30.6616)(-26.4614,-30.4616)(-26.3482,-30.2616)(-26.235,-30.0616)(-26.1219,-29.8616)(-26.0088,-29.6616)(-25.8958,-29.4616)(-25.7827,-29.2616)(-25.6698,-29.0616)(-25.5568,-28.8616)(-25.4439,-28.6616)(-25.331,-28.4616)(-25.2182,-28.2616)(-25.1054,-28.0616)(-24.9926,-27.8616)(-24.8799,-27.6616)(-24.7672,-27.4616)(-24.6545,-27.2616)(-24.5419,-27.0616)(-24.4294,-26.8616)(-24.3169,-26.6616)(-24.2044,-26.4616)(-24.092,-26.2616)(-23.9796,-26.0616)(-23.8673,-25.8616)(-23.755,-25.6616)(-23.6428,-25.4616)(-23.5306,-25.2616)(-23.4185,-25.0616)(-23.3064,-24.8616)(-23.1944,-24.6616)(-23.0824,-24.4616)(-22.9705,-24.2616)(-22.8587,-24.0616)(-22.7469,-23.8616)(-22.6352,-23.6616)(-22.5236,-23.4616)(-22.412,-23.2616)(-22.3005,-23.0616)(-22.189,-22.8616)(-22.0777,-22.6616)(-21.9664,-22.4616)(-21.8551,-22.2616)(-21.744,-22.0616)(-21.6329,-21.8616)(-21.5219,-21.6616)(-21.411,-21.4616)(-21.3002,-21.2616)(-21.1895,-21.0616)(-21.0789,-20.8616)(-20.9683,-20.6616)(-20.8579,-20.4616)(-20.7476,-20.2616)(-20.6373,-20.0616)(-20.5272,-19.8616)(-20.4172,-19.6616)(-20.3073,-19.4616)(-20.1975,-19.2616)(-20.0878,-19.0616)(-19.9782,-18.8616)(-19.8688,-18.6616)(-19.7595,-18.4616)(-19.6503,-18.2616)(-19.5413,-18.0616)(-19.4324,-17.8616)(-19.3237,-17.6616)(-19.2151,-17.4616)(-19.1067,-17.2616)(-18.9985,-17.0616)(-18.8904,-16.8616)(-18.7825,-16.6616)(-18.6748,-16.4616)(-18.5672,-16.2616)(-18.4599,-16.0616)(-18.3528,-15.8616)(-18.2458,-15.6616)(-18.1391,-15.4616)(-18.0327,-15.2616)(-17.9264,-15.0616)(-17.8204,-14.8616)(-17.7147,-14.6616)(-17.6092,-14.4616)(-17.5041,-14.2616)(-17.3992,-14.0616)(-17.2945,-13.8616)(-17.1903,-13.6616)(-17.0863,-13.4616)(-16.9827,-13.2616)(-16.8794,-13.0616)(-16.7765,-12.8616)(-16.674,-12.6616)(-16.5719,-12.4616)(-16.4703,-12.2616)(-16.369,-12.0616)(-16.2683,-11.8616)(-16.168,-11.6616)(-16.0683,-11.4616)(-15.9691,-11.2616)(-15.8705,-11.0616)(-15.7725,-10.8616)(-15.6751,-10.6616)(-15.5784,-10.4616)(-15.4824,-10.2616)(-15.3871,-10.0616)(-15.2926,-9.86156)(-15.199,-9.66156)(-15.1062,-9.46156)(-15.0144,-9.26156)(-14.9236,-9.06156)(-14.8339,-8.86156)(-14.7453,-8.66156)(-14.6579,-8.46156)(-14.5718,-8.26156)(-14.4871,-8.06156)(-14.4039,-7.86156)(-14.3223,-7.66156)(-14.2424,-7.46156)(-14.1643,-7.26156)(-14.0883,-7.06156)(-14.0144,-6.86156)(-13.9428,-6.66156)(-13.8738,-6.46156)(-13.8075,-6.26156)(-13.7442,-6.06156)(-13.6843,-5.86156)(-13.628,-5.66156)(-13.5757,-5.46156)(-13.5277,-5.26156)(-13.4847,-5.06156)(-13.4471,-4.86156)(-13.4156,-4.66156)(-13.3909,-4.46156)(-13.3739,-4.26156)(-13.3656,-4.06156)(-13.3672,-3.86156)(-13.3802,-3.66156)(-13.4063,-3.46156)(-13.4478,-3.26156)(-13.5072,-3.06156)(-13.5879,-2.86156)(-13.6941,-2.66156)(-13.8313,-2.46156)(-14.0066,-2.26156)(-14.2295,-2.06156)(-14.5135,-1.86156)(-14.8772,-1.66156)(-15.3486,-1.46156)(-15.9707,-1.26156)(-16.8144,-1.06156)(-18.0068,-0.861558)(-19.8031,-0.661558)(-22.8203,-0.461558)(-29.1411,-0.261558)(-57.6626,-0.0615576)(-60,-30)
		\psline[linewidth=2pt, fillstyle=solid, fillcolor=gray](26.8011,-31.0616)(26.6878,-30.8616)(26.5746,-30.6616)(26.4614,-30.4616)(26.3482,-30.2616)(26.235,-30.0616)(26.1219,-29.8616)(26.0088,-29.6616)(25.8958,-29.4616)(25.7827,-29.2616)(25.6698,-29.0616)(25.5568,-28.8616)(25.4439,-28.6616)(25.331,-28.4616)(25.2182,-28.2616)(25.1054,-28.0616)(24.9926,-27.8616)(24.8799,-27.6616)(24.7672,-27.4616)(24.6545,-27.2616)(24.5419,-27.0616)(24.4294,-26.8616)(24.3169,-26.6616)(24.2044,-26.4616)(24.092,-26.2616)(23.9796,-26.0616)(23.8673,-25.8616)(23.755,-25.6616)(23.6428,-25.4616)(23.5306,-25.2616)(23.4185,-25.0616)(23.3064,-24.8616)(23.1944,-24.6616)(23.0824,-24.4616)(22.9705,-24.2616)(22.8587,-24.0616)(22.7469,-23.8616)(22.6352,-23.6616)(22.5236,-23.4616)(22.412,-23.2616)(22.3005,-23.0616)(22.189,-22.8616)(22.0777,-22.6616)(21.9664,-22.4616)(21.8551,-22.2616)(21.744,-22.0616)(21.6329,-21.8616)(21.5219,-21.6616)(21.411,-21.4616)(21.3002,-21.2616)(21.1895,-21.0616)(21.0789,-20.8616)(20.9683,-20.6616)(20.8579,-20.4616)(20.7476,-20.2616)(20.6373,-20.0616)(20.5272,-19.8616)(20.4172,-19.6616)(20.3073,-19.4616)(20.1975,-19.2616)(20.0878,-19.0616)(19.9782,-18.8616)(19.8688,-18.6616)(19.7595,-18.4616)(19.6503,-18.2616)(19.5413,-18.0616)(19.4324,-17.8616)(19.3237,-17.6616)(19.2151,-17.4616)(19.1067,-17.2616)(18.9985,-17.0616)(18.8904,-16.8616)(18.7825,-16.6616)(18.6748,-16.4616)(18.5672,-16.2616)(18.4599,-16.0616)(18.3528,-15.8616)(18.2458,-15.6616)(18.1391,-15.4616)(18.0327,-15.2616)(17.9264,-15.0616)(17.8204,-14.8616)(17.7147,-14.6616)(17.6092,-14.4616)(17.5041,-14.2616)(17.3992,-14.0616)(17.2945,-13.8616)(17.1903,-13.6616)(17.0863,-13.4616)(16.9827,-13.2616)(16.8794,-13.0616)(16.7765,-12.8616)(16.674,-12.6616)(16.5719,-12.4616)(16.4703,-12.2616)(16.369,-12.0616)(16.2683,-11.8616)(16.168,-11.6616)(16.0683,-11.4616)(15.9691,-11.2616)(15.8705,-11.0616)(15.7725,-10.8616)(15.6751,-10.6616)(15.5784,-10.4616)(15.4824,-10.2616)(15.3871,-10.0616)(15.2926,-9.86156)(15.199,-9.66156)(15.1062,-9.46156)(15.0144,-9.26156)(14.9236,-9.06156)(14.8339,-8.86156)(14.7453,-8.66156)(14.6579,-8.46156)(14.5718,-8.26156)(14.4871,-8.06156)(14.4039,-7.86156)(14.3223,-7.66156)(14.2424,-7.46156)(14.1643,-7.26156)(14.0883,-7.06156)(14.0144,-6.86156)(13.9428,-6.66156)(13.8738,-6.46156)(13.8075,-6.26156)(13.7442,-6.06156)(13.6843,-5.86156)(13.628,-5.66156)(13.5757,-5.46156)(13.5277,-5.26156)(13.4847,-5.06156)(13.4471,-4.86156)(13.4156,-4.66156)(13.3909,-4.46156)(13.3739,-4.26156)(13.3656,-4.06156)(13.3672,-3.86156)(13.3802,-3.66156)(13.4063,-3.46156)(13.4478,-3.26156)(13.5072,-3.06156)(13.5879,-2.86156)(13.6941,-2.66156)(13.8313,-2.46156)(14.0066,-2.26156)(14.2295,-2.06156)(14.5135,-1.86156)(14.8772,-1.66156)(15.3486,-1.46156)(15.9707,-1.26156)(16.8144,-1.06156)(18.0068,-0.861558)(19.8031,-0.661558)(22.8203,-0.461558)(29.1411,-0.261558)(57.6626,-0.0615576)(60,-30)
		\psline(66.2747, -100)(-49.1953, 100)
		\psline(57.735, 114.791)(-57.735, -85.2088)
		\psline(-124.01, 0)(106.93, 0)
		\psline[linestyle=dotted](118.786, 9.04788)(-112.154, 9.04788)
		\psline[linestyle=dotted](121.7, 4)(-109.24, 4)
		\psline[linestyle=dotted](114.464, 16.5346)(-116.477, 16.5346)
		\psline[linestyle=dotted](-69.62, 94.2058)(45.85, -105.794)
		\psline[linestyle=dotted](-42.5047, -100)(72.9654, 100)
		\psline[linestyle=dotted](-61.4514, 108.354)(54.0186, -91.6458)
		\psline[linestyle=dotted](-59.085, 112.453)(56.385, -87.5471)
		\psline[linestyle=dotted](-58.842, -100)(56.6281, 100)
		\psline[linestyle=dotted](-63.5747, -100)(51.8954, 100)
		\put(-29.5,0.5){$s=0$}
		\put(-29.5,4.75){$s=4M_\pi^2$}
		\put(-29.5,10){$s=(M_K-\sqrt{s_\ell})^2$}
		\put(-29.5,17.5){$s=(M_K+\sqrt{s_\ell})^2$}
		\rput{-60}(16,-15){$t=0$}
		\rput{-60}(11,-20){$t=(M_K-M_\pi)^2$}
		\rput{-60}(-2.5,-25){$t=(M_K+M_\pi)^2$}
		\rput{60}(3.5,23){$u=0$}
		\rput{60}(9.5,21){$u=(M_K-M_\pi)^2$}
		\rput{60}(21,12.5){$u=(M_K+M_\pi)^2$}
		\put(-4,6){decay region}
		\put(-3.5,27){$s$-channel}
		\put(-27,-10){$t$-channel}
		\put(20,-10){$u$-channel}
		\put(-5,-10){real amplitude}
	\end{pspicture*}
	\caption{Mandelstam diagram for $K_{\ell4}$ for the case $s_\ell > 0$}
	\label{img:MandelstamDiagram2}
\end{figure}

Let us first study the general properties of matrix elements of the
hadronic axial vector current. It is instructive to draw a Mandelstam
diagram for the process (see figures~\ref{img:MandelstamDiagram1} and
\ref{img:MandelstamDiagram2}): since $s+t+u = M_K^2 + 2 M_\pi^2 + s_\ell =:
\Sigma_0$ is constant (for a fixed value of $s_\ell$), the Mandelstam
variables can be represented in one plane, using the fact that the sum of
distances of a point to the sides of an equilateral triangle is constant.

The same amplitude describes four processes:
\begin{itemize}
	\item the decay $K^+(k) \to \pi^+(p_1) \pi^-(p_2) A_\mu^\dagger(L)$,
	\item the $s$-channel scattering $K^+(k) A_\mu(-L) \to \pi^+(p_1) \pi^-(p_2)$,
	\item the $t$-channel scattering $K^+(k) \pi^-(-p_1) \to \pi^-(p_2) A_\mu^\dagger(L)$,
	\item the $u$-channel scattering $K^+(k) \pi^+(-p_2) \to \pi^+(p_1) A_\mu^\dagger(L)$.
\end{itemize}

The physical region of the decay starts at $s = 4 M_\pi^2$ and ends at $s =
(M_K - \sqrt{s_\ell})^2$. The $s$-channel scattering starts at $s = (M_K +
\sqrt{s_\ell})^2$. If $s_\ell = 0$ is assumed, the two regions touch at $s
= M_K^2$ (figure~\ref{img:MandelstamDiagram1}). 

The sub-threshold region $s < s_0 := 4 M_\pi^2$, $t < t_0 := (M_K +
M_\pi)^2$, $u < u_0 := (M_K + M_\pi)^2$ forms a triangle in the Mandelstam
plane where the amplitude is real. Branch cuts of the amplitude start at
each threshold $s_0$, $t_0$ and $u_0$. There, physical intermediate states
are possible ($\pi\pi$ intermediate states in the $s$-channel, $K\pi$
states in the $t$- and $u$-channel). 

\clearpage

\clearpage

\subsection{Isospin Decomposition}

Let us study the isospin properties of the $K_{\ell4}$ matrix element of
the hadronic axial vector current in the different channels: we decompose the
physical amplitude into amplitudes with definite isospin. 

\subsubsection{$s$-Channel}

We consider the matrix element
\begin{align}
	\mathcal{A}_\mu^{+-} = \< \pi^+(p_1) \pi^-(p_2) \big| A_\mu(0) \big| K^+(k) \> .
\end{align}
As the weak current satisfies $\Delta I = \frac{1}{2}$, the initial and final states can be decomposed as
\begin{align}
	\begin{split}
		A_\mu(0) \big| K^+(k) \> &= \frac{1}{\sqrt{2}} \big| 1, 0 \> + \frac{1}{\sqrt{2}} \big| 0, 0 \> , \\
		\< \pi^+(p_1) \pi^-(p_2) \big| &= \frac{1}{\sqrt{6}} \< 2, 0 \big| + \frac{1}{\sqrt{2}} \< 1, 0 \big| + \frac{1}{\sqrt{3}} \< 0, 0 \big| , \\
		\< \pi^-(p_1) \pi^+(p_2) \big| &= \frac{1}{\sqrt{6}} \< 2, 0 \big| - \frac{1}{\sqrt{2}} \< 1, 0 \big| + \frac{1}{\sqrt{3}} \< 0, 0 \big| .
	\end{split}
\end{align}
Hence, we can write the following decomposition of the matrix element into pure isospin amplitudes:
\begin{align}
	\begin{split}
		\mathcal{A}_\mu^{+-} = \frac{1}{2} \mathcal{A}_\mu^{(1)} + \frac{1}{\sqrt{6}} \mathcal{A}_\mu^{(0)} , \\
		\mathcal{A}_\mu^{-+} = -\frac{1}{2} \mathcal{A}_\mu^{(1)} + \frac{1}{\sqrt{6}} \mathcal{A}_\mu^{(0)} .
	\end{split}
\end{align}
Using $\mathcal{A}_\mu^{+-}(k, -L \to p_1, p_2) = \mathcal{A}_\mu^{-+}(k, -L \to p_2, p_1)$, we find the following relations:
\begin{align}
	\begin{split}
		\mathcal{A}_\mu^{0}(k, -L \to p_1, p_2) &= \sqrt{\frac{3}{2}} \left( \mathcal{A}_\mu^{+-}(k, -L \to p_1, p_2) + \mathcal{A}_\mu^{+-}(k, -L \to p_2, p_1) \right) , \\
		\mathcal{A}_\mu^{1}(k, -L \to p_1, p_2) &= \left( \mathcal{A}_\mu^{+-}(k, -L \to p_1, p_2) - \mathcal{A}_\mu^{+-}(k, -L \to p_2, p_1) \right) . \\
	\end{split}
\end{align}
The pure isospin form factors are related to the physical ones by
\begin{align}
	\begin{split}
		\label{eq:sChannelIsospinFormFactors}
		F^{(0)}(s,t,u) &= \sqrt{\frac{3}{2}}\left( F(s,t,u) + F(s,u,t) \right) , \\
		G^{(0)}(s,t,u) &= \sqrt{\frac{3}{2}}\left( G(s,t,u) - G(s,u,t) \right) , \\
		R^{(0)}(s,t,u) &= \sqrt{\frac{3}{2}}\left( R(s,t,u) + R(s,u,t) \right) , \\
		F^{(1)}(s,t,u) &= F(s,t,u) - F(s,u,t) , \\
		G^{(1)}(s,t,u) &= G(s,t,u) + G(s,u,t) , \\
		R^{(1)}(s,t,u) &= R(s,t,u) - R(s,u,t) .
	\end{split}
\end{align}
We further note that
\begin{align}
	\begin{split}
		\mathcal{A}_\mu^{(0)}(k, -L \to p_1, p_2) &= \mathcal{A}_\mu^{(0)}(k, -L \to p_2, p_1), \\
		\mathcal{A}_\mu^{(1)}(k, -L \to p_1, p_2) &= - \mathcal{A}_\mu^{(1)}(k, -L \to p_2, p_1) ,
	\end{split}
\end{align}
and that the form factors of the pure isospin amplitudes satisfy
\begin{align}
	\begin{split}
		F^{(0)}(s,t,u) &= F^{(0)}(s,u,t) , \\
		G^{(0)}(s,t,u) &= -G^{(0)}(s,u,t) , \\
		R^{(0)}(s,t,u) &= R^{(0)}(s,u,t) ,
	\end{split}
	\begin{split}
		F^{(1)}(s,t,u) &= -F^{(1)}(s,u,t), \\
		G^{(1)}(s,t,u) &= G^{(1)}(s,u,t), \\
		R^{(1)}(s,t,u) &= -R^{(1)}(s,u,t) .
	\end{split}
\end{align}

\subsubsection{$t$- and $u$-Channel}

In the crossed $t$-channel, we are concerned with the matrix element
\begin{align}
	\mathcal{A}_\mu^{+-} = \< \pi^-(p_2) \big| A_\mu(0) \big| K^+(k) \pi^-(-p_1)\> .
\end{align}
In the $u$-channel, we analogously look at
\begin{align}
	\mathcal{A}_\mu^{+-} = \< \pi^+(p_1) \big| A_\mu(0) \big| K^+(k) \pi^+(-p_2)\> .
\end{align}
Note that due to crossing, these matrix elements are described by the same
function -- or its analytic continuation -- as the corresponding
$s$-channel matrix element. 

The $t$-channel initial and final states have the isospin decompositions
\begin{align}
	\begin{split}
		\big| K^+(k) \pi^-(-p_1) \> &= \sqrt{\frac{2}{3}} \left| \frac{1}{2}, -\frac{1}{2} \right\> + \sqrt{\frac{1}{3}} \left| \frac{3}{2}, -\frac{1}{2} \right\> , \\
		\< \pi^-(p_2) \big| A_\mu(0) &= \sqrt{\frac{2}{3}} \left\< \frac{1}{2}, -\frac{1}{2} \right| + \sqrt{\frac{1}{3}} \left\< \frac{3}{2}, -\frac{1}{2} \right| ,
	\end{split}
\end{align}
whereas in the $u$-channel, we are concerned with a pure isospin 3/2 scattering:
\begin{align}
	\begin{split}
		\big| K^+(k) \pi^+(-p_2) \> &= \left| \frac{3}{2}, \frac{3}{2} \right\> , \\
		\< \pi^+(p_1) \big| A_\mu(0) &= \left\< \frac{3}{2}, \frac{3}{2} \right| .
	\end{split}
\end{align}
We find the following isospin relation:
\begin{align}
	\begin{split}
		\mathcal{A}_\mu^{(3/2)}(k, -p_2 \to L, p_1) &= \mathcal{A}_\mu^{+-}(k, -L \to p_1, p_2) \\
			&= \frac{2}{3} \mathcal{A}_\mu^{(1/2)}(k, -p_1 \to L, p_2) + \frac{1}{3} \mathcal{A}_\mu^{(3/2)}(k, -p_1 \to L, p_2) .
	\end{split}	
\end{align}
Note that the third component of the isospin does not alter the amplitude:
just insert an isospin rotation matrix together with its inverse between
in- and out-state to rotate the third component.


The amplitude that describes pure isospin 1/2 scattering in the $t$-channel is then
\begin{align}
	\mathcal{A}_\mu^{(1/2)}(k, -p_1 \to L, p_2) = \frac{3}{2} \mathcal{A}_\mu^{(3/2)}(k, -p_2 \to L, p_1) - \frac{1}{2} \mathcal{A}_\mu^{(3/2)}(k, -p_1 \to L, p_2) .
\end{align}
Defining analogous form factors for the isospin 1/2 amplitude, we find
\begin{align}
	\begin{split}
		\label{eq:Isospin12FormFactors}
		F^{(1/2)}(s,t,u) &= \frac{3}{2} F(s,t,u) - \frac{1}{2} F(s,u,t) , \\
		G^{(1/2)}(s,t,u) &= \frac{3}{2} G(s,t,u) + \frac{1}{2} G(s,u,t) , \\
		R^{(1/2)}(s,t,u) &= \frac{3}{2} R(s,t,u) - \frac{1}{2} R(s,u,t) .
	\end{split}
\end{align}

In the case $s_\ell = 0$, it may be convenient to look at a certain linear
combination of the form factors $F$ and $G$, as we did in
\cite{Stoffer2010, Colangelo2012, Stoffer2013}: 
\begin{align}
	F_1 := X F + (u-t) \frac{PL}{2X} G ,
\end{align}
where $X := \frac{1}{2}\lambda^{1/2}(M_K^2, s, s_\ell)$, $PL :=
\frac{1}{2}(M_K^2 - s - s_\ell)$ and $\lambda(a,b,c) := a^2 + b^2 + c^2 -
2(ab + bc + ca)$ is the K\"all\'en triangle function. 

Here, too, we can define the corresponding isospin $1/2$ form factor:
\begin{align}
	\begin{split}
		F_1^{(1/2)}(s,t,u) &:= X F^{(1/2)}(s,t,u) + (u-t) \frac{PL}{2X} G^{(1/2)}(s,t,u) \\
			&= \frac{3}{2} \left( X F(s,t,u) + (u-t) \frac{PL}{2X} G(s,t,u) \right) \\
			&\quad - \frac{1}{2} \left( X F(s,u,t) + (t-u) \frac{PL}{2X} G(s,u,t) \right) \\
			&= \frac{3}{2} F_1(s,t,u) - \frac{1}{2} F_1(s,u,t) .
	\end{split}
\end{align}

\clearpage

\subsection{Unitarity and Partial-Wave Expansion}

In this section, we will investigate the unitarity relations in the
different channels and work out expansions of the form factors into partial
waves with `nice' properties with respect to unitarity and analyticity: the
partial waves shall satisfy Watson's final-state theorem. As we will need
analytic continuations of the partial waves, we must also be careful not to
introduce kinematic singularities. 

The derivation of the partial-wave expansion has been done for the
$s$-channel in \cite{Riggenbach1992}. We now apply the same method to all
channels. 

\subsubsection{Helicity Amplitudes}

The quantities that have a simple expansion into partial waves are not the
form factors but the helicity amplitudes of the $2\to2$ scattering process
\cite{Jacob1959}. However, helicity partial waves contain kinematic
singularities. In order to determine them, we use the prescriptions of
\cite{Martin1970}.  

We obtain the helicity amplitudes by contracting the axial-vector-current
matrix element with the polarisation vectors of the off-shell $W$ boson. In
the $W$ rest frame, the polarisation vectors are given by: 
\begin{align}
	\begin{split}
		\varepsilon_t^\mu &= \left( 1, 0, 0, 0 \right) , \\
		\varepsilon_{\pm}^\mu &= \frac{1}{\sqrt{2}} \left( 0, 0, \pm 1, i \right) , \\
		\varepsilon_0^\mu &= \left( 0, 1, 0, 0 \right) .
	\end{split}
\end{align}
They are eigenvectors of the spin matrices $S^2$ and $S_1$, defined by
\begin{align}
	\begin{split}
		S_1 = \left(
				\begin{array}{cccc}
				 0 & 0 & 0 & 0 \\
				 0 & 0 & 0 & 0 \\
				 0 & 0 & 0 & -i \\
				 0 & 0 & i & 0 \\
				\end{array}
				\right) , \quad 
		S_2 &= \left(
				\begin{array}{cccc}
				 0 & 0 & 0 & 0 \\
				 0 & 0 & 0 & i \\
				 0 & 0 & 0 & 0 \\
				 0 & -i & 0 & 0 \\
				\end{array}
				\right) , \quad
		S_3 = \left(
				\begin{array}{cccc}
				 0 & 0 & 0 & 0 \\
				 0 & 0 & -i & 0 \\
				 0 & i & 0 & 0 \\
				 0 & 0 & 0 & 0 \\
				\end{array}
				\right) , \\
		S^2 = S_1^2 &+ S_2^2 + S_3^2 = \left(
				\begin{array}{cccc}
				 0 & 0 & 0 & 0 \\
				 0 & 2 & 0 & 0 \\
				 0 & 0 & 2 & 0 \\
				 0 & 0 & 0 & 2 \\
				\end{array}
				\right) .
	\end{split}
\end{align}
The eigenvalues $s(s+1)$ and $s_1$ of $S^2$ and $S_1$ are listed below:
\begin{center}
	\begin{tabular}{l r r r }
		\toprule
		 & $\varepsilon_t^\mu$ & $\varepsilon_{\pm}^\mu$ & $\varepsilon_0^\mu$ \\	
		 \hline
		$s$ & 0 & 1 & 1 \\
		$s_1$ & 0 & $\pm1$ & 0 \\
		\bottomrule
	\end{tabular}
\end{center}
If we boost the polarisation vectors into the frame where the $W$ momentum
is given by $L = ( L^0, L^1, 0, 0)$, $L^2 = s_\ell$, we obtain: 
\begin{align}
	\begin{split}
		\varepsilon_t^\mu &= \frac{1}{\sqrt{s_\ell}} \left( L^0, L^1, 0, 0 \right) , \\
		\varepsilon_{\pm}^\mu &= \frac{1}{\sqrt{2}} \left( 0, 0, \pm 1, i \right) , \\
		\varepsilon_0^\mu &= \frac{1}{\sqrt{s_\ell}} \left( L^1, L^0, 0, 0 \right) .
	\end{split}
\end{align}
The contractions of these basis vectors with $\mathcal{A}_\mu$ give the
different helicity amplitudes: 
\begin{align}
	\mathcal{A}_i := \mathcal{A}_\mu \varepsilon_i^\mu .
\end{align}

We extract the kinematic singularities by applying the recipe of
\cite{Martin1970}, chapter~7.3.5, to these helicity amplitudes.

\subsubsection{Partial-Wave Unitarity in the $s$-Channel}

\paragraph{Helicity Partial Waves}

The unitarity relation for the axial vector current matrix element reads
\begin{align}
	\begin{split}
		\Im\left( i \mathcal{A}_i^{(I)}(k,-L \to p_1, p_2) \right) &= \frac{1}{4} \int \widetilde{dq_1} \widetilde{dq_2} (2\pi)^4 \delta^{(4)}(p_1+p_2-q_1-q_2) \\
		&\qquad {\mathcal{T}^{(I)}}^*(q_1,q_2 \to p_1,p_2) \, i \mathcal{A}_i^{(I)}(k,-L\to q_1,q_2) ,
	\end{split}
\end{align}
where $\widetilde{dq} := \frac{d^3q}{(2\pi)^3 2 q^0}$ is the
Lorentz-invariant measure and where a symmetry factor $1/2$ for the pions
is included. $\mathcal{T}^{(I)}$ denotes the elastic isospin $I$
$\pi\pi$-scattering amplitude. Note that this relation is valid in the
physical region and that kinematic singularities have to be removed before
an analytic continuation. 

We perform the integrals:
\begin{align}
	\begin{split}
		\label{eq:sChannelUnitarity}
		\Im \Big( i \mathcal{A}_i^{(I)}(k,-L \to p_1, p_2) \Big)
		&= \frac{1}{16} \frac{1}{(2\pi)^2} \frac{1}{2} \sigma_\pi(s) \int d\Omega^\dprime \; {\mathcal{T}^{(I)}}^*(s,\cos\theta^\prime) \, i \mathcal{A}_i^{(I)}(s,\cos\theta^\dprime,\phi^\dprime) ,
	\end{split}
\end{align}
where $\sigma_\pi(s) = \sqrt{1 - 4M_\pi^2/s}$ and of course
$\cos\theta^\prime$ has to be understood as a function of
$\cos\theta^\dprime$ and $\phi^\dprime$ through the relation 
\begin{align}
	\cos\theta^\prime = \sin\theta \sin\theta^\dprime \cos\phi^\dprime + \cos\theta \cos\theta^\dprime .
\end{align}
If we expand $\mathcal{T}$ and $\mathcal{A}_i$ into appropriate partial
waves, we can perform the remaining angular integrals 
and find the unitarity relations for the $K_{\ell4}$ partial waves.

We expand the $\pi\pi$-scattering matrix element in the usual way:
\begin{align}
	\mathcal{T}^{(I)}(s, \cos\theta^\prime) = \sum_{l=0}^\infty P_l(\cos\theta^\prime) \, t_l^I(s) 
\end{align}
with
\begin{align}
	t_l^{I}(s) = \left| t_l^{I}(s) \right| e^{i \delta_l^{I}(s)} .
\end{align}

The $K_{\ell4}$ helicity amplitudes are expanded into appropriate Wigner
$d$-functions, which satisfy $d_{00}^{(l)}(\theta)=P_l(\cos\theta)$ and
$d_{10}^{(l)}(\theta) = -[l(l+1)]^{-1/2} \sin\theta \,
P_l^\prime(\cos\theta)$. We have to take care of the kinematic
singularities of the helicity amplitudes \cite{Jacob1959, Martin1970}: 
\begin{align}
	\begin{split}
		i \mathcal{A}_t^{(I)}(s,\cos\theta) &= \sum_{l=0}^\infty P_l(\cos\theta) \left( \frac{\lambda^{1/2}_{K\ell}(s) \sigma_\pi(s)}{M_K^2} \right)^l \; a^{(I)}_{t, l}(s) , \\
		i \mathcal{A}_0^{(I)}(s,\cos\theta) &= i \mathcal{\tilde A}_0^{(I)} \frac{\lambda^{1/2}_{K\ell}(s)}{M_K^2}  = \frac{\lambda^{1/2}_{K\ell}(s)}{M_K^2} \sum_{l=0}^\infty P_l(\cos\theta) \left( \frac{\lambda^{1/2}_{K\ell}(s) \sigma_\pi(s)}{M_K^2} \right)^l \; a^{(I)}_{0, l}(s) , \\
		i \mathcal{A}_2^{(I)}(s,\cos\theta,\phi) &= i \mathcal{\tilde A}_2^{(I)} \sin\theta = \sin\theta \sum_{l=1}^\infty P_l^\prime(\cos\theta) \left( \frac{\lambda^{1/2}_{K\ell}(s) \sigma_\pi(s)}{M_K^2} \right)^{l-1} \cos\phi \; a_{2,l}^{(I)}(s) ,
	\end{split}
\end{align}
where $\mathcal{A}_{2}^{(I)} := \mathcal{A}_{+}^{(I)} -
\mathcal{A}_-^{(I)}$. The square roots of the K\"all\'en function cancel
exactly the square root branch cuts in the Legendre polynomials between
$(M_K - \sqrt{s_\ell})^2$ and $(M_K + \sqrt{s_\ell})^2$. The factor $M_K^2$
in the denominators appears only for dimensional reasons. All the defined
partial waves $a^{(I)}_{i, l}$ are free of kinematic singularities and can
be used for an analytic continuation from the decay region through the
unphysical to the scattering region. 

If we insert the partial-wave expansions into the unitarity relation, the
remaining angular integrals can be performed and the unitarity relation for
the $K_{\ell4}$ partial waves emerges ($i=t,0,2$): 
\begin{align}
	\Im \left( a^{(I)}_{i,l}(s) \right) =  \frac{1}{2l+1} \frac{1}{32\pi} \sigma_\pi(s) \, {t_l^I}^*(s)\; a^{(I)}_{i,l}(s) .
\end{align}
In particular, we find that the phases of the $K_{\ell4}$ $s$-channel
partial waves are given by the elastic $\pi\pi$-scattering phases (this is
Watson's theorem) for all $s$ between $4M_\pi^2$ and some inelastic
threshold: 
\begin{align}
	a^{(I)}_{i,l}(s) = \left| a^{(I)}_{i,l}(s) \right| e^{i \delta_l^I(s)} .
\end{align}

\paragraph{Partial-Wave Expansion of the Form Factors in the $s$-Channel}

In order to find the partial-wave expansions of the form factors, we write explicitly
%
the helicity amplitudes (generalised to a generic $\phi$):
\begin{align}
	\begin{split}
		i \mathcal{A}_t^{(I)} &= i \mathcal{A}_\mu^{(I)} \varepsilon_t^\mu = i \mathcal{A}_\mu^{(I)} \frac{1}{\sqrt{s_\ell}} L^\mu \\
			&= \frac{1}{M_K \sqrt{s_\ell}} \bigg( \frac{1}{2} ( M_K^2 - s - s_\ell ) \; F^{(I)} + \frac{1}{2} \sigma_\pi(s) \lambda^{1/2}_{K\ell}(s) \cos\theta \; G^{(I)} + s_\ell \; R^{(I)} \bigg) , \\
		i \mathcal{A}_0^{(I)} &= i \mathcal{A}_\mu^{(I)} \varepsilon_0^\mu \\
			&= \frac{-1}{M_K \sqrt{s_\ell}} \bigg( \frac{1}{2} \lambda^{1/2}_{K\ell}(s) \; F^{(I)} + \frac{1}{2} (M_K^2 - s - s_\ell) \sigma_\pi(s) \cos\theta \; G^{(I)} \bigg) , \\
		i \mathcal{A}_2^{(I)} &= i \mathcal{A}_\mu^{(I)} \varepsilon_+^\mu - i \mathcal{A}_\mu^{(I)} \varepsilon_-^\mu \\
			&= \frac{-\sqrt{2}}{M_K} \bigg( \sqrt{s} \sigma_\pi(s) \sin\theta \cos\phi \; G^{(I)} \bigg) .
	\end{split}
\end{align}

Since the contribution of the form factor $R$ to the decay rate is
suppressed by $m_\ell^2$, it is invisible in the electron mode and we do
not have any data on it. We therefore look only for linear combinations of
the form factors $F$ and $G$ that possess a simple partial-wave
expansion. We find: 
\begin{align}
	\begin{split}
		F^{(I)} &+ \frac{\sigma_\pi(s) PL(s)}{X(s)} \cos\theta \; G^{(I)}  = F^{(I)} + \frac{(M_K^2-s-s_\ell) (u-t)}{\lambda_{K\ell}(s)} \; G^{(I)} \\
			&= - \frac{2 \sqrt{s_\ell}}{M_K} \sum_{l=0}^\infty P_l(\cos\theta) \left( \frac{\lambda^{1/2}_{K\ell}(s) \sigma_\pi(s)}{M_K^2} \right)^l a_{0, l}^{(I)}(s) , \\
		G^{(I)} &= - \frac{M_K}{\sqrt{2s} \sigma_\pi(s)} \sum_{l=1}^\infty P_l^\prime(\cos\theta) \left( \frac{\lambda^{1/2}_{K\ell}(s) \sigma_\pi(s)}{M_K^2} \right)^{l-1} a_{2,l}^{(I)}(s) .
	\end{split}
\end{align}

We write the partial-wave expansions of $F$ and $G$ in the form:
\begin{align}
	\begin{split}
		\label{eq:sChannelFormFactorPartialWaveExpansion}
		F^{(I)} &= \sum_{l=0}^\infty P_l(\cos\theta) \left( \frac{\lambda^{1/2}_{K\ell}(s) \sigma_\pi(s)}{M_K^2}\right)^l f_l^{(I)}(s) - \frac{\sigma_\pi PL}{X} \cos\theta \; G^{(I)}, \\
		G^{(I)} &= \sum_{l=1}^\infty P_l^\prime(\cos\theta) \left( \frac{\lambda^{1/2}_{K\ell}(s) \sigma_\pi(s)}{M_K^2}\right)^{l-1} g_l^{(I)}(s) ,
	\end{split}
\end{align}
where the partial waves $f_l^{(I)}$ and $g_l^{(I)}$ satisfy Watson's
theorem in the region $s>4M_\pi^2$: 
\begin{align}
	f_l^{(I)}(s) = \left| f_l^{(I)}(s) \right| e^{i \delta_l^I(s)} \quad , \quad g_l^{(I)}(s) = \left| g_l^{(I)}(s) \right| e^{i \delta_l^I(s)} .
\end{align}

\subsubsection{Partial-Wave Unitarity in the $t$-Channel}

\paragraph{Helicity Partial Waves}

The discussion in the crossed channels is a bit simpler because we are
interested in partial-wave expansions only in the region $t >
(M_K+M_\pi)^2$ or $u > (M_K + M_\pi)^2$, i.e.~above all initial and final
state thresholds and pseudo-thresholds. Therefore, we do not have to worry
about kinematic singularities, since we will not perform analytic
continuations into the critical regions. 

In the crossed channels, we consider $K\pi$ intermediate states in the
unitarity relation: 
\begin{align}
	\begin{split}
		\Im\left( i \mathcal{A}_i^{(1/2)}(k,-p_1 \to L, p_2) \right) &= \frac{1}{2} \int \widetilde{dq_K} \widetilde{dq_\pi} (2\pi)^4 \delta^{(4)}(k-p_1-q_K-q_\pi) \\
		& {\mathcal{T}^{(1/2)}}^*(q_K,q_\pi \to k,-p_1) \, i \mathcal{A}_i^{(1/2)}(q_K,q_\pi \to L,p_2) ,
	\end{split}
\end{align}
where $\mathcal{T}^{(1/2)}$ is the isospin $1/2$ elastic $K\pi$-scattering
amplitude. By performing the integrals we obtain: 
\begin{align}
	\begin{split}
		\label{eq:tChannelUnitarity}
		\Im \Big( i \mathcal{A}_i^{(1/2)}(k,-p_1 \to L, p_2) \Big)
		&= \frac{1}{8} \frac{1}{(2\pi)^2} \frac{\lambda^{1/2}_{K\pi}(t)}{2t} \int d\Omega_t^\dprime \; {\mathcal{T}^{(1/2)}}^*(t,\cos\theta_t^\prime) \, i \mathcal{A}_i^{(1/2)}(t,\cos\theta_t^\dprime,\phi_t^\dprime) .
	\end{split}
\end{align}

The $K\pi$ scattering matrix element is expanded in the usual way:
\begin{align}
	\mathcal{T}^{(1/2)}(t,\cos\theta_t) = \sum_{l=0}^\infty P_l(\cos\theta_t) t_l^{1/2}(t)
\end{align}
with
\begin{align}
	t_l^{1/2}(t) = \left| t_l^{1/2}(t) \right| e^{i \delta_l^{1/2}(t)} .
\end{align}

We expand the $K_{\ell4}$ helicity amplitudes as follows:
\begin{align}
	\begin{split}
		i \mathcal{A}_t^{(1/2)}(t,\cos\theta_t) &= \sum_{l=0}^\infty P_l(\cos\theta_t) \left( \frac{\lambda^{1/2}_{K\pi}(t) \lambda^{1/2}_{\ell\pi}(t)}{M_K^4} \right)^l a_{t,l}^{(1/2)}(t) , \\
		i \mathcal{A}_0^{(1/2)}(t,\cos\theta_t) &= \sum_{l=0}^\infty P_l(\cos\theta_t) \left( \frac{\lambda^{1/2}_{K\pi}(t) \lambda^{1/2}_{\ell\pi}(t)}{M_K^4} \right)^l a_{0,l}^{(1/2)}(t) , \\
		i \mathcal{A}_2^{(1/2)}(t,\cos\theta_t,\phi_t) &= i \mathcal{A}_+^{(1/2)}(t,\cos\theta_t,\phi_t) - i \mathcal{A}_-^{(1/2)}(t,\cos\theta_t,\phi_t) \\
			&= \sin\theta_t \cos\phi_t \sum_{l=1}^\infty P_l^\prime(\cos\theta_t) \left( \frac{\lambda^{1/2}_{K\pi}(t) \lambda^{1/2}_{\ell\pi}(t)}{M_K^4} \right)^{l-1} a_{2,l}^{(1/2)}(t) .
	\end{split}
\end{align}
By inserting these expansions into the unitarity relation
(\ref{eq:tChannelUnitarity}), we find that 
all the partial waves satisfy Watson's theorem ($i=t,0,2$):
\begin{align}
	\begin{split}
		\Im\left( a_{i,l}^{(1/2)}(t) \right) &= \frac{1}{2l+1}\frac{1}{16\pi} \frac{\lambda^{1/2}_{K\pi}(t)}{t} \; {t_l^{1/2}}^*(t) a_{i,l}^{(1/2)}(t) , \\
		a_{i,l}^{(1/2)}(t) &= \left| a_{i,l}^{(1/2)}(t) \right| e^{i\delta_l^{1/2}(t)} .
	\end{split}
\end{align}

\paragraph{Partial-Wave Expansion of the Form Factors in the $t$-Channel}

By contracting the axial vector current matrix element in the $t$-channel
with the polarisation vectors, we find the helicity amplitudes (for a
generic $\phi_t$). As we are not interested in $R$, we do not need the
$\mathcal{A}_t^{(1/2)}$ component: 
\begin{align}
	\begin{split}
		i \mathcal{A}_0^{(1/2)} &= i \mathcal{A}^{(1/2)}_\mu \varepsilon_0^\mu
			= \begin{aligned}[t]
			& \frac{-1}{M_K \sqrt{s_\ell}} \Bigg( \frac{1}{4t} \left( \lambda^{1/2}_{K\pi}(t) (M_\pi^2-s_\ell-t) \cos\theta_t + \lambda^{1/2}_{\ell\pi}(t) (M_K^2 - M_\pi^2 + t) \right) F^{(1/2)} \\
			& \quad + \frac{1}{4t} \left( \lambda^{1/2}_{K\pi}(t) (M_\pi^2-s_\ell-t) \cos\theta_t + \lambda^{1/2}_{\ell\pi}(t) (M_K^2 - M_\pi^2 - 3t) \right) G^{(1/2)} \Bigg) , \end{aligned} \\
		i \mathcal{A}_2^{(1/2)} &= i \mathcal{A}^{(1/2)}_\mu \varepsilon_+^\mu - i \mathcal{A}^{(1/2)}_\mu \varepsilon_-^\mu
			= \frac{1}{\sqrt{2}M_K} \left( \frac{\lambda^{1/2}_{K\pi}(t)}{\sqrt{t}} \sin\theta_t \cos\phi_t \; \left( F^{(1/2)} + G^{(1/2)} \right) \right) .
	\end{split}
\end{align}
This results in the following partial-wave expansions of the form factors: 
\begin{align}
	\begin{split}
		\label{eq:tChannelFormFactorPartialWaveExpansion2}
		F^{(1/2)} &= \sum_{l=0}^\infty P_l(\cos\theta_t) \left( \frac{\lambda^{1/2}_{K\pi}(t) \lambda^{1/2}_{\ell\pi}(t)}{M_K^4} \right)^l f_l^{(1/2)}(t) \\
			& - \frac{1}{2t} \left( M_K^2 - M_\pi^2 - 3 t + (M_\pi^2 - s_\ell - t) \frac{\lambda^{1/2}_{K\pi}(t)}{\lambda^{1/2}_{\ell\pi}(t)} \cos\theta_t \right) \sum_{l=1}^\infty P_l^\prime(\cos\theta_t) \left( \frac{\lambda^{1/2}_{K\pi}(t) \lambda^{1/2}_{\ell\pi}(t)}{M_K^4} \right)^{l-1} g_l^{(1/2)}(t) , \\
		G^{(1/2)} &= -\sum_{l=0}^\infty P_l(\cos\theta_t) \left( \frac{\lambda^{1/2}_{K\pi}(t) \lambda^{1/2}_{\ell\pi}(t)}{M_K^4} \right)^l f_l^{(1/2)}(t) \\
			&+ \frac{1}{2t} \left( M_K^2 - M_\pi^2 + t + (M_\pi^2 - s_\ell - t) \frac{\lambda^{1/2}_{K\pi}(t)}{\lambda^{1/2}_{\ell\pi}(t)} \cos\theta_t \right) \sum_{l=1}^\infty P_l^\prime(\cos\theta_t) \left( \frac{\lambda^{1/2}_{K\pi}(t) \lambda^{1/2}_{\ell\pi}(t)}{M_K^4} \right)^{l-1} g_l^{(1/2)}(t) ,
	\end{split}
\end{align}
where also the new partial waves $f_l^{(1/2)}$ and $g_l^{(1/2)}$ satisfy Watson's theorem in the region $t>(M_K+M_\pi)^2$:
\begin{align}
	f_l^{(1/2)}(t) = \left| f_l^{(1/2)}(t) \right| e^{i \delta_l^{1/2}(t)} \quad , \quad g_l^{(1/2)}(t) = \left| g_l^{(1/2)}(t) \right| e^{i \delta_l^{1/2}(t)} .
\end{align}

\subsubsection{Partial-Wave Unitarity in the $u$-Channel}

\paragraph{Helicity Partial Waves}

The $u$-channel (i.e.~the isospin $3/2$ case) can be treated in complete
analogy to the $t$-channel. We start with the unitarity relation: 
\begin{align}
	\begin{split}
		\label{eq:uChannelUnitarity}
		\Im\left( i \mathcal{A}_i^{(3/2)}(k,-p_2 \to L, p_1) \right) &= \frac{1}{2} \int \widetilde{dq_K} \widetilde{dq_\pi} (2\pi)^4 \delta^{(4)}(k-p_2-q_K-q_\pi) \\
		& \quad \cdot {\mathcal{T}^{(3/2)}}^*(q_K,q_\pi \to k,-p_2) \, i \mathcal{A}_i^{(3/2)}(q_K,q_\pi \to L,p_1) \\
		&= \frac{1}{8} \frac{1}{(2\pi)^2} \frac{\lambda^{1/2}_{K\pi}(u)}{2u} \int d\Omega_u^\dprime \; {\mathcal{T}^{(3/2)}}^*(u,\cos\theta_u^\prime) \, i \mathcal{A}_i^{(3/2)}(u,\cos\theta_u^\dprime,\phi_u^\dprime) .
	\end{split}
\end{align}
The $K\pi$-scattering matrix element is expanded as
\begin{align}
	\begin{split}
		\mathcal{T}^{(3/2)}(u,\cos\theta_u) &= \sum_{l=0}^\infty P_l(\cos\theta_u) t_l^{3/2}(u) , \\
		t_l^{3/2}(u) &= \left| t_l^{3/2}(u) \right| e^{i \delta_l^{3/2}(u)}
	\end{split}
\end{align}
and the $K_{\ell4}$ helicity amplitudes according to
\begin{align}
	\begin{split}
		i \mathcal{A}_t^{(3/2)}(u,\cos\theta_u) &= \sum_{l=0}^\infty P_l(\cos\theta_u)  \left( \frac{\lambda^{1/2}_{K\pi}(u) \lambda^{1/2}_{\ell\pi}(u)}{M_K^4} \right)^l  a_{t,l}^{(3/2)}(u) , \\
		i \mathcal{A}_0^{(3/2)}(u,\cos\theta_u) &= \sum_{l=0}^\infty P_l(\cos\theta_u) \left( \frac{\lambda^{1/2}_{K\pi}(u) \lambda^{1/2}_{\ell\pi}(u)}{M_K^4} \right)^l a_{0,l}^{(3/2)}(u) , \\
		i \mathcal{A}_2^{(3/2)}(u,\cos\theta_u,\phi_u) &= i \mathcal{A}_+^{(3/2)}(u,\cos\theta_u,\phi_u) - i \mathcal{A}_-^{(3/2)}(u,\cos\theta_u,\phi_u) \\
			&= \sin\theta_u \cos\phi_u \sum_{l=1}^\infty P_l^\prime(\cos\theta_u) \left( \frac{\lambda^{1/2}_{K\pi}(u) \lambda^{1/2}_{\ell\pi}(u)}{M_K^4} \right)^{l-1} a_{2,l}^{(3/2)}(u) .
	\end{split}
\end{align}
Performing the angular integrals in the unitarity relation, we find that
the partial waves satisfy Watson's theorem ($i=t,0,2$): 
\begin{align}
	\begin{split}
		\Im\left( a_{i,l}^{(3/2)}(u) \right) &= \frac{1}{2l+1}\frac{1}{16\pi} \frac{\lambda^{1/2}_{K\pi}(u)}{u} \; {t_l^{3/2}}^*(u) a_{i,l}^{(3/2)}(u) , \\
		a_{i,l}^{(3/2)}(u) &= \left| a_{i,l}^{(3/2)}(u) \right| e^{i\delta_l^{3/2}(u)} .
	\end{split}
\end{align}

\paragraph{Partial-Wave Expansion of the Form Factors in the $u$-Channel}

The contraction of the axial vector current matrix element with the
polarisation vectors yields:
\begin{align}
	\begin{split}
		i \mathcal{A}_0^{(3/2)} &= i \mathcal{A}_\mu^{(3/2)} \varepsilon_0^\mu = \begin{aligned}[t]
			& \frac{-1}{M_K \sqrt{s_\ell}} \Bigg( \frac{1}{4u} \left( \lambda^{1/2}_{K\pi}(u) (M_\pi^2-s_\ell-u) \cos\theta_u + \lambda^{1/2}_{\ell\pi}(u) (M_K^2 - M_\pi^2 + u) \right) F \\
			& \quad - \frac{1}{4u} \left( \lambda^{1/2}_{K\pi}(u) (M_\pi^2-s_\ell-u) \cos\theta_u + \lambda^{1/2}_{\ell\pi}(u) (M_K^2 - M_\pi^2 - 3u) \right) G \Bigg) , \end{aligned} \\
		i \mathcal{A}_2^{(3/2)} &= i \mathcal{A}_\mu^{(3/2)} \varepsilon_+^\mu - i \mathcal{A}_\mu^{(3/2)} \varepsilon_-^\mu = \frac{1}{\sqrt{2}M_K} \Bigg( \frac{\lambda^{1/2}_{K\pi}(u)}{\sqrt{u}} \sin\theta_u \cos\phi_u \left( F - G \right) \Bigg) .
	\end{split}
\end{align}
Hence, the partial-wave expansion of the form factors is given by
\begin{align}
	\begin{split}
		\label{eq:uChannelFormFactorPartialWaveExpansion2}
		F &= \sum_{l=0}^\infty P_l(\cos\theta_u) \left( \frac{\lambda^{1/2}_{K\pi}(u) \lambda^{1/2}_{\ell\pi}(u)}{M_K^4} \right)^l f_l^{(3/2)}(u) \\
			& - \frac{1}{2u} \left( M_K^2 - M_\pi^2 - 3u + (M_\pi^2 - s_\ell - u ) \frac{\lambda^{1/2}_{K\pi}(u)}{\lambda^{1/2}_{\ell\pi}(u)} \cos\theta_u \right) \sum_{l=1}^\infty P_l^\prime(\cos\theta_u)  \left( \frac{\lambda^{1/2}_{K\pi}(u) \lambda^{1/2}_{\ell\pi}(u)}{M_K^4} \right)^{l-1} g_l^{(3/2)}(u) , \\
		G &= \sum_{l=0}^\infty P_l(\cos\theta_u) \left( \frac{\lambda^{1/2}_{K\pi}(u) \lambda^{1/2}_{\ell\pi}(u)}{M_K^4} \right)^l f_l^{(3/2)}(u) \\
			& - \frac{1}{2u} \left( M_K^2 - M_\pi^2 + u + (M_\pi^2 - s_\ell - u ) \frac{\lambda^{1/2}_{K\pi}(u)}{\lambda^{1/2}_{\ell\pi}(u)} \cos\theta_u \right) \sum_{l=1}^\infty P_l^\prime(\cos\theta_u)  \left( \frac{\lambda^{1/2}_{K\pi}(u) \lambda^{1/2}_{\ell\pi}(u)}{M_K^4} \right)^{l-1} g_l^{(3/2)}(u) ,
	\end{split}
\end{align}
where the partial waves $f_l^{(3/2)}$ and $g_l^{(3/2)}$ satisfy Watson's
theorem in the region $u>(M_K+M_\pi)^2$: 
\begin{align}
	f_l^{(3/2)}(u) = \left| f_l^{(3/2)}(u) \right| e^{i \delta_l^{3/2}(u)} \quad , \quad g_l^{(3/2)}(u) = \left| g_l^{(3/2)}(u) \right| e^{i \delta_l^{3/2}(u)} .
\end{align}

\subsubsection{Projection and Analytic Structure of the Partial Waves}

\label{sec:ProjectionAnalyticStructurePartialWaves}

The several partial waves $f_l^{(I)}$ and $g_l^{(I)}$ can be calculated by
angular projections: 
\begin{align}
	\begin{split}
		\label{eqn:SChannelPartialWaveProjection}
		f_l^{(I)}(s) &= \left( \frac{M_K^2}{\lambda^{1/2}_{K\ell}(s) \sigma_\pi(s)} \right)^{l} \frac{2l+1}{2} \int_{-1}^1 dz \, P_l(z) \left( F^{(I)}(s,z) + \frac{\sigma_\pi(s) PL(s)}{X(s)} z G^{(I)}(s,z) \right) , \\
		g_l^{(I)}(s) &= \left( \frac{M_K^2}{\lambda^{1/2}_{K\ell}(s) \sigma_\pi(s)} \right)^{l-1} \int_{-1}^1 dz \, \frac{P_{l-1}(z) - P_{l+1}(z)}{2} \, G^{(I)}(s,z) ,
	\end{split}
\end{align}
where $X^{(I)}(s,z) := X^{(I)}(s,t(s,z),u(s,z))$, $X\in\{F,G\}$, $I\in\{0,1\}$ and
\begin{align}
	\begin{split}
		t(s,z) &= \frac{1}{2} \left( \Sigma_0 - s - 2 X \sigma_\pi z \right) , \\
		u(s,z) &= \frac{1}{2} \left( \Sigma_0 - s + 2 X \sigma_\pi z \right) .
	\end{split}
\end{align}
Since $t(s,-z)=u(s,z)$, the definition of the pure isospin form factors
(\ref{eq:sChannelIsospinFormFactors}) implies 
\begin{align}
	\begin{split}
		f_l^{(0)}(s) &= g_l^{(0)}(s) = 0 \quad \forall \; l \text{ odd} , \\
		f_l^{(1)}(s) &= g_l^{(1)}(s) = 0 \quad \forall \; l \text{ even} .
	\end{split}
\end{align}
Hence, we can as well directly use the partial waves of the physical form
factors: 
\begin{align}
	\begin{split}
		\label{eq:PartialWaveProjectionSChannelPhysicalFF}
		f_l(s) &= \left( \frac{M_K^2}{\lambda^{1/2}_{K\ell}(s) \sigma_\pi(s)} \right)^{l} \frac{2l+1}{2} \int_{-1}^1 dz \, P_l(z) \left( F(s,z) + \frac{\sigma_\pi(s) PL(s)}{X(s)} z G(s,z) \right) , \\
		g_l(s) &= \left( \frac{M_K^2}{\lambda^{1/2}_{K\ell}(s) \sigma_\pi(s)} \right)^{l-1} \int_{-1}^1 dz \, \frac{P_{l-1}(z) - P_{l+1}(z)}{2} \, G(s,z) ,
	\end{split}
\end{align}
which still fulfil Watson's theorem
\begin{align}
	f_l(s) = \left| f_l(s) \right| e^{i \delta_l^I(s)} \quad , \quad g_l(s) = \left| g_l(s) \right| e^{i \delta_l^I(s)} ,
\end{align}
where $I = (l\mod2)$.

In the crossed channels, the partial wave projections are given by
\begin{align}
	\begin{split}
		\label{eq:PartialWaveProjectionTUChannel}
		f_l^{(1/2)}(t) &= \left( \frac{M_K^4}{\lambda^{1/2}_{K\pi}(t) \lambda^{1/2}_{\ell\pi}(t)} \right)^l \frac{2l+1}{2} \int_{-1}^1 dz_t \, P_l(z_t) \Bigg( \frac{F^{(1/2)}(t,z_t) - G^{(1/2)}(t,z_t)}{2} \\
			+ \frac{1}{2t} \bigg( &M_K^2 - M_\pi^2 - t + (M_\pi^2 - s_\ell - t) \frac{\lambda^{1/2}_{K\pi}(t)}{\lambda^{1/2}_{\ell\pi}(t)} z_t \bigg) \frac{F^{(1/2)}(t,z_t) + G^{(1/2)}(t,z_t)}{2} \Bigg) , \\
		g_l^{(1/2)}(t) &= \left( \frac{M_K^4}{\lambda^{1/2}_{K\pi}(t) \lambda^{1/2}_{\ell\pi}(t)} \right)^{l-1} \int_{-1}^1 dz_t \, \frac{P_{l-1}(z_t) - P_{l+1}(z_t)}{2} \, \frac{F^{(1/2)}(t,z_t) + G^{(1/2)}(t,z_t)}{2} , \\
		f_l^{(3/2)}(u) &= \left( \frac{M_K^4}{\lambda^{1/2}_{K\pi}(u) \lambda^{1/2}_{\ell\pi}(u)} \right)^l \frac{2l+1}{2} \int_{-1}^1 dz_u \, P_l(z_u) \Bigg( \frac{F(u,z_u) + G(u,z_u)}{2} \\
			 + \frac{1}{2u} &\left( M_K^2 - M_\pi^2 - u + (M_\pi^2 - s_\ell - u) \frac{\lambda^{1/2}_{K\pi}(u)}{\lambda^{1/2}_{\ell\pi}(u)} z_u \right) \frac{F(u,z_u) - G(u,z_u)}{2} \Bigg) , \\
		g_l^{(3/2)}(u) &= \left( \frac{M_K^4}{\lambda^{1/2}_{K\pi}(u) \lambda^{1/2}_{\ell\pi}(u)} \right)^{l-1} \int_{-1}^1 dz_u \, \frac{P_{l-1}(z_u) - P_{l+1}(z_u)}{2} \, \frac{F(u,z_u) - G(u,z_u)}{2} ,
	\end{split}
\end{align}
where $X^{(I)}(t,z_t) := X^{(I)}(s(t,z_t),t,u(t,z_t))$, $X^{(I)}(u,z_u) := X^{(I)}(s(u,z_u),t(u,z_u),u)$, $X\in\{F,G\}$ and
\begin{align}
	\begin{split}
		s(t,z_t) &= \frac{1}{2} \left( \Sigma_0 - t + \frac{1}{t} \left( z_t \, \lambda^{1/2}_{K\pi}(t) \lambda^{1/2}_{\ell\pi}(t) - \Delta_{K\pi}\Delta_{\ell\pi} \right) \right) , \\
		u(t,z_t) &= \frac{1}{2} \left( \Sigma_0 - t - \frac{1}{t} \left( z_t \, \lambda^{1/2}_{K\pi}(t) \lambda^{1/2}_{\ell\pi}(t) - \Delta_{K\pi}\Delta_{\ell\pi} \right) \right)  , \\
		s(u,z_u) &= \frac{1}{2} \left( \Sigma_0 - u + \frac{1}{u} \left( z_u \, \lambda^{1/2}_{K\pi}(u) \lambda^{1/2}_{\ell\pi}(u) - \Delta_{K\pi}\Delta_{\ell\pi} \right) \right) , \\
		t(u,z_u) &= \frac{1}{2} \left( \Sigma_0 - u - \frac{1}{u} \left( z_u \, \lambda^{1/2}_{K\pi}(u) \lambda^{1/2}_{\ell\pi}(u) - \Delta_{K\pi}\Delta_{\ell\pi} \right) \right) .
	\end{split}
\end{align}

The construction of the partial waves has been done in a way that excludes
kinematic singularities for $s > 4 M_\pi^2$ and $t,u>(M_K+M_\pi)^2$. There
may still be kinematic singularities present below these regions, but they
do not bother us. But also the analytic structure of the partial waves
with respect to dynamic singularities is not trivial. 

For the $s$-channel partial waves, there is of course the right-hand cut at
$s>4M_\pi^2$. Further cuts can appear through the angular integration,
i.e.~at points where the integration contour in the $t$- or $u$-plane
touches the crossed channel cuts. If $s$ lies in the physical decay region,
the integration path is just a horizontal line from one end of the decay
region to the other (see the Mandelstam diagram in
figure~\ref{img:MandelstamDiagram2}). When we continue analytically into
the region $(M_K-\sqrt{s_\ell})^2 < s < (M_K + \sqrt{s_\ell})^2$, the
integration path moves into the complex $t$- and $u$-plane and crosses the
real Mandelstam plane at $t=u$: the square root of the K\"all\'en function $X
= \frac{1}{2}\lambda^{1/2}_{K\ell}(s)$ is purely imaginary in this
region. One has to know which branch of the square root should be
taken. The correct sign is found by taking $s$ real and shifting $M_K \to
M_K + i \epsilon$ (see \cite{Kacser1963}). With this prescription, the
K\"all\'en function turns counterclockwise around $\lambda_{K\ell}=0$ when
$s$ runs from $s<(M_K-\sqrt{s_\ell})^2$ to $s>(M_K+\sqrt{s_\ell})^2$. The
square root of the K\"all\'en function therefore takes the following
values:
\begin{align}
	\lambda^{1/2}_{K\ell}(s) = \left\{
	\begin{array}{r c}
		+ | \lambda^{1/2}_{K\ell}(s) | & s < (M_K-\sqrt{s_\ell})^2 , \\
		+ i | \lambda^{1/2}_{K\ell}(s) | & (M_K-\sqrt{s_\ell})^2 < s < (M_K+\sqrt{s_\ell})^2 , \\
		- | \lambda^{1/2}_{K\ell}(s) | & (M_K+\sqrt{s_\ell})^2 < s .
	\end{array} \right. 
\end{align}
In the region $s>(M_K+\sqrt{s_\ell})^2$, the integration path again lies in
the real Mandelstam plane from one to the other end of the scattering
region. 

As we are away from the $t$- and $u$-channel unitarity cuts at
$t,u>(M_K+M_\pi)^2$, this extension of the integration path into the
complex $t$- and $u$-plane is the only subtlety that has to be taken into
account. 

In the region $s<4M_\pi^2$, there is a left-hand cut at $s\in(-\infty,0)$:
the integration path extends again into the complex $t$- and $u$-plane in
the region $0<s<4M_\pi^2$ (due to the second square root). It diverges at
$s=0$ and returns to the real axis at $s<0$, but this time it touches the
$t$- and $u$-channel unitarity cuts at $t,u>(M_K+M_\pi)^2$ which produces
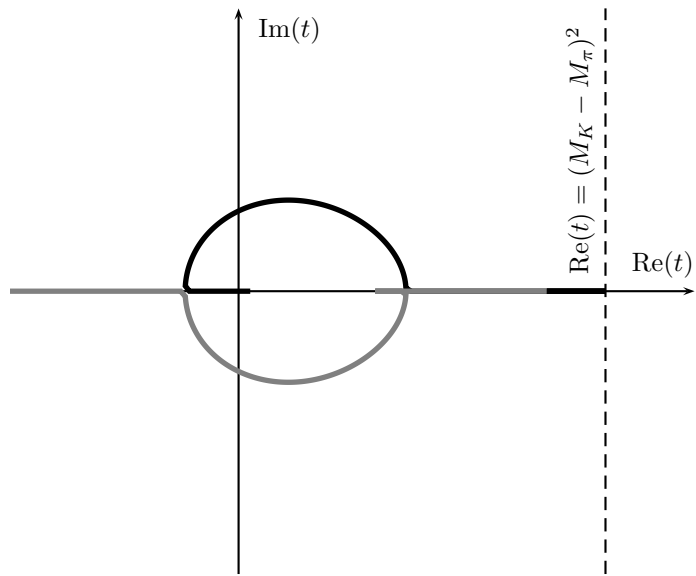
the left-hand cut of the $s$-channel partial waves. 

This left-hand cut can be most easily found by looking at the end-points of
the integration paths: solving the equation 
\begin{align}
	\begin{split}
		t_\pm &= u_\mp = \frac{1}{2} \left( \Sigma_0 - s \mp 2 X(s) \sigma_\pi(s) \right)
	\end{split}
\end{align}
for $t_\pm > (M_K+M_\pi)^2$ gives the left-hand cut $s\in(-\infty,0)$.

Let us consider the crossed $t$-channel (the situation in the $u$-channel
is analogous). We have defined the partial-wave expansion in the scattering
region $t>(M_K+M_\pi)^2$. Therefore, we also define the square root
branches of the K\"all\'en functions $\lambda^{1/2}_{K\pi}$ and
$\lambda^{1/2}_{\ell\pi}$ in this region. The sign of the square root
branch can be absorbed into the definition of the partial waves. 

The right-hand $t$-channel unitarity cut at $t>(M_K+M_\pi)^2$ also shows up
in the partial waves. A second possibility for singularities in the
$t$-channel partial waves arises when the integration path touches the $s$-
or $u$-channel unitarity cuts. For $t>(M_K+M_\pi)^2$, the integration path
lies  on the negative real axis of the $s$- and $u$-planes (this can be
seen in the Mandelstam diagram in figure~\ref{img:MandelstamDiagram2}). In
the region $(M_K-M_\pi)^2 < t < (M_K+M_\pi)^2$, the integration path
extends into the complex $s$- and $u$-plane. For the value of $t$
fulfilling $\frac{1}{2} \left( \Sigma_0 - t - \frac{1}{t}
  \Delta_{K\pi}\Delta_{\ell\pi} \right) = 4 M_\pi^2$, the integration path
in the $s$-plane touches the $s$-channel branch cut. From this point on
towards smaller values of $t$, the integration path has to be deformed in
the $s$-plane. Since the $u$-channel cut appears only at $u>(M_K+M_\pi)^2$,
such a deformation is not needed in the $u$-plane. At $t=(M_K-M_\pi)^2$,
the integration path in the $s$-plane has then the shape of a horseshoe
wrapped around the $s$-channel cut. For even smaller values of $t$, the
path unwraps itself in a continuous way, such that for $t <
\frac{1}{2}(M_K^2 - 2M_\pi^2 + s_\ell)$, the integration path lies
completely on the upper side of the $s$-channel cut. 

The cut structure in the $t$-channel partial wave is rather complicated, at
least for $s_\ell > 0$: The left-hand cuts can be found by solving the
equations 
\begin{align}
	\begin{split}
		s_\pm &= \frac{1}{2} \left( \Sigma_0 - t + \frac{1}{t} \left( \pm \, \lambda^{1/2}_{K\pi}(t) \lambda^{1/2}_{\ell\pi}(t) - \Delta_{K\pi}\Delta_{\ell\pi} \right) \right) , \\
		u_\pm &= \frac{1}{2} \left( \Sigma_0 - t - \frac{1}{t} \left( \pm \, \lambda^{1/2}_{K\pi}(t) \lambda^{1/2}_{\ell\pi}(t) - \Delta_{K\pi}\Delta_{\ell\pi} \right) \right) ,
	\end{split}
\end{align}
for $s_\pm > 4 M_\pi^2$ and $u_\pm > (M_K+M_\pi)^2$. While the second
equation results in a cut along the real axis, the first equation produces
an egg-shaped cut structure in the complex $t$-plane with $\Re(t) <
(M_K-M_\pi)^2$, shown in figure~\ref{img:tChannelPartialWaveCuts}. The
exact shape depends on the value of $s_\ell$. 

\input{sections/Singularities}

\subsubsection{Simplifications for $s_\ell \to 0$}

\label{sec:PartialWaveSimplificationsZeroSL}

In the experiment, a dependence on $s_\ell$ has been observed only in the
first partial wave of the form factor $F$ \cite{Batley2010,Batley2012}. If
we neglect this dependence on $s_\ell$ and assume that $s_\ell = 0$, the
treatment can be significantly simplified. 

\begin{itemize}
	\item The square root of the K\"all\'en function simplifies to
          \[\lim\limits_{s_\ell\to0}\lambda^{1/2}_{K\ell}(s) = M_K^2 - s
          ,\] the square root branch cut disappears. Hence, the integration
          path for the angular integrals in the $s$-channel always lies on
          the real axis. 
	\item The left-hand cut structure of $t$- and $u$-channel partial
          waves simplifies to a straight line along the real axis. The
          egg-shaped branch cuts disappear in the limit $s_\ell\to0$. 
	\item From (\ref{eq:sChannelFormFactorPartialWaveExpansion}), we
          see that the quantity 
	\begin{align}
		\begin{split}
			\lim_{s_\ell\to0} F_1^{(I)} &= \lim_{s_\ell\to0} \left(\frac{1}{2} \lambda^{1/2}_{K\ell}(s) F^{(I)} + \frac{1}{2} \frac{(M_K^2-s-s_\ell) (u-t)}{\lambda^{1/2}_{K\ell}(s)} \; G^{(I)} \right) \\
				&= \frac{M_K^2 - s}{2} F^{(I)} + \frac{u-t}{2} G^{(I)}
		\end{split}
	\end{align}
	has a simple $s$-channel partial-wave expansion into Legendre
        polynomials. If we consider
        (\ref{eq:tChannelFormFactorPartialWaveExpansion2}) in the limit
        $s_\ell \to 0$, we find that exactly the same linear combination of
        the form factors $F^{(1/2)}$ and $G^{(1/2)}$ has a simple
        $t$-channel partial-wave expansion into Legendre polynomials. The
        same follows from
        (\ref{eq:uChannelFormFactorPartialWaveExpansion2}) for the
        $u$-channel. In this limit, the form factor $F_1$ can therefore be
        treated independently from the other form factors. This is the
        procedure that has been followed in \cite{Stoffer2010,
          Colangelo2012, Stoffer2013}. 
\end{itemize}

There are several reasons why we abstain here from taking the limit
$s_\ell\to0$, which would result in a substantial simplification of the
whole treatment. The experiments provide data on both form factors $F$ and
$G$. In order to include all the available information, we deal with both
form factors at the same time. There is also some data available on the
dependence on $s_\ell$, which we include in this treatment. And finally,
the matching to \ChPT{} becomes much cleaner if it is performed with $F$
and $G$ directly, since these are the form factors with the simplest chiral
representation. 


\subsection{Reconstruction Theorem}

Since the form factors $F$ and $G$ describe a hadronic four-`particle'
process, they depend on the three Mandelstam variables $s$, $t$ and $u$ and
therefore possess a rather complicated analytic structure. However, it is
possible to decompose the form factors into a sum of functions that depend
only on a single Mandelstam variable -- a procedure known under the name of
`reconstruction theorem' \cite{Stern1993, Ananthanarayan2001}. Such a
decomposition provides a major simplification of the problem and leads to a
powerful dispersive description.  

\subsubsection{Decomposition of the Form Factors}

The explicit derivation of the decomposition of the form factors $F$ and
$G$ into functions of a single Mandelstam variable can be found in
\cite{Stoffer2014a}. It is based on fixed-$s$, fixed-$t$ and fixed-$u$
dispersion relations. We have to assume a certain asymptotic behaviour of
the form factors, e.g.~for fixed $u$, we assume 
\begin{align}
	\label{eq:FroissartInspiredAsymptotics}
	\lim_{|s|\to\infty} \frac{X_s^u(s)}{s^n} = \lim_{|t|\to\infty} \frac{X_t^u(t)}{t^n} = 0 ,
\end{align}
where the Froissart bound \cite{Froissart1961} suggests $n=2$. However, we
are also interested in the case $n=3$ in order to meet the asymptotic
behaviour of the NNLO \ChPT{} form factors. We therefore write down either
a twice- or thrice-subtracted dispersion relation for the form
factors. Then, we use the partial-wave expansions derived in the previous
section. We neglect the imaginary parts of $D$- and higher waves, an
approximation that is violated only at $\O(p^8)$ in the chiral power
counting. It implements the case $s_\ell \neq 0$. 

The result of the decomposition is the following:
\begin{align}
	\label{eq:FormFactorDecomposition}
	\begin{split}
		F(s,t,u) &= M_0(s) + \frac{u-t}{M_K^2} M_1(s) \\
			&\quad + \frac{2}{3} N_0(t) + \frac{2}{3} \frac{t(s-u) + \Delta_{K\pi}\Delta_{\ell\pi}}{M_K^4} N_1(t) - \frac{2}{3} \frac{\Delta_{K\pi}-3t}{2M_K^2} \tilde N_1(t) \\
			&\quad + \frac{1}{3} R_0(t) + \frac{1}{3} \frac{t(s-u) + \Delta_{K\pi}\Delta_{\ell\pi}}{M_K^4} R_1(t) - \frac{1}{3} \frac{\Delta_{K\pi}-3t}{2M_K^2} \tilde R_1(t) \\
			&\quad + R_0(u) + \frac{u(s-t) + \Delta_{K\pi}\Delta_{\ell\pi}}{M_K^4} R_1(u) - \frac{\Delta_{K\pi}-3u}{2M_K^2} \tilde R_1(u) \\
			&\quad + \O(p^8) , \\
		G(s,t,u) &= \tilde M_1(s) \\
			&\quad - \frac{2}{3} N_0(t) - \frac{2}{3} \frac{t(s-u) + \Delta_{K\pi}\Delta_{\ell\pi}}{M_K^4} N_1(t) + \frac{2}{3} \frac{\Delta_{K\pi}+t}{2M_K^2} \tilde N_1(t) \\
			&\quad - \frac{1}{3} R_0(t) - \frac{1}{3} \frac{t(s-u) + \Delta_{K\pi}\Delta_{\ell\pi}}{M_K^4} R_1(t) + \frac{1}{3} \frac{\Delta_{K\pi}+t}{2M_K^2} \tilde R_1(t) \\
			&\quad + R_0(u) + \frac{u(s-t) + \Delta_{K\pi}\Delta_{\ell\pi}}{M_K^4} R_1(u) - \frac{\Delta_{K\pi} + u}{2M_K^2} \tilde R_1(u) \\
			&\quad + \O(p^8) .
	\end{split}
\end{align}
In the case $n=2$, the various functions of one variable are given by
\begin{align}
	\begin{split}
		\label{eq:FunctionsOfOneVariable}
		M_0(s) &= m_0^0 + m_0^1 \frac{s}{M_K^2} + \frac{s^2}{\pi} \int_{s_0}^\infty \frac{\Im f_0(s^\prime)}{(s^\prime - s - i\epsilon){s^\prime}^2} ds^\prime , \\
		M_1(s) &= m_1^0 + \frac{s}{\pi} \int_{s_0}^\infty \frac{1}{(s^\prime - s - i\epsilon) s^\prime} \Im \left( f_1(s^\prime) - \frac{2PL(s^\prime) M_K^2}{\lambda_{K\ell}(s^\prime)} g_1(s^\prime) \right) ds^\prime , \\
		\tilde M_1(s) &= \tilde m_1^0 + \tilde m_1^1 \frac{s}{M_K^2} + \frac{s^2}{\pi} \int_{s_0}^\infty \frac{\Im g_1(s^\prime)}{(s^\prime - s - i\epsilon){s^\prime}^2} ds^\prime , \\
		N_0(t) &= n_0^1 \frac{t}{M_K^2} + \frac{t^2}{\pi} \int_{t_0}^\infty \frac{\Im  f_0^{(1/2)}(t^\prime)}{(t^\prime-t-i\epsilon){t^\prime}^2} dt^\prime , \\
		N_1(t) &= \frac{1}{\pi} \int_{t_0}^\infty \frac{1}{t^\prime - t - i\epsilon} \Im \left( f_1^{(1/2)}(t^\prime) + \frac{(\Delta_{\ell\pi}+t^\prime)M_K^4}{2t^\prime \lambda_{\ell\pi}(t^\prime)} g_1^{(1/2)}(t^\prime) \right) dt^\prime , \\
		\tilde N_1(t) &= \frac{t}{\pi} \int_{t_0}^\infty \frac{M_K^2}{t^\prime} \frac{\Im g_1^{(1/2)}(t^\prime)}{(t^\prime-t-i\epsilon)t^\prime} dt^\prime , \\
		R_0(t) &= \frac{t^2}{\pi} \int_{t_0}^\infty \frac{\Im  f_0^{(3/2)}(t^\prime)}{(t^\prime-t-i\epsilon){t^\prime}^2} dt^\prime , \\
		R_1(t) &= \frac{1}{\pi} \int_{t_0}^\infty \frac{1}{t^\prime - t - i\epsilon} \Im \left( f_1^{(3/2)}(t^\prime) + \frac{(\Delta_{\ell\pi}+t^\prime)M_K^4}{2t^\prime \lambda_{\ell\pi}(t^\prime)} g_1^{(3/2)}(t^\prime) \right) dt^\prime , \\
		\tilde R_1(t) &= \frac{t}{\pi} \int_{t_0}^\infty \frac{M_K^2}{t^\prime} \frac{\Im g_1^{(3/2)}(t^\prime)}{(t^\prime-t-i\epsilon)t^\prime} dt^\prime ,
	\end{split}
\end{align}
while for $n=3$, the functions of one variable are
\begin{align}
	\begin{split}
		\label{eq:FunctionsOfOneVariable3Subtr}
		M_0(s) &= m_0^0 + m_0^1 \frac{s}{M_K^2} + m_0^2 \frac{s^2}{M_K^4} + \frac{s^3}{\pi} \int_{s_0}^\infty \frac{\Im f_0(s^\prime)}{(s^\prime - s - i\epsilon){s^\prime}^3} ds^\prime , \\
		M_1(s) &= m_1^0 + m_1^1 \frac{s}{M_K^2} + \frac{s^2}{\pi} \int_{s_0}^\infty \frac{1}{(s^\prime - s - i\epsilon) {s^\prime}^2} \Im \left( f_1(s^\prime) - \frac{2PL(s^\prime) M_K^2}{\lambda_{K\ell}(s^\prime)} g_1(s^\prime) \right) ds^\prime , \\
		\tilde M_1(s) &= \tilde m_1^0 + \tilde m_1^1 \frac{s}{M_K^2} + \tilde m_1^2 \frac{s^2}{M_K^4} + \frac{s^3}{\pi} \int_{s_0}^\infty \frac{\Im g_1(s^\prime)}{(s^\prime - s - i\epsilon){s^\prime}^3} ds^\prime , \\
		N_0(t) &= n_0^1 \frac{t}{M_K^2} + n_0^2 \frac{t^2}{M_K^4} + \frac{t^3}{\pi} \int_{t_0}^\infty \frac{\Im  f_0^{(1/2)}(t^\prime)}{(t^\prime-t-i\epsilon){t^\prime}^3} dt^\prime , \\
		N_1(t) &= n_1^0 + \frac{t}{\pi} \int_{t_0}^\infty \frac{1}{(t^\prime - t - i\epsilon)t^\prime} \Im \left( f_1^{(1/2)}(t^\prime) + \frac{(\Delta_{\ell\pi}+t^\prime)M_K^4}{2t^\prime \lambda_{\ell\pi}(t^\prime)} g_1^{(1/2)}(t^\prime) \right) dt^\prime , \\
		\tilde N_1(t) &= \tilde n_1^1 \frac{t}{M_K^2} + \frac{t^2}{\pi} \int_{t_0}^\infty \frac{M_K^2}{t^\prime} \frac{\Im g_1^{(1/2)}(t^\prime)}{(t^\prime-t-i\epsilon){t^\prime}^2} dt^\prime , \\
		R_0(t) &= \frac{t^3}{\pi} \int_{t_0}^\infty \frac{\Im  f_0^{(3/2)}(t^\prime)}{(t^\prime-t-i\epsilon){t^\prime}^3} dt^\prime , \\
		R_1(t) &= \frac{t}{\pi} \int_{t_0}^\infty \frac{1}{(t^\prime - t - i\epsilon){t^\prime}} \Im \left( f_1^{(3/2)}(t^\prime) + \frac{(\Delta_{\ell\pi}+t^\prime)M_K^4}{2t^\prime \lambda_{\ell\pi}(t^\prime)} g_1^{(3/2)}(t^\prime) \right) dt^\prime , \\
		\tilde R_1(t) &= \frac{t^2}{\pi} \int_{t_0}^\infty \frac{M_K^2}{t^\prime} \frac{\Im g_1^{(3/2)}(t^\prime)}{(t^\prime-t-i\epsilon){t^\prime}^2} dt^\prime .
	\end{split}
\end{align}

Actually, since the $P$-wave of isospin $I=3/2$ $K\pi$ scattering is real
at $\O(p^6)$, so are the partial waves $f_1^{(3/2)}$ and
$g_1^{(3/2)}$. Hence, the functions $R_1(t)$ and $\tilde R_1(t)$ could be
dropped altogether in the decomposition. The phase $\delta_1^{3/2}$ is also
known to be tiny in phenomenology.

\subsubsection{Ambiguity of the Decomposition}

We have decomposed the form factors $F$ and $G$ into functions of one
variable $M_0(s)$, $\ldots$. However, while the form factors are observable
quantities, these functions are not. It is possible to redefine the
functions $M_0(s)$, $\ldots$ without changing the form factors and hence
without changing the physics. 

Therefore, let us study this ambiguity of the decomposition of the form
factors. We require the form factors to be invariant under a change of the
functions of one variable: 
\begin{align}
	\begin{split}
		M_0(s) &\mapsto M_0(s) + \delta M_0(s) , \\
		M_1(s) &\mapsto M_1(s) + \delta M_1(s) , \\
			&\ldots ,
	\end{split}
\end{align}
which we call `gauge transformation'. The shifts have to satisfy
\begin{align}
	\begin{split}
		0 &= \delta M_0(s) + \frac{u-t}{M_K^2} \delta M_1(s) \\
			&\quad + \frac{2}{3} \delta N_0(t) + \frac{2}{3} \frac{t(s-u) + \Delta_{K\pi}\Delta_{\ell\pi}}{M_K^4} \delta N_1(t) - \frac{2}{3} \frac{\Delta_{K\pi}-3t}{2M_K^2} \delta \tilde N_1(t) \\
			&\quad + \frac{1}{3} \delta R_0(t) + \frac{1}{3} \frac{t(s-u) + \Delta_{K\pi}\Delta_{\ell\pi}}{M_K^4} \delta R_1(t) - \frac{1}{3} \frac{\Delta_{K\pi}-3t}{2M_K^2} \delta \tilde R_1(t) \\
			&\quad + \delta R_0(u) + \frac{u(s-t) + \Delta_{K\pi}\Delta_{\ell\pi}}{M_K^4} \delta R_1(u) - \frac{\Delta_{K\pi}-3u}{2M_K^2} \delta \tilde R_1(u) ,
	\end{split} \\
	\begin{split}
		0 &= \delta \tilde M_1(s) \\
			&\quad - \frac{2}{3} \delta N_0(t) - \frac{2}{3} \frac{t(s-u) + \Delta_{K\pi}\Delta_{\ell\pi}}{M_K^4} \delta N_1(t) + \frac{2}{3} \frac{\Delta_{K\pi}+t}{2M_K^2} \delta \tilde N_1(t) \\
			&\quad - \frac{1}{3} \delta R_0(t) - \frac{1}{3} \frac{t(s-u) + \Delta_{K\pi}\Delta_{\ell\pi}}{M_K^4} \delta R_1(t) + \frac{1}{3} \frac{\Delta_{K\pi}+t}{2M_K^2} \delta \tilde R_1(t) \\
			&\quad + \delta R_0(u) + \frac{u(s-t) + \Delta_{K\pi}\Delta_{\ell\pi}}{M_K^4} \delta R_1(u) - \frac{\Delta_{K\pi}+u}{2M_K^2} \delta \tilde R_1(u) .
	\end{split}
\end{align}
The solution to these equations is found in the following way: we
substitute one of the three kinematic variables by means of $s+t+u =
\Sigma_0$. Then, we take the derivative with respect to one of the two
remaining variables and substitute back $\Sigma_0 = s + t + u$. After four
or five such differentiations, one gets differential equations for single
functions $\delta M_0$, $\ldots$ with the following solution: 
\begin{align}
	\begin{split}
		\delta M_0(s) &= c_0^{M_0} + c_1^{M_0} s + c_2^{M_0} s^2 , \\
		\delta M_1(s) &= c_0^{M_1} + c_1^{M_1} s + c_2^{M_1} s^2 , \\
		\delta \tilde M_1(s) &= c_0^{\tilde M_1} + c_1^{\tilde M_1} s + c_2^{\tilde M_1} s^2 + c_3^{\tilde M_1} s^3 , \\
		\delta N_0(t) &= c_{-1}^{N_0} t^{-1} + c_0^{N_0} + c_1^{N_0} t + c_2^{N_0} t^2 , \\
		\delta N_1(t) &= c_{-1}^{N_1} t^{-1} + c_0^{N_1} + c_1^{N_1} t , \\
		\delta \tilde N_1(t) &= c_{-1}^{\tilde N_1} t^{-1} + c_0^{\tilde N_1} + c_1^{\tilde N_1} t + c_2^{\tilde N_1} t^2 , \\
		\delta R_0(t) &= c_{-1}^{R_0} t^{-1} + c_0^{R_0} + c_1^{R_0} t + c_2^{R_0} t^2 , \\
		\delta R_1(t) &= c_{-1}^{R_1} t^{-1} + c_0^{R_1} + c_1^{R_1} t , \\
		\delta \tilde R_1(t) &= c_{-1}^{\tilde R_1} t^{-1} + c_0^{\tilde R_1} + c_1^{\tilde R_1} t + c_2^{\tilde R_1} t^2 .
	\end{split}
\end{align}
Inserting these solutions into the various differential equations results
in algebraic equations for the diverse coefficients. In the end, there
remain 13 independent parameters. In complete generality, we therefore have
a gauge freedom of 13 parameters in the decomposition
(\ref{eq:FormFactorDecomposition}). The gauge can be fixed by imposing
constraints on the Taylor expansion or the asymptotic behaviour of the
functions $M_0(s)$, $\ldots$. 

First, let us restrict the gauge freedom by imposing the same vanishing
Taylor coefficients as in (\ref{eq:FunctionsOfOneVariable}), i.e.~we
exclude all the pole terms, the constants in $N_0$, $\tilde N_1$, $R_0$,
$\tilde R_1$ and even a linear term in $R_0$. Then, we further demand that
asymptotically the functions behave at most as in
(\ref{eq:FunctionsOfOneVariable3Subtr}), i.e.~like $M_1(s) = \O(s)$,
$\tilde M_1(s) = \O(s^2)$, $N_1(t) = \O(1)$, $\tilde N_1(t) = \O(t)$,
$R_1(t) = \O(1)$ and $\tilde R_1(t) = \O(t)$. After imposing these
constraints, we are left with a restricted gauge freedom of 3 parameters,
which we call $C^{R_0}$, $A^{R_1}$ and $B^{\tilde R_1}$: 
\begin{align}
	\begin{split}
		\label{eq:GaugeTransformation}
		\delta M_0(s) &= \left(2 A^{R_1} - B^{\tilde R_1} + 2 C^{R_0}\right) \frac{\left( \Sigma_0 - s \right)^2 - \Delta_{K\pi}\Delta_{\ell\pi}}{2 M_K^4} , \\
		\delta M_1(s) &= -\left(A^{R_1} + B^{\tilde R_1} + 2 C^{R_0}\right)\frac{\Sigma_0}{M_K^2} + B^{\tilde R_1} \frac{\Delta_{K\pi}}{2 M_K^2} + \left(B^{\tilde R_1} + 2 C^{R_0}\right) \frac{s}{M_K^2} , \\
		\delta \tilde M_1(s) &= \left( \left(B^{\tilde R_1} - 2 C^{R_0}\right) \Sigma_0^2 - \left(2 A^{R_1} + B^{\tilde R_1} - 2 C^{R_0} \right) \Delta_{K\pi} \Delta_{\ell\pi} + B^{\tilde R_1} \Sigma_0 \Delta_{K\pi} \right) \frac{1}{2 M_K^4} \\
			& - \left(B^{\tilde R_1} \Delta_{K\pi} + \left(A^{R_1} + B^{\tilde R_1} - 2 C^{R_0}\right)  2 \Sigma_0 \right) \frac{s}{2 M_K^4} + \left(2 A^{R_1} + B^{\tilde R_1} - 2 C^{R_0}\right) \frac{s^2}{2 M_K^4} , \\
		\delta N_0(t) &= - \left(2 A^{R_1} - B^{\tilde R_1} + 2 C^{R_0} \right) \frac{3 t (\Delta_{K\pi} + 2 \Sigma_0)}{8 M_K^4} + \left(6 A^{R_1} - 3 B^{\tilde R_1} -10 C^{R_0} \right) \frac{t^2}{8 M_K^4} , \\
		\delta N_1(t) &= - \frac{1}{4} \left(2 A^{R_1} + 3 B^{\tilde R_1} - 6 C^{R_0} \right) , \\
		\delta \tilde N_1(t) &= -\left(6 A^{R_1} + 5 B^{\tilde R_1} + 6 C^{R_0} \right)\frac{t}{4 M_K^2} , \\
		\delta R_0(t) &= C^{R_0} \frac{t^2}{M_K^4} , \\
		\delta R_1(t) &= A^{R_1} , \\
		\delta \tilde R_1(t) &= B^{\tilde R_1} \frac{t}{M_K^2}.
	\end{split}
\end{align}
In order to fix the gauge completely, we have to impose further
conditions. We will use two different gauges. The first one corresponds to
the case of an asymptotic behaviour that needs $n=2$ subtractions. It is
most suitable for our numerical dispersive representation of the form
factors and for the NLO chiral result. In this case, the asymptotic
behaviour excludes quadratic terms in $\delta M_0(s)$ and $\delta \tilde
M_1(s)$ or a linear term in $\delta M_1(s)$. Hence, in the representation
(\ref{eq:FunctionsOfOneVariable}), the gauge is completely fixed. 

The chiral representation, being an expansion in the masses and momenta,
does not necessarily reproduce the correct asymptotic behaviour. The
$\O(p^6)$ chiral expressions show an asymptotic behaviour that needs $n=3$
subtractions. In this case, one has to fix the gauge rather with the Taylor
coefficients, e.g.~by excluding a quadratic term in $R_0$, a constant term
in $R_1$ and a linear term in $\tilde R_1$. Therefore, also in the
representation (\ref{eq:FunctionsOfOneVariable3Subtr}), the gauge is
completely fixed. 

Note that the second representation (\ref{eq:FunctionsOfOneVariable3Subtr})
makes less restrictive assumptions about the asymptotic
behaviour. Therefore, the first representation
(\ref{eq:FunctionsOfOneVariable}) is a special case of the second
(\ref{eq:FunctionsOfOneVariable3Subtr}). One can easily switch from the
first to the second representation with the help of the gauge
transformation (\ref{eq:GaugeTransformation}). In this case, the additional
subtraction constants will be given by sum rules. 


\subsubsection{Simplifications for $s_\ell\to0$}

As a test of the decomposition, let us study the linear combination
\begin{align}
	\begin{split}
		F_1(s,t,u) = \frac{1}{2}(M_K^2 - s) F(s,t,u) + \frac{1}{2}(u-t) G(s,t,u)
	\end{split}
\end{align}
in the limit $s_\ell\to0$. We neglect the contribution of the isospin $3/2$ $P$-wave:
\begin{align}
	\begin{split}
		\lim_{s_\ell\to0} &F_1(s,t,u) = \lim_{s_\ell\to0} \left( \frac{M_K^2 - s}{2} F(s,t,u) + \frac{u-t}{2} G(s,t,u) \right) \\
			&= \frac{M_K^2 - s}{2} M_0(s) \\
			&\quad + \frac{u-t}{2} \left[ \frac{M_K^2 - s}{M_K^2} M_1(s) +  \tilde M_1(s) \right] \\
			&\quad + \frac{2}{3} \left[ (t-M_\pi^2) N_0(t) \right] + \frac{1}{3} \left[ (t-M_\pi^2) R_0(t) \right] + (u-M_\pi^2) R_0(u) \\
			&\quad - \frac{2}{3} \left(t(u-s)+(M_K^2-M_\pi^2)M_\pi^2\right) \left[ \frac{t-M_\pi^2}{M_K^4} N_1(t) - \frac{1}{2 M_K^2} \tilde N_1(t) \right] .
	\end{split}
\end{align}

By identifying
\begin{align}
	\begin{split}
		M_0^{F_1}(s) &= \frac{M_K^2-s}{2} M_0(s) , \\
		M_1^{F_1}(s) &= \frac{1}{2} \left( \frac{M_K^2-s}{M_K^2} M_1(s) + \tilde M_1(s) \right) , \\
		N_0^{F_1}(t) &= (t-M_\pi^2) N_0(t) , \\
		R_0^{F_1}(t) &= (t-M_\pi^2) R_0(t) , \\
		N_1^{F_1}(t) &= \frac{t-M_\pi^2}{M_K^4} N_1(t) - \frac{1}{2M_K^2} \tilde N_1(t) ,
	\end{split}
\end{align}
we recover the decomposition of the form factor $F_1$ used in
\cite{Stoffer2010, Colangelo2012, Stoffer2013}. We further note that the
imaginary parts of the functions of one variable in this decomposition are
given by 
\begin{align}
	\begin{split}
		\Im M_0^{F_1}(s) &= \frac{M_K^2-s}{2} \Im f_0(s) , \\
		\Im M_1^{F_1}(s) &= \frac{M_K^2-s}{2M_K^2} \Im f_1(s) , \\
		\Im N_0^{F_1}(t) &= (t-M_\pi^2) \Im f_0^{(1/2)}(t) , \\
		\Im R_0^{F_1}(t) &= (t-M_\pi^2) \Im f_0^{(3/2)}(t) , \\
		\Im N_1^{F_1}(t) &= \frac{t-M_\pi^2}{M_K^4} \Im f_1^{(1/2)}(t) ,
	\end{split}
\end{align}
and repeat the observation of
section~\ref{sec:PartialWaveSimplificationsZeroSL} that in the limit
$s_\ell\to0$, these partial waves are given by projections of $F_1$ in all
three channels. Hence, the form factor $F_1$ decouples in this limit and
can be treated independently in the above decomposition. 


\subsection{Integral Equations}

\subsubsection{Omn\`es Representation}

The decomposition of the form factors (\ref{eq:FormFactorDecomposition})
signifies a major simplification, since we only have to deal with functions
of a single Mandelstam variable. These functions
(\ref{eq:FunctionsOfOneVariable}, \ref{eq:FunctionsOfOneVariable3Subtr})
are constructed in such a way that they only contain the right-hand cut of
the corresponding partial wave. Their imaginary part on the upper rim of
their cut is given by 
\begin{align}
	\begin{split}
		\Im M_0(s) &= \Im f_0(s) , \\
		\Im M_1(s) &= \Im \left( f_1(s) - \frac{2PL(s) M_K^2}{\lambda_{K\ell}(s)} g_1(s) \right) , \\
		\Im \tilde M_1(s) &= \Im g_1(s) , \\
		\Im N_0(t) &= \Im  f_0^{(1/2)}(t) , \\
		\Im N_1(t) &= \Im \left( f_1^{(1/2)}(t) + \frac{(\Delta_{\ell\pi}+t)M_K^4}{2t \lambda_{\ell\pi}(t)} g_1^{(1/2)}(t) \right) , \\
		\Im \tilde N_1(t) &= \Im \left(\frac{M_K^2}{t} g_1^{(1/2)}(t) \right) , \\
		\Im R_0(t) &= \Im  f_0^{(3/2)}(t) , \\
		\Im R_1(t) &= \Im \left( f_1^{(3/2)}(t) + \frac{(\Delta_{\ell\pi}+t)M_K^4}{2t \lambda_{\ell\pi}(t)} g_1^{(3/2)}(t) \right) , \\
		\Im \tilde R_1(t) &= \Im \left(\frac{M_K^2}{t} g_1^{(3/2)}(t) \right) .
	\end{split}
\end{align}
Therefore, we can write
\begin{align}
	\begin{split}
		\label{eq:HatFunctionPartialWavesRelation}
		M_0(s) + \hat M_0(s) &= f_0(s) , \\
		M_1(s) + \hat M_1(s) &= f_1(s) - \frac{2PL(s) M_K^2}{\lambda_{K\ell}(s)} g_1(s) , \\
		 \tilde M_1(s) + \hat{\tilde M}_1(s) &= g_1(s) , \\
		N_0(t) + \hat N_0(t) &=  f_0^{(1/2)}(t) , \\
		N_1(t) + \hat N_1(t) &= f_1^{(1/2)}(t) + \frac{(\Delta_{\ell\pi}+t)M_K^4}{2t \lambda_{\ell\pi}(t)} g_1^{(1/2)}(t) , \\
		\tilde N_1(t) + \hat{\tilde N}_1(t) &= \frac{M_K^2}{t} g_1^{(1/2)}(t) , \\
		R_0(t) + \hat R_0(t) &=  f_0^{(3/2)}(t) , \\
		R_1(t) + \hat R_1(t) &= f_1^{(3/2)}(t) + \frac{(\Delta_{\ell\pi}+t)M_K^4}{2t \lambda_{\ell\pi}(t)} g_1^{(3/2)}(t) , \\
		\tilde R_1(t) + \hat{\tilde R}_1(t) &= \frac{M_K^2}{t} g_1^{(3/2)}(t) ,
	\end{split}
\end{align}
where the `hat functions' $\hat M_0(s)$, $\ldots$ are real on the cut:
indeed, they do not possess a right-hand cut, but contain the (possibly
complicated) left-hand cut structure of the partial waves (see
section~\ref{sec:ProjectionAnalyticStructurePartialWaves}). Writing $\Im
f_0(s) = f_0(s) e^{-i\delta_0^0(s)} \sin\delta_0^0(s)$, $\ldots$ leads
directly to the following equations: 
\begin{align}
	\begin{split}
		\Im M_0(s) &= ( M_0(s) + \hat M_0(s) ) e^{-i\delta_0^0(s)} \sin\delta_0^0(s) , \\
		\Im M_1(s) &= ( M_1(s) + \hat M_1(s) ) e^{-i\delta_1^1(s)} \sin\delta_1^1(s) , \\
		\Im \tilde M_1(s) &= ( \tilde M_1(s) + \hat{\tilde M}_1(s) ) e^{-i\delta_1^1(s)} \sin\delta_1^1(s) , \\
		\Im N_0(t) &= ( N_0(t) + \hat N_0(t) ) e^{-i\delta_0^{1/2}(t)} \sin\delta_0^{1/2}(t) , \\
		\Im N_1(t) &= ( N_1(t) + \hat N_1(t) ) e^{-i\delta_1^{1/2}(t)} \sin\delta_1^{1/2}(t) , \\
		\Im \tilde N_1(t) &= ( \tilde N_1(t) + \hat{\tilde N}_1(t) ) e^{-i\delta_1^{1/2}(t)} \sin\delta_1^{1/2}(t) , \\
		\Im R_0(t) &= ( R_0(t) + \hat R_0(t) ) e^{-i\delta_0^{3/2}(t)} \sin\delta_0^{3/2}(t) , \\
		\Im R_1(t) &= ( R_1(t) + \hat R_1(t) ) e^{-i\delta_1^{3/2}(t)} \sin\delta_1^{3/2}(t) , \\
		\Im \tilde R_1(t) &= ( \tilde R_1(t) + \hat{\tilde R}_1(t) ) e^{-i\delta_1^{3/2}(t)} \sin\delta_1^{3/2}(t) ,
	\end{split}
\end{align}
where, below some inelastic threshold, the phases $\delta_l^I$ agree with
the elastic $\pi\pi$- or $K\pi$-scattering phase shifts. Therefore, the
functions $M_0$, $\ldots$ are given by the solution to the inhomogeneous
Omn\`es problem. The minimal number of subtractions appearing in the Omn\`es
representation is determined by the asymptotic behaviour of the functions
$M_0$, $\ldots$ and the phases $\delta_l^I$. 

Let us extend these equations even to the region above inelastic thresholds
by replacing $\delta \mapsto \omega$, 
\begin{align}
	\begin{split}
		\Im M_0(s) &= ( M_0(s) + \hat M_0(s) ) e^{-i\omega_0^0(s)} \sin\omega_0^0(s) , \\
		&\ldots ,
	\end{split}
\end{align}
where $\omega_l^I(s) = \delta_l^I(s) + \eta_l^I(s)$ and $\eta_l^I(s) = 0$
below the inelastic threshold $s = \Lambda^2$. 

We define the usual once-subtracted Omn\`es function
\begin{align}
	\label{eq:OmnesFunction}
	\Omega(s) := \exp\left( \frac{s}{\pi} \int_{s_0}^\infty \frac{\delta(s^\prime)}{(s^\prime - s - i\epsilon) s^\prime} ds^\prime \right) .
\end{align}
If the asymptotic behaviour of the phase is $\lim\limits_{s\to\infty}
\delta(s) = m \pi$, the Omn\`es function behaves asymptotically as
$\O(s^{-m})$. Provided that the function $M(s)$ behaves asymptotically as
$\O(s^k)$, we can write a dispersion relation for $M/\Omega$ that leads to
a modified Omn\`es solution 
\begin{align}
	\begin{split}
		\label{eq:ModifiedOmnesSolution}
		M(s) = \Omega(s) &\left\{ P_{n-1}(s) + \frac{s^n}{\pi} \int_{s_0}^{\Lambda^2} \frac{\hat M(s^\prime) \sin\delta(s^\prime)}{|\Omega(s^\prime)| (s^\prime - s - i\epsilon) {s^\prime}^n} ds^\prime  \right. \\
			&\quad + \frac{s^n}{\pi} \int_{\Lambda^2}^{\infty} \frac{\hat M(s^\prime) \sin\delta(s^\prime)}{|\Omega(s^\prime)| (s^\prime - s - i\epsilon) {s^\prime}^n} ds^\prime \\
			&\quad + \left. \frac{s^n}{\pi} \int_{\Lambda^2}^\infty \frac{(\hat M(s^\prime) + \Re M(s^\prime)) \sin\eta(s^\prime)}{|\Omega(s^\prime)| \cos(\delta(s^\prime) + \eta(s^\prime)) (s^\prime - s - i\epsilon) {s^\prime}^n} ds^\prime \right\} ,
	\end{split}
\end{align}
where the order of the subtraction polynomial is $n-1=k+m$.

Actually, we do not know the phase $\delta$ at high
energies. Inelasticities due to multi-Goldstone boson intermediate states,
i.e.~more than two Goldstone bosons, appear only at $\O(p^8)$
\cite{Stern1993}, hence the most important inelastic contribution would
certainly be a $K\bar K$ intermediate state in the $s$-channel. This could
be included by using experimental input on $\eta$ up to $s\approx(1.4 \,
\mathrm{GeV})^2$. 

We could make a Taylor expansion of the inelasticity integral and neglect
terms that only contribute at $\O(p^8)$ to the form factors by applying the
power counting $\frac{s}{\Lambda^2} \sim p^2$. This would introduce quite a
lot of unknown Taylor coefficients. Here, we follow another strategy: we
set $\eta = 0$ and assign a large error to the phases $\delta$ at high
energies. We assume further that $\delta$ reaches a multiple of $\pi$ above
a certain $s = \Lambda^2$. The two high-energy integrals in
(\ref{eq:ModifiedOmnesSolution}) drop in this case.

Assuming that the phases behave asymptotically like $\delta_0^0 \to \pi$ ,
$\delta_1^1 \to \pi$ and all other $\delta_l^I \to 0$, we find the
following solution for the case of $n=2$ subtractions: 
\begin{align}
	\begin{alignedat}{2}
		\label{eq:FunctionsOfOneVariableOmnes}
		M_0(s) &= \Omega_0^0(s) & &\bigg\{ a^{M_0} + b^{M_0} \frac{s}{M_K^2} + c^{M_0} \frac{s^2}{M_K^4} + \frac{s^3}{\pi} \int_{s_0}^{\Lambda^2} \frac{\hat M_0(s^\prime) \sin\delta_0^0(s^\prime)}{|\Omega_0^0(s^\prime)| (s^\prime - s - i\epsilon) {s^\prime}^3} ds^\prime \bigg\} , \\
		M_1(s) &= \Omega_1^1(s) & &\bigg\{ a^{M_1} + b^{M_1}  \frac{s}{M_K^2} + \frac{s^2}{\pi} \int_{s_0}^{\Lambda^2} \frac{\hat M_1(s^\prime) \sin\delta_1^1(s^\prime)}{|\Omega_1^1(s^\prime)| (s^\prime - s - i\epsilon) {s^\prime}^2} ds^\prime  \bigg\} , \\
		 \tilde M_1(s) &= \Omega_1^1(s) & &\bigg\{ a^{\tilde M_1} + b^{\tilde M_1}  \frac{s}{M_K^2} + c^{\tilde M_1}  \frac{s^2}{M_K^4} + \frac{s^3}{\pi} \int_{s_0}^{\Lambda^2} \frac{\hat{\tilde M}_1(s^\prime) \sin\delta_1^1(s^\prime)}{|\Omega_1^1(s^\prime)| (s^\prime - s - i\epsilon) {s^\prime}^3} ds^\prime \bigg\} , \\
		N_0(t) &=  \Omega_0^{1/2}(t) & &\bigg\{ b^{N_0} \frac{t}{M_K^2} + \frac{t^2}{\pi} \int_{t_0}^{\Lambda^2} \frac{\hat N_0(t^\prime) \sin\delta_0^{1/2}(t^\prime)}{|\Omega_0^{1/2}(t^\prime)| (t^\prime - t - i\epsilon) {t^\prime}^2} dt^\prime  \bigg\} , \\
		N_1(t) &= \Omega_1^{1/2}(t) & &\bigg\{ \frac{1}{\pi} \int_{t_0}^{\Lambda^2} \frac{\hat N_1(t^\prime) \sin\delta_1^{1/2}(t^\prime)}{|\Omega_1^{1/2}(t^\prime)| (t^\prime - t - i\epsilon)} dt^\prime  \bigg\} , \\
		\tilde N_1(t) &= \Omega_1^{1/2}(t) & &\bigg\{ \frac{t}{\pi} \int_{t_0}^{\Lambda^2} \frac{\hat{\tilde N}_1(t^\prime) \sin\delta_1^{1/2}(t^\prime)}{|\Omega_1^{1/2}(t^\prime)| (t^\prime - t - i\epsilon) t^\prime} dt^\prime  \bigg\} , \\
		R_0(t) &=  \Omega_0^{3/2}(t) & &\bigg\{ \frac{t^2}{\pi} \int_{t_0}^{\Lambda^2} \frac{\hat R_0(t^\prime) \sin\delta_0^{3/2}(t^\prime)}{|\Omega_0^{3/2}(t^\prime)| (t^\prime - t - i\epsilon) {t^\prime}^2} dt^\prime  \bigg\} , \\
		R_1(t) &=  \Omega_1^{3/2}(t) & &\bigg\{ \frac{1}{\pi} \int_{t_0}^{\Lambda^2} \frac{\hat R_1(t^\prime) \sin\delta_1^{3/2}(t^\prime)}{|\Omega_1^{3/2}(t^\prime)| (t^\prime - t - i\epsilon)} dt^\prime  \bigg\} , \\
		\tilde R_1(t) &=   \Omega_1^{3/2}(t) & &\bigg\{ \frac{t}{\pi} \int_{t_0}^{\Lambda^2} \frac{\hat{\tilde R}_1(t^\prime) \sin\delta_1^{3/2}(t^\prime)}{|\Omega_1^{3/2}(t^\prime)| (t^\prime - t - i\epsilon) {t^\prime}} dt^\prime  \bigg\} ,
	\end{alignedat}
\end{align}
where we have fixed some of the subtraction constants in $N_0$, $\tilde
N_1$, $R_0$ and $\tilde R_1$ to zero by imposing the same Taylor expansion
as in the defining equation (\ref{eq:FunctionsOfOneVariable}). 

Note that driving the $K\pi$ phases to 0 is somehow artificial. They are
rather supposed to reach $\pi$ at high energies. However, this would
introduce three more subtraction constants in our framework. Since the
high-energy behaviour of the phases does not have an important influence on
our results, we abstain from introducing more subtractions and take these
effects into account in the systematic uncertainty. 

In the case of $n=3$ subtractions, six additional subtraction constants
appear in the Omn\`es representation. The conversion of a solution for
$n=2$ into a solution for $n=3$ requires a gauge transformation in the
Omn\`es representation, as explained in
appendix~\ref{sec:AppendixOmnes3Subtractions}. 

\subsubsection{Hat Functions}

The hat functions appearing in the Omn\`es solution to the functions of one
variable (\ref{eq:FunctionsOfOneVariableOmnes}) can be computed through
partial-wave projections of the form factors:
(\ref{eq:HatFunctionPartialWavesRelation}) should be understood as the
defining equation of the hat functions. One has to compute the partial-wave
projections of the decomposed form factors $F$ and $G$
(\ref{eq:FormFactorDecomposition}) and subtract the function of one
variable ($M_0$, $\ldots$). Finally, one obtains an expression for the hat
functions in terms of angular averages of the single-variable functions. The
explicit expressions for the hat functions are given in
appendix~\ref{sec:AppendixHatFunctions}.


\section{Numerical Solution of the Dispersion Relation}

\label{sec:Kl4NumericalSolutionDR}

\subsection{Iterative Solution of the Dispersion Relation}

The reconstruction theorem has allowed us to decompose the form factors
into functions of one variable, (\ref{eq:FormFactorDecomposition}). The
nine functions of one variable are given unambiguously by the Omn\`es
solutions (\ref{eq:FunctionsOfOneVariableOmnes}). The hat functions
appearing in the dispersive integrals are given by angular integrals of the
nine functions of one variable and link these functions
together. Therefore, we face a set of coupled integral equations,
parametrised by the nine subtraction constants $a^{M_0}$, $b^{M_0}$, $\ldots$
as defined in~(\ref{eq:FunctionsOfOneVariableOmnes}). In this section we
discuss a method for solving these equations numerically. We will
assume an asymptotic behaviour of the form factors that requires only $n=2$
subtractions. 

A crucial property of this set of equations is that they are linear in the
subtraction constants. Any solution can be written as a linear combination
of nine basis solutions. Our main task is therefore to determine
numerically these nine basis solutions.

So far, the invariant squared energy of the dilepton system, $s_\ell$, has
been treated as an external parameter. On the one hand, it appears in
the definition of the hat functions. On the other hand, the subtraction
constants have an implicit dependence on $s_\ell$. To make this
dependence explicit we write the form factors as: 
\begin{align}
	\begin{split}
		X(s,t,u) &= \sum_{i=1}^9 a_i(s_\ell) X_i(s,t,u) ,
	\end{split}
\end{align}
where $X\in\{F,G\}$, $\{a_i\}_i = \{ a^{M_0}, b^{M_0}, \ldots, b^{N_0} \}$
and where $X_i$ denotes the basis solution with $a_k = \delta_{ik}$,
$k\in\{1,\ldots,9\}$. If $s_\ell$ is allowed to vary, the `functions of one
variable' become actually functions of two variables, $M_0(s,s_\ell)$,
$\ldots$ 

Our strategy is as follows. We determine the basis solutions by a
numerical iteration of the coupled integral equations. Each basis solution
is a function of $s$, $t$ and $u$, where $s+t+u = \Sigma_0 = M_K^2 +
2M_\pi^2 + s_\ell$, or equivalently a function of $s$, $s_\ell$ and
$\cos\theta$. Since $s_\ell$ is a fixed external parameter in the integral
equations, the iterative solution has to be performed for each value of
$s_\ell$ separately. Once the basis solutions are computed, the subtraction
constants (or rather functions) have to be determined by suitable means,
such as a fit to data, the soft-pion theorem and \ChPT{} input. As the
dependence on $s_\ell$ has been found to be rather weak in experiments, the
subtraction functions can be well approximated by a low-order polynomial in
$s_\ell$. 

In summary, we need the nine basis solutions for a set of values of
$s_\ell$, so as to allow us to calculate them for any value of $s_\ell$ by
interpolation. Again, since the dependence on $s_\ell$ appears to be rather
weak, we will need only a low number of values of $s_\ell$.


To calculate each of the basis solutions we use the following iterative procedure:
\begin{enumerate}
	\item set the initial value of the functions $M_0$, $\ldots$ to
          Omn\`es function $\times$ subtraction polynomial (the polynomial
          is in fact either 0 or a simple monomial with coefficient 1 for a
          particular basis solution); 
	\item calculate the hat functions $\hat M_0$, $\ldots$ by means of
          angular integrals of the functions $M_0$, $\ldots$ ; 
	\item calculate the new values of the functions $M_0$, $\ldots$ as
          Omn\`es function $\times$ (polynomial + dispersive part), where
          in the dispersion integral the hat function calculated in step 2
          appears; 
	\item go to step 2 and iterate this procedure until convergence.
\end{enumerate}
It turns out that this iteration converges quickly. After five or six
iterations, the relative changes are of order $10^{-6}$.

\subsection{Phase Input}

\subsubsection{$\pi\pi$ Phase Shifts}

\label{sec:Kl4pipiPhaseShifts}

\begin{figure}[H]
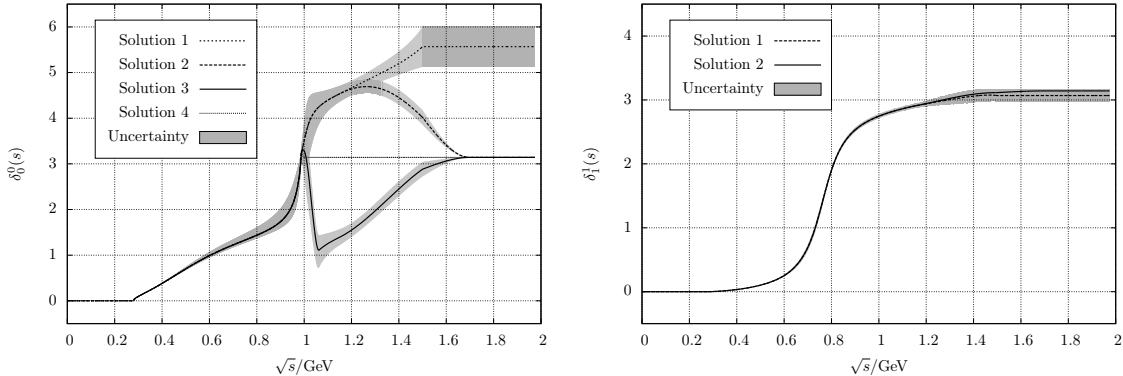

	\centering
	\scalebox{0.59}{
		\input{plots/d00.tex}
		\input{plots/d11.tex}
		}
	\caption{$\pi\pi$ phase shift inputs}
	\label{plot:PiPiPhaseShifts}
\end{figure}

For the pion scattering phase shifts, we use the parametrisation of
\cite{Ananthanarayan2001a, Caprini2012}. The solution depends on 28
para\-meters that can vary within a certain range. The curve labeled as
Solution~1 in figure~\ref{plot:PiPiPhaseShifts} shows the central solution
for the phase shifts as well as the error band due to uncertainty in the
parameters (summed in quadrature). 

Two aspects deserve special attention. First, the phase for Solution~1 is
just taken constant above $\sqrt{s} \approx 1.5$~GeV. Our derivation of the
dispersion relation, however, relies on the assumption that $\delta_0^0(s)
\to \pi$, $\delta_1^1(s) \to \pi$ for $s\to\infty$. We should therefore
change the high-energy behaviour of the phases such that they reach $\pi$
at $s=\Lambda^2$. The exact way how this is achieved should not have an
influence on the result at low energies, especially in the physical
region of the decay. We choose to interpolate smoothly between the value of
Solution~1  and $\pi$: 
\begin{align}
	\begin{split}
		\delta_0^0(s)_\mathrm{sol.2} &:= \left(1 -
                f_\mathrm{int}(s_1,s_2,s) \right)
                \delta_0^0(s)_\mathrm{sol.1} + f_\mathrm{int}(s_1,s_2,s)
                \pi , \\ 
		\delta_1^1(s)_\mathrm{sol.2} &:= \left(1 -
                f_\mathrm{int}(s_1,s_2,s) \right)
                \delta_1^1(s)_\mathrm{sol.1} + f_\mathrm{int}(s_1,s_2,s)
                \pi , \\ 
		f_\mathrm{int}(s_1,s_2,s) &:= \left\{ \begin{matrix} 0 &
                  \text{if } s < s_1 , \\ \frac{(s-s_1)^2 (3 s_2 - 2 s -
                    s_1)}{(s_2 - s_1)^3}  & \text{if } s_1 \le s < s_2 ,
                  \\ 1 & \text{if } s_2 \le s . \end{matrix} \right.  
	\end{split}
\end{align}
Figure~\ref{plot:PiPiPhaseShifts} shows Solution~2 with $s_1 = 68 M_\pi^2$
and $s_2 = 148 M_\pi^2$. These values can be varied and should not have an
important influence. 

The second subtlety is the problem of the behaviour around the $K\bar K$
threshold \cite{Oller2007}: are the $K_{\ell4}$ partial waves expected to
have a peak or a dip in the vicinity of the $K \bar K$ threshold, i.e.~do
they rather behave like the strange or the non-strange scalar form factor
of the pion? The answer to this question could be obtained from a
coupled-channel analysis of the $K_{\ell4}$ amplitude, which however goes
beyond the scope of this paper. In case of a dip we have to modify the
phase such that it follows $\delta_0^0(s)-\pi$ above the $K \bar K$
threshold. The third solution shown in figure~\ref{plot:PiPiPhaseShifts} is
given by 
\begin{align}
	\begin{split}
		\delta_0^0(s)_\mathrm{sol.3} &:= \left(1 -
                f_\mathrm{int}(s_1,s_2,s) \right) \left(
                \delta_0^0(s)_\mathrm{sol.1} - f_\mathrm{int}(\tilde s_1,
                \tilde s_2, s) \pi \right) + f_\mathrm{int}(s_1,s_2,s) \pi
                ,  
	\end{split}
\end{align}
with $\tilde s_1 = 4 M_K^2$ and $\tilde s_2 = \tilde s_1 + 8 M_\pi^2$.

The Solution~4 in figure~\ref{plot:PiPiPhaseShifts} corresponds to
Solution~2 but with $s_1 = 4 M_K^2$ and $s_2 = s_1 + M_\pi^2$. 

As the question of the correct behaviour around the $K\bar K$ threshold is
not easy to answer, we declare Solution~3 as the `central' one and use
all the other solutions to determine the systematic uncertainty. 

\begin{figure}[H]
	\centering
	\scalebox{0.59}{
		\input{plots/d012.tex}
		\input{plots/d112.tex}
		}
	\caption{$K\pi$ phase shift inputs, isospin $I=1/2$}
	\label{plot:KPiPhaseShifts12}
\end{figure}

\begin{figure}[H]
	\centering
	\scalebox{0.59}{
		\input{plots/d032.tex}
		\input{plots/d132.tex}
		}
	\caption{$K\pi$ phase shift inputs, isospin $I=3/2$}
	\label{plot:KPiPhaseShifts32}
\end{figure}

%


\subsubsection{$K\pi$ Phase Shifts}

\label{sec:Kl4KpiPhaseShifts}

For the crossed channels, we also need the $K\pi$ phase shifts as an
input. We use the phase shifts and uncertainties of \cite{Buettiker2004,
  Boito2010}, but add by hand a more conservative uncertainty that reaches
$20^\circ$ at $t=150M_\pi^2$. For the very small phase $\delta_1^{3/2}$, we
just assume a 100\% uncertainty. These phase solutions are shown in
figures~\ref{plot:KPiPhaseShifts12} and \ref{plot:KPiPhaseShifts32} as
`Solution~1'. 

In the derivation of the dispersion relation, we assume that the $K\pi$
phases go to zero at high energies. We implement this by interpolating
smoothly between Solution~1 and zero with
$f_\mathrm{int}(t_1,t_2,t)$. These modified phase shifts with $t_1 = 150
M_\pi^2$ and $t_2 = 250 M_\pi^2$ are displayed as `Solution~2' in
figures~\ref{plot:KPiPhaseShifts12} and \ref{plot:KPiPhaseShifts32}. The
difference between `Solution~1' and `Solution~2' is taken as a measure of
the systematic uncertainty due to the high-energy behaviour of the $K\pi$
phases. 

\subsection{Omn\`es Functions}

\begin{figure}[H]
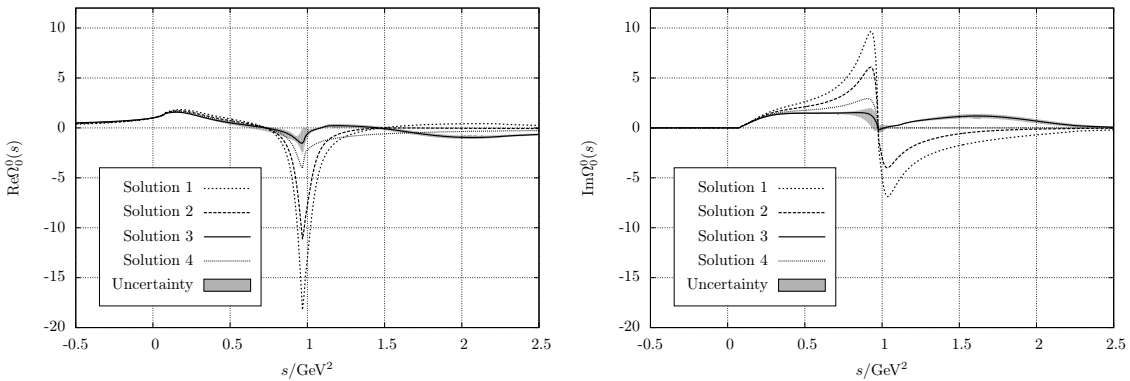

	\centering
	\scalebox{0.59}{
		\input{plots/ReOm00.tex}
		\input{plots/ImOm00.tex}
		}
	\caption{$\pi\pi$ $S$-wave Omn\`es function, isospin $I=0$}
	\label{plot:OmnesFunction00}
\end{figure}

\begin{figure}[H]
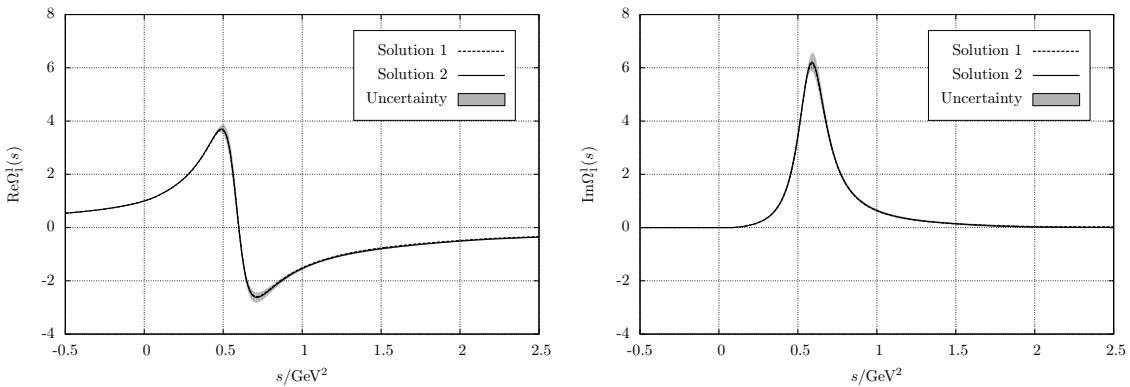

	\centering
	\scalebox{0.59}{
		\input{plots/ReOm11.tex}
		\input{plots/ImOm11.tex}
		}
	\caption{$\pi\pi$ $P$-wave Omn\`es function, isospin $I=1$}
	\label{plot:OmnesFunction11}
\end{figure}

In a first step, the six Omn\`es functions are computed, defined by
\begin{align}
	\begin{split}
		\Omega_l^I(s) := \exp\left( \frac{s}{\pi} \int_{s_0}^\infty
                \frac{\delta_l^I(s^\prime)}{(s^\prime - s - i\epsilon)
                  s^\prime} ds^\prime \right) , 
	\end{split}
\end{align}
where $s_0$ denotes the respective threshold. We show only the results for
the $\pi\pi$ Omn\`es functions, see figures~\ref{plot:OmnesFunction00} and
\ref{plot:OmnesFunction11}. 
	In the case of $\Omega_0^0$, the Omn\`es function is computed for
        the phase Solution~3 -- the corresponding uncertainty is obtained
        by summing in quadrature the variations generated by the
        uncertainties of all 28 parameters. The differences, appropriately
        weighted, are summed up in quadrature to give the error band. For
        the phase Solutions~1, 2 and 4, only the central curve is shown. 
Note that the differences between the various high-energy phase solutions
are much larger than the error band due to the phase parameters. However,
at low energy these differences are well described by polynomials and can
be absorbed at low energies by the subtraction constants of the dispersion
relation. This implies that the uncertainty generated by the unknown
high-energy behaviour of the phase shifts will be moderate.

%
%
%


\subsection{Hat Functions and Angular Projection}

During the iterative solution of the dispersion relation, the hat functions
have to be computed by means of angular averages. Since the hat functions
appear in the integrand of the dispersive integrals, they have to be known
just on the real axis above the threshold of the respective channel. 

In the $s_\ell = 0$ case, the calculation of the angular integrals is
straightforward. The functions $M_0$, $\ldots$ need to be computed on the
real axis, also for negative values of their argument. As described in
section~\ref{sec:ProjectionAnalyticStructurePartialWaves}, a subtlety
arises in the case $s_\ell \neq 0$: in the calculation of the $s$-channel
hat functions, we have to know the angular integrals of the $t$- and
$u$-channel functions $N_0$, $\ldots$. In the region $(M_K -
\sqrt{s_\ell})^2 < s < (M_K + \sqrt{s_\ell})^2$, the angular integration
path extends into the complex $t$- or $u$-plane. Therefore, the $t$- and
$u$-channel functions $N_0$, $\ldots$ have to be computed not only on the
real axis but also in the complex plane. Since the region where this
happens is much below the $t$- or $u$-channel cut, we have two options how
to perform this: 
\begin{itemize}
	\item integrate on a straight line in the complex $t$- or
          $u$-plane. The functions $N_0(t)$, $\ldots$ have to be known in
          an egg-shaped region of $s_\ell$-dependent size. The egg lies
          within $M_\pi^2 - M_K \sqrt{s_\ell} < \Re(t) < M_\pi^2 + M_K
          \sqrt{s_\ell}$. The functions $N_0(t)$, $\ldots$ can be computed
          on a two-dimensional grid covering this egg and then
          e.g.~interpolated with a 2D spline. 
	\item since the functions $N_0(t)$, $\ldots$ are analytic in the
          region of the egg, the angular integration path can be deformed
          to lie always on the border of the egg. Therefore, the functions
          $N_0(t)$,~$\ldots$ only have to be computed on points lying on
          this border (in addition to the real axis) and 1D interpolation
          methods can be applied. 
\end{itemize}
The first method is more straightforward, the second needs less computing
time. The second one requires a change of variable that we briefly
describe.

\begin{figure}[t]
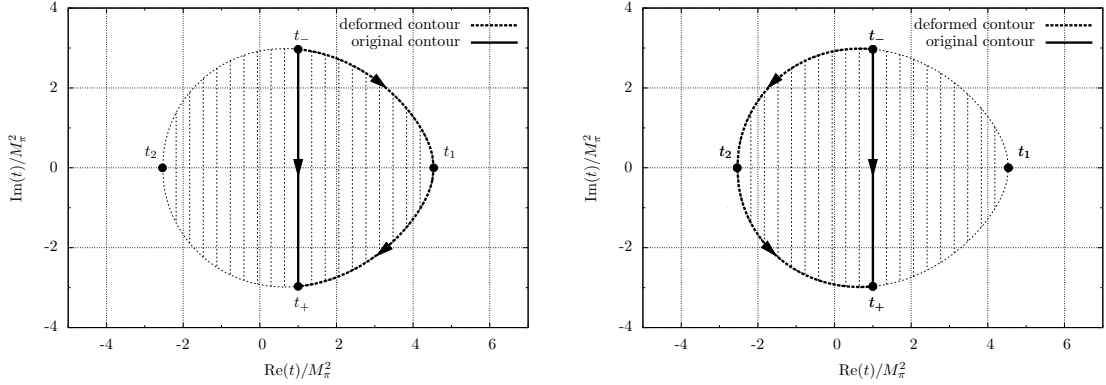

	\centering
	\scalebox{0.59}{
		\input{plots/egg.tex}
		\input{plots/egg2.tex}
		}
	\caption{Angular integration contours for $s_\ell = M_\pi^2$}
	\label{fig:Egg}
\end{figure}

We want to compute the angular integral
\begin{align}
	\begin{split}
		\< z^n X \>_{t_s}(s) = \frac{1}{2} \int_{-1}^1 dz \, z^n X(t(s,z)) ,
	\end{split}
\end{align}
where e.g.~$X = N_0$ and
\begin{align}
	\begin{split}
		t(s,z) = \frac{1}{2} \left( \Sigma_0 - s - \lambda^{1/2}_{K\ell}(s) \sigma_\pi(s) z \right) .
	\end{split}
\end{align}
The square root of the K\"all\'en function is defined by
\begin{align}
	\lambda^{1/2}_{K\ell}(s) = \left\{
	\begin{array}{r c}
		+ | \lambda^{1/2}_{K\ell}(s) | & s < (M_K-\sqrt{s_\ell})^2 , \\
		+ i | \lambda^{1/2}_{K\ell}(s) | & (M_K-\sqrt{s_\ell})^2 < s < (M_K+\sqrt{s_\ell})^2 , \\
		- | \lambda^{1/2}_{K\ell}(s) | & (M_K+\sqrt{s_\ell})^2 < s 
	\end{array} \right. 
\end{align}
and the critical region is $s_- < s < s_+$, where we define
\begin{align}
	\begin{split}
		s_\pm := (M_K \pm \sqrt{s_\ell})^2 .
	\end{split}
\end{align}
In this region, the angular integration path in the complex $t$-plane runs
from $t_- := t(s,-1)$ to $t_+ := t(s,1)$. Due to the analyticity of the
function $X(t)$, the straight contour can be deformed along the border of
the egg, either to pass $t_1 := t(s_-,z)$ or $t_2 := t(s_+,z)$, see the two
plots in figure~\ref{fig:Egg}. 
Defining
\begin{align}
	\begin{split}
		z_s(t) = \frac{1}{\lambda^{1/2}_{K\ell}(s) \sigma_\pi(s)} \left( \Sigma_0 - s - 2 t \right) ,
	\end{split}
\end{align}
we rewrite the angular integral as a complex integral:
\begin{align}
	\begin{split}
		\< z^n X \>_{t_s} &= \frac{1}{2} \int_{t_-}^{t_+} \frac{dz_s}{dt} dt \, z_s^n(t) X(t) \\
			&= - \frac{1}{\lambda_{K\ell}^{1/2}(s) \sigma_\pi(s)} \int_{t_-}^{t_+} dt \, z_s^n(t) X(t) \\
			&=  - \frac{1}{\lambda_{K\ell}^{1/2}(s) \sigma_\pi(s)} \left(  \int_{t_-}^{t_1} dt \, z_s^n(t) X(t) -  \int_{t_+}^{t_1} dt \, z_s^n(t) X(t) \right) ,
	\end{split}
\end{align}
or equivalently
\begin{align}
	\begin{split}
		\< z^n X \>_{t_s} &=  - \frac{1}{\lambda_{K\ell}^{1/2}(s) \sigma_\pi(s)} \left(  \int_{t_-}^{t_2} dt \, z_s^n(t) X(t) -  \int_{t_+}^{t_2} dt \, z_s^n(t) X(t) \right) .
	\end{split}
\end{align}
We parametrise the border of the egg by the following curves:
\begin{align}
	\begin{split}
		t_s^\pm(\xi) &:= t(\xi,\pm1) = \frac{1}{2} \left( \Sigma_0 - \xi \mp \lambda_{K\ell}^{1/2}(\xi)\sigma_\pi(\xi) \right) , \quad \xi \in [ s_-, s_+] ,
	\end{split}
\end{align}
hence
\begin{align}
	\begin{split}
		\< z^n X \>_{t_s} &= - \frac{1}{\lambda_{K\ell}^{1/2}(s) \sigma_\pi(s)} \begin{aligned}[t]
			& \Bigg( \int_s^{s_-} d\xi \frac{d t_s^-(\xi)}{d\xi} z_s^n(t_s^-(\xi)) X(t_s^-(\xi)) \\
			& - \int_s^{s_-} d\xi \frac{dt_s^+(\xi)}{d\xi} z_s^n(t_s^+(\xi)) X(t_s^+(\xi)) \Bigg) , \end{aligned}
	\end{split}
\end{align}
or
\begin{align}
	\begin{split}
		\< z^n X \>_{t_s} &= - \frac{1}{\lambda_{K\ell}^{1/2}(s) \sigma_\pi(s)} \begin{aligned}[t]
			& \Bigg( \int_s^{s_+} d\xi \frac{d t_s^-(\xi)}{d\xi} z_s^n(t_s^-(\xi)) X(t_s^-(\xi)) \\
			&  - \int_s^{s_+} d\xi \frac{dt_s^+(\xi)}{d\xi} z_s^n(t_s^+(\xi)) X(t_s^+(\xi)) \Bigg) , \end{aligned}
	\end{split}
\end{align}
where
\begin{align}
	\begin{split}
		\frac{d t_s^\pm(\xi)}{d\xi} &= \frac{1}{2} \left( -1 \mp \frac{d(\lambda_{K\ell}^{1/2}(\xi) \sigma_\pi(\xi))}{d\xi} \right) \\
			&= \frac{1}{2} \left(-1 \mp \frac{2 M_K^4 M_\pi^2 - M_K^2 \left(4 M_\pi^2 s_\ell + \xi^2\right)+(s_\ell-\xi) \left(2 M_\pi^2 (\xi+s_\ell)-\xi^2\right)}{\xi^2 \lambda_{K\ell}^{1/2}(\xi) \sigma_\pi(\xi)} \right) , \\
		z_s(t_s^\pm(\xi)) &= \frac{1}{\lambda_{K\ell}^{1/2}(s) \sigma_\pi(s)} \left( \xi - s \pm \lambda_{K\ell}^{1/2}(\xi) \sigma_\pi(\xi) \right) .
	\end{split}
\end{align}
Note that
\begin{align}
	\begin{split}
		z_s(t_s^+(\xi)) &= - z_s(t_s^-(\xi))^* , \\
		t_s^+(\xi) &= t_s^-(\xi)^* , \\
		\frac{d t_s^+(\xi)}{d\xi} &= \left( \frac{dt_s^-(\xi)}{d\xi} \right)^*
	\end{split}
\end{align}
and hence, due to the Schwarz reflection principle
\begin{align}
	\begin{split}
		X(t_s^+(\xi)) = X(t_s^-(\xi))^* .
	\end{split}
\end{align}
Therefore, the function $X$ has to be computed only on the `upper half-egg':
\begin{align}
	\begin{split}
		\< z^n X \>_{t_s} &= \frac{1}{\lambda_{K\ell}^{1/2}(s) \sigma_\pi(s)} \int_{s_-}^s d\xi \begin{aligned}[t]
			& \Bigg( \frac{dt_s^-(\xi)}{d\xi} z_s^n(t_s^-(\xi)) X(t_s^-(\xi)) \\
			& - (-1)^n \left( \frac{dt_s^-(\xi)}{d\xi} z_s^n(t_s^-(\xi)) X(t_s^-(\xi)) \right)^* \Bigg) \end{aligned}
	\end{split}
\end{align}
or
\begin{align}
	\begin{split}
		\< z^n X \>_{t_s} &= -\frac{1}{\lambda_{K\ell}^{1/2}(s) \sigma_\pi(s)} \int_s^{s_+} d\xi \begin{aligned}[t]
			& \Bigg( \frac{dt_s^-(\xi)}{d\xi} z_s^n(t_s^-(\xi)) X(t_s^-(\xi)) \\
			& - (-1)^n \left( \frac{dt_s^-(\xi)}{d\xi} z_s^n(t_s^-(\xi)) X(t_s^-(\xi)) \right)^* \Bigg) . \end{aligned}
	\end{split}
\end{align}
Although both descriptions are valid in the range $s_- < s < s_+$, one may
choose to use the first in the region $s_- < s < s_m$ and the second in the
region $s_m < s < s_+$, where $s_m$ lies somewhere in the middle of $s_-$
and $s_+$, e.g.~$s_m = (s_-+s_+)/2$. The motivation to do so is to avoid
numerical instabilities: the integral from $s_-$ to $s$ with $s \to s_+$
must tend to zero to give a finite value for the hat function. The integral
over the whole half-egg, however, accumulates a numerical uncertainty. 

\subsection{Results for the Basis Solutions}

\begin{figure}[H]
	\centering
	\scalebox{0.6}{
		\input{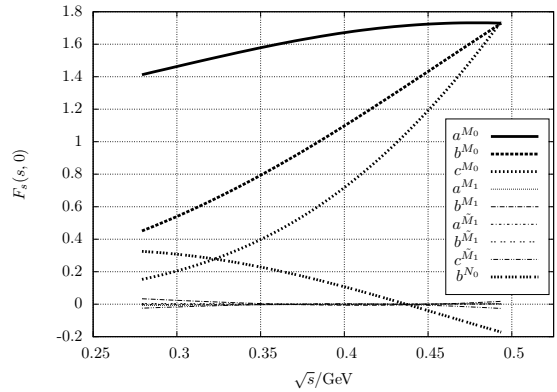}
		}
	\caption{$S$-wave of the form factor $F$ for the different basis solutions for $s_\ell = 0$}
	\label{fig:BasisSolutionsSwave}
\end{figure}

\begin{figure}[H]
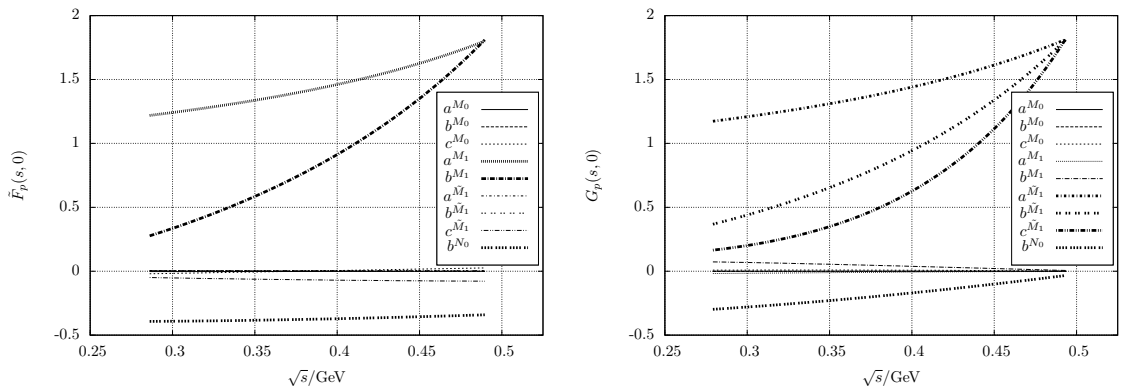

	\centering
	\scalebox{0.59}{
		\input{plots/Fp.tex}
		\input{plots/Gp.tex}
		}
	\caption{$P$-waves of the form factors $F$ and $G$ for the different basis solutions for $s_\ell = 0$}
	\label{fig:BasisSolutionsPwaves}
\end{figure}

We have now all the ingredients to compute the nine basis solutions of the dispersion relation. The final result will be a linear combination thereof. In section~\ref{sec:Kl4SubtractionConstantsDetermination}, we will describe how to determine this linear combination. We will fit experimental data on the partial waves defined by
\begin{align}
	\begin{split}
		\label{eqn:PartialWavesFuncOneVariableRelations1}
		F_s(s,s_\ell) &= \left( M_0(s,s_\ell) + \hat M_0(s,s_\ell) \right) e^{-i \delta_0^0(s)} , \\
		\tilde F_p(s,s_\ell) &= \left( M_1(s,s_\ell) + \hat M_1(s,s_\ell) \right) e^{-i \delta_1^1(s)} , \\
		G_p(s,s_\ell) &= \left( \tilde M_1(s,s_\ell) + \hat{\tilde M}_1(s,s_\ell) \right) e^{-i \delta_1^1(s)} .
	\end{split}
\end{align}
The figures~\ref{fig:BasisSolutionsSwave} and \ref{fig:BasisSolutionsPwaves} show the partial waves of the basis solutions in the case $s_\ell = 0$. They are computed with the phase solutions that reach the asymptotic values of $\pi$ in the case of the $\pi\pi$ phases and 0 in the case of the $K\pi$ phases. For $\delta_0^0$, the solution with the drop around the $K\bar K$ threshold is used. The figures illustrate what can be learnt also from the definitions (\ref{eq:FunctionsOfOneVariableOmnes}) and (\ref{eqn:PartialWavesFuncOneVariableRelations1}): the data on the partial wave $F_s$ will constrain mainly the subtraction constants appearing in $M_0$, the data on $F_p$ mainly the constants in $M_1$ and the data on $G_p$ mainly the constants in $\tilde M_1$. An exception is the constant $b^{N_0}$: through the hat functions, it is constrained by the data on all partial waves.

Besides experimental data on the partial waves, we will also use two soft-pion theorems as an additional source of information. Table~\ref{tab:BasisSolutionsSPT} shows the values of $(F-G)(M_\pi^2, M_K^2, M_\pi^2)$ and $(F+G)(M_\pi^2, M_\pi^2, M_K^2)$ for the basis solutions. The first soft-pion theorem, which implies $(F-G)(M_\pi^2, M_K^2, M_\pi^2) \approx 0$, constrains mainly a linear combination of $a^{M_0}$, $a^{M_1}$, $a^{\tilde M_1}$ and $b^{N_0}$.

\begin{table}[H]
	\centering
	\begin{tabular}{c c c}
		\toprule
		basis solution & $(F-G)_\mathrm{SPP1}$ & $(F+G)_\mathrm{SPP2}$ \\[0.1cm]
		\hline \\[-0.3cm]
		$a^{M_0}$ &		$\m1.06$	& $\m1.05$ \\
		$b^{M_0}$ &		$\m0.08$	& $\m0.09$ \\
		$c^{M_0}$ &		$\m0.03$	& $-0.01$ \\
		$a^{M_1}$ &		$-1.03$	& $\m0.93$ \\
		$b^{M_1}$ &		$\m0.05$	& $\m0.11$ \\
		$a^{\tilde M_1}$ &	$-1.07$	& $\m1.02$ \\
		$b^{\tilde M_1}$ &	$-0.05$	& $\m0.09$ \\
		$c^{\tilde M_1}$ &	$-0.10$	& $-0.01$ \\
		$b^{N_0}$ &		$\m1.62$	& $-0.01$ \\
		\bottomrule
	\end{tabular}
	\caption{Values of the two relevant combinations of the form factors $F$ and $G$ at the soft-pion points, computed for the basis solutions.}
	\label{tab:BasisSolutionsSPT}
\end{table}


\section{Determination of the Subtraction Constants}

\label{sec:Kl4SubtractionConstantsDetermination}

In the previous chapter, we have described how to solve numerically the Omn\`es dispersion relation for the form factors $F$ and $G$. The solution is parametrised in terms of the subtraction constants $a^{M_0}$,~$\ldots$. The next task is now to determine these subtraction constants in order to fix the parametric freedom. We use three different sources of information for the determination of the subtraction constants:
\begin{itemize}
	\item experimental data on the $K_{\ell4}$ form factors,
	\item the soft-pion theorem, providing relations between $F$, $G$ and the $K_{\ell3}$ vector form factor,
	\item input from \ChPT{}.
\end{itemize}

The soft-pion theorem (SPT) is valid up to corrections of $\O(M_\pi^2)$ and hence can be considered as a strong constraint. From the two high-statistics experiments NA48/2 and E865 we have data on the $S$- and $P$-waves of the form factors. Although these experiments have achieved impressive results, the data alone does not determine all the subtraction constants with satisfactory precision. Therefore, we use chiral input to fix some of the subtraction constants that are not well determined by the fit to data.

In the following, we describe in more detail what data we use for the fits and how these fits are performed. We were provided with additional unpublished data from the E865 experiment and include the data sets of NA48/2 that became available only recently as an addendum to the original publication \cite{Batley2012}. Therefore, our fits include the maximal amount of experimental information on the $K_{\ell4}$ form factors $F$ and $G$ that is currently available.

\subsection{Experimental Data}

\label{sec:ExperimentalKl4Data}

The NA48/2 experiment defines the partial wave expansion of the form factors as
\begin{align}
	\begin{split}
		F &= F_s e^{i\delta_s} + F_p e^{i\delta_p} \cos\theta + \ldots , \\
		G &= G_p e^{i\delta_p} + \ldots
	\end{split}
\end{align}
and further defines the linear combination
\begin{align}
	\begin{split}
		\label{eq:Gptilde}
		\tilde G_p := G_p + \frac{X}{\sigma_\pi PL} F_p .
	\end{split}
\end{align}
For us, it is convenient to define the partial wave
\begin{align}
	\begin{split}
		\label{eq:FptildeDefinition}
		\tilde F_p := \frac{M_K^2}{2X\sigma_\pi} F_p .
	\end{split}
\end{align}
In our former treatment of the form factor $F_1$ \cite{Stoffer2010,
  Colangelo2012, Stoffer2013}, it was most convenient to use the data on
$F_s$ and $\tilde G_p$ (which corresponds to the $P$-wave of $F_1$). Now
that we describe both form factors $F$ and $G$, we prefer to fit
the three partial waves $F_s$, $\tilde F_p$ and $G_p$.

The comparison with our definition of the $s$-channel partial-wave expansions
\begin{align}
	\begin{split}
		F &= \sum_{l=0}^\infty P_l(\cos\theta) \left( \frac{2 X \sigma_\pi}{M_K^2}\right)^l f_l - \frac{\sigma_\pi PL}{X} \cos\theta \; G , \\
		G &= \sum_{l=1}^\infty P_l^\prime(\cos\theta) \left( \frac{2 X \sigma_\pi}{M_K^2}\right)^{l-1} g_l ,
	\end{split}
\end{align}
allows us to identify
\begin{align}
	\begin{split}
		\label{eq:Kl4PartialWavesIdentification}
		F_s e^{i \delta_s} &= f_0 , \\
		F_p e^{i \delta_p} &= \frac{2X\sigma_\pi}{M_K^2} f_1 - \frac{\sigma_\pi PL}{X} g_1 , \\
		G_p e^{i \delta_p} &= g_1 .
	\end{split}
\end{align}
The phase shifts are just given by the $\pi\pi$ phases that we use as input. With (\ref{eq:HatFunctionPartialWavesRelation}), we find the fitting equations:
\begin{align}
	\begin{split}
		\label{eqn:PartialWavesFuncOneVariableRelations}
		F_s(s,s_\ell) &= \left( M_0(s,s_\ell) + \hat M_0(s,s_\ell) \right) e^{-i \delta_0^0(s)} , \\
		\tilde F_p(s,s_\ell) &= \left( M_1(s,s_\ell) + \hat M_1(s,s_\ell) \right) e^{-i \delta_1^1(s)} , \\
		G_p(s,s_\ell) &= \left( \tilde M_1(s,s_\ell) + \hat{\tilde M}_1(s,s_\ell) \right) e^{-i \delta_1^1(s)} .
	\end{split}
\end{align}

The NA48/2 collaboration has performed phenomenological fits of the form \cite{Batley2010,Batley2012}
\begin{align}
	\begin{split}
		\label{eq:NA48PhenomenologicalFit}
		\frac{F_s(s,s_\ell)}{f_s} &= 1 + \frac{f_s^\prime}{f_s} q^2 + \frac{f_s^\dprime}{f_s} q^4 + \frac{f_e^\prime}{f_s} \frac{s_\ell}{4M_\pi^2} , \\
		\frac{F_p(s,s_\ell)}{f_s} &= \frac{f_p}{f_s} , \\
		\frac{G_p(s,s_\ell)}{f_s} &= \frac{g_p}{f_s} + \frac{g_p^\prime}{f_s} q^2 ,
	\end{split}
\end{align}
where $q^2 = \frac{s}{4M_\pi^2}-1$. In a first step, only the normalised
coefficients were measured \cite{Batley2010}. In a second step, the
normalisation $f_s$ was determined from the branching ratio measurement and
a phase-space integration, using the parametrisation
(\ref{eq:NA48PhenomenologicalFit}) and the fitted normalised coefficients \cite{Batley2012}.

However, one should note that from (\ref{eq:Kl4PartialWavesIdentification})
it follows that $F_p$ has to vanish at the $\pi\pi$ threshold like $\sim
\sqrt{q^2}$. The phenomenological fit (\ref{eq:NA48PhenomenologicalFit}) of
\cite{Batley2010,Batley2012}, which assumes $F_p$ to be constant in $q^2$,
gives a wrong threshold behaviour. We have not tried to estimate its
influence on the determination of the normalisation $f_s$. For our purpose,
we find it convenient to work with $\tilde F_p$, which does not contain
kinematic prefactors. 

Because all the basis solutions use the same $\pi\pi$ phase as input, the real quantities $F_s$, $\tilde F_p$ and $G_p$ are still linear combinations of the corresponding quantities computed with the basis solutions. Note that the partial waves can be negative, i.e.~one really has to rotate the $\pi\pi$ phase away and not just take the absolute value.

For our fits, we use the experimental values of NA48/2 \cite{Batley2010, Batley2012} and E865 \cite{Pislak2001, Pislak2003} on the partial waves. Some remarks on these numbers are appropriate.
\begin{itemize}
	\item Originally, the published NA48/2 data consisted of 10 bins in $s$-direction. Very recently, a two-dimensional data set on $F_s(s,s_\ell)$ has become available (addendum to \cite{Batley2012}): in this set, not only a single bin but up to 10 bins are used in $s_\ell$-direction.
	\item The barycentre values of $s_\ell$ for the original 10 bins of NA48/2 also became available in the addendum to \cite{Batley2012}. A value of $s_\ell$ could also be extracted from the relation (\ref{eq:Gptilde}) between $F_p$, $G_p$ and $\tilde G_p$ \cite{Stoffer2014}. However, this value does not agree with the barycentre.
	\item We compute the value of $\tilde F_p$ with (\ref{eq:FptildeDefinition}) using the values of $F_p$ and the barycentre values of $s$ and $s_\ell$.
	\item There is a discrepancy between \cite{Batley2010} and
          \cite{Batley2012}. The statistical and systematic uncertainties
          for $F_s$ in the NA48/2 data have to be calculated from the
          normalised coefficients in \cite{Batley2010}. The correct
          uncertainties are also listed in the addendum to
          \cite{Batley2012}. 
	\item The published values of $F_s$ in the 10 bins of NA48/2 have been normalised in such a way that a fit of the form (\ref{eq:NA48PhenomenologicalFit}) with $f_e^\prime = 0$ results in $F_s(0,0)/f_s = 1$, although a non-zero value of $f_e^\prime$ has been obtained from a fit to the two-dimensional data set. In order to take the $s_\ell$-dependence consistently into account, we have to increase the values of $F_s$ by 0.77\%.
	\item The E865 experiment has assumed in the analysis that the form factors do not depend on $s_\ell$. The values of $s_\ell$ for each bin were not published.\footnote{We thank Peter Truöl and Andries van der Schaaf, who performed a new analysis of the Brookhaven data in order to extract the barycentre values of $s_\ell$.}
	\item The E865 experiment only provides data on the first partial waves $F_s$ and $G_p$.
	\item The E865 papers \cite{Pislak2001, Pislak2003} include the fully correlated error of the normalisation of $1.2\%$ in their systematic errors (added in quadrature).\footnote{We thank Stefan Pislak and Peter Truöl for this additional unpublished information.} It needs a special treatment for unbiased fitting.
\end{itemize}

In the data analysis of both experiments, radiative corrections have been
applied to some extent. More reliable radiative corrections based on a
fixed-order calculation \cite{Stoffer2014} can be applied a posteriori at
least to the NA48/2 data. Furthermore, neither the E865 nor the NA48/2
experiment has corrected the isospin-breaking effects due to the quark and
meson mass differences. The calculation of \cite{Stoffer2014} also allows
for their correction. The resulting numbers are given in
appendix~\ref{sec:AppendixData}. We add the uncertainties of the isospin
corrections (without the higher order estimate) in quadrature to the
systematic errors. The one-dimensional NA48/2 values also include the
mentioned correction of the normalisation of $F_s$ by $0.77\%$ due to the
$s_\ell$-dependence.

In addition to the statistical and systematic errors, we take into account
the correlations between $F_p$ and $G_p$ of the NA48/2 data, which also
became available with the addendum to \cite{Batley2012}. There are however
several correlations that we neglect, either because they are not available
or because we assume them to play a minor role. These include the
bin-to-bin correlations of the $P$-waves and the correlations with the
$S$-wave. We also neglect the correlation due to the isospin-breaking
corrections and correlations between the two experiments due to external
input. We do not expect any of these neglected correlations to
significantly affect our fits, but of course it would be better to
check. If the complete set of experimental correlations will become
available, it will be possible to do that.


\subsection{Soft-Pion Theorem}

In addition to the experimental input on the partial waves, we use the
soft-pion theorem (SPT) \cite{Treiman1972, DeAlfaro1973} as a second source
of information to determine the subtraction constants. 

There are two different soft-pion theorems for $K_{\ell4}$, depending on
which pion is taken to be soft. If the momentum $p_1$ of the positively
charged pion is sent to zero, the Mandelstam variables become $s =
M_\pi^2$, $t = M_K^2$, $u = s_\ell$. Since the SPT is valid only at
$\mathcal{O}(M_\pi^2)$, we set $u = s_\ell + M_\pi^2$, such that the
relation $s+t+u = M_K^2 + 2 M_\pi^2 + s_\ell$ remains valid and one does
not need to worry about defining an off-shell form factor. 

The first SPT states \cite{Stoffer2010}:
\begin{align}
	\label{eqn:SPT1}
	F(M_\pi^2, M_K^2, M_\pi^2+s_\ell) - G(M_\pi^2, M_K^2, M_\pi^2+s_\ell) = \mathcal{O}(M_\pi^2) .
\end{align}

If the momentum $p_2$ of the negatively charged pion is sent to zero, the
Mandelstam variables become $s = M_\pi^2$, $t = s_\ell$, $u = M_K^2$. We
set $t = s_\ell + M_\pi^2$. 

The second SPT gives a relation to the $K_{\ell3}$ vector form factor:
\begin{align}
	\label{eqn:SPT2}
	F(M_\pi^2, M_\pi^2+s_\ell, M_K^2) + G(M_\pi^2, M_\pi^2+s_\ell, M_K^2) - \frac{\sqrt{2} M_K}{F_\pi} f_+(M_\pi^2+s_\ell) = \mathcal{O}(M_\pi^2) .
\end{align}

At leading order in \ChPT{}, the SPTs are fulfilled exactly, i.e.~the
right-hand sides of the equations (\ref{eqn:SPT1}) and (\ref{eqn:SPT2})
vanish, at NLO and NNLO, there appear $\O(M_\pi^2)$ corrections. 

Numerically, it turns out that the first SPT is fulfilled to a higher
precision than the second SPT. At NLO, the correction to the first SPT is
about $0.4\%$ for $s_\ell=0$, the second SPT gets a correction of $2.0\%$
if $f_+(M_\pi^2)$ is used. If we make the arbitrary replacement
$f_+(M_\pi^2) \mapsto f_+(0)$, again an $\O(M_\pi^2)$ effect, the deviation
in the second SPT increases to $4.9\%$. This confirms that the size of the
observed deviations from the SPT is natural.

At NNLO, the corrections become slightly larger.\footnote{We thank Johan
  Bijnens and Ilaria Jemos for providing the C++ implementation of the NNLO
  expressions.} If the $\O(p^6)$ LECs $C_i^r$ are all put to zero and
$s_\ell = 0$ as well, the first SPT is fulfilled at $1.0\%$, the second at
$4.4\%$ with $f_+(M_\pi^2)$ or $7.6\%$ with $f_+(0)$. If the $C_i^r$ parts
are replaced by the estimates of \cite{Amoros2000,Bijnens2003} (resonances
estimates in the case of $K_{\ell4}$), the accuracy of the first SPT is
$1.5\%$, the one of the second SPT $5.4\%$ using $f_+(M_\pi^2)$ or again
$7.6\%$ using $f_+(0)$. 

We use the size of the NNLO corrections to the SPT as an estimate of the
tolerance that we allow in the fits when using the SPTs as constraints.

\subsection{Fitting Method}

In the following, we describe how we perform the fit. Basically, we have to
deal with a simple linear fit. The only subtlety is the fact that the data
contains a fully correlated uncertainty of the normalisation, which is a
multiplicative quantity. The fact that we use two experiments with
different normalisation errors asks for a special fitting method to avoid a
bias \cite{DAgostini1994, Ball2010}. We apply the `$t_0$-method' of
\cite{Ball2010}. 

First, we construct a covariance matrix for the observations as follows.
\begin{itemize}
	\item For all the partial-wave data that we want to fit we construct the covariance matrix with the squared statistical errors on the diagonal and the statistical covariance between the $P$-waves as off-diagonal elements.
	\item We add the uncorrelated systematic errors, which do not contain the error of the normalisation, in quadrature to the diagonal entries.
	\item We may or may not include the two soft-pion theorems as additional observations. If we do so, we take e.g.~$F-G$ at the first soft-pion point (SPP) and $F+G$ at the second SPP as observations. As uncertainties, we take a value typical for the deviation in \ChPT{} at NNLO, e.g.~$1\%$ or $2\%$ of the LO value of $F$ for the first SPT and a few percent of $\sqrt{2} M_K / F_\pi f_+(0)$ for the second SPT.
	\item We add the errors of the normalisation to the covariance matrix, which are in block-diagonal form for the data of the two experiments:
	\begin{align}
		\begin{split}
			(\mathrm{cov})_{ij} = (\mathrm{rel.cov.})_{ij} + (\mathrm{norm.cov.})_{ij}, \quad (\mathrm{norm.cov.})_{ij} = \Delta_I^2 \, f(s^i,s_\ell^i) f(s^j,s_\ell^j) \delta_{I_i, I_j} ,
		\end{split}
	\end{align}
	where $\Delta_I$ denotes the error of the normalisation for experiment $I$. $I_i$ is the index of the experiment (1 or 2) corresponding to the data point $i$ and $f(s^i,s_\ell^i)$ is the value of the \textit{fitted} partial wave. In a first step, this value has to be computed under the assumption of some starting values for the fit parameters.
\end{itemize}

The fit requires then an iteration. One has to minimise the error function defined by
\begin{align}
	\begin{split}
		\chi^2 = v^T P v ,
	\end{split}
\end{align}
where $v$ is the vector of the residues, i.e.~the differences between the observations and computed values. $P$ is the inverse of the covariance matrix constructed above: $P = (\mathrm{cov})^{-1}$. The minimum of the $\chi^2$ function can be either found with some minimisation routine or, since the fit is linear, directly with the explicit solution
\begin{align}
	\begin{split}
		\mathrm{par} = ( A^T P A )^{-1} A^T P O ,
	\end{split}
\end{align}
where $O$ is the vector of observations and
\begin{align}
	\begin{split}
		A_{ij} = \frac{\p f(s^i,s_\ell^i)}{\p \mathrm{param}_j}
	\end{split}
\end{align}
is the design matrix to be determined with the values of the basis solutions.

With these new values for the fit parameters, one again computes the new covariance matrix (the contribution for the normalisation changes) and iterates this procedure. It turns out that only very few iterations are needed to reach convergence.

If we do not want to determine a parameter through the fit but fix it beforehand to a non-zero value, we have to subtract the fixed contribution from the observations $O$, such that $O$ is purely linear in the parameters and contains no constant contributions.

In the above discussion, we have not specified what we use as fit parameters. One option is to fit the subtraction constants. Since we want to include an $s_\ell$-dependence in the subtraction constants, we write e.g.
\begin{align}
	\begin{split}
		\label{eqn:SlDependenceOmnesSubtractionConstants}
		a^{M_0}(s_\ell) = a^{M_0}_0 + a^{M_0}_1 \frac{s_\ell}{M_K^2} + \ldots ,
	\end{split}
\end{align}
where $a^{M_0}_0$, $a^{M_0}_1$, $\ldots$ are now the parameters collected in the above vector `par'. Another option is to use the matching equations to \ChPT{}, which provide a linear relation between the subtraction constants and the LECs we are interested in, and perform the fit directly with the LECs.


\subsection{Matching to \ChPT{}}

The final goal of this treatment is the determination of low-energy
constants of \ChPT{}. Instead of fitting directly the $K_{\ell4}$ data with
the chiral expressions, we use the dispersive representation as an
intermediate step. The dispersion relation provides a model-independent
resummation of final-state rescattering effects. Therefore, we expect that
even the most important effects beyond $\O(p^6)$ are included in the
dispersion relation. Of course, in order to extract values for the LECs,
one has to perform a matching of the dispersive and the chiral
representations. This can be done e.g.~on the level of the form factors
\cite{Stoffer2010, Colangelo2012, Stoffer2013}. Since the dispersion
relation describes the energy dependence, the matching point can be outside
the physical region, i.e.~even at lower energies, where \ChPT{} is expected
to converge better.

Here, we follow an improved strategy for the matching: we match the
dispersive and the chiral representations not on the level of form factors
but directly on the level of subtraction constants. Since the decomposition
(\ref{eq:FormFactorDecomposition}) is valid up to terms of $\O(p^8)$, the
one-loop and even the two-loop result can be written in this form, which
allows us to extract a chiral representation of the subtraction
constants. This procedure has the advantage that the matching is performed
for each function of one variable $M_0(s)$,~$\ldots$ at its subtraction
point, i.e.~at $s=0$, $t=0$ and $u=0$, where indeed the chiral
representation is expected to be reliable.

\subsubsection{Matching Equations at $\O(p^4)$}

\paragraph{Reconstruction of the \ChPT{} Form Factors}

Let us start by reconstructing the NLO form factors in the standard dispersive form (\ref{eq:FunctionsOfOneVariable}).

The LO \ChPT{} form factors are given by
\begin{align}
	\begin{split}
		F_\mathrm{LO} = G_\mathrm{LO} = \frac{M_K}{\sqrt{2} F_\pi} .
	\end{split}
\end{align}
With the partial wave projections (\ref{eq:PartialWaveProjectionSChannelPhysicalFF}), we find
\begin{align}
	\begin{split}
		f_0^\mathrm{LO}(s) &= \frac{M_K}{\sqrt{2} F_\pi} , \\
		f_1^\mathrm{LO}(s) &= \frac{M_K}{\sqrt{2} F_\pi} \frac{M_K^2 PL}{2 X^2} , \\
		g_1^\mathrm{LO}(s) &= \frac{M_K}{\sqrt{2} F_\pi} .
	\end{split}
\end{align}
The isospin 1/2 form factors (\ref{eq:Isospin12FormFactors}) are given by
\begin{align}
	\begin{split}
		F^{(1/2)}_\mathrm{LO} = \frac{M_K}{\sqrt{2} F_\pi} , \quad G^{(1/2)}_\mathrm{LO} = \frac{\sqrt{2} M_K}{F_\pi} .
	\end{split}
\end{align}
Hence, the partial waves in the crossed channels (\ref{eq:PartialWaveProjectionTUChannel}) are
\begin{align}
	\begin{split}
		f_{0,\mathrm{LO}}^{(1/2)}(t) &= \frac{M_K}{\sqrt{2} F_\pi} \frac{3 \Delta_{K\pi} - 5 t}{4t} , \\
		f_{1,\mathrm{LO}}^{(1/2)}(t) &= \frac{M_K}{\sqrt{2} F_\pi} \frac{3 M_K^4( M_\pi^2 - s_\ell - t)}{4t \lambda_{\ell\pi}(t)} , \\
		g_{1,\mathrm{LO}}^{(1/2)}(t) &= \frac{3 M_K}{2\sqrt{2} F_\pi} , \\
		f_{0,\mathrm{LO}}^{(3/2)}(u) &= \frac{M_K}{\sqrt{2} F_\pi} , \\
		f_{1,\mathrm{LO}}^{(3/2)}(u) &= 0 , \\
		g_{1,\mathrm{LO}}^{(3/2)}(u) &= 0 .
	\end{split}
\end{align}

The $\pi\pi$-scattering amplitude can be written as \cite{GasserLeutwyler1984}
\begin{align}
	\begin{split}
		T^{(0)}(s,t,u) &= 3 A(s,t,u) + A(t,u,s) + A(u,s,t) , \\
		T^{(1)}(s,t,u) &= A(t,u,s) - A(u,s,t) ,
	\end{split}
\end{align}
where at LO
\begin{align}
	\begin{split}
		A^\mathrm{LO}(s,t,u) = \frac{s - M_\pi^2}{F_\pi^2} .
	\end{split}
\end{align}
The Mandelstam variables for $\pi\pi$ scattering satisfy
\begin{align}
	\begin{split}
		s+t+u &= 4 M_\pi^2 , \\
		t &= -2 q^2( 1 - z ) ,
	\end{split}
\end{align}
where $q^2 = \frac{s}{4} - M_\pi^2$, $z = \cos\theta$. Hence, the $\pi\pi$ partial waves are
\begin{align}
	\begin{split}
		t_{0,\mathrm{LO}}^0(s) &= \frac{1}{2} \int_{-1}^1 dz \, T^{(0)}_\mathrm{LO}(s,z) = \frac{2s - M_\pi^2}{F_\pi^2} , \\
		t_{1,\mathrm{LO}}^1(s) &= \frac{3}{2} \int_{-1}^1 dz \, z T^{(1)}_\mathrm{LO}(s,z) = \frac{s - 4 M_\pi^2}{F_\pi^2} .
	\end{split}
\end{align}

The $K\pi$-scattering amplitude is given by \cite{Bernard1991}
\begin{align}
	\begin{split}
		T^{(1/2)}(s,t,u) &= \frac{3}{2} T^{(3/2)}(u,t,s) - \frac{1}{2} T^{(3/2)}(s,t,u) ,
	\end{split}
\end{align}
and at LO
\begin{align}
	\begin{split}
		T^{(3/2)}(s,t,u) &= \frac{1}{2 F_\pi^2} ( M_K^2 + M_\pi^2 - s ) .
	\end{split}
\end{align}
Of course, the Mandelstam variables satisfy here $s+t+u = 2 M_K^2 + 2 M_\pi^2$. The partial waves are given by
\begin{align}
	\begin{split}
		t_{0,\mathrm{LO}}^{1/2}(s) &= \frac{1}{8 s F_\pi^2} \left(5 s^2 - 2 s(M_K^2+M_\pi^2) - 3\Delta_{K\pi}^2 \right) , \\
		t_{1,\mathrm{LO}}^{1/2}(s) &= \frac{1}{8 s F_\pi^2} \left(3 s^2 - 6 s(M_K^2+M_\pi^2) + 3\Delta_{K\pi}^2 \right) , \\
		t_{0,\mathrm{LO}}^{3/2}(s) &= \frac{1}{2 F_\pi^2} (M_K^2 + M_\pi^2 - s ) , \\
		t_{1,\mathrm{LO}}^{3/2}(s) &= 0 .
	\end{split}
\end{align}

Using the unitarity relation for the $K_{\ell4}$ partial waves, we can now easily construct their imaginary parts at NLO:
\begin{align}
	\begin{split}
		\Im f_l^\mathrm{NLO}(s) &= \frac{1}{2l+1} \frac{1}{32\pi} \sigma_\pi(s) t_{l,\mathrm{LO}}^{I*}(s) f_l^\mathrm{LO}(s) , \\
		\Im g_l^\mathrm{NLO}(s) &= \frac{1}{2l+1} \frac{1}{32\pi} \sigma_\pi(s) t_{l,\mathrm{LO}}^{I*}(s) g_l^\mathrm{LO}(s) , \\
		\Im f_{l,\mathrm{NLO}}^{(I)}(t) &= \frac{1}{2l+1} \frac{1}{16\pi} \frac{\lambda_{K\pi}^{1/2}(t)}{t} t_{l,\mathrm{LO}}^{I*}(t) f_{l,\mathrm{LO}}^{(I)}(t) , \\
		\Im g_{l,\mathrm{NLO}}^{(I)}(t) &= \frac{1}{2l+1} \frac{1}{16\pi} \frac{\lambda_{K\pi}^{1/2}(t)}{t} t_{l,\mathrm{LO}}^{I*}(t) g_{l,\mathrm{LO}}^{(I)}(t) ,
	\end{split}
\end{align}
hence
\begin{align}
	\begin{split}
		\Im f_0^\mathrm{NLO}(s) &= \frac{1}{32\pi} \sigma_\pi(s) \frac{M_K( 2s - M_\pi^2)}{\sqrt{2} F_\pi^3} , \\
		\Im f_1^\mathrm{NLO}(s) &= \frac{1}{3} \frac{1}{32\pi} \sigma_\pi(s) \frac{M_K( s - 4 M_\pi^2)}{\sqrt{2} F_\pi^3}  \frac{M_K^2 PL}{2 X^2} , \\
		\Im g_1^\mathrm{NLO}(s) &= \frac{1}{3} \frac{1}{32\pi} \sigma_\pi(s) \frac{M_K(s - 4 M_\pi^2)}{\sqrt{2} F_\pi^3} , \\
		\Im f_{0,\mathrm{NLO}}^{(1/2)}(t) &= \frac{1}{16\pi} \frac{\lambda_{K\pi}^{1/2}(t)}{t} \frac{M_K}{32\sqrt{2} t^2 F_\pi^3} \left(5 t^2 - 2 t(M_K^2+M_\pi^2) - 3\Delta_{K\pi}^2 \right) (3 \Delta_{K\pi} - 5 t) , \\
		\Im f_{1,\mathrm{NLO}}^{(1/2)}(t) &= \frac{1}{16\pi} \frac{\lambda_{K\pi}^{1/2}(t)}{t} \frac{M_K}{8 \sqrt{2} t F_\pi^3} \left(3 t^2 - 6 t(M_K^2+M_\pi^2) + 3\Delta_{K\pi}^2 \right) \frac{M_K^4( M_\pi^2 - s_\ell - t)}{4t \lambda_{\ell\pi}(t)} , \\
		\Im g_{1,\mathrm{NLO}}^{(1/2)}(t) &= \frac{1}{16\pi} \frac{\lambda_{K\pi}^{1/2}(t)}{t} \frac{M_K}{16 \sqrt{2} t F_\pi^3} \left(3 t^2 - 6 t(M_K^2+M_\pi^2) + 3\Delta_{K\pi}^2 \right) , \\
		\Im f_{0,\mathrm{NLO}}^{(3/2)}(u) &= \frac{1}{16\pi} \frac{\lambda_{K\pi}^{1/2}(u)}{u} \frac{M_K}{2 \sqrt{2} F_\pi^3} (M_K^2 + M_\pi^2 - u ) , \\
		\Im f_{1,\mathrm{NLO}}^{(3/2)}(u) &= 0 , \\
		\Im g_{1,\mathrm{NLO}}^{(3/2)}(u) &= 0 .
	\end{split}
\end{align}
By inserting these imaginary parts into the dispersion integrals in (\ref{eq:FunctionsOfOneVariable}), we can reconstruct the NLO form factors.
For the comparison with the explicit loop calculation, we rewrite the dispersive integrals in terms of loop functions (see appendix~\ref{sec:AppendixScalarLoopFunctions}):
\begin{align}
	\begin{split}
		M_0^\mathrm{NLO}(s) &= m_{0,\mathrm{NLO}}^0 + m_{0,\mathrm{NLO}}^1 \frac{s}{M_K^2} + \frac{M_K}{2 \sqrt{2} F_\pi^3} \Big( (2s - M_\pi^2) \left( \bar B_{\pi\pi}(s) - \bar B_{\pi\pi}(0) \right) + M_\pi^2 s \, \bar B_{\pi\pi}^\prime(0) \Big) , \\
		M_1^\mathrm{NLO}(s) &= m_{1,\mathrm{NLO}}^0 , \\
		\tilde M_1^\mathrm{NLO}(s) &= \tilde m_{1,\mathrm{NLO}}^0 + \tilde m_{1,\mathrm{NLO}}^1 \frac{s}{M_K^2} + \frac{M_K}{6 \sqrt{2} F_\pi^3} \Big( (s-4M_\pi^2) \left( \bar B_{\pi\pi}(s) - \bar B_{\pi\pi}(0) \right) + 4 M_\pi^2  s \, \bar B_{\pi\pi}^\prime(0) \Big) , \\
		N_0^\mathrm{NLO}(t) &= n_{0,\mathrm{NLO}}^1 \frac{t}{M_K^2} + \frac{M_K}{32\sqrt{2} F_\pi^3} \bigg(  \left( - 25 t  + 5 (5 M_K^2 - M_\pi^2) \right) \left( \bar B_{K\pi}(t) - \bar B_{K\pi}(0) \right) \\
			& + \frac{3 \Delta_{K\pi}}{t^2} \left( t (3 M_K^2 - 7 M_\pi^2) - 3 \Delta_{K\pi}^2 \right) \left( \bar B_{K\pi}(t) - \bar B_{K\pi}(0) - t \,  \bar B_{K\pi}^\prime(0) - \frac{t^2}{2} \,  \bar B_{K\pi}^\dprime(0) \right) \\
			& - 5 t (5 M_K^2 - M_\pi^2) \bar B_{K\pi}^\prime(0) + \frac{3}{2} t \Delta_{K\pi}^3 \bar B_{K\pi}^\tprime(0) \bigg) , \\
		N_1^\mathrm{NLO}(t) &= 0 , \\
		\tilde N_1^\mathrm{NLO}(t) &= \frac{3 M_K^3}{16 \sqrt{2} F_\pi^3} \begin{aligned}[t]
			& \bigg( \frac{1}{t^2} \left( t^2 - 2t(M_K^2+M_\pi^2) + \Delta_{K\pi}^2 \right) \left( \bar B_{K\pi}(t) - \bar B_{K\pi}(0) \right) \\
			& + \frac{1}{t} \left( 2t(M_K^2+M_\pi^2) - \Delta_{K\pi}^2 \right) \bar B_{K\pi}^\prime(0) - \frac{\Delta_{K\pi}^2}{2} \, \bar B_{K\pi}^\dprime(0) \bigg) , \end{aligned} \\
		R_0^\mathrm{NLO}(u) &= \frac{M_K}{2 \sqrt{2} F_\pi^3} \Big(  \left(M_K^2 + M_\pi^2 - u \right) \left( \bar B_{K\pi}(u) - \bar B_{K\pi}(0) \right) - \left(M_K^2 + M_\pi^2 \right) u \, \bar B_{K\pi}^\prime(0) \Big) , \\
		R_1^\mathrm{NLO}(u) &= 0 , \\
		\tilde R_1^\mathrm{NLO}(u) &= 0 .
	\end{split}
\end{align}
We can now compare this expression with the one-loop calculation
\cite{Bijnens1990,Riggenbach1991,Bijnens1994}. As in our dispersive
treatment, we only consider $\pi\pi$ intermediate states in the $s$-channel
and $K\pi$ intermediate states in the crossed channels, the $K\bar K$ and
$\eta\eta$ loops in the $s$-channel and the $K\eta$ loops in the
$t$-channel have to be expanded in a Taylor series and absorbed by the
subtraction polynomial. The comparison of the dispersive representation
with the loop calculation then allows the extraction of the $\O(p^4)$
values for the subtraction constants.

Note that the only contributions that we neglect when writing the $\O(p^4)$
loop calculation in the dispersive form are the second and higher order
Taylor coefficients of the expanded loop functions of higher intermediate
states ($K\bar K$, $\eta\eta$ and $K\eta$). The result for the $\O(p^4)$
subtraction constants can be found in
appendix~\ref{sec:AppendixNLOSubtractionConstantsStandardRep}. 

\paragraph{\ChPT{} Form Factors in the Omn\`es Representation}

The reason why we do not use the standard dispersive form
(\ref{eq:FunctionsOfOneVariable}) for the numerical solution of the
dispersion relation but rather the Omn\`es representation
(\ref{eq:FunctionsOfOneVariableOmnes}) is mainly the separation of
final-state rescattering effects: the Omn\`es function resums the most
important rescattering effects. The remaining dispersive integrals take the
interplay of the different channels into account. 

It is therefore desirable to perform the matching to \ChPT{} not on the
level of the standard dispersive form but directly with the Omn\`es
representation. This should avoid mixing the final-state resummation with
the determination of the LECs. 

However, it is not possible to write directly the \ChPT{} representation in
the Omn\`es form, because the chiral expansion of the phase shifts does not
have the correct asymptotic behaviour. At LO, the phases grow linearly,
hence the Omn\`es dispersion integral (\ref{eq:OmnesFunction}) is
logarithmically divergent. Therefore, we subtract the dispersion integral
once more: 
\begin{align}
	\begin{split}
		\Omega(s) &= \exp\left( \frac{s}{\pi} \int_{s_0}^\infty \frac{\delta(s^\prime)}{(s^\prime - s - i \epsilon) s^\prime} ds^\prime \right) \\
			&= \exp\left(  \frac{s}{\pi} \int_{s_0}^\infty \frac{\delta(s^\prime)}{{s^\prime}^2} ds^\prime +  \frac{s^2}{\pi} \int_{s_0}^\infty \frac{\delta(s^\prime)}{(s^\prime - s - i \epsilon) {s^\prime}^2} ds^\prime \right) \\
			&=: \exp\left( \omega \frac{s}{M_K^2} +  \frac{s^2}{\pi} \int_{s_0}^\infty \frac{\delta(s^\prime)}{(s^\prime - s - i \epsilon) {s^\prime}^2} ds^\prime \right) .
	\end{split}
\end{align}
$\omega$ is divergent if evaluated in \ChPT{}. Let us postpone the determination of this constant for a moment.

Let us now use the Omn\`es representation to reconstruct the NLO result for
the form factors. At LO, the functions of one variable are simply given by 
\begin{align}
	\begin{split}
		M_0^\mathrm{LO}(s) &= \tilde M_1^\mathrm{LO}(s) = \frac{M_K}{\sqrt{2} F_\pi} , \\
		M_1^\mathrm{LO}(s) &= N_0^\mathrm{LO}(t) = N_1^\mathrm{LO}(t) = \tilde N_1^\mathrm{LO}(t) = R_0^\mathrm{LO}(u) = R_1^\mathrm{LO}(u) = \tilde R_1^\mathrm{LO}(u) = 0 .
	\end{split}
\end{align}
We start by calculating the hat functions at LO:
\begin{align}
	\begin{split}
		\hat M_0^\mathrm{LO}(s) &= \hat M_1^\mathrm{LO}(s) = \hat {\tilde M}_1^\mathrm{LO}(s) = \hat N_1^\mathrm{LO}(t) = \hat R_1^\mathrm{LO}(u) = \hat {\tilde R}_1^\mathrm{LO}(u) = 0 , \\
		\hat N_0^\mathrm{LO}(t) &= \frac{M_K}{\sqrt{2} F_\pi} \frac{3\Delta_{K\pi} - 5t}{4t} , \\
		\hat {\tilde N}_1^\mathrm{LO}(t) &= \frac{M_K}{\sqrt{2} F_\pi} \frac{3 M_K^2}{2 t} , \\
		\hat R_0^\mathrm{LO}(u) &= \frac{M_K}{\sqrt{2} F_\pi} .
	\end{split}
\end{align}
Further, we need the phase shifts at LO:
\begin{align}
	\begin{split}
		\delta_{0,\mathrm{LO}}^0(s) &= \frac{1}{32 \pi F_\pi^2} ( 2s - M_\pi^2 ) \sigma_\pi(s) , \\
		\delta_{1,\mathrm{LO}}^1(s) &= \frac{1}{96 \pi F_\pi^2} ( s - 4 M_\pi^2 ) \sigma_\pi(s) , \\
		\delta_{0,\mathrm{LO}}^{1/2}(t) &= \frac{1}{128 \pi  F_\pi^2} \left(5 t^2 - 2 t(M_K^2+M_\pi^2) - 3\Delta_{K\pi}^2 \right) \frac{\lambda_{K\pi}^{1/2}(t)}{t^2} , \\
		\delta_{1,\mathrm{LO}}^{1/2}(t) &= \frac{1}{384 \pi F_\pi^2} \left(3 t^2 - 6 t(M_K^2+M_\pi^2) + 3\Delta_{K\pi}^2 \right) \frac{\lambda_{K\pi}^{1/2}(t)}{t^2} , \\
		\delta_{0,\mathrm{LO}}^{3/2}(u) &= \frac{1}{32 \pi F_\pi^2} (M_K^2 + M_\pi^2 - u ) \frac{\lambda_{K\pi}^{1/2}(u)}{u} , \\
		\delta_{1,\mathrm{LO}}^{3/2}(u) &= 0 .
	\end{split}
\end{align}
We expand the Omn\`es representation (\ref{eq:FunctionsOfOneVariableOmnes}) at NLO:
\begin{align}
	\small
	\begin{split}
		M_0^\mathrm{NLO}(s) &= \left( 1 + \omega_0^0 \frac{s}{M_K^2} +  \frac{s^2}{\pi} \int_{s_0}^\infty \frac{\delta_{0,\mathrm{LO}}^0(s^\prime)}{(s^\prime - s - i \epsilon) {s^\prime}^2} ds^\prime \right) \left( a^{M_0} + b^{M_0} \frac{s}{M_K^2} + c^{M_0} \frac{s^2}{M_K^4} \right), \\
		M_1^\mathrm{NLO}(s) &= \left( 1 + \omega_1^1 \frac{s}{M_K^2} +  \frac{s^2}{\pi} \int_{s_0}^\infty \frac{\delta_{1,\mathrm{LO}}^1(s^\prime)}{(s^\prime - s - i \epsilon) {s^\prime}^2} ds^\prime \right) \left( a^{M_1} + b^{M_1}  \frac{s}{M_K^2}  \right) , \\
		 \tilde M_1^\mathrm{NLO}(s) &= \left( 1 + \omega_1^1 \frac{s}{M_K^2} +  \frac{s^2}{\pi} \int_{s_0}^\infty \frac{\delta_{1,\mathrm{LO}}^1(s^\prime)}{(s^\prime - s - i \epsilon) {s^\prime}^2} ds^\prime \right) \left( a^{\tilde M_1} + b^{\tilde M_1}  \frac{s}{M_K^2} + c^{\tilde M_1}  \frac{s^2}{M_K^4} \right) , \\
		N_0^\mathrm{NLO}(t) &= \left( 1 + \omega_0^{1/2} \frac{t}{M_K^2} +  \frac{t^2}{\pi} \int_{t_0}^\infty \frac{\delta_{0,\mathrm{LO}}^{1/2}(t^\prime)}{(t^\prime - t - i \epsilon) {t^\prime}^2} dt^\prime \right) \left( b^{N_0} \frac{t}{M_K^2} + \frac{t^2}{\pi} \int_{t_0}^\infty \frac{\hat N_0^\mathrm{LO}(t^\prime) \delta_{0,\mathrm{LO}}^{1/2}(t^\prime)}{(t^\prime - t - i\epsilon) {t^\prime}^2} dt^\prime  \right) , \\
		N_1^\mathrm{NLO}(t) &= 0 , \\
		\tilde N_1^\mathrm{NLO}(t) &= \left( 1 + \omega_1^{1/2} \frac{t}{M_K^2} +  \frac{t^2}{\pi} \int_{t_0}^\infty \frac{\delta_{1,\mathrm{LO}}^{1/2}(t^\prime)}{(t^\prime - t - i \epsilon) {t^\prime}^2} dt^\prime \right) \Bigg( \frac{t}{\pi} \int_{t_0}^\infty \frac{\hat{\tilde N}_1^\mathrm{LO}(t^\prime) \delta_{1,\mathrm{LO}}^{1/2}(t^\prime)}{(t^\prime - t - i\epsilon) t^\prime} dt^\prime  \Bigg) , \\
		R_0^\mathrm{NLO}(u) &= \left( 1 + \omega_0^{3/2} \frac{u}{M_K^2} +  \frac{u^2}{\pi} \int_{u_0}^\infty \frac{\delta_{0,\mathrm{LO}}^{3/2}(u^\prime)}{(u^\prime - u - i \epsilon) {u^\prime}^2} du^\prime \right) \left( \frac{u^2}{\pi} \int_{u_0}^\infty \frac{\hat R_0^\mathrm{LO}(u^\prime) \delta_{0,\mathrm{LO}}^{3/2}(u^\prime)}{(u^\prime - u - i\epsilon) {u^\prime}^2} du^\prime  \right) , \\
		R_1^\mathrm{NLO}(u) &= 0 , \\
		\tilde R_1^\mathrm{NLO}(u) &= 0 .
	\end{split}
\end{align}
If we further expand these expressions chirally and neglect higher orders, we obtain (note that only $a^{M_0}$ and $a^{\tilde M_1}$ do not vanish at LO):
\begin{align}
	\small
	\begin{split}
		M_0^\mathrm{NLO}(s) &= a^{M_0}_\mathrm{LO} \left( 1 + \omega_0^0 \frac{s}{M_K^2} +  \frac{s^2}{\pi} \int_{s_0}^\infty \frac{\delta_{0,\mathrm{LO}}^0(s^\prime)}{(s^\prime - s - i \epsilon) {s^\prime}^2} ds^\prime \right) + \Delta a^{M_0}_\mathrm{NLO} + b^{M_0}_\mathrm{NLO} \frac{s}{M_K^2} + c^{M_0}_\mathrm{NLO} \frac{s^2}{M_K^4} , \\
		M_1^\mathrm{NLO}(s) &= a^{M_1}_\mathrm{NLO} + b^{M_1}_\mathrm{NLO}  \frac{s}{M_K^2} , \\
		 \tilde M_1^\mathrm{NLO}(s) &= a^{\tilde M_1}_\mathrm{LO} \left( 1 + \omega_1^1 \frac{s}{M_K^2} +  \frac{s^2}{\pi} \int_{s_0}^\infty \frac{\delta_{1,\mathrm{LO}}^1(s^\prime)}{(s^\prime - s - i \epsilon) {s^\prime}^2} ds^\prime \right) + \Delta a^{\tilde M_1}_\mathrm{NLO} + b^{\tilde M_1}_\mathrm{NLO}  \frac{s}{M_K^2} + c^{\tilde M_1}_\mathrm{NLO}  \frac{s^2}{M_K^4} , \\
		N_0^\mathrm{NLO}(t) &= b^{N_0}_\mathrm{NLO} \frac{t}{M_K^2} + \frac{t^2}{\pi} \int_{t_0}^\infty \frac{\hat N_0^\mathrm{LO}(t^\prime) \delta_{0,\mathrm{LO}}^{1/2}(t^\prime)}{(t^\prime - t - i\epsilon) {t^\prime}^2} dt^\prime , \\
		N_1^\mathrm{NLO}(t) &= 0 , \\
		\tilde N_1^\mathrm{NLO}(t) &= \frac{t}{\pi} \int_{t_0}^\infty \frac{\hat{\tilde N}_1^\mathrm{LO}(t^\prime) \delta_{1,\mathrm{LO}}^{1/2}(t^\prime)}{(t^\prime - t - i\epsilon) t^\prime} dt^\prime , \\
		R_0^\mathrm{NLO}(u) &= \frac{u^2}{\pi} \int_{u_0}^\infty \frac{\hat R_0^\mathrm{LO}(u^\prime) \delta_{0,\mathrm{LO}}^{3/2}(u^\prime)}{(u^\prime - u - i\epsilon) {u^\prime}^2} du^\prime , \\
		R_1^\mathrm{NLO}(u) &= 0 , \\
		\tilde R_1^\mathrm{NLO}(u) &= 0 ,
	\end{split}
\end{align}
where
\begin{align}
	\begin{split}
		a^{M_0}_\mathrm{NLO} &= a^{M_0}_\mathrm{LO} + \Delta a^{M_0}_\mathrm{NLO}, \quad a^{M_0}_\mathrm{LO} = \frac{M_K}{\sqrt{2} F_\pi} , \\
		a^{\tilde M_1}_\mathrm{NLO} &= a^{\tilde M_1}_\mathrm{LO} + \Delta a^{\tilde M_1}_\mathrm{NLO}, \quad a^{\tilde M_1}_\mathrm{LO} = \frac{M_K}{\sqrt{2} F_\pi} .
	\end{split}
\end{align}
Next, we insert the LO phases and hat functions:
\begin{align}
	\begin{split}
		M_0^\mathrm{NLO}(s) &= a^{M_0}_\mathrm{NLO} + \left( b^{M_0}_\mathrm{NLO} + \frac{M_K}{\sqrt{2} F_\pi} \omega_0^0 \right) \frac{s}{M_K^2}  + c^{M_0}_\mathrm{NLO} \frac{s^2}{M_K^4} \\
			&\quad + \frac{s^2}{\pi} \int_{s_0}^\infty \frac{1}{(s^\prime - s - i \epsilon) {s^\prime}^2} \frac{\sigma_\pi(s^\prime)}{32 \pi} \frac{M_K (2s^\prime - M_\pi^2)}{\sqrt{2} F_\pi^3}  ds^\prime , \\
		M_1^\mathrm{NLO}(s) &= a^{M_1}_\mathrm{NLO} + b^{M_1}_\mathrm{NLO}  \frac{s}{M_K^2} , \\
		 \tilde M_1^\mathrm{NLO}(s) &= a^{\tilde M_1}_\mathrm{NLO} + \left( b^{\tilde M_1}_\mathrm{NLO} + \frac{M_K}{\sqrt{2} F_\pi} \omega_1^1 \right) \frac{s}{M_K^2} + c^{\tilde M_1}_\mathrm{NLO}  \frac{s^2}{M_K^4} \\
		 	&\quad + \frac{s^2}{\pi} \int_{s_0}^\infty \frac{1}{(s^\prime - s - i \epsilon) {s^\prime}^2} \frac{\sigma_\pi(s^\prime)}{32 \pi} \frac{M_K(s^\prime - 4 M_\pi^2)}{3\sqrt{2} F_\pi^3} ds^\prime , \\
		N_0^\mathrm{NLO}(t) &= b^{N_0}_\mathrm{NLO} \frac{t}{M_K^2} + \frac{t^2}{\pi} \int_{t_0}^\infty \begin{aligned}[t]
			& \frac{1}{(t^\prime - t - i\epsilon) {t^\prime}^2} \frac{\lambda_{K\pi}^{1/2}(t^\prime)}{16 \pi t^\prime} \frac{M_K(3\Delta_{K\pi} - 5t^\prime)}{32 \sqrt{2}{t^\prime}^2 F_\pi^3} \\
			& \cdot \left(5 {t^\prime}^2 - 2 t^\prime(M_K^2+M_\pi^2) - 3\Delta_{K\pi}^2 \right) dt^\prime , \end{aligned} \\
		N_1^\mathrm{NLO}(t) &= 0 , \\
		\tilde N_1^\mathrm{NLO}(t) &= \frac{t}{\pi} \int_{t_0}^\infty \frac{1}{(t^\prime - t - i\epsilon) t^\prime} \frac{\lambda_{K\pi}^{1/2}(t^\prime)}{16 \pi t^\prime} \frac{M_K^3}{16\sqrt{2} {t^\prime}^2 F_\pi^3} \left(3 {t^\prime}^2 - 6 t^\prime(M_K^2+M_\pi^2) + 3\Delta_{K\pi}^2 \right) dt^\prime , \\
		R_0^\mathrm{NLO}(u) &= \frac{u^2}{\pi} \int_{u_0}^\infty \frac{1}{(u^\prime - u - i\epsilon) {u^\prime}^2} \frac{\lambda_{K\pi}^{1/2}(u^\prime)}{16 \pi u^\prime} \frac{M_K}{2\sqrt{2} F_\pi^3} (M_K^2 + M_\pi^2 - u^\prime ) du^\prime , \\
		R_1^\mathrm{NLO}(u) &= 0 , \\
		\tilde R_1^\mathrm{NLO}(u) &= 0 .
	\end{split}
\end{align}
We see that the form of the Omn\`es representation is completely equivalent
to the standard representation, apart from the presence of the additional
subtraction constants $c^{M_0}$, $b^{M_1}$ and $c^{\tilde M_1}$, which also
need to be determined. 
We expand the $t$-channel $K\eta$ integrals up to linear terms in $t$ and find:
\begin{align}
	\begin{split}
		\label{eqn:NLORelationOmnesStandardSubtrConst}
		a^{M_0}_\mathrm{NLO} &= m_{0,\mathrm{NLO}}^0 , \\
		b^{M_0}_\mathrm{NLO} &= m_{0,\mathrm{NLO}}^1 - \frac{M_K}{\sqrt{2} F_\pi} \omega_0^0 , \\
		c^{M_0}_\mathrm{NLO} &= \frac{M_K^3}{\sqrt{2}F_\pi^3} \frac{15 M_\eta^4 + M_K^2 M_\pi^2}{1920 \pi^2 M_\eta^4} , \\
		a^{M_1}_\mathrm{NLO} &= m_{1,\mathrm{NLO}}^0 , \\
		b^{M_1}_\mathrm{NLO} &= 0 , \\
		a^{\tilde M_1}_\mathrm{NLO} &= \tilde m_{1,\mathrm{NLO}}^0 , \\
		b^{\tilde M_1}_\mathrm{NLO} &= \tilde m_{1,\mathrm{NLO}}^1 - \frac{M_K}{\sqrt{2} F_\pi} \omega_1^1 , \\
		c^{\tilde M_1}_\mathrm{NLO} &= \frac{M_K^3}{\sqrt{2} F_\pi^3} \frac{1}{1920 \pi^2} , \\
		b^{N_0}_\mathrm{NLO} &= n_{0,\mathrm{NLO}}^1 .
	\end{split}
\end{align}
The constants $\omega_0^0$ and $\omega_1^1$ cannot be evaluated with the
chiral phases. If we evaluate them with the physical phases, this leads to
exactly the same matching equations for the determination of the $L_i^r$ as
if we would match the Taylor expansion of the Omn\`es representation with
the Taylor expansion of the chiral result. Note, however, that the
expressions obtained for $c^{M_0}$, $b^{M_1}$ and $c^{\tilde M_1}$ are
different. E.g.~for $b^{M_1}$, the chiral expansion leads to
$b^{M_1}_\mathrm{NLO} = 0$ while a Taylor expansion of the dispersion
relation would require $b^{M_1} = - m_{1,\mathrm{NLO}}^0
{\Omega_1^1}^\prime(0) M_K^2$, where ${\Omega_1^1}^\prime$ is the
derivative of the Omn\`es function calculated with the physical phases. Of
course, the difference is a higher order effect in the chiral counting. As
higher order effects can be important if due to final state rescattering,
we would not like to intermingle them with the matching of the subtraction
constants. The matching on the basis of Taylor coefficients would require
the linear term of $M_1(s)$ to vanish exactly, while the matching based on
the chiral expansion of the dispersion relation gives a non-zero linear
term in $M_1(s)$ due to the Omn\`es function -- this is important
information which we wish to make use of in our fits.

\subsubsection{Matching Equations at $\O(p^6)$}

\paragraph{Decomposition of the NNLO Form Factors}

In the following, we describe the decomposition of the two-loop result such that the matching can be performed at NNLO. Since the NNLO chiral result has a different asymptotic behaviour than the NLO result and our numerical dispersive representation, we have to use the representation (\ref{eq:FunctionsOfOneVariable3Subtr}), which uses a different gauge and more subtractions than (\ref{eq:FunctionsOfOneVariable}).

The imaginary parts of the $K_{\ell4}$ partial waves at NNLO could again be reconstructed using the unitarity relations, e.g.
\begin{align}
	\begin{split}
		\Im f_l^\mathrm{NNLO}(s) &= \frac{1}{2l+1} \frac{1}{32\pi} \sigma_\pi(s) \left( t_{l,\mathrm{LO}}^{I*}(s) f_l^\mathrm{LO}(s) + \Delta t_{l,\mathrm{NLO}}^{I*}(s) f_l^\mathrm{LO}(s) + t_{l,\mathrm{LO}}^{I*}(s) \Delta f_l^\mathrm{NLO}(s) \right) .
	\end{split}
\end{align}
However, instead of proceeding as for NLO, it is more straightforward to decompose the two-loop result directly into functions of one variable, then to impose the gauge condition and extract the Taylor coefficients of the functions of one variable.

The two-loop result for the form factors $F$ and $G$ was computed in \cite{Amoros2000}. We have the full expressions in form of a C++ program at hand.\footnote{We thank Johan Bijnens and Ilaria Jemos for providing the C++ implementation of the NNLO expressions.} It has the following structure:
\begin{align}
	\begin{split}
		X^\mathrm{NNLO}(s,t,u) = X^\mathrm{LO} & + X^\mathrm{NLO}_L(s,t,u) + X^\mathrm{NLO}_R(s,t,u) \\
			& + X^\mathrm{NNLO}_{C}(s,t,u) + X^\mathrm{NNLO}_{L}(s,t,u) + X^\mathrm{NNLO}_{P}(s,t,u) \\
			& + X^\mathrm{NNLO}_{VS}(s,t,u) + X^\mathrm{NNLO}_{VT}(s,t,u) + X^\mathrm{NNLO}_{VU}(s,t,u) ,
	\end{split}
\end{align}
where $X\in\{F,G\}$ and the different parts denote the following:
\begin{itemize}
	\item $X^\mathrm{NLO}_L$: NLO polynomial containing the LECs $L_i^r$,
	\item $X^\mathrm{NLO}_R$: NLO loops,
	\item $X^\mathrm{NNLO}_C$: NNLO polynomial containing the LECs $C_i^r$,
	\item $X^\mathrm{NNLO}_L$: NNLO part containing $L_i^r \times L_i^r$ and $L_i^r \times \mathrm{loop}$,
	\item $X^\mathrm{NNLO}_P$: NNLO two-loop part without vertex integrals,
	\item $X^\mathrm{NNLO}_{VS}$: NNLO vertex integrals in the $s$-channel,
	\item $X^\mathrm{NNLO}_{VT}$: NNLO vertex integrals in the $t$-channel,
	\item $X^\mathrm{NNLO}_{VU}$: NNLO vertex integrals in the $u$-channel.
\end{itemize}

In appendix~\ref{sec:AppendixTwoLoopDecomposition}, we perform the explicit decomposition of the two-loop result into functions of one Mandelstam variable according to (\ref{eq:FormFactorDecomposition}) and (\ref{eq:FunctionsOfOneVariable3Subtr}) and evaluate numerically the subtraction constants.

\paragraph{NNLO Form Factors in the Omn\`es Representation}

As we already pointed out for the NLO matching, it is desirable to use the Omn\`es representation rather than the standard dispersion relation for the matching and the determination of the LECs. Let us therefore derive the matching equations at NNLO in the Omn\`es scheme.

We have to use the second gauge for the decomposition of the NNLO representation (\ref{eq:FunctionsOfOneVariable3Subtr}). As a starting point, let us find the NLO Omn\`es subtraction constants in the second gauge. In the first gauge, we found $R_1^\mathrm{NLO} = \tilde R_1^\mathrm{NLO} = 0$, hence
\begin{align}
	\begin{split}
		c_\mathrm{NLO}^{R_0} &= \frac{M_K}{\sqrt{2} F_\pi^3} \frac{M_K^4}{4} \left((M_K^2 + M_\pi^2) \bar B_{K\pi}^\dprime(0) - 2 \bar B_{K\pi}^\prime(0) \right) , \\
		a_\mathrm{NLO}^{R_1} &= b_\mathrm{NLO}^{\tilde R_1} = 0.
	\end{split}
\end{align}
The gauge-transformation (\ref{eq:GaugeTransformation}) is then defined by
\begin{align}
	\begin{split}
		C_\mathrm{NLO}^{R_0} &= \frac{M_K}{\sqrt{2} F_\pi^3} \frac{1}{32\pi^2} \frac{M_K^4}{\Delta_{K\pi}^4} \Bigg( \frac{(M_K^2 + M_\pi^2)(M_K^4 - 8 M_K^2 M_\pi^2 + M_\pi^4)}{3} + \frac{4 M_K^4 M_\pi^4 \ln\left(\frac{M_K^2}{M_\pi^2}\right)}{\Delta_{K\pi}} \Bigg) , \\
		A_\mathrm{NLO}^{R_1} &= B_\mathrm{NLO}^{\tilde R_1} = 0.
	\end{split}
\end{align}
At NLO, the shifts in the subtraction constants (\ref{eq:OmnesGaugeTransformationParameters}) are therefore given by
\begin{align}
	\begin{alignedat}{4}
		\label{eq:NLOGaugeTransformationSubtractionConstants}
		\delta a_\mathrm{NLO}^{M_0} &= \frac{\Sigma_0^2 - \Delta_{K\pi}\Delta_{\ell\pi}}{M_K^4} C_\mathrm{NLO}^{R_0} , \quad & \delta b_\mathrm{NLO}^{M_0} &= - \frac{2 \Sigma_0}{M_K^2} C_\mathrm{NLO}^{R_0} , \quad & \delta c_\mathrm{NLO}^{M_0} &= C_\mathrm{NLO}^{R_0} , \quad & \delta d_\mathrm{NLO}^{M_0} &= 0 , \\
		\delta a_\mathrm{NLO}^{M_1} &= - \frac{2 \Sigma_0}{M_K^2} C_\mathrm{NLO}^{R_0} , \quad & \delta b_\mathrm{NLO}^{M_1} &= 2 C_\mathrm{NLO}^{R_0} , \quad & \delta c_\mathrm{NLO}^{M_1} &= 0 , \\
		\delta a_\mathrm{NLO}^{\tilde M_1} &= - \frac{\Sigma_0^2 - \Delta_{K\pi} \Delta_{\ell\pi}}{M_K^4} C_\mathrm{NLO}^{R_0} , \quad & \delta b_\mathrm{NLO}^{\tilde M_1} &= \frac{2\Sigma_0}{M_K^2} C_\mathrm{NLO}^{R_0} , \quad & \delta c_\mathrm{NLO}^{\tilde M_1} &= - C_\mathrm{NLO}^{R_0} , \quad & \delta d_\mathrm{NLO}^{\tilde M_1} &= 0 , \\
		\delta b_\mathrm{NLO}^{N_0} &= - \frac{3 (\Delta_{K\pi} + 2 \Sigma_0)}{4 M_K^2} C_\mathrm{NLO}^{R_0} , \quad & \delta c_\mathrm{NLO}^{N_0} &= - \frac{5}{4} C_\mathrm{NLO}^{R_0} , \\
		\delta a_\mathrm{NLO}^{N_1} &= \frac{3}{2} C_\mathrm{NLO}^{R_0} , \quad & \delta b_\mathrm{NLO}^{\tilde N_1} &= -\frac{3}{2} C_\mathrm{NLO}^{R_0} , \\
		\delta c_\mathrm{NLO}^{R_0} &= C_\mathrm{NLO}^{R_0} , \quad & \delta a_\mathrm{NLO}^{R_1} &= 0 , \quad & \delta b_\mathrm{NLO}^{\tilde R_1} &= 0 .
	\end{alignedat}
\end{align}

When studying now the Omn\`es representation at NNLO, we notice that the asymptotic behaviour of the phases at NNLO is even worse than at NLO, hence we have to subtract the Omn\`es function three times:
\begin{align}
	\begin{split}
		\label{eqn:ThriceSubtractedOmnesFunction}
		\Omega(s) &= \exp\left( \frac{s}{\pi} \int_{s_0}^\infty \frac{\delta(s^\prime)}{(s^\prime - s - i \epsilon) s^\prime} ds^\prime \right) \\
			&= \exp\left(  \frac{s}{\pi} \int_{s_0}^\infty \frac{\delta(s^\prime)}{{s^\prime}^2} ds^\prime +  \frac{s^2}{\pi} \int_{s_0}^\infty \frac{\delta(s^\prime)}{{s^\prime}^3} ds^\prime +  \frac{s^3}{\pi} \int_{s_0}^\infty \frac{\delta(s^\prime)}{(s^\prime - s - i \epsilon) {s^\prime}^3} ds^\prime \right) \\
			&=: \exp\left( \omega \frac{s}{M_K^2} +  \bar\omega \frac{s^2}{M_K^4} +  \frac{s^3}{\pi} \int_{s_0}^\infty \frac{\delta(s^\prime)}{(s^\prime - s - i \epsilon) {s^\prime}^3} ds^\prime \right) .
	\end{split}
\end{align}
$\omega$ and $\bar\omega$ are both divergent if evaluated in \ChPT{} at NNLO, hence we will use the physical phases to determine them.

In the case of the NLO matching, we have derived the relation between the standard and the Omn\`es subtraction constants (\ref{eqn:NLORelationOmnesStandardSubtrConst}) by comparing the Taylor coefficients of the chirally expanded Omn\`es representation with the Taylor coefficients of the standard dispersive representation. Although it is instructive to understand the chiral expansion of the Omn\`es representation, a shortcut can be taken. Note that the chiral expansion and the Taylor expansion are interchangeable. Therefore, we easily obtain the relations between the standard subtraction constants $m_0^0$,~$\ldots$ and the Omn\`es subtraction constants $a^{M_0}$,~$\ldots$ by chirally expanding the Taylor coefficients of the Omn\`es representation (\ref{eq:FunctionsOfOneVariableOmnes3Subtr}) and comparing it with the Taylor coefficients of (\ref{eq:FunctionsOfOneVariable3Subtr}).

This leads to the following relations between the relevant subtraction constants:
\begin{align}
	\begin{split}
		\label{eqn:NNLORelationOmnesStandardSubtrConst}
		m_{0}^{0,\mathrm{NNLO}} &= a_\mathrm{NNLO}^{M_0} , \\
		m_{0}^{1,\mathrm{NNLO}} &= 
		b_\mathrm{NNLO}^{M_0} + \omega_0^0 a_\mathrm{NLO}^{M_0} , \\
		m_{0}^{2,\mathrm{NNLO}} 
			&= c_\mathrm{NNLO}^{M_0} + \omega_0^0 b_\mathrm{NLO}^{M_0} + \frac{1}{2} {\omega_0^0}^2 a_\mathrm{LO}^{M_0} + a_\mathrm{NLO}^{M_0} \bar\omega_0^0 + h.o. , \\
		m_{1}^{0,\mathrm{NNLO}} &= a_\mathrm{NNLO}^{M_1} , \\
		m_{1}^{1,\mathrm{NNLO}} &= b_\mathrm{NNLO}^{M_1} + \omega_1^1 a_\mathrm{NLO}^{M_1} , \\
		\tilde m_{1}^{0,\mathrm{NNLO}} &= a_\mathrm{NNLO}^{\tilde M_1} , \\
		\tilde m_{1}^{1,\mathrm{NNLO}} &= b_\mathrm{NNLO}^{\tilde M_1} + \omega_1^1 a_\mathrm{NLO}^{\tilde M_1} , \\
		\tilde m_{1}^{2,\mathrm{NNLO}} &= c_\mathrm{NNLO}^{\tilde M_1} + \omega_1^1 b_\mathrm{NLO}^{\tilde M_1} + \frac{1}{2} {\omega_1^1}^2 a_\mathrm{LO}^{\tilde M_1} + a_\mathrm{NLO}^{\tilde M_1} \bar \omega_1^1 + h.o. , \\
		n_{0}^{1,\mathrm{NNLO}} &= b_\mathrm{NNLO}^{N_0} , \\
		n_{0}^{2,\mathrm{NNLO}} &= c_\mathrm{NNLO}^{N_0} + \omega_0^{1/2} b_\mathrm{NLO}^{N_0} , \\
		n_{1}^{0,\mathrm{NNLO}} &= a_\mathrm{NNLO}^{N_1} , \\
		\tilde n_{1}^{1,\mathrm{NNLO}} &= b_\mathrm{NNLO}^{\tilde N_1} .
	\end{split}
\end{align}
The NNLO chiral expansion of the full Omn\`es representation can be found in appendix~\ref{sec:AppendixNNLOChiralExpansionOmnes} and leads to the same result. It can be used to identify all the imaginary parts and to connect the different dispersive integrals with the discontinuities of the loop diagrams.


\section{Results}

\label{sec:Kl4Results}

In this chapter, we discuss the results for the low-energy constants that
we determine by fitting the dispersive representation to data and matching
it to \ChPT{}. In order to understand the differences between the results
at NLO and NNLO and the source of complications that appear at NNLO, it is
useful to study in a first step the results of direct \ChPT{} fits. We
perform direct fits at NLO and NNLO and compare our results with the
literature before using the whole machinery of the dispersive framework
matched to \ChPT{} at NLO and finally at NNLO. 

\subsection{Comparison of Direct \ChPT{} Fits}

The most recent fits to $K_{\ell4}$ data performed in the literature are \cite{Bijnens2014}. There, a global fit is performed, taking into account the threshold expansion parameters of the $K_{\ell4}$ form factor measurement of NA48/2 \cite{Batley2012}:
\begin{align}
	\begin{alignedat}{3}
		\label{eqn:NA48ThresholdParameters}
		F &= f_s + f_s^\prime q^2 + \ldots, \quad & f_s &= 5.705 \pm 0.035, \quad & f_s^\prime &= 0.867 \pm 0.050 \\
		G &= g_p + g_p^\prime q^2 + \ldots, \quad & g_p &= 4.952 \pm 0.086, \quad & g_p^\prime &= 0.508 \pm 0.122,
	\end{alignedat}
\end{align}
where $q^2 = \frac{s}{4M_\pi^2} - 1$. In \cite{Bijnens2014}, the above quantities are fitted with the form factors at $\cos\theta = 0$ instead of the first partial wave. In addition to the $K_{\ell4}$ form factor data, the global fit of \cite{Bijnens2014} uses many other inputs, like data on the different decay constants and masses, $\pi\pi$- and $K\pi$-scattering parameters, quark mass ratios etc.

We compare now different strategies for direct fits with the results of \cite{Bijnens2014}. We use only $K_{\ell4}$ data for our fits and therefore are only sensitive to the LECs $L_1^r$, $L_2^r$ and $L_3^r$ \cite{Stoffer2010}. The other LECs are taken as a fixed input.

\subsubsection{Direct Fits at $\O(p^4)$}

\paragraph{Fits of Threshold Parameters}

In order to make the connection to \cite{Bijnens2014}, we first perform a direct NLO fit to the NA48/2 threshold parameters in (\ref{eqn:NA48ThresholdParameters}). Using $\cos\theta=0$, i.e.~the first Taylor coefficient of an expansion in $z=\cos\theta$, and the LEC inputs $L_4^r=0$ and the fitted value for $L_5^r$ of \cite{Bijnens2014}, we reproduce almost exactly the result of \cite{Bijnens2014} for $L_1^r$, $L_2^r$ and $L_3^r$, see the second and third column in table~\ref{tab:NLOThresholdFitsNA48}. If we use instead the partial-wave projection (\ref{eqn:SChannelPartialWaveProjection}), the fit results for $L_1^r$ and $L_2^r$ change a bit, as shown in the fourth column of table~\ref{tab:NLOThresholdFitsNA48}. The last column uses lattice results \cite{MILC2009,Aoki2013} for the input LECs.

\begin{table}[H]
	\centering
	\begin{tabular}{c c c c c}
		\toprule
						&	Ref.~\cite{Bijnens2014}	& 	Taylor		 &	PWE		&	PWE		 \\[0.1cm]
		\hline \\[-0.3cm]	
		$10^3 \cdot L_1^r$	&	$\m0.98(09)$		&	$\m0.99(09)$	&	$\m1.15(09)$	&	$\m1.17(09)$	\\
		$10^3 \cdot L_2^r$	&	$\m1.56(09)$		&	$\m1.57(09)$	&	$\m1.48(08)$	&	$\m1.50(08)$	\\
		$10^3 \cdot L_3^r$	&	$-3.82(30)$		&	$-3.83(30)$	&	$-3.82(30)$	&	$-3.87(30)$	\\
		$10^3 \cdot L_4^r$	&	$\equiv 0$		&	$\equiv 0$	&	$\equiv 0$	&	$\equiv0.04$	\\
		$10^3 \cdot L_5^r$	&	$\m1.23(06)$		&	$\equiv1.23$	&	$\equiv1.23$	&	$\equiv0.84$	\\
		\hline \\[-0.3cm]
		$\chi^2$			&	16				&	0.3			&	0.3			&	0.3			\\
		dof				& 	5				&	1	 		&	1			&	1			\\
		$\chi^2/$dof		& 	3.2				&	0.3	 		&	0.3			&	0.3			\\
		\bottomrule
	\end{tabular}
	\caption{Comparison of direct NLO fits to the NA48/2 threshold parameters \cite{Batley2012}. The renormalisation scale is $\mu = 770$~MeV. The last column uses the lattice determination of \cite{MILC2009,Aoki2013} for the input LECs. The uncertainties are purely statistical.}
	\label{tab:NLOThresholdFitsNA48}
\end{table}

\paragraph{Fits of the Complete Form Factor Data}

In a next step, we no longer fit the threshold expansion parameters (\ref{eqn:NA48ThresholdParameters}) of the form factors, but the form factor data of NA48/2 \cite{Batley2010,Batley2012} and E865 \cite{Pislak2001, Pislak2003}, discussed in section~\ref{sec:ExperimentalKl4Data}. The second column of table~\ref{tab:NLODirectFormFactorFits} shows the result of the NLO fit to the one-dimensional NA48/2 data without isospin corrections (but with the corrected normalisation of $F_s$ to account for the $s_\ell$-dependence). In the third column, isospin corrections are applied to the fitted data (table~\ref{tab:NA48DataIsoCorr}). The fourth and fifth column show the results of a combined fit to NA48/2 and E865 data (table~\ref{tab:E865DataIsoCorr}). The smaller $\chi^2$ value in the fits to the data with isospin-breaking corrections is due to the fact that the isospin corrections introduce an additional uncertainty in the data.

\begin{table}[H]
	\centering
	\begin{tabular}{c c c c c c}
		\toprule
						&	NA48/2		 &	NA48/2, \cancel{iso}	&	NA48/2 \& E865	&	NA48/2 \& E865, \cancel{iso}	 \\[0.1cm]
		\hline \\[-0.3cm]	
		$10^3 \cdot L_1^r$	&	$\m0.69(03)$	&	$\m0.71(04)$		&	$\m0.62(03)$		&	$\m0.64(04)$	\\
		$10^3 \cdot L_2^r$	&	$\m1.88(07)$	&	$\m1.80(08)$		&	$\m1.79(06)$		&	$\m1.70(06)$	\\
		$10^3 \cdot L_3^r$	&	$-3.89(13)$	&	$-3.93(14)$		&	$-3.62(11)$		&	$-3.60(12)$	\\
		$10^3 \cdot L_4^r$	&	$\equiv 0.04$	&	$\equiv 0.04$		&	$\equiv 0.04$		&	$\equiv 0.04$	\\
		$10^3 \cdot L_5^r$	&	$\equiv 0.84$	&	$\equiv 0.84$		&	$\equiv 0.84$		&	$\equiv 0.84$	\\
		$10^3 \cdot L_9^r$	&	$\equiv 5.93$	&	$\equiv 5.93$		&	$\equiv 5.93$		&	$\equiv 5.93$	\\
		\hline																					 \\[-0.3cm]
		$\chi^2$			&	159.4		&	67.5				&	199.9			&	117.1		\\
		dof				& 	27	 		&	27				&	39				&	39			\\
		$\chi^2/$dof		& 	5.9	 		&	2.5				&	5.1				&	3.0			\\
		\bottomrule
	\end{tabular}
	\caption{Comparison of direct NLO fits to the NA48/2 and E865 form factor measurements. The renormalisation scale is $\mu = 770$~MeV. For $L_4^r$ and $L_5^r$, we use lattice input \cite{MILC2009,Aoki2013}, for $L_9^r$ the determination of \cite{BijnensTalavera2002}. The uncertainties are purely statistical.}
	\label{tab:NLODirectFormFactorFits}
\end{table}

\begin{figure}[H]
	\centering
	\scalebox{0.75}{
		\small
		\input{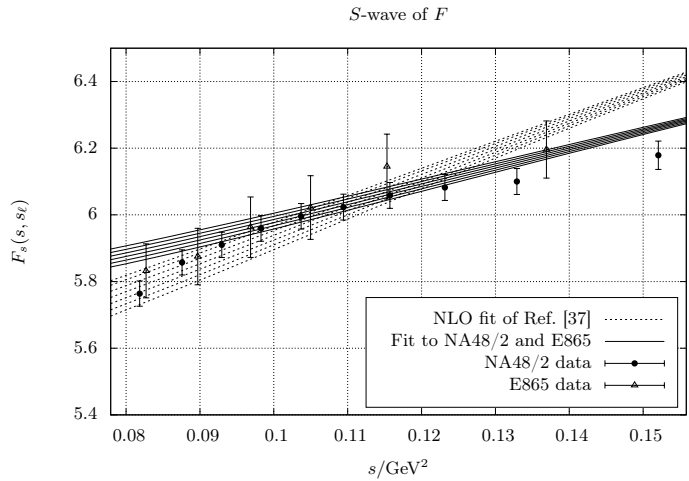}
		}
	\caption{Comparison of different fits for the $S$-wave of the form factor $F$: NA48/2 threshold parameter fit of \cite{Bijnens2014} and a fit to the full data set. The $(s,s_\ell)$ phase space is projected on the $s$-axis. No isospin corrections are applied.}
	\label{fig:FsDirectFits}
\end{figure}

Figure~\ref{fig:FsDirectFits} shows a comparison of the NA48/2 threshold parameter fit of \cite{Bijnens2014} with the result of the fit to the whole form factor data set (forth column of table~\ref{tab:NLODirectFormFactorFits}). It helps to understand the difference between the fitted LECs in the two procedures: in the fit to the threshold parameters, the curvature of the form factor is neglected. Since the NLO chiral representation cannot reproduce the curvature, the data points at higher energies reduce the slope in a fit to the whole data set.

\subsubsection{Direct Fits at $\O(p^6)$}

\begin{table}[H]
	\centering
	\tabcolsep=0.15cm
	\begin{tabular}{c c c c c c c}
		\toprule
						&	Ref.~\cite{Bijnens2014}	&	Ref.~\cite{Bijnens2014}	&	NA48/2	&	NA48/2 \& E865	&	NA48/2	&	NA48/2 \& E865	 \\[0.1cm]
		\hline \\[-0.3cm]	
		$C_i^r$			&	$\equiv0$		&	BE14		&	$\equiv0$			&	$\equiv0$			&	BE14		&	BE14		\\
		\hline \\[-0.3cm]	
		$10^3 \cdot L_1^r$	&	$\m0.67(06)$	&	$\m0.53(06)$	&	$\m0.34(03)$		&	$\m0.28(02)$		&	$\m0.33(03)$	&	$\m0.27(02)$	\\
		$10^3 \cdot L_2^r$	&	$\m0.17(04)$	&	$\m0.81(04)$	&	$\m0.42(06)$		&	$\m0.35(05)$		&	$\m0.95(06)$	&	$\m0.89(05)$	\\
		$10^3 \cdot L_3^r$	&	$-1.76(21)$	&	$-3.07(20)$	&	$-1.54(14)$		&	$-1.25(11)$		&	$-3.06(14)$	&	$-2.80(11)$	\\
		$10^3 \cdot L_4^r$	&	$\m0.73(10)$	&	$\equiv0.3$	&	$\equiv 0.04$		&	$\equiv 0.04$		&	$\equiv 0.04$	&	$\equiv 0.04$	\\
		$10^3 \cdot L_5^r$	&	$\m0.65(05)$	&	$\m1.01(06)$	&	$\equiv 0.84$		&	$\equiv 0.84$		&	$\equiv 0.84$	&	$\equiv 0.84$	\\
		$10^3 \cdot L_6^r$	&	$\m0.25(09)$	&	$\m0.14(05)$	&	$\equiv 0.07$		&	$\equiv 0.07$		&	$\equiv 0.07$	&	$\equiv 0.07$	\\
		$10^3 \cdot L_7^r$	&	$-0.17(06)$	&	$-0.34(09)$	&	$\equiv -0.34$		&	$\equiv -0.34$		&	$\equiv -0.34$	&	$\equiv -0.34$	\\
		$10^3 \cdot L_8^r$	&	$\m0.22(08)$	&	$\m0.47(10)$	&	$\equiv 0.36$		&	$\equiv 0.36$		&	$\equiv 0.36$	&	$\equiv 0.36$	\\
		$10^3 \cdot L_9^r$	&				&				&	$\equiv 5.93$		&	$\equiv 5.93$		&	$\equiv 5.93$	&	$\equiv 5.93$	\\
		\hline																													 \\[-0.3cm]
		$\chi^2$			&	26			&	1.0			&	81.3				&	128.7			&	52.5			&	91.2			\\
		dof				& 	9	 		&		 		&	27				&	39				&	27			&	39			\\
		$\chi^2/$dof		& 	2.9	 		&		 		&	3.0				&	3.3				&	1.9			&	2.3			\\
		\bottomrule
	\end{tabular}
	\caption{Direct NNLO fits for different choices of the $C_i^r$. The
          results of the fits of \cite{Bijnens2014} are shown for
          comparison. The renormalisation scale is $\mu = 770$~MeV. Our
          results are fits to the entire form factor data including isospin
          corrections. The uncertainties are purely statistical. The NLO
          input LECs $L_4^r$, $L_5^r$, $L_6^r$ and $L_8^r$ are lattice
          determinations \cite{MILC2009,Aoki2013}, $L_7^r$ is the BE14
          value \cite{Bijnens2014} and $L_9^r$ is taken from
          \cite{BijnensTalavera2002}.} 
	\label{tab:NNLOThresholdFitsBE14}
\end{table}

\ChPT{} at NNLO suffers from the problem that many new low-energy constants $C_i^r$ appear in the $\O(p^6)$ Lagrangian. In $K_{\ell4}$, in total 24 linearly independent combinations of the $C_i^r$ enter in the NNLO chiral representation of the form factors $F$ and $G$. A fit of so many parameters seems out of question. We would rather like to use some input values for the $C_i^r$. Unfortunately, only very few of the NNLO LECs are known reliably. We could either use determinations of the $C_i^r$ with models like the chiral quark model \cite{Jiang2010}, a resonance estimate \cite{Amoros2000,Bijnens2012} or the educated guess of \cite{Bijnens2014}. These different estimates, however, do not lead to compatible results \cite{Bijnens2014}.

In table~\ref{tab:NNLOThresholdFitsBE14}, we display the results of our
direct \ChPT{} fits at NNLO in comparison with the results of
\cite{Bijnens2014}. In contrast to \cite{Bijnens2014}, we do not use the
threshold parameters but the whole form factor data sets of NA48/2 and E865
corrected by isospin-breaking effects \cite{Stoffer2014}. It turns out that
even at NNLO, \ChPT{} has trouble to reproduce the curvature of the $F_s$
data. We also note that the results for the fitted LECs at NNLO differ
quite significantly from the results at NLO.

\subsection{Matching the Dispersion Relation to \ChPT{}}

With the direct \ChPT{} fits, we have seen a number of problems: First, at
NLO and even at NNLO, the energy dependence of the $F_s$ form factor is not
very well described. Second, at $\O(p^6)$, the appearance of quite a large
number of additional LECs reduces the predictive power of \ChPT{}. Some
input values for the $C_i^r$ have to be assumed, as a fit of $K_{\ell4}$
data alone cannot determine all these LECs.

We now turn to the results using the dispersive representation as an
intermediate step in the determination of the LECs: we fit the $K_{\ell4}$
form factor data with the dispersion relation. The matching to \ChPT{}
relates the subtraction constants of the dispersion relation to the
LECs. As the dispersion relation provides a resummation of final-state
rescattering effects, we trust that we will obtain a better description of the
energy dependence of the form factors. However, it is clear that the
matching of the dispersion relation to NNLO \ChPT{} will not be free of the
problem related to the large number of LECs. We will alleviate the
situation by including additional constraints on the chiral convergence in
the fit. This will enable us to fit partially the contribution of the NNLO
LECs to the subtraction constants.

\subsubsection{Matching at $\O(p^4)$}

Our numerical solution of the dispersion relation
(\ref{eq:FunctionsOfOneVariableOmnes}) is parametrised by nine subtraction
constants, which in fact are functions of $s_\ell$. If we use the matching
at NLO to provide a chiral representation of the subtraction constants, we 
see that $a^{M_0}_\mathrm{NLO}$ and $a^{\tilde M_1}_\mathrm{NLO}$ are
linear in $s_\ell$, while the other subtraction constants do not depend on
$s_\ell$. We therefore introduce this $s_\ell$-dependence according to
(\ref{eqn:SlDependenceOmnesSubtractionConstants}) and have to determine in
total 11 parameters.

We fit our dispersive representation to the data of both experiments, shown
in appendix~\ref{sec:AppendixData}. In the case of NA48/2, the use of the
two-dimensional instead of the one-dimensional data set has basically no
effect on the determination of the LECs $L_1^r$, $L_2^r$ and $L_3^r$ but
gives us the option to fit the $s_\ell$-dependence and therefore to
determine also $L_9^r$. In order to test the influence of the
isospin-breaking corrections, we also perform fits to data without isospin
corrections.

An unconstrained fit with the 11 subtraction parameters leads to a low
relative $\chi^2$ of 0.77 (with 94 degrees of freedom, dof) for the NA48/2
data alone or 0.74 (106 dof) for the combined data set of NA48/2 and
E865. However, the soft-pion theorems in such a fit are not well
reproduced. Therefore, we chose to use the soft-pion theorems as
constraints in the fit: the first soft-pion theorem (\ref{eqn:SPT1}) with a
tolerance of 2\% and the second soft-pion theorem (\ref{eqn:SPT2}) of
5\%. These numbers are inspired by the typical NNLO deviation. In these
fits the relative $\chi^2$ slightly increases to 0.79 (96
dof) for the NA48/2 fit and 0.77 (108 dof) for the combined fit. This shows
that in a fit with all 11 parameters, the soft-pion theorems are not
fulfilled automatically but are not a strong additional constraint.

In an unconstrained fit, the result for the subtraction constants turns out
to be rather unstable: the statistical uncertainties are large and some of
the subtraction constants change drastically if the E865 data is
included. We consider these fits of little interest and fix to an a
priori value those subtraction constants that have the largest statistical
uncertainty: these are the subtraction constants of highest order in each
function and the one parametrising the $s_\ell$-dependence in $G_p$,
i.e. $c^{M_0}$, $b^{M_1}$, $c^{\tilde M_1}$ and $a^{\tilde M_1}_1$. We fix
these subtraction constants to the NLO chiral prediction in the matching
(\ref{eqn:NLORelationOmnesStandardSubtrConst}): while $c^{M_0}$, $b^{M_1}$
and $c^{\tilde M_1}$ are purely numerical, $a^{\tilde M_1}_1$ depends on
$L_2^r$ and $L_9^r$. We take those two LECs as input and iterate the fit
after the matching to reach self-consistency for $L_2^r$ (and $L_9^r$ if
this LEC is determined in the matching as well).

Seven subtraction constants $a^{M_0}_0$, $a^{M_0}_1$, $b^{M_0}$, $a^{M_1}$,
$a^{\tilde M_1}_0$, $b^{\tilde M_1}$ and $b^{N_0}$ remain to be fitted to
data. In the matching equations
(i.e.~(\ref{eqn:NLORelationOmnesStandardSubtrConst}) together with
appendix~\ref{sec:AppendixNLOSubtractionConstantsStandardRep}), the LECs
$L_1^r$, $L_2^r$, $L_3^r$ and $L_9^r$ are overdetermined. Hence, we have to
use a second $\chi^2$ minimisation to fix these LECs. As an alternative to
this two-step procedure (first fit to data, then matching to \ChPT{}), we
can directly use the NLO chiral representation of the subtraction constants
and perform the fit of the dispersive representation to data with the LECs as
fitting parameters. As expected, these two strategies lead to almost
identical numerical results for the LECs.

In table~\ref{tab:ResultsDispersionNLOMatching}, we show the results of the
fits of the dispersion relation matched to NLO \ChPT{}. For the input LECs,
we use lattice results \cite{MILC2009,Aoki2013}: 
\begin{align}
	\begin{split}
		\label{eq:Kl4InputLECsNLOMatching}
		10^3 \cdot L_4^r &= 0.04(14) , \\
		10^3 \cdot L_5^r &= 0.84(38) .
	\end{split}
\end{align}

The $\chi^2$ and degrees of freedom correspond to the strategy of using the
LECs as fitting parameters. If we use the two-step fitting/matching
strategy instead, the $\chi^2/$dof of the fit of the subtraction constants
to data is good: around 0.8 for the fit to NA48/2 and around 1.0 for the
fit to both experiments. At the same time, the relative $\chi^2/$dof of the
matching is bad (between 2.9 and 6.1). This is not surprising because the
sum of the total $\chi^2$ of the two steps is approximately equal to the
total $\chi^2$ in the one-step procedure, while the dof in the second step
are drastically reduced.

The first bracket indicates the statistical uncertainty due to the fitted
data. The second bracket gives the systematic uncertainty. In
section~\ref{sec:ErrorAnalysis}, we will discuss in more detail the
different sources of uncertainty.
\begin{table}[H]
	\small
	\centering
	\tabcolsep=0.08cm
	\begin{tabular}{c c c c c c c}
		\toprule
						&	NA48/2			&	NA48/2 \& E865	&	NA48/2			&	NA48/2 \& E865	&	NA48/2			&	NA48/2 \& E865	\\[0.1cm]
		\hline																																	\\[-0.3cm]	
		Isospin corr.		&	\xmark			&	\xmark			&	\cmark			&	\cmark			&	\cmark			&	\cmark			\\
		\hline																																	\\[-0.3cm]	
		$\sigma_\mathrm{SPT1}$	&	---			&	---				&	---				&	---				&	2\%				&	2\%				\\
		$\sigma_\mathrm{SPT2}$	&	---			&	---				&	---				&	---				&	5\%				&	5\%				\\
		\hline																																	\\[-0.3cm]	
		$10^3 \cdot L_1^r$	&	$\m0.52(02)(05)$	&	$\m0.48(02)(05)$	&	$\m0.54(02)(05)$	&	$\m0.50(02)(05)$	&	$\m0.54(02)(05)$	&	$\m0.50(02)(05)$	\\
		$10^3 \cdot L_2^r$	&	$\m1.00(05)(07)$	&	$\m0.94(04)(07)$	&	$\m0.94(05)(07)$	&	$\m0.88(05)(07)$	&	$\m0.94(05)(07)$	&	$\m0.88(05)(07)$	\\
		$10^3 \cdot L_3^r$	&	$-3.03(11)(07)$	&	$-2.83(09)(07)$	&	$-2.99(11)(07)$	&	$-2.79(10)(07)$	&	$-2.99(11)(07)$	&	$-2.80(10)(07)$	\\
		\hline																																	\\[-0.3cm]	
		$10^3 \cdot L_9^r$	&	$\m4.70(40)(63)$	&	$\m4.64(39)(61)$	&	$\m4.51(43)(63)$	&	$\m4.44(43)(61)$	&	$\m4.52(43)(63)$	&	$\m4.45(43)(61)$	\\
		\hline																																	\\[-0.3cm]	
		$\chi^2$			&	100.9			&	133.3			&	86.1				&	116.8			&	98.0				&	128.8			\\
		dof				& 	101				&	113				&	101				&	113				&	103				&	115				\\
		$\chi^2/$dof		& 	1.0				&	1.2				&	0.9				&	1.0				&	1.0				&	1.1				\\
		\bottomrule
	\end{tabular}
	\caption{Fit results for the dispersion relation matched to \ChPT{} at NLO. The renormalisation scale is $\mu = 770$~MeV.}
	\label{tab:ResultsDispersionNLOMatching}
\end{table}

The fit results for $L_9^r$ are not in agreement with the determination of \cite{BijnensTalavera2002}, 
\begin{align}
	\label{eq:Kl4InputL9}
	10^3 \cdot L_9^r = 5.93(43).
\end{align}
Note that the influence of $L_9^r$ on $L_1^r$, $L_2^r$ and $L_3^r$ is
minimal: if $L_9^r$ is fixed to (\ref{eq:Kl4InputL9}), we find $10^3 \cdot
L_1^r = 0.51(02)(06)$, $10^3 \cdot L_2^r = 0.89(05)(07)$ and $10^3 \cdot
L_3^r = -2.82(10)(07)$.

While the final results for the LECs do not differ significantly in the
one-step and two-step strategies, a difference can be observed concerning
the soft-pion theorems. If we use the two-step matching strategy, the
soft-pion theorems are not automatically satisfied, but if they are imposed
as a fitting constraint, they can be perfectly satisfied with only a slight
increase of the $\chi^2$. In contrast, in the one-step strategy, where the
subtraction constants have to fulfil the chiral constraints, the accuracy
of the soft-pion theorems lies at $\sim4\%$ and $\sim10\%$
respectively. This does not change with the soft-pion constraints added to
the fit, which only increases the $\chi^2$ a bit.

The influence of the isospin-breaking corrections of \cite{Stoffer2014} is
about the size of the statistical uncertainty in the case of $L_1^r$ and
$L_2^r$, while $L_3^r$ is less sensitive to the isospin effects.

A plot of the data points indicates that the two experiments NA48/2 and
E865 are in agreement, which is confirmed by the fit results. We find it
worthwhile to stress that this is only the case if the normalisation of the
$F_s$ data points of NA48/2 is determined including the $s_\ell$-dependence
(for the values in the 10 published bins, this requires the normalisation
to be increased by $0.77\%$). If the published values are used, which are
normalised neglecting the $s_\ell$-dependence, a quite strong tension
between the two experiments is observed, resulting in higher $\chi^2$
values for combined fits.

\begin{figure}[H]
	\centering
	\scalebox{0.75}{
		\small
		\input{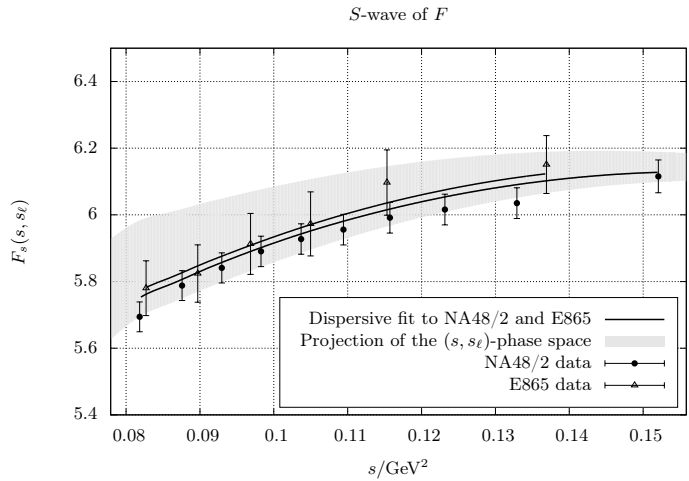}
		}
	\caption{Fit result for the $S$-wave of the form factor $F$. The dispersive description reproduces beautifully the curvature of the form factor. The $(s,s_\ell)$-phase space is projected on the $s$-axis, the plotted lines correspond to splines through the $(s,s_\ell)$-values of the two data sets.}
	\label{fig:FsDispersionRelation}
\end{figure}

\begin{figure}[H]
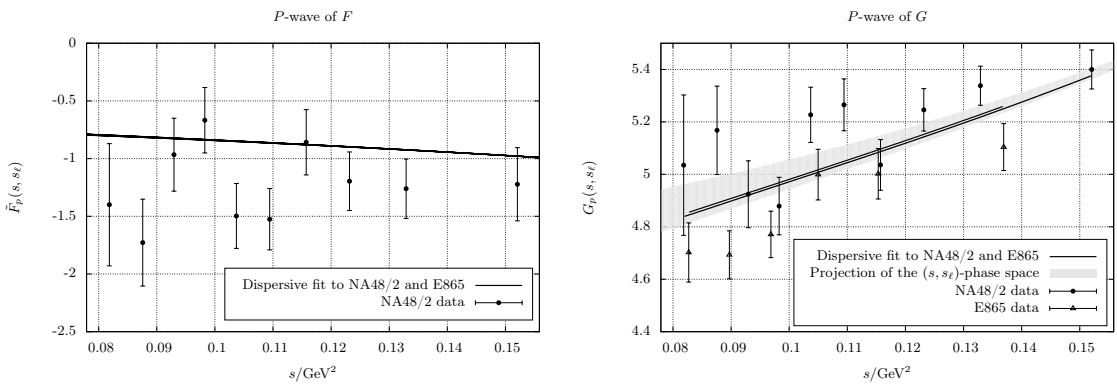

	\centering
	\scalebox{0.59}{
		\small
		\input{plots/FptDRMatching}
		\input{plots/GpDRMatching}
		}
	\caption{Fit results for the $P$-waves of the form factors $F$ and $G$. The $(s,s_\ell)$-phase space is again projected on the $s$-axis.}
	\label{fig:FptGpDispersionRelation}
\end{figure}

We note that the $\chi^2$ in the dispersive treatment is clearly improved
compared to the direct fit with \ChPT{} at NLO: in a fit to the
one-dimensional data in appendix~\ref{sec:AppendixData}, the $\chi^2$ of a
dispersive fit is 1.2 instead of 2.5 for the direct chiral fit (both with
27 dof). This is illustrated in figure~\ref{fig:FsDispersionRelation}: in
contrast to a pure chiral treatment, the dispersion relation allows to
describe the curvature of the $S$-wave of the form factor $F$. We interpret
this as the result of the resummation of final-state rescattering
effects. Figure~\ref{fig:FptGpDispersionRelation} shows the fitted
$P$-waves of $F$ and $G$.

\subsubsection{Matching at $\O(p^6)$}

We have seen that when using one-loop \ChPT{}, the dispersive treatment
clearly exhibits its powers, and the advantage over a pure chiral treatment
is evident: the dispersive representation is able to describe the
energy-dependence of the form factors, hence the $\chi^2$ of the fit to the
whole form factor data is much better. Due to the resummation of
final-state rescattering effects, we expect the dispersive representation
to capture the most important higher-order contributions and to render the
determination of the LECs more robust.

In combination with two-loop \ChPT{}, the treatment becomes more
difficult. The matching equations at NNLO relate the subtraction constants
to chiral expressions that contain the $\O(p^6)$ LECs $C_i^r$. The largest
obstacle in a chiral treatment at NNLO is the large number of poorly known
$C_i^r$. In the dispersive treatment with NNLO matching, the same problem
occurs. It turns out that the determination of the NLO LECs is still
strongly affected by the choice of the $C_i^r$, a situation known from
direct \ChPT{} fits \cite{Bijnens2012, Bijnens2014}.

In order to alleviate this problem, we note that not all choices of the
input $C_i^r$ lead to a good convergence of the chiral expansion. In our
dispersion relation, there appear nine subtraction constants, which are
gauge-dependent quantities. Since the gauge transformation
(\ref{eq:GaugeTransformation}) is described by three parameters, we can
find six gauge-invariant linear combinations of subtraction constants. For
these linear combinations, we require a good chiral convergence. We obtain
this by modifying the fitting procedure as follows.
\begin{itemize}
	\item We introduce 9 additional fitting parameters, corresponding
          to the contribution of the $C_i^r$ to the subtraction constants.
	\item We add to the $\chi^2$ nine observations of these parameters
          corresponding to the input values of the $C_i^r$ with a 50\%
          tolerance for the linear combinations of the $C_i^r$.
	\item We add to the $\chi^2$ six observations of the total
          $\O(p^6)$ correction to the gauge-invariant linear combinations
          of subtraction constants. The observation is zero $\pm5.6\%$ of
          the $\O(p^4)$ contribution (5.6\% corresponds to $M_\eta^2/(4\pi
          F_\pi)^2$).
\end{itemize}
With this setup, we are able to perform the NNLO matching with a reduced
dependence on the input values of the $C_i^r$. In
table~\ref{tab:ResultsDispersionNNLOMatching}, we present the matching
results at NNLO, using the `preferred values' of \cite{Bijnens2014} as
input for the $C_i^r$.
\begin{table}[H]
	\centering
	\begin{tabular}{c c c c c}
		\toprule
						&	NA48/2				&	NA48/2 \& E865	&	NA48/2						&	NA48/2 \& E865				\\[0.1cm]
		\hline																														\\[-0.3cm]	
		$10^3 \cdot L_1^r$	&	$\m0.82(16)(09)$		&	$\m0.69(16)(08)$	&	$\m0.93(17)(04)$		&	$\m0.78(17)(03)$		\\
		$10^3 \cdot L_2^r$	&	$\m0.71(10)(10)$		&	$\m0.63(09)(10)$	&	$\m1.11(17)(08)$		&	$\m0.97(17)(08)$		\\
		$10^3 \cdot L_3^r$	&	$-3.10(40)(27)$		&	$-2.63(39)(24)$	&	$-3.96(49)(14)$		&	$-3.38(48)(10)$		\\
		\hline																														\\[-0.3cm]	
		$10^3 \cdot L_9^r$	&	$\equiv5.93$			&	$\equiv5.93$		&	$\m8.36(87)(48)$				&	$\m8.05(86)(39)$				\\
		\hline																														\\[-0.3cm]	
		$\chi^2$			&	91.8				&	123.9				&	83.1							&	115.3						\\
		dof				& 	110				&	122					&	109							&	121							\\
		$\chi^2/$dof		& 	0.8				&	1.0					&	0.8							&	1.0							\\
		\bottomrule
	\end{tabular}
	\caption{Fit results for the dispersion relation matched to \ChPT{}
          at NNLO. The renormalisation scale is $\mu = 770$~MeV. As in
          table \ref{tab:NNLOThresholdFitsBE14}, we use lattice input for
          $L_4^r$, $L_5^r$, $L_6^r$ and $L_8^r$ \cite{MILC2009,Aoki2013},
          $L_7^r$ is the BE14 value \cite{Bijnens2014} and the input value
          for $L_9^r$ is taken from \cite{BijnensTalavera2002}.}
	\label{tab:ResultsDispersionNNLOMatching}
\end{table}
The fit results with $L_9^r$ taken as input are shown in the second and
third column of table~\ref{tab:ResultsDispersionNNLOMatching}. Here, the
corrections from NLO to NNLO matching for all three LECs are smaller than
the corrections between NLO and NNLO observed in direct \ChPT{} fits. The
larger uncertainties with respect to the NLO matching are explained by the
additional fitting parameters for the $C_i^r$ contribution to the
subtraction constants. If we take as input for the $C_i^r$ the
resonance estimate of \cite{Bijnens2012}, we obtain
$\{L_1^r, L_2^r, L_3^r\} = \{ 0.65, 0.26, -1.79 \} \cdot 10^{-3}$. With the $C_i^r$
input taken from the chiral quark model \cite{Jiang2010},
we find $\{L_1^r, L_2^r, L_3^r\} = \{ 0.49, 0.65, -2.44 \} \cdot 10^{-3}$.
We prefer the BE14 input values for the $C_i^r$, because they lead to the best
chiral convergence and the best $\chi^2$ of the fit.

The fit results change quite drastically if we include $L_9^r$ in the
fit. These fit results are shown in the fourth and fifth column of
table~\ref{tab:ResultsDispersionNNLOMatching}. In the matching equations at
NNLO, a stronger correlation between $L_9^r$ and the other LECs is
introduced due to their appearance in the $s_\ell$-dependence. At present,
alternative determinations of $L_9^r$ are clearly more reliable than this
one, and we therefore prefer here the fits with $L_9^r$ taken as
input. However, if the $s_\ell$-dependence of the form factors will be
measured in forthcoming experiments with even higher statistics, this could
provide a new reliable way to determine $L_9^r$.

\subsection{Error Analysis}

\label{sec:ErrorAnalysis}

In the following, we analyse the different sources of uncertainties in the
determination of the LECs using the NLO and NNLO matching. Let us give once
more the NLO and NNLO values for the LECs, obtained from the combined fits
to the NA48/2 and E865 data, where $L_9^r$ is taken as a fixed input
\cite{BijnensTalavera2002}:
\begin{table}[H]
	\centering
	\tabcolsep=0.15cm
	\begin{tabular}{c c c c c}
		\toprule
							&	NLO				&	NNLO			\\[0.1cm]
		\hline														\\[-0.3cm]	
		$10^3 \cdot L_1^r(\mu)$	&	$\m0.51(02)(06)$	&	$\m0.69(16)(08)$		\\
		$10^3 \cdot L_2^r(\mu)$	&	$\m0.89(05)(07)$	&	$\m0.63(09)(10)$		\\
		$10^3 \cdot L_3^r(\mu)$	&	$-2.82(10)(07)$	&	$-2.63(39)(24)$		\\
		\bottomrule
	\end{tabular}
	\caption{Matching results for the LECs at NLO and NNLO. The scale is $\mu=770$~MeV.}
	\label{tab:LirNLOvsNNLOResults}
\end{table}
The first error indicates the statistical one, i.e.~the error calculated in
the linear fit of the parameters. This error is due to the uncertainty of
the fitted data including isospin corrections. In the case of NNLO
matching, it includes also the uncertainty introduced with the 50\%
tolerance of the $C_i^r$ contribution to the subtraction constants. The
second error is due to the systematics of our approach. The corresponding
statistical and systematic correlations are shown in
tables~\ref{tab:LirCorrelationsNLO} and \ref{tab:LirCorrelationsNNLO}. 

\begin{figure}[H]
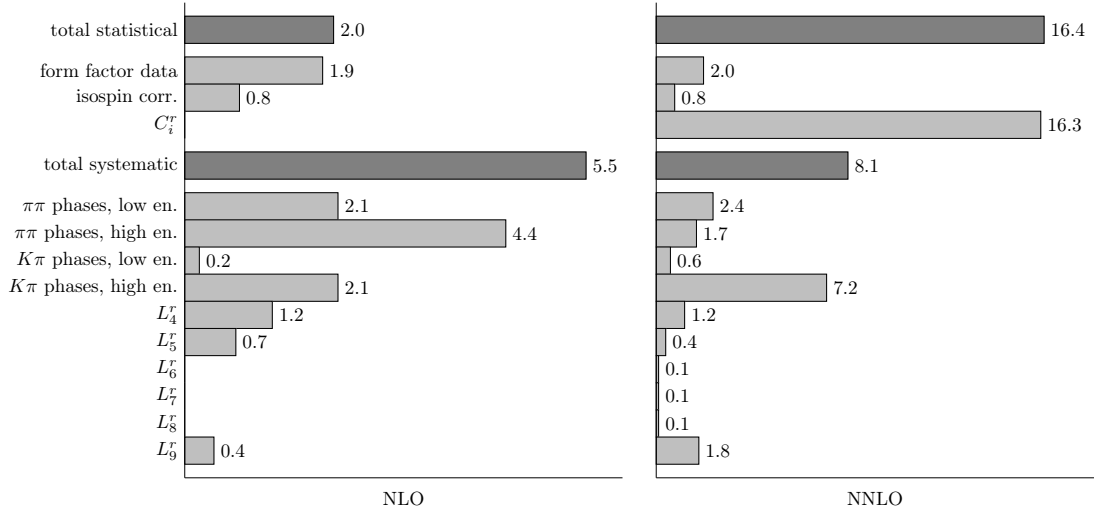

	\centering
	\renewcommand{\bcfontstyle}{\rmfamily}
	\scalebox{0.72}{
	\begin{bchart}[max=6.0,plain]
		\bcbar[label={total statistical},color=gray,value=2.0]{2.04}
		\smallskip
		\bcbar[label={form factor data},color=lightgray,value=1.9]{1.89}
		\bcbar[label={isospin corr.},color=lightgray,value=0.8]{0.75}
		\bcbar[label={$C_i^r$},color=lightgray]{}
		\smallskip
		\bcbar[label=total systematic,color=gray]{5.5}
		\smallskip
		\bcbar[label={$\pi\pi$ phases, low en.},color=lightgray]{2.1}
		\bcbar[label={$\pi\pi$ phases, high en.},color=lightgray]{4.4}
		\bcbar[label={$K\pi$ phases, low en.},color=lightgray]{0.2}
		\bcbar[label={$K\pi$ phases, high en.},color=lightgray]{2.1}
		\bcbar[label=$L_4^r$,color=lightgray]{1.2}
		\bcbar[label=$L_5^r$,color=lightgray]{0.7}
		\bcbar[label=$L_6^r$,color=lightgray]{}
		\bcbar[label=$L_7^r$,color=lightgray]{}
		\bcbar[label=$L_8^r$,color=lightgray]{}
		\bcbar[label=$L_9^r$,color=lightgray]{0.4}
		\bcxlabel{NLO}
	\end{bchart} }%
	\scalebox{0.72}{
	\begin{bchart}[max=18.5,plain]
		\bcbar[color=gray,value=16.4]{16.39}
		\smallskip
		\bcbar[color=lightgray]{2.0}
		\bcbar[color=lightgray,value=0.8]{0.78}
		\bcbar[color=lightgray,value=16.3]{16.25}
		\smallskip
		\bcbar[color=gray]{8.1}
		\smallskip
		\bcbar[color=lightgray]{2.4}
		\bcbar[color=lightgray]{1.7}
		\bcbar[color=lightgray]{0.6}
		\bcbar[color=lightgray]{7.2}
		\bcbar[color=lightgray]{1.2}
		\bcbar[color=lightgray]{0.4}
		\bcbar[color=lightgray]{0.1}
		\bcbar[color=lightgray]{0.1}
		\bcbar[color=lightgray]{0.1}
		\bcbar[color=lightgray]{1.8}
		\bcxlabel{NNLO}
	\end{bchart} }%
	\caption{Contributions to the uncertainty of $L_1^r$ in the $\O(p^4)$ and $\O(p^6)$ matching in units of $10^{-5}$.}
	\label{fig:ErrorsL1r}
\end{figure}

\begin{figure}[H]
	\centering
	\renewcommand{\bcfontstyle}{\rmfamily}
	\scalebox{0.72}{
	\begin{bchart}[max=7.5,plain]
		\bcbar[label={total statistical},color=gray,value=4.6]{4.61}
		\smallskip
		\bcbar[label={form factor data},color=lightgray,value=4.3]{4.27}
		\bcbar[label={isospin corr.},color=lightgray,value=1.7]{1.73}
		\bcbar[label={$C_i^r$},color=lightgray]{}
		\smallskip
		\bcbar[label=total systematic,color=gray]{6.5}
		\smallskip
		\bcbar[label={$\pi\pi$ phases, low en.},color=lightgray]{1.8}
		\bcbar[label={$\pi\pi$ phases, high en.},color=lightgray]{4.1}
		\bcbar[label={$K\pi$ phases, low en.},color=lightgray]{0.3}
		\bcbar[label={$K\pi$ phases, high en.},color=lightgray]{1.9}
		\bcbar[label=$L_4^r$,color=lightgray]{3.7}
		\bcbar[label=$L_5^r$,color=lightgray]{2.1}
		\bcbar[label=$L_6^r$,color=lightgray]{}
		\bcbar[label=$L_7^r$,color=lightgray]{}
		\bcbar[label=$L_8^r$,color=lightgray]{}
		\bcbar[label=$L_9^r$,color=lightgray]{0.3}
		\bcxlabel{NLO}
	\end{bchart} }%
	\scalebox{0.72}{
	\begin{bchart}[max=11.0,plain]
		\bcbar[color=gray]{9.5}
		\smallskip
		\bcbar[color=lightgray,value=3.7]{3.69}
		\bcbar[color=lightgray,value=1.4]{1.42}
		\bcbar[color=lightgray,value=8.6]{8.64}
		\smallskip
		\bcbar[color=gray]{9.6}
		\smallskip
		\bcbar[color=lightgray]{0.5}
		\bcbar[color=lightgray]{4.9}
		\bcbar[color=lightgray]{0.1}
		\bcbar[color=lightgray]{0.5}
		\bcbar[color=lightgray]{0.8}
		\bcbar[color=lightgray]{4.5}
		\bcbar[color=lightgray]{0.1}
		\bcbar[color=lightgray]{0.2}
		\bcbar[color=lightgray]{0.1}
		\bcbar[color=lightgray]{6.9}
		\bcxlabel{NNLO}
	\end{bchart} }%
	\caption{Contributions to the uncertainty of $L_2^r$ in the $\O(p^4)$ and $\O(p^6)$ matching in units of $10^{-5}$.}
	\label{fig:ErrorsL2r}
\end{figure}

\begin{figure}[H]
	\centering
	\renewcommand{\bcfontstyle}{\rmfamily}
	\scalebox{0.72}{
	\begin{bchart}[max=11.0,plain]
		\bcbar[label={total statistical},color=gray,value=9.5]{9.55}
		\smallskip
		\bcbar[label={form factor data},color=lightgray,value=9.0]{8.99}
		\bcbar[label={isospin corr.},color=lightgray,value=3.2]{3.23}
		\bcbar[label={$C_i^r$},color=lightgray]{}
		\smallskip
		\bcbar[label=total systematic,color=gray]{7.0}
		\smallskip
		\bcbar[label={$\pi\pi$ phases, low en.},color=lightgray]{1.5}
		\bcbar[label={$\pi\pi$ phases, high en.},color=lightgray]{1.3}
		\bcbar[label={$K\pi$ phases, low en.},color=lightgray]{0.4}
		\bcbar[label={$K\pi$ phases, high en.},color=lightgray]{5.0}
		\bcbar[label=$L_4^r$,color=lightgray]{0.7}
		\bcbar[label=$L_5^r$,color=lightgray]{4.2}
		\bcbar[label=$L_6^r$,color=lightgray]{}
		\bcbar[label=$L_7^r$,color=lightgray]{}
		\bcbar[label=$L_8^r$,color=lightgray]{}
		\bcbar[label=$L_9^r$,color=lightgray]{0.9}
		\bcxlabel{NLO}
	\end{bchart} }%
	\scalebox{0.72}{
	\begin{bchart}[max=45.0,plain]
		\bcbar[color=gray]{39.3}
		\smallskip
		\bcbar[color=lightgray,value=11.0]{11.04}
		\bcbar[color=lightgray,value=3.6]{3.57}
		\bcbar[color=lightgray,value=37.6]{37.55}
		\smallskip
		\bcbar[color=gray]{23.7}
		\smallskip
		\bcbar[color=lightgray]{2.8}
		\bcbar[color=lightgray]{5.9}
		\bcbar[color=lightgray]{1.2}
		\bcbar[color=lightgray]{14.9}
		\bcbar[color=lightgray]{2.9}
		\bcbar[color=lightgray]{8.7}
		\bcbar[color=lightgray]{0.2}
		\bcbar[color=lightgray]{0.0}
		\bcbar[color=lightgray]{0.3}
		\bcbar[color=lightgray]{14.7}
		\bcxlabel{NNLO}
	\end{bchart} }%
	\caption{Contributions to the uncertainty of $L_3^r$ in the $\O(p^4)$ and $\O(p^6)$ matching in units of $10^{-5}$.}
	\label{fig:ErrorsL3r}
\end{figure}

\clearpage

\begin{table}[H]
	\centering
	\tabcolsep=0.15cm
	\begin{tabular}{r | c c}
		stat.~corr. & $L_2^r$ & $\hphantom{-}L_3^r$ \\
		\hline
		$L_1^r$ & $0.49$ & $-0.72$ \\
		$L_2^r$ & & $-0.95$ \\	
	\end{tabular}
	\hspace{0.5cm}
	\begin{tabular}{r | c c}
		syst.~corr. & $\hphantom{-}L_2^r$ & $\hphantom{-}L_3^r$ \\
		\hline
		$L_1^r$ & $-0.69$ & $-0.31$ \\
		$L_2^r$ & & $-0.29$ \\	
	\end{tabular}
	\caption{Statistical and systematic correlations of the fitted LECs at NLO.}
	\label{tab:LirCorrelationsNLO}
\end{table}

\begin{table}[H]
	\centering
	\tabcolsep=0.15cm
	\begin{tabular}{r | c c}
		stat.~corr. & $L_2^r$ & $\hphantom{-}L_3^r$ \\
		\hline
		$L_1^r$ & $0.31$ & $-0.32$ \\
		$L_2^r$ & & $-0.84$ \\	
	\end{tabular}
	\hspace{0.5cm}
	\begin{tabular}{r | c c}
		syst.~corr. & $\hphantom{-}L_2^r$ & $\hphantom{-}L_3^r$ \\
		\hline
		$L_1^r$ & $0.23$ & $-0.83$ \\
		$L_2^r$ & & $-0.70$ \\	
	\end{tabular}
	\caption{Statistical and systematic correlations of the fitted LECs at NNLO.}
	\label{tab:LirCorrelationsNNLO}
\end{table}

Figures~\ref{fig:ErrorsL1r}, \ref{fig:ErrorsL2r} and \ref{fig:ErrorsL3r} show bar charts of the uncertainties of the LECs. The fractional uncertainties are summed in squares and determined as follows.
\begin{itemize}
	\item The uncertainty due to the $K_{\ell4}$ form factor data is the statistical uncertainty of a fit where no isospin corrections are included and the $C_i^r$ contributions are fixed to the fitted values.
	\item The uncertainty due to isospin corrections is the difference in squares of the statistical uncertainties of fits to data with and without isospin corrections, again with the $C_i^r$ contributions fixed to the fitted values.
	\item The uncertainty due to the $C_i^r$ is the difference in squares of the statistical uncertainties of fits to isospin corrected data with the $C_i^r$ contributions either fitted or fixed to the fitted values.
	\item For the $\pi\pi$ phases \cite{Ananthanarayan2001a, Caprini2012}, we vary all the 28 parameters and sum the variations of the LECs in squares. In the bar charts, this is the uncertainty labelled by `$\pi\pi$ phases, low energy'.
	\item The next fractional uncertainty is due to the high-energy behaviour of the $\pi\pi$ phases. We sum in squares the differences between the high-energy solutions explained in section~\ref{sec:Kl4pipiPhaseShifts}.
	\item The $K\pi$ phases are simply varied between the centre and upper/lower limit of the error bands. This influence is labelled as `$K\pi$ phases, low energy'.
	\item The uncertainty due to the high-energy behaviour of the $K\pi$ phases is estimated with the two solutions for each of the $K\pi$ phases as explained in section~\ref{sec:Kl4KpiPhaseShifts}.
	\item The input LECs are varied by their uncertainties given in (\ref{eq:Kl4InputLECsNLOMatching}) and (\ref{eq:Kl4InputL9}).
	\item We have checked that the numerical uncertainties due to the discretisation, interpolation and numerical integration of the functions as well as the iteration procedure are completely negligible.
\end{itemize}

We note that at NLO, the largest contribution to the systematic errors
comes from the high-energy behaviour of the phase shifts, either from the
$\pi\pi$ phases in the case of $L_1^r$ and $L_2^r$ or the $K\pi$ phases in
the case of $L_3^r$. The uncertainties due to the low-energy
parametrisation of the phases are small. The uncertainty due to the input
LEC $L_9^r$ is very small as well.

At NNLO, the high-energy behaviour of the phases is again a large
contribution to the uncertainty. $L_9^r$ has now a large impact on the
uncertainty of $L_2^r$. The additional LECs $L_6^r$, $L_7^r$ and $L_8^r$
have almost no influence on the uncertainty. The largest uncertainty is due
to the fitted contribution of the $C_i^r$, which is part of the statistical
uncertainty.

\section{Conclusion and Outlook}

\label{sec:ConclusionOutlook}

We have presented a new dispersive treatment of $K_{\ell4}$ decays, which
provides a very accurate description of the hadronic form factors $F$ and
$G$. The dispersion relation is valid up to and including $\O(p^6)$ in the
chiral counting. Furthermore, it provides a resummation of final-state
$\pi\pi$- and $K\pi$-rescattering effects, which we believe to be the most
important contribution beyond $\O(p^6)$.

Our dispersion relation for $K_{\ell4}$ is written in the form of an Omnès
representation. It consists of a set of coupled integral equations. We have
solved this system numerically with an iterative procedure. The solutions
are parametrised by subtraction constants, which we have determined in a
fit to data and by using the soft-pion theorem as well as chiral input. In
contrast to a pure chiral description, the dispersion relation describes
perfectly the experimentally observed curvature of the $S$-wave of the form
factor $F$, which we interpret as a result of significant
$\pi\pi$-rescattering effects. This is yet another case in which
high-precision data clearly call for effects which go even beyond NNLO in
\ChPT{}. These effects only concern the momentum dependence of the form
factors: we see no sign that quark mass dependence beyond NNLO is required
by data.

By using the matching equations to \ChPT{} we have extracted the values of
the low-energy constants $L_1^r$, $L_2^r$ and $L_3^r$. The correction from
NLO to NNLO, when matching the chiral and dispersive representations and
fitting the latter to the data are smaller than the corrections from NLO to
NNLO observed in direct \ChPT{} fits. Constraints on the chiral convergence
of the subtraction constants allow us to reduce the dependence on the input
values for the $C_i^r$. Still, the poorly known values of the $C_i^r$ are
responsible for the larger uncertainties in the matching at NNLO.

Our results for the LECs obtained by matching \ChPT{} at NLO are:
\begin{align}
L_1^r = 0.51(06)\cdot 10^{-3} , \qquad L_2^r = 0.89(09) \cdot 10^{-3} ,\qquad
L_3^r = -2.82(12)\cdot 10^{-3} , \end{align}
whereas the matching at NNLO gives
\begin{align}
L_1^r = 0.69(18) \cdot 10^{-3} , \qquad L_2^r = 0.63(13) \cdot 10^{-3} , \qquad
L_3^r = -2.63(46) \cdot 10^{-3} . \end{align}
The two-dimensional NA48/2 data set for the $S$-wave of $F$, which shows
both the $s$- as well as the $s_\ell$-dependence, has allowed us
to extract a value for $L_9^r$, which is roughly compatible with previous
determinations. In accuracy, however, it cannot yet compete, as it reflects
the low precision in the measurement of the $s_\ell$-dependence of $F$. The
determination of $L_9^r$ is also quite strongly dependent on whether the
matching is done at NLO or NNLO.

\section*{Acknowledgements}
\addcontentsline{toc}{section}{Acknowledgements}

We cordially thank Brigitte Bloch-Devaux, Stefan Pislak, Peter Truöl and Andries van der Schaaf for providing additional data from the NA48/2 and E865 experiments and for many helpful discussions on the experiments and the data analysis.
We are grateful to Hans Bijnens and Ilaria Jemos for their support with the two-loop implementation of the form factors.
We thank Jürg Gasser, Bastian Kubis, Stefan Lanz and Heiri Leutwyler for many interesting and valuable discussions and Gerhard Ecker for useful comments on the manuscript.
PS thanks the Swiss National Science Foundation for a mobility grant. EP and PS are grateful to the Los Alamos National Laboratory, where part of this work was carried out.

Financial support by the Swiss National Science Foundation, the DFG (CRC 16, ``Subnuclear Structure of Matter'') and the U.S.~Department of Energy (contract DEAC05-06OR23177) is gratefully acknowledged.

\begin{appendices}
	\numberwithin{equation}{section}


\section{Scalar Loop Functions}

\label{sec:AppendixScalarLoopFunctions}

We use the following conventions for the scalar one-loop functions in $n$ dimensions:
\begin{align}
	\begin{split}
		A_0(m^2) &:= \frac{1}{i} \int \frac{d^nq}{(2\pi)^n} \frac{1}{[ q^2 - m^2 ]} , \\
		B_0(p^2, m_1^2, m_2^2) &:= \frac{1}{i} \int \frac{d^nq}{(2\pi)^n} \frac{1}{[ q^2 - m_1^2 ] [ (q+p)^2 - m_2^2 ]} .
	\end{split}
\end{align}
These loop functions are UV-divergent. We define the renormalised loop functions in the $\overline{MS}$ scheme:
\begin{align}
	\begin{split}
		A_0(m^2) &= -2 m^2 \lambda + \bar A_0(m^2) + \O(4-n) , \\
		B_0(p^2, m_1^2, m_2^2) &= -2\lambda + \bar B_0(p^2, m_1^2, m_2^2) + \O(4-n) ,
	\end{split}
\end{align}
where
\begin{align}
	\begin{split}
		\lambda = \frac{\mu^{n-4}}{16\pi^2} \left( \frac{1}{n-4} - \frac{1}{2} \left( \ln(4\pi) + 1 - \gamma_E \right) \right) .
	\end{split}
\end{align}
$\mu$ denotes the renormalisation scale.

The renormalised loop functions are given by \cite{Amoros2000}
\begin{align}
	\begin{split}
		\bar A_0(m^2) &= -\frac{m^2}{16\pi^2} \ln\left( \frac{m^2}{\mu^2} \right) , \\
		\bar B_0(p^2, m_1^2, m_2^2) &= -\frac{1}{16\pi^2} \frac{m_1^2 \ln\left(\frac{m_1^2}{\mu^2}\right) - m_2^2 \ln\left(\frac{m_2^2}{\mu^2}\right)}{m_1^2 - m_2^2} \\
			&+ \frac{1}{32\pi^2} \left( 2 + \left( -\frac{\Delta}{p^2} + \frac{\Sigma}{\Delta} \right) \ln\left( \frac{m_1^2}{m_2^2} \right) - \frac{\nu}{p^2} \ln\left( \frac{(p^2+\nu)^2 - \Delta^2}{(p^2-\nu)^2 - \Delta^2} \right) \right) ,
	\end{split}
\end{align}
where
\begin{align}
	\begin{split}
		\Delta &:= m_1^2 - m_2^2 , \\
		\Sigma &:= m_1^2 + m_2^2 , \\
		\nu &:= \lambda^{1/2}(s,m_1^2,m_2^2) .
	\end{split}
\end{align}

The renormalised two-point function fulfils a once-subtracted dispersion relation:
\begin{align}
	\begin{split}
		\bar B_0(s, m_1^2, m_2^2) &= \bar B_0(0, m_1^2, m_2^2) + \frac{s}{\pi} \int_{(m_1+m_2)^2}^\infty \frac{\Im \bar B_0(s^\prime,m_1^2,m_2^2)}{(s^\prime - s - i\epsilon)s^\prime} ds^\prime ,
	\end{split}
\end{align}
where the imaginary part is given by
\begin{align}
	\begin{split}
		\Im \bar B_0(s, m_1^2, m_2^2) = \frac{1}{16 \pi} \frac{\lambda^{1/2}(s,m_1^2,m_2^2)}{s}
	\end{split}
\end{align}
and the value at $s=0$ is
\begin{align}
	\begin{split}
		B_0(0,m_1^2,m_2^2) =  -\frac{1}{16\pi^2} \frac{m_1^2 \ln\left(\frac{m_1^2}{\mu^2}\right) - m_2^2 \ln\left(\frac{m_2^2}{\mu^2}\right)}{m_1^2 - m_2^2} .
	\end{split}
\end{align}
The first and second derivative at $s=0$ are
\begin{align}
	\begin{split}
		B_0^\prime(0,m_1^2,m_2^2) &= \frac{1}{32\pi^2} \frac{\Delta \Sigma - 2 m_1^2 m_2^2 \ln\left(\frac{m_1^2}{m_2^2}\right)}{\Delta^3} , \\
		B_0^\dprime(0,m_1^2,m_2^2) &= \frac{1}{48\pi^2} \frac{\Delta(m_1^4 + 10 m_1^2 m_2^2 + m_2^4) - 6 m_1^2 m_2^2 \Sigma \ln\left(\frac{m_1^2}{m_2^2}\right)}{\Delta^5} .
	\end{split}
\end{align}


\section{Kinematics}

For each channel, the partial-wave expansion is performed in the corresponding rest frame, i.e.~in the $\pi\pi$ centre-of-mass frame for the $s$-channel and in one of the $K\pi$ centre-of-mass frames for the $t$- and $u$-channel. Therefore, we work out explicitly the kinematics in the three different frames.

\subsection{Legendre Polynomials and Spherical Harmonics}

\label{sec:LegendrePolynomials}

For the partial-wave-expansions, we make use of several relations between spherical harmonics and Legendre polynomials.

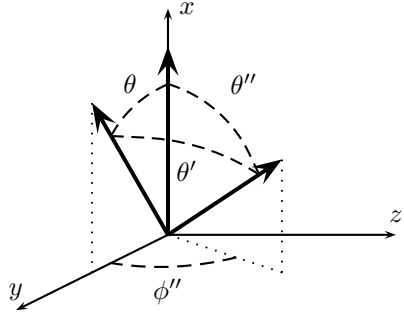
\begin{figure}[ht]
	\centering
	\psset{unit=1cm}
	\begin{pspicture}(-3,-3)(3,3)
		\psline{->}(0,0)(0,3)
		\psline{->}(0,0)(3,0)
		\psline{->}(0,0)(-2,-1)
		\psline[linewidth=1.5pt, arrowsize=8pt]{->}(0,0)(0,2.5)
		\psline[linewidth=1.5pt, arrowsize=8pt]{->}(0,0)(-1,1.75)
		\psline[linewidth=1.5pt, arrowsize=8pt]{->}(0,0)(1.5,1)
		\psline[linestyle=dotted](1.5,1)(1.5,-0.5)
		\psline[linestyle=dotted](0,0)(1.5,-0.5)
		\psline[linestyle=dotted](-1,1.75)(-1,-0.5)
		\psline[linearc=1.75,linestyle=dashed](0,2)(0.75,1.75)(1.2,0.8)
		\psline[linearc=1.5,linestyle=dashed](0,2)(-0.5,1.75)(-0.75,1.31)
		\psline[linearc=3,linestyle=dashed](-0.75,1.31)(0.5,1.25)(1.2,0.8)
		\psline[linearc=4,linestyle=dashed](-0.75,-0.375)(0,-0.5)(0.9,-0.3)
		\rput(0.25,3){$x$}
		\rput(-2,-0.75){$y$}
		\rput(3,0.25){$z$}
		\rput(-0.5,2){$\theta$}
		\rput(0.25,0.85){$\theta^\prime$}
		\rput(1,2){$\theta^\dprime$}
		\rput(0,-0.75){$\phi^\dprime$}
	\end{pspicture}
	\caption{Vectors and angles appearing in the addition theorem for spherical harmonics}
	\label{img:AnglesSphericalHarmonics}
\end{figure}

We use the addition theorem for the spherical harmonics and the relations between Legendre polynomials or derivatives of Legendre polynomials to spherical harmonics:
\begin{align}
	P_l(\cos\theta^\prime) &= \frac{4\pi}{2l+1} \sum_{m=-l}^l Y_l^m(\theta,0) {Y_l^m}^*(\theta^\dprime,\phi^\dprime) , \\
	P_{l^\prime}(\cos\theta^\dprime) &= \sqrt{\frac{4\pi}{2l^\prime+1}} Y_{l^\prime}^0(\theta^\dprime,\phi^\dprime) \qquad \text{(for any $\phi^\dprime$)} , \\
	P_{l^\prime}^\prime(\cos\theta^\dprime) \sin\theta^\dprime &= (-1)\sqrt{\frac{4\pi}{2l^\prime+1}} \sqrt{\frac{(l^\prime+1)!}{(l^\prime-1)!}} \, {Y_{l^\prime}^1}^*(\theta^\dprime,\phi^\dprime) e^{i\phi^\dprime} ,
\end{align}
where $P_l^\prime(z) := \frac{d}{dz}P_l(z)$. The different angles are defined in figure~\ref{img:AnglesSphericalHarmonics}.

We can now easily derive the addition theorem for the Legendre polynomials:
\begin{align}
	\begin{split}
		&\int d\Omega^\dprime P_l(\cos\theta^\prime) P_{l^\prime}(\cos\theta^\dprime) \\
		 &= \int d\Omega^\dprime \frac{4\pi}{2l+1}\sum_{m=-l}^l Y_l^m(\theta,0) {Y_l^m}^*(\theta^\dprime,\phi^\dprime) \sqrt{\frac{4\pi}{2l^\prime+1}} Y_{l^\prime}^0(\theta^\dprime,\phi^\dprime) \\
		 &= \sum_{m=-l}^l Y_l^m(\theta,0) \frac{4\pi}{2l+1} \sqrt{\frac{4\pi}{2l^\prime+1}} \underbrace{\int d\Omega^\dprime {Y_l^m}^*(\theta^\dprime,\phi^\dprime) Y_{l^\prime}^0(\theta^\dprime,\phi^\dprime)}_{\delta_{ll^\prime}\delta_{m0}} \\
		 &= \delta_{ll^\prime} \frac{4\pi}{2l+1} \sqrt{\frac{4\pi}{2l+1}} \, Y_l^0(\theta,0) = \delta_{ll^\prime} \frac{4\pi}{2l+1} P_l(\cos\theta) ,
	\end{split}
\end{align}
as well as the following relation:
\begin{align}
	\begin{split}
		&\int d\Omega^\dprime P_l(\cos\theta^\prime) P_{l^\prime}^\prime(\cos\theta^\dprime) \sin\theta^\dprime e^{-i \phi^\dprime} \\
		 &= \int d\Omega^\dprime \frac{4\pi}{2l+1}\sum_{m=-l}^l {Y_l^m}^*(\theta,0) Y_l^m(\theta^\dprime,\phi^\dprime) (-1) \sqrt{\frac{4\pi}{2l^\prime+1}} \sqrt{\frac{(l^\prime+1)!}{(l^\prime-1)!}} \, {Y_{l^\prime}^1}^*(\theta^\dprime,\phi^\dprime) \\
		 &= \sum_{m=-l}^l {Y_l^m}^*(\theta,0) \frac{4\pi}{2l+1} (-1) \sqrt{\frac{4\pi}{2l^\prime+1}} \sqrt{\frac{(l^\prime+1)!}{(l^\prime-1)!}} \underbrace{\int d\Omega^\dprime Y_l^m(\theta^\dprime,\phi^\dprime) {Y_{l^\prime}^1}^*(\theta^\dprime,\phi^\dprime)}_{\delta_{ll^\prime}\delta_{m1}} \\
		 &= \delta_{ll^\prime} \frac{4\pi}{2l+1} (-1) \sqrt{\frac{4\pi}{2l+1}} \sqrt{\frac{(l+1)!}{(l-1)!}} \, {Y_l^1}^*(\theta,0) = \delta_{ll^\prime} \frac{4\pi}{2l+1} P_l^\prime(\cos\theta) \sin\theta .
	\end{split}
\end{align}
Since the right-hand side is real, we conclude that
\begin{align}
	\begin{split}
		\int d\Omega^\dprime P_l(\cos\theta^\prime) P_{l^\prime}^\prime(\cos\theta^\dprime) \sin\theta^\dprime \cos\phi^\dprime &= \delta_{ll^\prime} \frac{4\pi}{2l+1} P_l^\prime(\cos\theta) \sin\theta , \\
		\int d\Omega^\dprime P_l(\cos\theta^\prime) P_{l^\prime}^\prime(\cos\theta^\dprime) \sin\theta^\dprime \sin\phi^\dprime &= 0 .
	\end{split}
\end{align}

\subsection{Kinematics in the $s$-Channel}

In the $\pi\pi$ centre-of-mass frame, the four-momenta of the different particles take the following values:
\begin{align}
	\begin{aligned}
		k &= \left( \sqrt{M_K^2 + \vec k^2}, \vec k \right) , \; & q_1 &= \left( \sqrt{ M_\pi^2 + \vec q^2 }, \vec q \right) , \; & p_1 &= \left( \sqrt{ M_\pi^2 + \vec p^2 }, \vec p \right) , \\
		-L &= \left( -\sqrt{ s_\ell + \vec k^2}, -\vec k \right) , \; & q_2 &= \left( \sqrt{ M_\pi^2 + \vec q^2 }, -\vec q \right) , \; & p_2 &= \left( \sqrt{ M_\pi^2 + \vec p^2 }, -\vec p \right) ,
	\end{aligned}
\end{align}
where $q_1$ and $q_2$ will be the momenta of intermediate pions. Note that we choose here the decay region ($L^0$ is positive), but could have equally well chosen the scattering region.

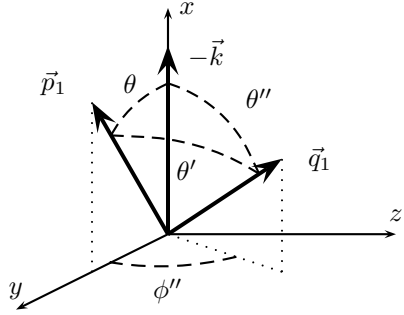
\begin{figure}[ht]
	\centering
	\psset{unit=1cm}
	\begin{pspicture}(-3,-3)(3,3)
		\psline{->}(0,0)(0,3)
		\psline{->}(0,0)(3,0)
		\psline{->}(0,0)(-2,-1)
		\psline[linewidth=1.5pt, arrowsize=8pt]{->}(0,0)(0,2.5)
		\psline[linewidth=1.5pt, arrowsize=8pt]{->}(0,0)(-1,1.75)
		\psline[linewidth=1.5pt, arrowsize=8pt]{->}(0,0)(1.5,1)
		\psline[linestyle=dotted](1.5,1)(1.5,-0.5)
		\psline[linestyle=dotted](0,0)(1.5,-0.5)
		\psline[linestyle=dotted](-1,1.75)(-1,-0.5)
		\psline[linearc=1.75,linestyle=dashed](0,2)(0.75,1.75)(1.2,0.8)
		\psline[linearc=1.5,linestyle=dashed](0,2)(-0.5,1.75)(-0.75,1.31)
		\psline[linearc=3,linestyle=dashed](-0.75,1.31)(0.5,1.25)(1.2,0.8)
		\psline[linearc=4,linestyle=dashed](-0.75,-0.375)(0,-0.5)(0.9,-0.3)
		\rput(0.25,3){$x$}
		\rput(-2,-0.75){$y$}
		\rput(3,0.25){$z$}
		\rput(-0.5,2){$\theta$}
		\rput(0.25,0.85){$\theta^\prime$}
		\rput(1.2,1.8){$\theta^\dprime$}
		\rput(0,-0.75){$\phi^\dprime$}
		\rput(0.5,2.4){$-\vec k$}
		\rput(-1.5,2){$\vec p_1$}
		\rput(2,1){$\vec q_1$}
	\end{pspicture}
	\caption{Vectors and angles in the $s$-channel centre-of-mass frame}
	\label{img:AnglesSChannel}
\end{figure}

Inserting these expressions into $s = (k - L)^2 = (q_1 + q_2)^2 = (p_1 + p_2)^2$ gives the values of $\vec k^2$, $\vec q^2$ and $\vec p^2$. We choose the directions of the three-vectors according to figure~\ref{img:AnglesSChannel}, i.e.~the angles are defined as $\theta := \angle(-\vec k, \vec p_1)$, $\theta^\prime := \angle(\vec p_1, \vec q_1)$, $\theta^\dprime := \angle(-\vec k, \vec q_1)$.

We end up with the following explicit expressions for the four-vectors:
\begin{align}
	\begin{aligned}
		k &= \left( \frac{M_K^2+s-s_\ell}{2\sqrt{s}}, -\frac{\lambda^{1/2}_{K\ell}(s)}{2\sqrt{s}}, 0, 0 \right) , \\
		L &= \left( \frac{M_K^2-s-s_\ell}{2\sqrt{s}}, -\frac{\lambda^{1/2}_{K\ell}(s)}{2\sqrt{s}}, 0, 0 \right) , \\
		q_1 &= \left( \frac{\sqrt{s}}{2}, \sqrt{\frac{s}{4}-M_\pi^2} \cos\theta^\dprime, \sqrt{\frac{s}{4}-M_\pi^2} \sin\theta^\dprime \cos\phi^\dprime, \sqrt{\frac{s}{4}-M_\pi^2} \sin\theta^\dprime \sin\phi^\dprime \right) , \\
		q_2 &= \left( \frac{\sqrt{s}}{2}, -\sqrt{\frac{s}{4}-M_\pi^2} \cos\theta^\dprime, -\sqrt{\frac{s}{4}-M_\pi^2} \sin\theta^\dprime \cos\phi^\dprime, -\sqrt{\frac{s}{4}-M_\pi^2} \sin\theta^\dprime \sin\phi^\dprime \right) , \\
		p_1 &= \left( \frac{\sqrt{s}}{2}, \sqrt{\frac{s}{4}-M_\pi^2} \cos\theta, \sqrt{\frac{s}{4}-M_\pi^2} \sin\theta, 0 \right) , \\
		p_2 &= \left( \frac{\sqrt{s}}{2}, -\sqrt{\frac{s}{4}-M_\pi^2} \cos\theta, -\sqrt{\frac{s}{4}-M_\pi^2} \sin\theta, 0 \right) ,
	\end{aligned}
\end{align}
where $\lambda_{K\ell}(s) := \lambda(M_K^2, s_\ell, s)$. Note that $\lambda_{K\ell}^{1/2}(s)$ has a square root branch cut in the $s$-plane between $(M_K-\sqrt{s_\ell})^2$ and $(M_K+\sqrt{s_\ell})^2$ and changes sign when we continue it analytically to the scattering region. We will have to pay attention that we do not introduce this kinematic singularity into the partial-wave expansion.

In order to express the $s$-channel scattering angle $\theta$ with the Mandelstam variables, we calculate:
\begin{align}
	\begin{split}
		t - u &= (k-p_1)^2 - (k-p_2)^2 = k^2 + p_1^2 - 2kp_1 - k^2 - p_2^2 + 2kp_2 \\
			&= 2k(p_2-p_1) = 2(k^0 (p_2^0 - p_1^0) - \vec k \cdot (\vec p_2 - \vec p_1) ) \\
			&= \frac{\lambda^{1/2}_{K\ell}(s)}{\sqrt{s}} \left(-2\sqrt{\frac{s}{4}-M_\pi^2}\cos\theta \right) = -\lambda_{K\ell}^{1/2}(s) \sqrt{1-\frac{4M_\pi^2}{s}} \cos\theta \\
			&= - 2 X(s) \sigma_\pi(s) \cos\theta ,
	\end{split}
\end{align}
hence
\begin{align}
	\cos\theta = \frac{u-t}{2X\sigma_\pi} ,
\end{align}
where $\sigma_\pi(s) := \sqrt{1-4M_\pi^2/s}$ and $X(s) = \frac{1}{2}\lambda^{1/2}_{K\ell}(s)$ as before.

\subsection{Kinematics in the $t$-Channel}

In the $t$-channel, we are in the $K\pi$ centre-of-mass frame and look at the $t$-channel scattering region:
\begin{align}
	\begin{aligned}
		k &= \left( \sqrt{M_K^2 + \vec k^2}, \vec k \right) , \; & q_K &= \left( \sqrt{ M_K^2 + \vec q_K^2 }, \vec q_K \right) , \; & p_2 &= \left( \sqrt{ M_\pi^2 + \vec p_2^2 }, \vec p_2 \right) , \\
		-p_1 &= \left( \sqrt{ M_\pi^2 + \vec k^2}, -\vec k \right) , \; & q_\pi &= \left( \sqrt{ M_\pi^2 + \vec q_K^2 }, -\vec q_K \right) , \; & L &= \left( \sqrt{ s_\ell + \vec p_2^2 }, -\vec p_2 \right) .
	\end{aligned}
\end{align}

Inserting these expressions into $t = (k - p_1)^2 = (q_K + q_\pi)^2 = (p_2 + L)^2$ gives the values of $\vec k^2$, $\vec q_K^2$ and $\vec p_2^2$. We choose the directions of the three-vectors according to figure~\ref{img:AnglesTChannel}, i.e.~the angles are defined as $\theta_t := \angle(-\vec k, \vec p_2)$, $\theta_t^\prime := \angle(\vec k, \vec q_K)$, $\theta_t^\dprime := \angle(-\vec q_K, \vec p_2)$.

\begin{figure}[ht]
	\centering
	\psset{unit=1cm}
	\begin{pspicture}(-3,-3)(3,3)
		\psline{->}(0,0)(0,3)
		\psline{->}(0,0)(3,0)
		\psline{->}(0,0)(-2,-1)
		\psline[linewidth=1.5pt, arrowsize=8pt]{->}(0,0)(0,2.5)
		\psline[linewidth=1.5pt, arrowsize=8pt]{->}(0,0)(-1,1.75)
		\psline[linewidth=1.5pt, arrowsize=8pt]{->}(0,0)(1.5,1)
		\psline[linestyle=dotted](1.5,1)(1.5,-0.5)
		\psline[linestyle=dotted](0,0)(1.5,-0.5)
		\psline[linestyle=dotted](-1,1.75)(-1,-0.5)
		\psline[linearc=1.75,linestyle=dashed](0,2)(0.75,1.75)(1.2,0.8)
		\psline[linearc=1.5,linestyle=dashed](0,2)(-0.5,1.75)(-0.75,1.31)
		\psline[linearc=3,linestyle=dashed](-0.75,1.31)(0.5,1.25)(1.2,0.8)
		\psline[linearc=4,linestyle=dashed](-0.75,-0.375)(0,-0.5)(0.9,-0.3)
		\rput(0.25,3){$x$}
		\rput(-2,-0.75){$y$}
		\rput(3,0.25){$z$}
		\rput(-0.5,2){$\theta_t$}
		\rput(0.25,0.85){$\theta_t^\prime$}
		\rput(1.2,1.8){$\theta_t^\dprime$}
		\rput(0,-0.75){$\phi_t^\dprime$}
		\rput(0.5,2.4){$\vec p_2$}
		\rput(-1.5,2){$-\vec k$}
		\rput(2,1){$-\vec q_K$}
	\end{pspicture}
	\caption{Vectors and angles in the $t$-channel centre-of-mass frame}
	\label{img:AnglesTChannel}
\end{figure}

We find the following results:
\begin{align}
	\begin{aligned}
		k &= \left( \frac{t + M_K^2-M_\pi^2}{2\sqrt{t}}, -\frac{\lambda^{1/2}_{K\pi}(t)}{2\sqrt{t}} \cos\theta_t, -\frac{\lambda^{1/2}_{K\pi}(t)}{2\sqrt{t}} \sin\theta_t, 0 \right) , \\
		p_1 &= \left( \frac{M_K^2-M_\pi^2-t}{2\sqrt{t}}, -\frac{\lambda^{1/2}_{K\pi}(t)}{2\sqrt{t}} \cos\theta_t, -\frac{\lambda^{1/2}_{K\pi}(t)}{2\sqrt{t}} \sin\theta_t, 0 \right) , \\
		q_K &= \left( \frac{t + M_K^2-M_\pi^2}{2\sqrt{t}}, -\frac{\lambda^{1/2}_{K\pi}(t)}{2\sqrt{t}} \cos\theta_t^\dprime, -\frac{\lambda^{1/2}_{K\pi}(t)}{2\sqrt{t}} \sin\theta_t^\dprime \cos\phi_t^\dprime, -\frac{\lambda^{1/2}_{K\pi}(t)}{2\sqrt{t}} \sin\theta_t^\dprime \sin\phi_t^\dprime \right) , \\
		q_\pi &= \left( \frac{t-M_K^2+M_\pi^2}{2\sqrt{t}}, \frac{\lambda^{1/2}_{K\pi}(t)}{2\sqrt{t}} \cos\theta_t^\dprime, \frac{\lambda^{1/2}_{K\pi}(t)}{2\sqrt{t}} \sin\theta_t^\dprime \cos\phi_t^\dprime, \frac{\lambda^{1/2}_{K\pi}(t)}{2\sqrt{t}} \sin\theta_t^\dprime \sin\phi_t^\dprime \right) , \\
		p_2 &= \left( \frac{t-s_\ell+M_\pi^2}{2\sqrt{t}}, \frac{\lambda^{1/2}_{\ell\pi}(t)}{2\sqrt{t}}, 0, 0 \right) , \\
		L &= \left( \frac{t+s_\ell-M_\pi^2}{2\sqrt{t}}, -\frac{\lambda^{1/2}_{\ell\pi}(t)}{2\sqrt{t}}, 0, 0 \right) ,
	\end{aligned}
\end{align}
where $\lambda_{K\pi}(t) := \lambda(M_K^2, M_\pi^2, t)$ and  $\lambda_{\ell\pi}(t) := \lambda(s_\ell, M_\pi^2, t)$. Again, the square root of the first of these Källén functions has in the $t$-plane a branch cut between $(M_K-M_\pi)^2$ and $(M_K+M_\pi)^2$, the second between $(M_\pi-\sqrt{s_\ell})^2$ and $(M_\pi+\sqrt{s_\ell})^2$. Since we need the partial-wave expansion only in the scattering region $t>(M_K+M_\pi)^2$, these branch cuts are not relevant.

We calculate the $t$-channel scattering angle $\theta_t$ as a function of the Mandelstam variables:
\begin{align}
	\begin{split}
		s - u &= (p_1+p_2)^2 - (k-p_2)^2 = p_1^2 + p_2^2 + 2p_1 p_2 - k^2 - p_2^2 + 2kp_2 \\
			&= M_\pi^2 - M_K^2 + 2 p_2 ( k + p_1 ) \\
			&= M_\pi^2 - M_K^2 + 2 \left( \frac{t-s_\ell+M_\pi^2}{2\sqrt{t}} \, \frac{M_K^2 - M_\pi^2}{\sqrt{t}} + \frac{\lambda^{1/2}_{\ell\pi}(t)}{2\sqrt{t}} \frac{\lambda^{1/2}_{K\pi}(t)}{\sqrt{t}} \cos\theta_t \right) ,
	\end{split}
\end{align}
hence
\begin{align}
	\cos\theta_t = \frac{t(s-u) + \Delta_{K\pi} \Delta_{\ell\pi}}{\lambda^{1/2}_{K\pi}(t) \lambda^{1/2}_{\ell\pi}(t)}.
\end{align}

\subsection{Kinematics in the $u$-Channel}

The $u$-channel is completely analogous to the $t$-channel:
\begin{align}
	\begin{aligned}
		k &= \left( \sqrt{M_K^2 + \vec k^2}, \vec k \right) , \; & q_K &= \left( \sqrt{ M_K^2 + \vec q_K^2 }, \vec q_K \right) , \; & p_1 &= \left( \sqrt{ M_\pi^2 + \vec p_1^2 }, \vec p_1 \right) , \\
		-p_2 &= \left( \sqrt{ M_\pi^2 + \vec k^2}, -\vec k \right) , \; & q_\pi &= \left( \sqrt{ M_\pi^2 + \vec q_K^2 }, -\vec q_K \right) , \; & L &= \left( \sqrt{ s_\ell + \vec p_1^2 }, -\vec p_1 \right) .
	\end{aligned}
\end{align}

Inserting these expressions into $u = (k - p_2)^2 = (q_K + q_\pi)^2 = (p_1 + L)^2$ gives the values of $\vec k^2$, $\vec q_K^2$ and $\vec p_1^2$. We choose the directions of the three-vectors according to figure~\ref{img:AnglesUChannel}, i.e.~the angles are defined as $\theta_u := \angle(-\vec k, \vec p_1)$, $\theta_u^\prime := \angle(\vec k, \vec q_K)$, $\theta_u^\dprime := \angle(-\vec q_K, \vec p_1)$.

\begin{figure}[ht]
	\centering
	\psset{unit=1cm}
	\begin{pspicture}(-3,-3)(3,3)
		\psline{->}(0,0)(0,3)
		\psline{->}(0,0)(3,0)
		\psline{->}(0,0)(-2,-1)
		\psline[linewidth=1.5pt, arrowsize=8pt]{->}(0,0)(0,2.5)
		\psline[linewidth=1.5pt, arrowsize=8pt]{->}(0,0)(-1,1.75)
		\psline[linewidth=1.5pt, arrowsize=8pt]{->}(0,0)(1.5,1)
		\psline[linestyle=dotted](1.5,1)(1.5,-0.5)
		\psline[linestyle=dotted](0,0)(1.5,-0.5)
		\psline[linestyle=dotted](-1,1.75)(-1,-0.5)
		\psline[linearc=1.75,linestyle=dashed](0,2)(0.75,1.75)(1.2,0.8)
		\psline[linearc=1.5,linestyle=dashed](0,2)(-0.5,1.75)(-0.75,1.31)
		\psline[linearc=3,linestyle=dashed](-0.75,1.31)(0.5,1.25)(1.2,0.8)
		\psline[linearc=4,linestyle=dashed](-0.75,-0.375)(0,-0.5)(0.9,-0.3)
		\rput(0.25,3){$x$}
		\rput(-2,-0.75){$y$}
		\rput(3,0.25){$z$}
		\rput(-0.5,2.1){$\theta_u$}
		\rput(0.25,0.85){$\theta_u^\prime$}
		\rput(1.2,1.8){$\theta_u^\dprime$}
		\rput(0,-0.75){$\phi_u^\dprime$}
		\rput(0.5,2.4){$\vec p_1$}
		\rput(-1.5,2){$-\vec k$}
		\rput(2,1){$-\vec q_K$}
	\end{pspicture}
	\caption{Vectors and angles in the $u$-channel centre-of-mass frame}
	\label{img:AnglesUChannel}
\end{figure}

The results for the $u$-channel are then:
\begin{align}
	\begin{aligned}
		k &= \left( \frac{u + M_K^2-M_\pi^2}{2\sqrt{u}}, -\frac{\lambda^{1/2}_{K\pi}(u)}{2\sqrt{u}} \cos\theta_u, -\frac{\lambda^{1/2}_{K\pi}(u)}{2\sqrt{u}} \sin\theta_u, 0 \right) , \\
		p_2 &= \left( \frac{M_K^2-M_\pi^2-u}{2\sqrt{u}}, -\frac{\lambda^{1/2}_{K\pi}(u)}{2\sqrt{u}} \cos\theta_u, -\frac{\lambda^{1/2}_{K\pi}(u)}{2\sqrt{u}} \sin\theta_u, 0 \right) , \\
		q_K &= \left( \frac{u + M_K^2-M_\pi^2}{2\sqrt{u}}, -\frac{\lambda^{1/2}_{K\pi}(u)}{2\sqrt{u}} \cos\theta_u^\dprime, -\frac{\lambda^{1/2}_{K\pi}(u)}{2\sqrt{u}} \sin\theta_u^\dprime \cos\phi_u^\dprime, -\frac{\lambda^{1/2}_{K\pi}(u)}{2\sqrt{u}} \sin\theta_u^\dprime \sin\phi_u^\dprime \right) , \\
		q_\pi &= \left( \frac{u-M_K^2+M_\pi^2}{2\sqrt{u}}, \frac{\lambda^{1/2}_{K\pi}(u)}{2\sqrt{u}} \cos\theta_u^\dprime, \frac{\lambda^{1/2}_{K\pi}(u)}{2\sqrt{u}} \sin\theta_u^\dprime \cos\phi_u^\dprime, \frac{\lambda^{1/2}_{K\pi}(u)}{2\sqrt{u}} \sin\theta_u^\dprime \sin\phi_u^\dprime \right) , \\
		p_1 &= \left( \frac{u-s_\ell+M_\pi^2}{2\sqrt{u}}, \frac{\lambda^{1/2}_{\ell\pi}(u)}{2\sqrt{u}}, 0, 0 \right) , \\
		L &= \left( \frac{u+s_\ell-M_\pi^2}{2\sqrt{u}}, -\frac{\lambda^{1/2}_{\ell\pi}(u)}{2\sqrt{u}}, 0, 0 \right) .
	\end{aligned}
\end{align}

Let us calculate the $u$-channel scattering angle $\theta_u$ as a function of the Mandelstam variables:
\begin{align}
	\begin{split}
		s - t &= (p_1+p_2)^2 - (k-p_1)^2 = p_1^2 + p_2^2 + 2 p_1 p_2 - k^2 - p_1^2 + 2 k p_1 \\
			&= M_\pi^2 - M_K^2 + 2 p_1 ( k + p_2 ) \\
			&= M_\pi^2 - M_K^2 + 2 \left( \frac{u-s_\ell+M_\pi^2}{2\sqrt{u}} \, \frac{M_K^2 - M_\pi^2}{\sqrt{u}} + \frac{\lambda^{1/2}_{\ell\pi}(u)}{2\sqrt{u}} \frac{\lambda^{1/2}_{K\pi}(u)}{\sqrt{u}} \cos\theta_u \right) ,
	\end{split}
\end{align}
hence
\begin{align}
	\cos\theta_u = \frac{u(s-t) + \Delta_{K\pi} \Delta_{\ell\pi}}{\lambda^{1/2}_{K\pi}(u) \lambda^{1/2}_{\ell\pi}(u)}.
\end{align}


\section{Omnès Solution to the Dispersion Relation}

\subsection{Solution for $n=3$ Subtractions}

\label{sec:AppendixOmnes3Subtractions}

For $n=3$ subtractions, the Omnès representation reads
\begin{align}
	\begin{alignedat}{2}
		\label{eq:FunctionsOfOneVariableOmnes3Subtr}
		M_0(s) &= \Omega_0^0(s) & &\bigg\{ a^{M_0} + b^{M_0} \frac{s}{M_K^2} + c^{M_0} \frac{s^2}{M_K^4} + d^{M_0} \frac{s^3}{M_K^6} + \frac{s^4}{\pi} \int_{s_0}^{\Lambda^2} \frac{\hat M_0(s^\prime) \sin\delta_0^0(s^\prime)}{|\Omega_0^0(s^\prime)| (s^\prime - s - i\epsilon) {s^\prime}^4} ds^\prime \bigg\} , \\
		M_1(s) &= \Omega_1^1(s) & &\bigg\{ a^{M_1} + b^{M_1}  \frac{s}{M_K^2} + c^{M_1} \frac{s^2}{M_K^4} + \frac{s^3}{\pi} \int_{s_0}^{\Lambda^2} \frac{\hat M_1(s^\prime) \sin\delta_1^1(s^\prime)}{|\Omega_1^1(s^\prime)| (s^\prime - s - i\epsilon) {s^\prime}^3} ds^\prime  \bigg\} , \\
		 \tilde M_1(s) &= \Omega_1^1(s) & &\bigg\{ a^{\tilde M_1} + b^{\tilde M_1}  \frac{s}{M_K^2} + c^{\tilde M_1}  \frac{s^2}{M_K^4} + d^{\tilde M_1} \frac{s^3}{M_K^6} + \frac{s^4}{\pi} \int_{s_0}^{\Lambda^2} \frac{\hat{\tilde M}_1(s^\prime) \sin\delta_1^1(s^\prime)}{|\Omega_1^1(s^\prime)| (s^\prime - s - i\epsilon) {s^\prime}^4} ds^\prime \bigg\} , \\
		N_0(t) &=  \Omega_0^{1/2}(t) & &\bigg\{ b^{N_0} \frac{t}{M_K^2} + c^{N_0} \frac{t^2}{M_K^4} + \frac{t^3}{\pi} \int_{t_0}^{\Lambda^2} \frac{\hat N_0(t^\prime) \sin\delta_0^{1/2}(t^\prime)}{|\Omega_0^{1/2}(t^\prime)| (t^\prime - t - i\epsilon) {t^\prime}^3} dt^\prime  \bigg\} , \\
		N_1(t) &= \Omega_1^{1/2}(t) & &\bigg\{ a^{N_1} + \frac{t}{\pi} \int_{t_0}^{\Lambda^2} \frac{\hat N_1(t^\prime) \sin\delta_1^{1/2}(t^\prime)}{|\Omega_1^{1/2}(t^\prime)| (t^\prime - t - i\epsilon)t^\prime} dt^\prime  \bigg\} , \\
		\tilde N_1(t) &= \Omega_1^{1/2}(t) & &\bigg\{ b^{\tilde N_1} \frac{t}{M_K^2} + \frac{t^2}{\pi} \int_{t_0}^{\Lambda^2} \frac{\hat{\tilde N}_1(t^\prime) \sin\delta_1^{1/2}(t^\prime)}{|\Omega_1^{1/2}(t^\prime)| (t^\prime - t - i\epsilon) {t^\prime}^2} dt^\prime  \bigg\} , \\
		R_0(t) &=  \Omega_0^{3/2}(t) & &\bigg\{ \frac{t^3}{\pi} \int_{t_0}^{\Lambda^2} \frac{\hat R_0(t^\prime) \sin\delta_0^{3/2}(t^\prime)}{|\Omega_0^{3/2}(t^\prime)| (t^\prime - t - i\epsilon) {t^\prime}^3} dt^\prime  \bigg\} , \\
		R_1(t) &=  \Omega_1^{3/2}(t) & &\bigg\{ \frac{t}{\pi} \int_{t_0}^{\Lambda^2} \frac{\hat R_1(t^\prime) \sin\delta_1^{3/2}(t^\prime)}{|\Omega_1^{3/2}(t^\prime)| (t^\prime - t - i\epsilon)t^\prime} dt^\prime  \bigg\} , \\
		\tilde R_1(t) &=   \Omega_1^{3/2}(t) & &\bigg\{ \frac{t^2}{\pi} \int_{t_0}^{\Lambda^2} \frac{\hat{\tilde R}_1(t^\prime) \sin\delta_1^{3/2}(t^\prime)}{|\Omega_1^{3/2}(t^\prime)| (t^\prime - t - i\epsilon) {t^\prime}^2} dt^\prime  \bigg\} .
	\end{alignedat}
\end{align}

Let us work out how to transform the Omnès representation (\ref{eq:FunctionsOfOneVariableOmnes}) into the one with more subtractions (\ref{eq:FunctionsOfOneVariableOmnes3Subtr}).
We start by subtracting all the dispersive integrals once more, using the relation
\begin{align}
	\begin{split}
		\frac{1}{s^\prime-s} = \frac{1}{s^\prime} + \frac{s}{(s^\prime-s)s^\prime} .
	\end{split}
\end{align}
This generates nine additional subtraction constants:
\begin{align}
	\footnotesize
	\begin{alignedat}{2}
		M_0(s) &= \Omega_0^0(s) & &\bigg\{ a^{M_0} + b^{M_0} \frac{s}{M_K^2} + c^{M_0} \frac{s^2}{M_K^4} + d^{M_0} \frac{s^3}{M_K^6} + \frac{s^4}{\pi} \int_{s_0}^{\Lambda^2} \frac{\hat M_0(s^\prime) \sin\delta_0^0(s^\prime)}{|\Omega_0^0(s^\prime)| (s^\prime - s - i\epsilon) {s^\prime}^4} ds^\prime \bigg\} , \\
		M_1(s) &= \Omega_1^1(s) & &\bigg\{ a^{M_1} + b^{M_1}  \frac{s}{M_K^2} + c^{M_1} \frac{s^2}{M_K^4} + \frac{s^3}{\pi} \int_{s_0}^{\Lambda^2} \frac{\hat M_1(s^\prime) \sin\delta_1^1(s^\prime)}{|\Omega_1^1(s^\prime)| (s^\prime - s - i\epsilon) {s^\prime}^3} ds^\prime  \bigg\} , \\
		 \tilde M_1(s) &= \Omega_1^1(s) & &\bigg\{ a^{\tilde M_1} + b^{\tilde M_1}  \frac{s}{M_K^2} + c^{\tilde M_1}  \frac{s^2}{M_K^4} + d^{\tilde M_1} \frac{s^3}{M_K^6} + \frac{s^4}{\pi} \int_{s_0}^{\Lambda^2} \frac{\hat{\tilde M}_1(s^\prime) \sin\delta_1^1(s^\prime)}{|\Omega_1^1(s^\prime)| (s^\prime - s - i\epsilon) {s^\prime}^4} ds^\prime \bigg\} , \\
		N_0(t) &=  \Omega_0^{1/2}(t) & &\bigg\{ b^{N_0} \frac{t}{M_K^2} + c^{N_0} \frac{t^2}{M_K^4} + \frac{t^3}{\pi} \int_{t_0}^{\Lambda^2} \frac{\hat N_0(t^\prime) \sin\delta_0^{1/2}(t^\prime)}{|\Omega_0^{1/2}(t^\prime)| (t^\prime - t - i\epsilon) {t^\prime}^3} dt^\prime  \bigg\} , \\
		N_1(t) &= \Omega_1^{1/2}(t) & &\bigg\{ a^{N_1} + \frac{t}{\pi} \int_{t_0}^{\Lambda^2} \frac{\hat N_1(t^\prime) \sin\delta_1^{1/2}(t^\prime)}{|\Omega_1^{1/2}(t^\prime)| (t^\prime - t - i\epsilon){t^\prime}} dt^\prime  \bigg\} , \\
		\tilde N_1(t) &= \Omega_1^{1/2}(t) & &\bigg\{ b^{\tilde N_1} \frac{t}{M_K^2} + \frac{t^2}{\pi} \int_{t_0}^{\Lambda^2} \frac{\hat{\tilde N}_1(t^\prime) \sin\delta_1^{1/2}(t^\prime)}{|\Omega_1^{1/2}(t^\prime)| (t^\prime - t - i\epsilon) {t^\prime}^2} dt^\prime  \bigg\} , \\
		R_0(t) &=  \Omega_0^{3/2}(t) & &\bigg\{ c^{R_0} \frac{t^2}{M_K^4} + \frac{t^3}{\pi} \int_{t_0}^{\Lambda^2} \frac{\hat R_0(t^\prime) \sin\delta_0^{3/2}(t^\prime)}{|\Omega_0^{3/2}(t^\prime)| (t^\prime - t - i\epsilon) {t^\prime}^3} dt^\prime  \bigg\} , \\
		R_1(t) &=  \Omega_1^{3/2}(t) & &\bigg\{ a^{R_1} +  \frac{t}{\pi} \int_{t_0}^{\Lambda^2} \frac{\hat R_1(t^\prime) \sin\delta_1^{3/2}(t^\prime)}{|\Omega_1^{3/2}(t^\prime)| (t^\prime - t - i\epsilon) t^\prime} dt^\prime  \bigg\} , \\
		\tilde R_1(t) &=   \Omega_1^{3/2}(t) & &\bigg\{ b^{\tilde R_1} \frac{t}{M_K^2} + \frac{t^2}{\pi} \int_{t_0}^{\Lambda^2} \frac{\hat{\tilde R}_1(t^\prime) \sin\delta_1^{3/2}(t^\prime)}{|\Omega_1^{3/2}(t^\prime)| (t^\prime - t - i\epsilon) {t^\prime}^2} dt^\prime  \bigg\} .
	\end{alignedat}
\end{align}
To get rid of the subtraction constants in the $R$-functions, we apply a gauge transformation (\ref{eq:GaugeTransformation}). To this end, let us write the gauge transformation in the Omnès representation:
\begin{align}
	\footnotesize
	\begin{alignedat}{2}
		\label{eq:OmnesGaugeTransformation}
		\delta M_0(s) &= \Omega_0^0(s) & &\bigg\{ \delta a^{M_0} + \delta b^{M_0} \frac{s}{M_K^2} + \delta c^{M_0} \frac{s^2}{M_K^4} + \delta d^{M_0} \frac{s^3}{M_K^6} + \frac{s^4}{\pi} \int_{s_0}^{\Lambda^2} \frac{\delta \hat M_0(s^\prime) \sin\delta_0^0(s^\prime)}{|\Omega_0^0(s^\prime)| (s^\prime - s - i\epsilon) {s^\prime}^4} ds^\prime \bigg\} , \\
		\delta M_1(s) &= \Omega_1^1(s) & &\bigg\{ \delta a^{M_1} + \delta b^{M_1}  \frac{s}{M_K^2} + \delta c^{M_1} \frac{s^2}{M_K^4} + \frac{s^3}{\pi} \int_{s_0}^{\Lambda^2} \frac{\delta \hat M_1(s^\prime) \sin\delta_1^1(s^\prime)}{|\Omega_1^1(s^\prime)| (s^\prime - s - i\epsilon) {s^\prime}^3} ds^\prime  \bigg\} , \\
		\delta  \tilde M_1(s) &= \Omega_1^1(s) & &\bigg\{ \delta a^{\tilde M_1} + \delta b^{\tilde M_1}  \frac{s}{M_K^2} + \delta c^{\tilde M_1}  \frac{s^2}{M_K^4} + \delta d^{\tilde M_1} \frac{s^3}{M_K^6} + \frac{s^4}{\pi} \int_{s_0}^{\Lambda^2} \frac{\delta \hat{\tilde M}_1(s^\prime) \sin\delta_1^1(s^\prime)}{|\Omega_1^1(s^\prime)| (s^\prime - s - i\epsilon) {s^\prime}^4} ds^\prime \bigg\} , \\
		\delta N_0(t) &=  \Omega_0^{1/2}(t) & &\bigg\{ \delta b^{N_0} \frac{t}{M_K^2} + \delta c^{N_0} \frac{t^2}{M_K^4} + \frac{t^3}{\pi} \int_{t_0}^{\Lambda^2} \frac{\delta \hat N_0(t^\prime) \sin\delta_0^{1/2}(t^\prime)}{|\Omega_0^{1/2}(t^\prime)| (t^\prime - t - i\epsilon) {t^\prime}^3} dt^\prime  \bigg\} , \\
		\delta N_1(t) &= \Omega_1^{1/2}(t) & &\bigg\{ \delta a^{N_1} + \frac{t}{\pi} \int_{t_0}^{\Lambda^2} \frac{\delta \hat N_1(t^\prime) \sin\delta_1^{1/2}(t^\prime)}{|\Omega_1^{1/2}(t^\prime)| (t^\prime - t - i\epsilon){t^\prime}} dt^\prime  \bigg\} , \\
		\delta \tilde N_1(t) &= \Omega_1^{1/2}(t) & &\bigg\{ \delta b^{\tilde N_1} \frac{t}{M_K^2} + \frac{t^2}{\pi} \int_{t_0}^{\Lambda^2} \frac{\delta \hat{\tilde N}_1(t^\prime) \sin\delta_1^{1/2}(t^\prime)}{|\Omega_1^{1/2}(t^\prime)| (t^\prime - t - i\epsilon) {t^\prime}^2} dt^\prime  \bigg\} , \\
		\delta R_0(t) &=  \Omega_0^{3/2}(t) & &\bigg\{ \delta c^{R_0} \frac{t^2}{M_K^4} + \frac{t^3}{\pi} \int_{t_0}^{\Lambda^2} \frac{\delta \hat R_0(t^\prime) \sin\delta_0^{3/2}(t^\prime)}{|\Omega_0^{3/2}(t^\prime)| (t^\prime - t - i\epsilon) {t^\prime}^3} dt^\prime  \bigg\} , \\
		\delta R_1(t) &=  \Omega_1^{3/2}(t) & &\bigg\{ \delta a^{R_1} +  \frac{t}{\pi} \int_{t_0}^{\Lambda^2} \frac{\delta \hat R_1(t^\prime) \sin\delta_1^{3/2}(t^\prime)}{|\Omega_1^{3/2}(t^\prime)| (t^\prime - t - i\epsilon) t^\prime} dt^\prime  \bigg\} , \\
		\delta \tilde R_1(t) &=   \Omega_1^{3/2}(t) & &\bigg\{ \delta b^{\tilde R_1} \frac{t}{M_K^2} + \frac{t^2}{\pi} \int_{t_0}^{\Lambda^2} \frac{\delta \hat{\tilde R}_1(t^\prime) \sin\delta_1^{3/2}(t^\prime)}{|\Omega_1^{3/2}(t^\prime)| (t^\prime - t - i\epsilon) {t^\prime}^2} dt^\prime  \bigg\} .
	\end{alignedat}
\end{align}
Since the gauge transformation is a polynomial and has no discontinuity, the changes in the hat functions are given by $\delta \hat M_0 = - \delta M_0$ etc., which assures that the partial waves are unchanged. The shifts in the subtraction constants are most easily found by comparing the Taylor expansion of (\ref{eq:OmnesGaugeTransformation}) with (\ref{eq:GaugeTransformation}):
\begin{align}
\label{eq:OmnesGaugeTransformationParameters}
\scalebox{0.8}{
\begin{minipage}{1.1\textwidth}
$	\begin{split}
		\delta a^{M_0} &= \left(2 A^{R_1} - B^{\tilde R_1} + 2 C^{R_0}\right) \frac{\Sigma_0^2 - \Delta_{K\pi}\Delta_{\ell\pi}}{2M_K^4} , \\
		\delta b^{M_0} &= - \left(2 A^{R_1} - B^{\tilde R_1} + 2 C^{R_0}\right) \left( \frac{\Sigma_0}{M_K^2} + \omega_0^0 \frac{\Sigma_0^2 - \Delta_{K\pi}\Delta_{\ell\pi}}{2 M_K^4} \right) , \\
		\delta c^{M_0} &= \left(2 A^{R_1} - B^{\tilde R_1} + 2 C^{R_0}\right) \left(\frac{1}{2} + \omega_0^0 \frac{\Sigma_0}{M_K^2} + \left( \frac{{\omega_0^0}^2}{2} - \bar \omega_0^0 \right) \frac{\Sigma_0^2 - \Delta_{K\pi}\Delta_{\ell\pi}}{2M_K^4} \right) , \\
		\delta d^{M_0} &= - \left(2 A^{R_1} - B^{\tilde R_1} + 2 C^{R_0}\right) \left( \frac{\omega_0^0}{2} - \left( \bar\omega_0^0 - \frac{{\omega_0^0}^2}{2} \right) \frac{\Sigma_0}{M_K^2} + \left( {\omega_0^0}^3 - 6 \omega_0^0 \bar\omega_0^0 + 6 \bar{\bar\omega}_0^0 \right) \frac{\Sigma_0^2 - \Delta_{K\pi} \Delta_{\ell\pi}}{12 M_K^4} \right) , \\
		\delta a^{M_1} &= -\left(A^{R_1} + B^{\tilde R_1} + 2 C^{R_0}\right)\frac{\Sigma_0}{M_K^2} + B^{\tilde R_1} \frac{\Delta_{K\pi}}{2 M_K^2} , \\
		\delta b^{M_1} &=  \left(B^{\tilde R_1} + 2 C^{R_0}\right) + \omega_1^1 \left( \left(A^{R_1} + B^{\tilde R_1} + 2 C^{R_0}\right)\frac{\Sigma_0}{M_K^2} - B^{\tilde R_1} \frac{\Delta_{K\pi}}{2 M_K^2} \right) , \\
		\delta c^{M_1} &= - \left( \left(A^{R_1} + B^{\tilde R_1} + 2 C^{R_0}\right)\frac{\Sigma_0}{M_K^2} - B^{\tilde R_1} \frac{\Delta_{K\pi}}{2 M_K^2}\right) \left(\frac{{\omega_1^1}^2}{2} - \bar\omega_1^1 \right) - \omega_1^1 \left(B^{\tilde R_1} + 2 C^{R_0} \right) , \\
		\delta a^{\tilde M_1} &= \left(B^{\tilde R_1} - 2 C^{R_0}\right) \frac{\Sigma_0^2}{2 M_K^4} - \left(2 A^{R_1} + B^{\tilde R_1} - 2 C^{R_0} \right) \frac{\Delta_{K\pi} \Delta_{\ell\pi}}{2 M_K^4} + B^{\tilde R_1} \frac{\Sigma_0 \Delta_{K\pi}}{2 M_K^4} , \\
		\delta b^{\tilde M_1} &=  - \left(B^{\tilde R_1} \frac{\Delta_{K\pi}}{2 M_K^2} + \left(A^{R_1} + B^{\tilde R_1} - 2 C^{R_0}\right) \frac{\Sigma_0}{M_K^2} \right) \\
			&\quad - \omega_1^1 \left(\left(B^{\tilde R_1} - 2 C^{R_0}\right) \frac{\Sigma_0^2}{2 M_K^4} - \left(2 A^{R_1} + B^{\tilde R_1} - 2 C^{R_0} \right) \frac{\Delta_{K\pi} \Delta_{\ell\pi}}{2 M_K^4} + B^{\tilde R_1} \frac{\Sigma_0 \Delta_{K\pi}}{2 M_K^4} \right) , \\
		\delta c^{\tilde M_1} &= \frac{1}{2} \left(2 A^{R_1} + B^{\tilde R_1} - 2 C^{R_0}\right) + \omega_1^1 \left(B^{\tilde R_1} \frac{\Delta_{K\pi}}{2 M_K^2} + \left(A^{R_1} + B^{\tilde R_1} - 2 C^{R_0}\right) \frac{\Sigma_0}{M_K^2} \right) \\
			&\quad + \left( \frac{{\omega_1^1}^2}{2} - \bar \omega_1^1 \right) \left( \left(B^{\tilde R_1} - 2 C^{R_0}\right) \frac{\Sigma_0^2}{2 M_K^4} - \left(2 A^{R_1} + B^{\tilde R_1} - 2 C^{R_0} \right) \frac{\Delta_{K\pi} \Delta_{\ell\pi}}{2 M_K^4} + B^{\tilde R_1} \frac{\Sigma_0 \Delta_{K\pi}}{2 M_K^4} \right) , \\
		\delta d^{\tilde M_1} &= - \frac{1}{2} \omega_1^1 \left(2 A^{R_1} + B^{\tilde R_1} - 2 C^{R_0}\right) - \left( \frac{{\omega_1^1}^2}{2} - \bar\omega_1^1 \right) \left(B^{\tilde R_1} \frac{\Delta_{K\pi}}{2 M_K^2} + \left(A^{R_1} + B^{\tilde R_1} - 2 C^{R_0}\right) \frac{\Sigma_0}{M_K^2} \right) \\
			&\quad - \frac{1}{6} \left( {\omega_1^1}^3 - 6 \omega_1^1 \bar\omega_1^1 + 6 \bar{\bar\omega}_1^1 \right) \left( \left(B^{\tilde R_1} - 2 C^{R_0}\right) \frac{\Sigma_0^2}{2 M_K^4} - \left(2 A^{R_1} + B^{\tilde R_1} - 2 C^{R_0} \right) \frac{\Delta_{K\pi} \Delta_{\ell\pi}}{2 M_K^4} + B^{\tilde R_1} \frac{\Sigma_0 \Delta_{K\pi}}{2 M_K^4} \right) , \\
		\delta b^{N_0} &= - \left(2 A^{R_1} - B^{\tilde R_1} + 2 C^{R_0} \right) \frac{3 (\Delta_{K\pi} + 2 \Sigma_0)}{8 M_K^2} , \\
		\delta c^{N_0} &= \frac{1}{8} \left(6 A^{R_1} - 3 B^{\tilde R_1} -10 C^{R_0} \right) + \omega_0^{1/2} \left(2 A^{R_1} - B^{\tilde R_1} + 2 C^{R_0} \right) \frac{3 (\Delta_{K\pi} + 2 \Sigma_0)}{8 M_K^2} , \\
		\delta a^{N_1} &= - \frac{1}{4} \left(2 A^{R_1} + 3 B^{\tilde R_1} - 6 C^{R_0} \right) , \\
		\delta b^{\tilde N_1} &= -\frac{1}{4} \left(6 A^{R_1} + 5 B^{\tilde R_1} + 6 C^{R_0} \right) , \\
		\delta c^{R_0} &= C^{R_0} , \\
		\delta a^{R_1} &= A^{R_1} , \\
		\delta b^{\tilde R_1} &= B^{\tilde R_1} ,
	\end{split}
	$
\end{minipage}}
\end{align}
where $\omega$, $\bar\omega$ and $\bar{\bar\omega}$ are defined by applying subtractions to the Omnès functions:
\begin{align}
	\footnotesize
	\begin{split}
		\Omega(s) &= \exp\left( \frac{s}{\pi} \int_{s_0}^\infty \frac{\delta(s^\prime)}{(s^\prime - s - i \epsilon) s^\prime} ds^\prime \right) \\
			&= \exp\left(  \frac{s}{\pi} \int_{s_0}^\infty \frac{\delta(s^\prime)}{{s^\prime}^2} ds^\prime +  \frac{s^2}{\pi} \int_{s_0}^\infty \frac{\delta(s^\prime)}{{s^\prime}^3} ds^\prime +  \frac{s^3}{\pi} \int_{s_0}^\infty \frac{\delta(s^\prime)}{{s^\prime}^4} ds^\prime +  \frac{s^4}{\pi} \int_{s_0}^\infty \frac{\delta(s^\prime)}{(s^\prime - s - i \epsilon) {s^\prime}^4} ds^\prime \right) \\
			&=: \exp\left( \omega \frac{s}{M_K^2} +  \bar\omega \frac{s^2}{M_K^4} + \bar{\bar\omega} \frac{s^3}{M_K^6} + \frac{s^4}{\pi} \int_{s_0}^\infty \frac{\delta(s^\prime)}{(s^\prime - s - i \epsilon) {s^\prime}^4} ds^\prime \right) .
	\end{split}
\end{align}
In order to obtain the form (\ref{eq:FunctionsOfOneVariableOmnes3Subtr}), the subtraction constants in the $R$-functions can now be removed with the gauge transformation
\begin{align}
	\begin{split}
		C^{R_0} &= -c^{R_0} , \\
		A^{R_1} &= -a^{R_1} , \\
		B^{\tilde R_1} &= -b^{\tilde R_1} .
	\end{split}
\end{align}

\subsection{Hat Functions}

\label{sec:AppendixHatFunctions}

In the following, we provide the explicit expressions for the hat functions that appear in the Omnès solution to the dispersion relation.
\begin{align}
	\footnotesize
	\begin{split}
		\hat M_0(s) &=  \frac{2}{3} \Big( \<N_0\>_{t_s}+2\<R_0\>_{t_s} \Big) - \Big( \<z N_0\>_{t_s} + 2 \<z R_0\>_{t_s} \Big) \frac{2 \sigma_\pi PL }{3 X} \\
			&\quad - \Big( \<N_1\>_{t_s} + 2 \<R_1\>_{t_s} \Big) \frac{3 s^2-4 s \Sigma_0+\Sigma_0^2-4 \Delta_{K\pi} \Delta_{\ell\pi}}{6 M_K^4} \\
			&\quad + \Big( \<z N_1\>_{t_s} + 2 \<z R_1\>_{t_s} \Big) \frac{\sigma _{\pi } \left(-4 PL \Delta_{K\pi} \Delta_{\ell\pi} + PL \left(3 s^2-4 s \Sigma_0+\Sigma_0^2\right)-4 s X^2\right)}{6 M_K^4 X} \\
			&\quad + \Big( \<z^2 N_1\>_{t_s} + 2 \<z^2 R_1\>_{t_s} \Big) \frac{2 \sigma_\pi^2 \left( PL s + X^2 \right)}{3 M_K^4} - \Big( \<z^3 N_1\>_{t_s} + 2 \<z^3 R_1\>_{t_s} \Big) \frac{2 \sigma_\pi^3 PL \, X}{3 M_K^4} \\
			&\quad - \Big( \<\tilde N_1\>_{t_s} + 2 \<\tilde R_1\>_{t_s} \Big) \frac{2 \Delta_{K\pi} + 3 s - 3 \Sigma_0}{6 M_K^2} \\
			&\quad + \Big( \<z \tilde N_1\>_{t_s} + 2 \<z \tilde R_1\>_{t_s} \Big) \frac{\sigma_\pi \left(PL \left(2 \Delta_{K\pi} - s + \Sigma_0\right)-6 X^2\right)}{6 M_K^2 X}  - \Big( \<z^2 \tilde N_1\>_{t_s} + 2 \<z^2 \tilde R_1\>_{t_s} \Big) \frac{\sigma_\pi^2 PL}{3 M_K^2} , \\
	\end{split}
\end{align}
\begin{align}
	\footnotesize
	\begin{split}
		\hat M_1(s) &= \Big( \<N_0\>_{t_s} - \<R_0\>_{t_s} \Big) \frac{M_K^2 PL}{2 X^2}  + \Big( \< z N_0 \>_{t_s} - \<z R_0\>_{t_s} \Big) \frac{M_K^2}{\sigma_\pi X} - \Big( \< z^2 N_0\>_{t_s} - \< z^2 R_0\>_{t_s} \Big) \frac{3 M_K^2 PL}{2 X^2} \\
			&\quad + \Big( \<N_1\>_{t_s} - \<R_1\>_{t_s} \Big) \frac{\left(4 \Delta_{K\pi} \Delta_{\ell\pi} -3 s^2+4 s \Sigma_0 - \Sigma_0^2\right) PL}{8 M_K^2 X^2} \\
			&\quad - \Big( \<z N_1\>_{t_s} - \<z R_1\>_{t_s} \Big) \frac{ 3 s^2 + 2 \sigma_\pi^2 PL s - 4 s \Sigma_0+\Sigma_0^2 - 4 \Delta_{K\pi} \Delta_{\ell\pi} }{4 M_K^2 \sigma_\pi X} \\
			&\quad + \Big( \<z^2 N_1\>_{t_s} - \<z^2 R_1\>_{t_s} \Big) \Bigg( \frac{ 3 PL \left(3 s^2 - 4 s \Sigma_0 + \Sigma_0^2\right) -12 PL \Delta_{K\pi} \Delta_{\ell\pi} }{8 M_K^2 X^2} +  \frac{ \sigma_\pi^2 PL - 2 s}{2 M_K^2} \Bigg) \\
			&\quad + \Big( \<z^3 N_1\>_{t_s} - \<z^3 R_1\>_{t_s} \Big) \frac{\sigma_\pi \left(3 s PL + 2 X^2\right)}{2 X M_K^2}  - \Big( \<z^4 N_1\>_{t_s} - \<z^4 R_1\>_{t_s} \Big) \frac{3 \sigma_\pi^2 PL}{2 M_K^2} \\
			&\quad - \Big( \<\tilde N_1\>_{t_s} - \<\tilde R_1\>_{t_s} \Big) \frac{PL \left(2 \Delta_{K\pi} - s + \Sigma_0\right)}{8 X^2} + \Big( \<z \tilde N_1\>_{t_s} - \<z \tilde R_1\>_{t_s} \Big) \frac{3 \Sigma_0 - 2 \Delta_{K\pi} + \sigma_\pi^2 PL - 3 s}{4 \sigma_\pi X} \\
			&\quad + \Big( \<z^2 \tilde N_1\>_{t_s} - \<z^2 \tilde R_1\>_{t_s} \Big) \left(\frac{3 PL \left(2 \Delta_{K\pi} - s + \Sigma_0 \right)}{8 X^2} - \frac{3}{2}\right) - \Big( \<z^3 \tilde N_1\>_{t_s} - \<z^3 \tilde R_1\>_{t_s} \Big) \frac{3 PL \sigma_\pi }{4 X} , \\
	\end{split}
\end{align}
\begin{align}
	\footnotesize
	\begin{split}
		\hat{\tilde M}_1(s) &= - \Big( \<N_0\>_{t_s} - \< R_0 \>_{t_s} \Big) + \Big( \<z^2 N_0\>_{t_s} - \< z^2 R_0 \>_{t_s} \Big) \\
			&\quad + \Big( \<N_1\>_{t_s} - \< R_1 \>_{t_s} \Big) \frac{3 s^2 - 4 s \Sigma_0 + \Sigma_0^2-4 \Delta_{K\pi} \Delta_{\ell\pi}}{4 M_K^4} + \Big( \<z N_1\>_{t_s} - \<z R_1 \>_{t_s} \Big) \frac{s \sigma_\pi X}{M_K^4} \\
			&\quad - \Big( \<z^2 N_1\>_{t_s} - \<z^2 R_1 \>_{t_s} \Big) \frac{3 s^2 - 4 s \Sigma_0 + \Sigma_0^2 + 4 \sigma_\pi^2 X^2 -4 \Delta_{K\pi} \Delta_{\ell\pi}}{4 M_K^4} \\
			&\quad - \Big( \<z^3 N_1\>_{t_s} - \<z^3 R_1 \>_{t_s} \Big) \frac{s \sigma_\pi X}{M_K^4} + \Big( \<z^4 N_1\>_{t_s} - \<z^4 R_1 \>_{t_s} \Big) \frac{\sigma_\pi^2 X^2}{M_K^4} \\
			&\quad + \Big( \<\tilde N_1\>_{t_s} - \<\tilde R_1 \>_{t_s} \Big) \frac{2 \Delta_{K\pi} - s + \Sigma_0}{4 M_K^2} - \Big( \<z \tilde N_1\>_{t_s} - \<z \tilde R_1 \>_{t_s} \Big) \frac{\sigma_\pi X}{2 M_K^2} \\
			&\quad - \Big( \<z^2 \tilde N_1\>_{t_s} - \<z^2 \tilde R_1 \>_{t_s} \Big) \frac{2 \Delta_{K\pi} - s + \Sigma_0}{4 M_K^2} + \Big( \<z^3 \tilde N_1\>_{t_s} - \<z^3 \tilde R_1 \>_{t_s} \Big) \frac{\sigma_\pi X}{2 M_K^2} , \\
	\end{split}
\end{align}
\begin{align}
	\footnotesize
	\begin{split}
		\hat N_0(t) &= \< M_0 \>_{s_t} \frac{\Delta_{K\pi}+t}{4 t}  - \< z M_0\>_{s_t} \frac{\lambda_{K\pi}^{1/2}(t) \left(\Delta_{\ell\pi} + t\right)}{4 t \lambda_{\ell\pi}^{1/2}(t)} + \< M_1 \>_{s_t} \frac{\left(\Delta_{K\pi} + t\right) \left(\Delta_{K\pi} \Delta_{\ell\pi} + t \left(\Sigma_0 - 3 t\right)\right)}{4 t^2 M_K^2} \\
			&\quad - \< z M_1 \>_{s_t} \frac{\lambda_{K\pi}^{1/2}(t) \left(\Delta_{K\pi} \left(\lambda_{\ell\pi}(t) + \Delta_{\ell\pi} \left(\Delta_{\ell\pi} + t\right)\right) + t \left(\lambda_{\ell\pi}(t) + \left(\Sigma_0-3 t\right) \left(\Delta_{\ell\pi}+t\right)\right)\right)}{4 t^2 M_K^2 \lambda_{\ell\pi}^{1/2}(t)} \\
			&\quad + \< z^2 M_1 \>_{s_t} \frac{\lambda_{K\pi}(t) \left(\Delta_{\ell\pi}+t\right)}{4 t^2 M_K^2} + \< \tilde M_1 \>_{s_t} \frac{\Delta_{K\pi}-3t}{2t}  - \< z \tilde M_1 \>_{s_t} \frac{\lambda_{K\pi}^{1/2}(t) \left(\Delta_{\ell\pi}+t\right)}{2 t \lambda_{\ell\pi}^{1/2}(t)} \\
			&\quad + \Big( \< N_0 \>_{u_t} - 4 \< R_0 \>_{u_t} \Big) \frac{t-\Delta_{K\pi}}{6 t}  + \Big( \< z N_0 \>_{u_t} - 4 \< z R_0 \>_{u_t} \Big) \frac{ \lambda_{K\pi}^{1/2}(t) \left(\Delta_{\ell\pi}+t\right)}{6 t \lambda_{\ell\pi}^{1/2}(t)} \\
			&\quad + \Big( \< N_1 \>_{u_t} - 4 \< R_1 \>_{u_t} \Big) \frac{\left(\Delta_{K\pi}-t\right) \left(\Delta_{K\pi} \Delta_{\ell\pi}+t \left(t-\Sigma_0\right)\right)\left(\Delta_{K\pi} \Delta_{\ell\pi}+t \left(\Sigma_0-3 t\right)\right)}{24 t^3 M_K^4} \\
			&\quad - \Big( \< z N_1 \>_{u_t} - 4 \< z R_1 \>_{u_t} \Big) \begin{aligned}[t]
				& \Bigg( \frac{\lambda_{K\pi}^{1/2}(t) \left(\Delta_{\ell\pi}+t\right) \left(\Delta_{K\pi} \Delta_{\ell\pi}+t \left(t-\Sigma_0\right)\right) \left(\Delta_{K\pi} \Delta_{\ell\pi}+t \left(\Sigma_0-3 t\right)\right)}{24 t^3 M_K^4 \lambda_{\ell\pi}^{1/2}(t) } \\
				& + \frac{\left(\Delta_{K\pi}-t\right) \lambda_{K\pi}^{1/2}(t) \lambda_{\ell\pi}^{1/2}(t) \left(\Delta_{K\pi} \Delta_{\ell\pi}+t^2\right)}{12 t^3 M_K^4} \Bigg) \end{aligned} \\
			&\quad + \Big( \< z^2 N_1 \>_{u_t} - 4 \< z^2 R_1 \>_{u_t} \Big) \Bigg( \frac{\left(\Delta_{K\pi}-t\right) \lambda_{K\pi}(t) \lambda_{\ell\pi}(t)}{24 t^3 M_K^4}+\frac{\lambda_{K\pi}(t) \left(\Delta_{\ell\pi}+t\right) \left(\Delta_{K\pi} \Delta_{\ell\pi}+t^2\right)}{12 t^3 M_K^4} \Bigg) \\
			&\quad - \Big( \< z^3 N_1 \>_{u_t} - 4 \< z^3 R_1 \>_{u_t} \Big) \frac{\lambda_{K\pi}^{3/2}(t) \lambda_{\ell\pi}^{1/2}(t) \left(\Delta_{\ell\pi}+t\right)}{24 t^3 M_K^4} \\
			&\quad - \Big( \< \tilde N_1 \>_{u_t} - 4 \< \tilde R_1 \>_{u_t} \Big) \frac{t \Delta_{K\pi} \left(3 \Delta_{\ell\pi}+\Sigma_0+t\right)+\Delta_{K\pi}^2 \left(\Delta_{\ell\pi}-2 t\right)+3 t^2 \left(\Sigma_0-t\right)}{24 t^2 M_K^2} \\
			&\quad + \Big( \< z \tilde N_1 \>_{u_t} - 4 \< z \tilde R_1 \>_{u_t} \Big) \begin{aligned}[t]
				& \Bigg( \frac{\lambda_{K\pi}^{1/2}(t) \left(\Delta_{\ell\pi}+t\right) \left(\Delta_{K\pi} \left(\Delta_{\ell\pi}-2 t\right)+t \left(\Sigma_0-t\right)\right)}{24 t^2 M_K^2 \lambda_{\ell\pi}^{1/2}(t)} \\
				& +\frac{\left(\Delta_{K\pi}+3 t\right) \lambda_{K\pi}^{1/2}(t) \lambda_{\ell\pi}^{1/2}(t)}{24 t^2 M_K^2} \Bigg) \end{aligned} \\
			&\quad - \Big( \< z^2 \tilde N_1 \>_{u_t} - 4 \< z^2 \tilde R_1 \>_{u_t} \Big) \frac{\lambda_{K\pi}(t) \left(\Delta_{\ell\pi}+t\right)}{24 t^2 M_K^2} , \\
	\end{split}
\end{align}

\begin{align}
	\footnotesize
	\begin{split}
		\hat N_1(t) &= \< M_0 \>_{s_t} \frac{3 M_K^4 \left(\Delta_{\ell\pi}+t\right)}{8 t \lambda_{\ell\pi}(t)} + \< z M_0 \>_{s_t} \frac{3 M_K^4 \left(\Delta_{K\pi}+t\right)}{4 t \lambda_{K\pi}^{1/2}(t) \lambda_{\ell\pi}^{1/2}(t)}  - \< z^2 M_0 \>_{s_t} \frac{9 M_K^4 \left(\Delta_{\ell\pi}+t\right)}{8 t \lambda_{\ell\pi}(t)} \\
			&\quad + \< M_1 \>_{s_t} \frac{3 M_K^2 \left(\Delta_{\ell\pi}+t\right) \left(\Delta_{K\pi} \Delta_{\ell\pi}+t \left(\Sigma_0-3 t\right)\right)}{8 t^2 \lambda_{\ell\pi}(t)} \\
			&\quad + \< z M_1 \>_{s_t} \Bigg( \frac{3 M_K^2 \left(\Delta_{K\pi}+t\right) \left(\Delta_{K\pi} \Delta_{\ell\pi}+t \left(\Sigma_0-3 t\right)\right)}{4 t^2 \lambda_{K\pi}^{1/2}(t) \lambda_{\ell\pi}^{1/2}(t)}-\frac{3 M_K^2 \lambda_{K\pi}^{1/2}(t) \left(\Delta_{\ell\pi}+t\right)}{8 t^2 \lambda_{\ell\pi}^{1/2}(t)} \Bigg) \\
			&\quad - \< z^2 M_1 \>_{s_t} \Bigg( \frac{9 M_K^2 \left(\Delta_{\ell\pi}+t\right) \left(\Delta_{K\pi} \Delta_{\ell\pi}+t \left(\Sigma_0-3t\right)\right)}{8 t^2 \lambda_{\ell\pi}(t)} + \frac{3 M_K^2 \left(\Delta_{K\pi}+t\right)}{4 t^2} \Bigg) \\
			&\quad + \< z^3 M_1 \>_{s_t} \frac{9 M_K^2 \lambda_{K\pi}^{1/2}(t) \left(\Delta_{\ell\pi}+t\right)}{8 t^2 \lambda_{\ell\pi}^{1/2}(t)} + \< \tilde M_1 \>_{s_t} \frac{3 M_K^4 \left(\Delta_{\ell\pi}+t\right)}{4 t \lambda_{\ell\pi}(t)} + \< z \tilde M_1 \>_{s_t} \frac{3 M_K^4 \left(\Delta_{K\pi}-3 t\right)}{2 t \lambda_{K\pi}^{1/2}(t) \lambda_{\ell\pi}^{1/2}(t)} \\
			&\quad - \< z^2 \tilde M_1 \>_{s_t} \frac{9 M_K^4 \left(\Delta_{\ell\pi}+t\right)}{4 t \lambda_{\ell\pi}(t)} - \Big( \< N_0 \>_{u_t} - 4 \< R_0 \>_{u_t} \Big) \frac{M_K^4 \left(\Delta_{\ell\pi}+t\right)}{4 t \lambda_{\ell\pi}(t)} \\
			&\quad + \Big( \< z N_0 \>_{u_t} - 4 \< z R_0 \>_{u_t} \Big) \frac{M_K^4 \left(t-\Delta_{K\pi}\right)}{2 t \lambda_{K\pi}^{1/2}(t) \lambda_{\ell\pi}^{1/2}(t) } + \Big( \< z^2 N_0 \>_{u_t} - 4 \< z^2 R_0 \>_{u_t} \Big) \frac{3 M_K^4 \left(\Delta_{\ell\pi}+t\right)}{4 t \lambda_{\ell\pi}(t)} \\
			&\quad + \Big( \< N_1 \>_{u_t} - 4 \< R_1 \>_{u_t} \Big) \frac{\left(\Delta_{\ell\pi}+t\right) \left(\Delta_{K\pi} \Delta_{\ell\pi}+t \left(t-\Sigma_0\right)\right)
   \left(\Delta_{K\pi} \Delta_{\ell\pi}+t \left(\Sigma_0-3 t\right)\right)}{16 t^3 \lambda_{\ell\pi}(t)} \\
			&\quad + \Big( \< z N_1 \>_{u_t} - 4 \< z R_1 \>_{u_t} \Big) \begin{aligned}[t]
				& \Bigg( \frac{\left(\Delta_{K\pi}-t\right) \left(\Delta_{K\pi} \Delta_{\ell\pi}+t \left(t-\Sigma_0\right)\right) \left(\Delta_{K\pi} \Delta_{\ell\pi}+t \left(\Sigma_0-3 t\right)\right)}{8 t^3 \lambda_{K\pi}^{1/2}(t) \lambda_{\ell\pi}^{1/2}(t)} \\
				& -\frac{\lambda_{K\pi}^{1/2}(t) \left(\Delta_{\ell\pi}+t\right) \left(\Delta_{K\pi} \Delta_{\ell\pi}+t^2\right)}{8 t^3 \lambda_{\ell\pi}^{1/2}(t) }  \Bigg) \end{aligned} \\
			&\quad - \Big( \< z^2 N_1 \>_{u_t} - 4 \< z^2 R_1 \>_{u_t} \Big) \begin{aligned}[t]
				& \Bigg( \frac{3 \left(\Delta_{\ell\pi}+t\right) \left(\Delta_{K\pi} \Delta_{\ell\pi}+t \left(t-\Sigma_0\right)\right) \left(\Delta_{K\pi} \Delta_{\ell\pi}+t \left(\Sigma_0-3 t\right)\right)}{16 t^3 \lambda_{\ell\pi}(t)} \\
				& - \frac{\lambda_{K\pi}(t) \left(\Delta_{\ell\pi}+t\right)}{16 t^3} + \frac{\left(\Delta_{K\pi}-t\right) \left(\Delta_{K\pi} \Delta_{\ell\pi}+t^2\right)}{4 t^3} \Bigg) \end{aligned} \\
			&\quad + \Big( \< z^3 N_1 \>_{u_t} - 4 \< z^3 R_1 \>_{u_t} \Big) \Bigg( \frac{\left(\Delta_{K\pi}-t\right) \lambda_{K\pi}^{1/2}(t) \lambda_{\ell\pi}^{1/2}(t) }{8 t^3}+\frac{3 \lambda_{K\pi}^{1/2}(t) \left(\Delta_{\ell\pi}+t\right) \left(\Delta_{K\pi} \Delta_{\ell\pi}+t^2\right)}{8 t^3 \lambda_{\ell\pi}^{1/2}(t)} \Bigg) \\
			&\quad - \Big( \< z^4 N_1 \>_{u_t} - 4 \< z^4 R_1 \>_{u_t} \Big) \frac{3 \lambda_{K\pi}(t) \left(\Delta_{\ell\pi}+t\right)}{16 t^3} \\
			&\quad + \Big( \< \tilde N_1 \>_{u_t} - 4 \< \tilde R_1 \>_{u_t} \Big) \frac{M_K^2 \left(\Delta_{\ell\pi}+t\right) \left(t^2+2 t \Delta_{K\pi} -\Sigma_0 t-\Delta_{K\pi} \Delta_{\ell\pi}\right)}{16 t^2 \lambda_{\ell\pi}(t)} \\
			&\quad + \Big( \< z \tilde N_1 \>_{u_t} - 4 \< z \tilde R_1 \>_{u_t} \Big) \begin{aligned}[t]
				& \Bigg( \frac{M_K^2 \lambda_{K\pi}^{1/2}(t) \left(\Delta_{\ell\pi}+t\right)}{16 t^2 \lambda_{\ell\pi}^{1/2}(t) } \\
				& -\frac{M_K^2 \left(t \Delta_{K\pi} \left(3 \Delta_{\ell\pi}+\Sigma_0+t\right)+\Delta_{K\pi}^2 \left(\Delta_{\ell\pi}-2 t\right)+3 t^2 \left(\Sigma_0-t\right)\right)}{8 t^2 \lambda_{K\pi}^{1/2}(t) \lambda_{\ell\pi}^{1/2}(t) } \Bigg) \end{aligned} \\
			&\quad + \Big( \< z^2 \tilde N_1 \>_{u_t} - 4 \< z^2 \tilde R_1 \>_{u_t} \Big) \begin{aligned}[t]
				& \Bigg( \frac{3 M_K^2 \left(\Delta_{\ell\pi}+t\right) \left(\Delta_{K\pi} \left(\Delta_{\ell\pi}-2 t\right)+t \left(\Sigma_0-t\right)\right)}{16 t^2 \lambda_{\ell\pi}(t)} \\
				& +\frac{M_K^2 \left(\Delta_{K\pi}+3 t\right)}{8 t^2} \Bigg) \end{aligned} \\
			&\quad - \Big( \< z^3 \tilde N_1 \>_{u_t} - 4 \< z^3 \tilde R_1 \>_{u_t} \Big) \frac{3 M_K^2 \lambda_{K\pi}^{1/2}(t) \left(\Delta_{\ell\pi}+t\right)}{16 t^2 \lambda_{\ell\pi}^{1/2}(t) } , \\
	\end{split}
\end{align}

\begin{align}
	\footnotesize
	\begin{split}
		\hat{\tilde N}_1(t) &= \< (1-z^2) M_0 \>_{s_t} \frac{3 M_K^2}{4 t} + \< (1-z^2) M_1 \>_{s_t} \frac{3 \left(\Delta_{K\pi} \Delta_{\ell\pi}+t \left(\Sigma_0 - 3 t\right)\right)}{4 t^2} \\
			&\quad - \< (1-z^2) z M_1 \>_{s_t} \frac{3 \lambda_{K\pi}^{1/2}(t) \lambda_{\ell\pi}^{1/2}(t)}{4 t^2} + \< (1-z^2) \tilde M_1 \>_{s_t} \frac{3 M_K^2}{2 t} \\
			&\quad - \Big( \< (1-z^2) N_0 \>_{u_t} - 4 \< (1-z^2) R_0 \>_{u_t} \Big) \frac{M_K^2}{2 t} \\
			&\quad + \Big( \< (1-z^2) N_1 \>_{u_t} - 4 \< (1-z^2) R_1 \>_{u_t} \Big) \frac{\left(\Delta_{K\pi} \Delta_{\ell\pi}+t \left(t-\Sigma_0\right)\right) \left(\Delta_{K\pi} \Delta_{\ell\pi}+t \left(\Sigma_0-3 t\right)\right)}{8 t^3 M_K^2} \\
			&\quad - \Big( \< (1-z^2) z N_1 \>_{u_t} - 4 \< (1-z^2) z R_1 \>_{u_t} \Big) \frac{\lambda_{K\pi}^{1/2}(t) \lambda_{\ell\pi}^{1/2}(t) \left(\Delta_{K\pi} \Delta_{\ell\pi}+t^2\right)}{4 t^3 M_K^2} \\
			&\quad + \Big( \< (1-z^2) z^2 N_1 \>_{u_t} - 4 \< (1-z^2) z^2 R_1 \>_{u_t} \Big) \frac{\lambda_{K\pi}(t) \lambda_{\ell\pi}(t)}{8 t^3 M_K^2} \\
			&\quad + \Big( \< (1-z^2) \tilde N_1 \>_{u_t} - 4 \< (1-z^2) \tilde R_1 \>_{u_t} \Big) \frac{t^2 + 2 t \Delta_{K\pi}-\Sigma_0 t-\Delta_{K\pi} \Delta_{\ell\pi}}{8 t^2} \\
			&\quad + \Big( \< (1-z^2) z \tilde N_1 \>_{u_t} - 4 \< (1-z^2) z \tilde R_1 \>_{u_t} \Big) \frac{\lambda_{K\pi}^{1/2}(t) \lambda_{\ell\pi}^{1/2}(t) }{8 t^2} , \\
	\end{split}
\end{align}

\begin{align}
	\footnotesize
	\begin{split}
		\hat R_0(u) &= \< M_0 \>_{s_u}  \frac{u+\Delta_{K\pi}}{4 u} - \< z M_0 \>_{s_u} \frac{\left(u+\Delta_{\ell\pi}\right) \lambda_{K\pi}^{1/2}(u)}{4 u \lambda_{\ell\pi}^{1/2}(u)} - \< M_1 \>_{s_u} \frac{\left(u+\Delta_{K\pi}\right) \left(u \left(\Sigma_0-3 u\right)+\Delta_{K\pi} \Delta_{\ell\pi}\right)}{8u^2 M_K^2} \\
			&\quad + \< z M_1 \>_{s_u} \Bigg( \frac{\lambda_{K\pi}^{1/2}(u) \lambda_{\ell\pi}^{1/2}(u) \left(u+\Delta_{K\pi}\right)}{8 u^2 M_K^2}+\frac{\left(u+\Delta_{\ell\pi}\right) \left(u \left(\Sigma_0-3 u\right)+\Delta_{K\pi} \Delta_{\ell\pi}\right) \lambda_{K\pi}^{1/2}(u)}{8 u^2 M_K^2 \lambda_{\ell\pi}^{1/2}(u)} \Bigg) \\
			&\quad - \< z^2 M_1 \>_{s_u} \frac{\left(u+\Delta_{\ell\pi}\right) \lambda_{K\pi}(u)}{8 u^2 M_K^2} + \< \tilde M_1 \>_{s_u}  \frac{3u-\Delta_{K\pi}}{4 u} + \< z \tilde M_1 \>_{s_u} \frac{\left(u+\Delta_{\ell\pi}\right) \lambda_{K\pi}^{1/2}(u)}{4 u \lambda_{\ell\pi}^{1/2}(u)} \\
			&\quad + \Big( 2 \< N_0 \>_{t_u} + \< R_0 \>_{t_u} \Big) \frac{\Delta_{K\pi} - u}{6u} - \Big( 2 \< z N_0 \>_{t_u} + \< z R_0 \>_{t_u} \Big) \frac{\left(u+\Delta_{\ell\pi}\right) \lambda_{K\pi}^{1/2}(u)}{6 u \lambda_{\ell\pi}^{1/2}(u)} \\
			&\quad - \Big( 2 \< N_1 \>_{t_u} + \< R_1 \>_{t_u} \Big) \frac{\left(\Delta_{K\pi}-u\right) \left(u \left(u-\Sigma_0\right)+\Delta_{K\pi} \Delta_{\ell\pi}\right) \left(u \left(\Sigma_0-3 u\right)+\Delta_{K\pi} \Delta_{\ell\pi}\right)}{24 u^3 M_K^4} \\
			&\quad + \Big( 2 \< z N_1 \>_{t_u} + \< z R_1 \>_{t_u} \Big) \begin{aligned}[t]
				&\Bigg( \frac{\left(\Delta_{K\pi}-u\right) \lambda_{K\pi}^{1/2}(u) \lambda_{\ell\pi}^{1/2}(u) \left(u^2+\Delta_{K\pi} \Delta_{\ell\pi}\right)}{12 u^3 M_K^4} \\
				& \hspace{-0.5cm} + \frac{\left(u+\Delta_{\ell\pi}\right) \left(u \left(u-\Sigma_0\right)+\Delta_{K\pi} \Delta_{\ell\pi}\right) \left(u \left(\Sigma_0-3 u\right)+\Delta_{K\pi} \Delta_{\ell\pi}\right) \lambda_{K\pi}^{1/2}(u)}{24 u^3 M_K^4 \lambda_{\ell\pi}^{1/2}(u)} \Bigg) \end{aligned} \\
			&\quad + \Big( 2 \< z^2 N_1 \>_{t_u} + \< z^2 R_1 \>_{t_u} \Big) \Bigg( \frac{\left(u-\Delta_{K\pi}\right) \lambda_{K\pi}(u) \lambda_{\ell\pi}(u)}{24 u^3 M_K^4}-\frac{\left(u+\Delta_{\ell\pi}\right) \left(u^2+\Delta_{K\pi} \Delta_{\ell\pi}\right) \lambda_{K\pi}(u)}{12 u^3 M_K^4} \Bigg) \\
			&\quad + \Big( 2 \< z^3 N_1 \>_{t_u} + \< z^3 R_1 \>_{t_u} \Big) \frac{\left(u+\Delta_{\ell\pi}\right) \lambda_{K\pi}^{3/2}(u) \lambda_{\ell\pi}^{1/2}(u)}{24 u^3 M_K^4} \\
			&\quad + \Big( 2 \< \tilde N_1 \>_{t_u} + \< \tilde R_1 \>_{t_u} \Big) \frac{3 \left(\Sigma_0-u\right) u^2+\Delta_{K\pi} \left(u+\Sigma_0+3 \Delta_{\ell\pi}\right) u+\Delta_{K\pi}^2 \left(\Delta_{\ell\pi}-2 u\right)}{24 u^2 M_K^2} \\
			&\quad + \Big( 2 \< z \tilde N_1 \>_{t_u} + \< z \tilde R_1 \>_{t_u} \Big) \begin{aligned}[t]
				& \Bigg( \frac{\left(u+\Delta_{\ell\pi}\right) \left(u^2-\Sigma_0 u+2 \Delta_{K\pi} u-\Delta_{K\pi} \Delta_{\ell\pi}\right) \lambda_{K\pi}^{1/2}(u)}{24 u^2 M_K^2 \lambda_{\ell\pi}^{1/2}(u)} \\
				& -\frac{\left(3 u+\Delta_{K\pi}\right) \lambda_{K\pi}^{1/2}(u) \lambda_{\ell\pi}^{1/2}(u)}{24 u^2 M_K^2} \Bigg) \end{aligned} \\
			&\quad + \Big( 2 \< z^2 \tilde N_1 \>_{t_u} + \< z^2 \tilde R_1 \>_{t_u} \Big) \frac{\left(u+\Delta_{\ell\pi}\right) \lambda_{K\pi}(u)}{24 u^2 M_K^2} , \\
	\end{split}
\end{align}

\begin{align}
	\footnotesize
	\begin{split}
		\hat R_1(u) &= \< M_0 \>_{s_u}  \frac{3 M_K^4 \left(u+\Delta_{\ell\pi}\right)}{8 u \lambda_{\ell\pi}(u)} + \< z M_0 \>_{s_u} \frac{3 M_K^4 \left(u+\Delta_{K\pi}\right)}{4 u \lambda_{K\pi}^{1/2}(u) \lambda_{\ell\pi}^{1/2}(u)} - \< z^2 M_0 \>_{s_u} \frac{9 M_K^4 \left(u+\Delta_{\ell\pi}\right)}{8 u \lambda_{\ell\pi}(u)} \\
			&\quad - \< M_1 \>_{s_u} \frac{3 M_K^2 \left(u+\Delta_{\ell\pi}\right) \left(u \left(\Sigma_0-3 u\right)+\Delta_{K\pi} \Delta_{\ell\pi}\right)}{16 u^2 \lambda_{\ell\pi}(u)} \\
			&\quad + \< z M_1 \>_{s_u} \Bigg( \frac{3 M_K^2 \lambda_{K\pi}^{1/2}(u) \left(\Delta_{\ell\pi}+u\right)}{16 u^2 \lambda_{\ell\pi}^{1/2}(u)}-\frac{3 M_K^2 \left(\Delta_{K\pi}+u\right) \left(\Delta_{K\pi} \Delta_{\ell\pi}+u \left(\Sigma_0-3 u\right)\right)}{8u^2 \lambda_{K\pi}^{1/2}(u) \lambda_{\ell\pi}^{1/2}(u) } \Bigg) \\
			&\quad + \< z^2 M_1 \>_{s_u} \Bigg( \frac{9 M_K^2 \left(\Delta_{\ell\pi}+u\right) \left(\Delta_{K\pi} \Delta_{\ell\pi}+u \left(\Sigma_0-3u\right)\right)}{16 u^2 \lambda_{\ell\pi}(u)}+\frac{3 M_K^2 \left(\Delta_{K\pi}+u\right)}{8 u^2} \Bigg) \\
			&\quad - \< z^3 M_1 \>_{s_u} \frac{9 M_K^2 \left(u+\Delta_{\ell\pi}\right) \lambda_{K\pi}^{1/2}(u)}{16 u^2 \lambda_{\ell\pi}^{1/2}(u)} - \< \tilde M_1 \>_{s_u} \frac{3 M_K^4 \left(u+\Delta_{\ell\pi}\right)}{8 u \lambda_{\ell\pi}(u)} - \< z \tilde M_1 \>_{s_u} \frac{3 M_K^4 \left(\Delta_{K\pi}-3 u\right)}{4 u \lambda_{K\pi}^{1/2}(u) \lambda_{\ell\pi}^{1/2}(u)} \\
			&\quad + \< z^2 \tilde M_1 \>_{s_u} \frac{9 M_K^4 \left(u+\Delta_{\ell\pi}\right)}{8 u \lambda_{\ell\pi}(u)} + \Big( 2 \< N_0 \>_{t_u} + \< R_0 \>_{t_u} \Big) \frac{M_K^4 \left(u+\Delta_{\ell\pi}\right)}{4 u \lambda_{\ell\pi}(u)} \\
			&\quad + \Big( 2 \< z N_0 \>_{t_u} + \< z R_0 \>_{t_u} \Big) \frac{M_K^4 \left(\Delta_{K\pi}-u\right)}{2 u \lambda_{K\pi}^{1/2}(u) \lambda_{\ell\pi}^{1/2}(u)} - \Big( 2 \< z^2 N_0 \>_{t_u} + \< z^2 R_0 \>_{t_u} \Big) \frac{3 M_K^4 \left(u+\Delta_{\ell\pi}\right)}{4 u \lambda_{\ell\pi}(u)} \\
			&\quad - \Big( 2 \< N_1 \>_{t_u} + \< R_1 \>_{t_u} \Big) \frac{\left(u+\Delta_{\ell\pi}\right) \left(u \left(u-\Sigma_0\right)+\Delta_{K\pi} \Delta_{\ell\pi}\right) \left(u \left(\Sigma_0-3 u\right)+\Delta_{K\pi} \Delta_{\ell\pi}\right)}{16 u^3 \lambda_{\ell\pi}(u)} \\
			&\quad + \Big( 2 \< z N_1 \>_{t_u} + \< z R_1 \>_{t_u} \Big) \begin{aligned}[t]
				& \Bigg( \frac{\lambda_{K\pi}^{1/2}(u) \left(\Delta_{\ell\pi}+u\right) \left(\Delta_{K\pi} \Delta_{\ell\pi}+u^2\right)}{8u^3 \lambda_{\ell\pi}^{1/2}(u)} \\
				& -\frac{\left(\Delta_{K\pi}-u\right) \left(\Delta_{K\pi} \Delta_{\ell\pi}+u \left(u-\Sigma_0\right)\right) \left(\Delta_{K\pi} \Delta_{\ell\pi}+u \left(\Sigma_0-3 u\right)\right)}{8 u^3 \lambda_{K\pi}^{1/2}(u) \lambda_{\ell\pi}^{1/2}(u)} \Bigg) \end{aligned} \\
			&\quad + \Big( 2 \< z^2 N_1 \>_{t_u} + \< z^2 R_1 \>_{t_u} \Big) \begin{aligned}[t]
				& \Bigg( \frac{3 \left(\Delta_{\ell\pi}+u\right) \left(\Delta_{K\pi} \Delta_{\ell\pi}+u \left(u-\Sigma_0\right)\right) \left(\Delta_{K\pi} \Delta_{\ell\pi}+u \left(\Sigma_0-3 u\right)\right)}{16 u^3 \lambda_{\ell\pi}(u)} \\
				& -\frac{\lambda_{K\pi}(u) \left(\Delta_{\ell\pi}+u\right)}{16 u^3}+\frac{\left(\Delta_{K\pi}-u\right) \left(\Delta_{K\pi} \Delta_{\ell\pi}+u^2\right)}{4 u^3} \Bigg) \end{aligned} \\
			&\quad + \Big( 2 \< z^3 N_1 \>_{t_u} + \< z^3 R_1 \>_{t_u} \Big) \begin{aligned}[t]
				& \Bigg( \frac{\left(u-\Delta_{K\pi}\right) \lambda_{K\pi}^{1/2}(u) \lambda_{\ell\pi}^{1/2}(u)}{8 u^3} \\
				& -\frac{3 \lambda_{K\pi}^{1/2}(u) \left(\Delta_{\ell\pi}+u\right) \left(\Delta_{K\pi} \Delta_{\ell\pi}+u^2\right)}{8 u^3 \lambda_{\ell\pi}^{1/2}(u)} \Bigg) \end{aligned} \\
			&\quad + \Big( 2 \< z^4 N_1 \>_{t_u} + \< z^4 R_1 \>_{t_u} \Big) \frac{3 \left(u+\Delta_{\ell\pi}\right) \lambda_{K\pi}(u)}{16 u^3} \\
			&\quad + \Big( 2 \< \tilde N_1 \>_{t_u} + \< \tilde R_1 \>_{t_u} \Big) \frac{M_K^2 \left(u+\Delta_{\ell\pi}\right) \left(u \left(\Sigma_0-u\right)+\Delta_{K\pi} \left(\Delta_{\ell\pi}-2u\right)\right)}{16 u^2 \lambda_{\ell\pi}(u)} \\
			&\quad + \Big( 2 \< z \tilde N_1 \>_{t_u} + \< z \tilde R_1 \>_{t_u} \Big) \begin{aligned}[t]
				&\Bigg( \frac{M_K^2 \left(u \Delta_{K\pi} \left(3 \Delta_{\ell\pi}+\Sigma_0+u\right)+\Delta_{K\pi}^2 \left(\Delta_{\ell\pi}-2 u\right)+3 u^2 \left(\Sigma_0-u\right)\right)}{8 u^2 \lambda_{K\pi}^{1/2}(u) \lambda_{\ell\pi}^{1/2}(u)} \\
				& -\frac{M_K^2 \lambda_{K\pi}^{1/2}(u) \left(\Delta_{\ell\pi}+u\right)}{16 u^2 \lambda_{\ell\pi}^{1/2}(u)} \Bigg) \end{aligned} \\
			&\quad + \Big( 2 \< z^2 \tilde N_1 \>_{t_u} + \< z^2 \tilde R_1 \>_{t_u} \Big) \begin{aligned}[t]
				& \Bigg( \frac{3 M_K^2 \left(\Delta_{\ell\pi}+u\right) \left(-\Delta_{K\pi} \Delta_{\ell\pi}+2 u \Delta_{K\pi}+u^2-\Sigma_0 u\right)}{16 u^2 \lambda_{\ell\pi}(u)} \\
				& -\frac{M_K^2 \left(\Delta_{K\pi}+3 u\right)}{8 u^2} \Bigg) \end{aligned} \\
			&\quad + \Big( 2 \< z^3 \tilde N_1 \>_{t_u} + \< z^3 \tilde R_1 \>_{t_u} \Big) \frac{3 M_K^2 \lambda_{K\pi}^{1/2}(u) \left(\Delta_{\ell\pi}+u\right)}{16 u^2 \lambda_{\ell\pi}^{1/2}(u)} , \\
	\end{split}
\end{align}

\begin{align}
	\footnotesize
	\begin{split}
		\hat{\tilde R}_1(u) &= \< (1-z^2) M_0 \>_{s_u} \frac{3 M_K^2}{4 u} - \< (1-z^2) M_1 \>_{s_u} \frac{3 \left(\Delta_{K\pi} \Delta_{\ell\pi}+u \left(\Sigma_0-3 u\right)\right)}{8 u^2} \\
			&\quad + \< (1-z^2) z M_1 \>_{s_u} \frac{3 \lambda_{K\pi}^{1/2}(u) \lambda_{\ell\pi}^{1/2}(u) }{8 u^2} - \< (1-z^2) \tilde M_1 \>_{s_u} \frac{3 M_K^2}{4 u} \\
			&\quad + \Big( 2 \< (1-z^2) N_0 \>_{t_u} + \< (1-z^2) R_0 \>_{t_u} \Big) \frac{M_K^2}{2 u} \\
			&\quad - \Big( 2 \< (1-z^2) N_1 \>_{t_u} + \< (1-z^2) R_1 \>_{t_u} \Big) \frac{\left(\Delta_{K\pi} \Delta_{\ell\pi}+u \left(u-\Sigma_0\right)\right) \left(\Delta_{K\pi} \Delta_{\ell\pi}+u \left(\Sigma_0-3 u\right)\right)}{8 u^3 M_K^2} \\
			&\quad + \Big( 2 \< (1-z^2) z N_1 \>_{t_u} + \< (1-z^2) z R_1 \>_{t_u} \Big) \frac{\lambda_{K\pi}^{1/2}(u) \lambda_{\ell\pi}^{1/2}(u) \left(\Delta_{K\pi} \Delta_{\ell\pi}+u^2\right)}{4 u^3 M_K^2} \\
			&\quad - \Big( 2 \< (1-z^2) z^2 N_1 \>_{t_u} + \< (1-z^2) z^2 R_1 \>_{t_u} \Big) \frac{\lambda_{K\pi}(u) \lambda_{\ell\pi}(u)}{8 u^3 M_K^2} \\
			&\quad + \Big( 2 \< (1-z^2) \tilde N_1 \>_{t_u} + \< (1-z^2) \tilde R_1 \>_{t_u} \Big) \frac{\Delta_{K\pi} \left(\Delta_{\ell\pi}-2 u\right)+u \left(\Sigma_0-u\right)}{8 u^2} \\
			&\quad - \Big( 2 \< (1-z^2) z \tilde N_1 \>_{t_u} + \< (1-z^2) z \tilde R_1 \>_{t_u} \Big) \frac{\lambda_{K\pi}^{1/2}(u) \lambda_{\ell\pi}^{1/2}(u)}{8 u^2} , \\
	\end{split}
\end{align}
where
\begin{align}
	\begin{split}
		\< z^n X \>_{t_s} := \frac{1}{2} \int_{-1}^1 z^n X(t(s,z)) dz , \\
		\< z^n X \>_{s_t} := \frac{1}{2} \int_{-1}^1 z^n X(s(t,z)) dz , \\
		\< z^n X \>_{u_t} := \frac{1}{2} \int_{-1}^1 z^n X(u(t,z)) dz , \\
		\< z^n X \>_{s_u} := \frac{1}{2} \int_{-1}^1 z^n X(s(u,z)) dz , \\
		\< z^n X \>_{t_u} := \frac{1}{2} \int_{-1}^1 z^n X(t(u,z)) dz , \\
	\end{split}
\end{align}
and
\begin{align}
	\begin{split}
		t(s,z) &= \frac{1}{2} \left( \Sigma_0 - s - 2 X \sigma_\pi z \right) , \\
		s(t,z) &= \frac{1}{2} \left( \Sigma_0 - t + \frac{1}{t} \left( z \, \lambda^{1/2}_{K\pi}(t) \lambda^{1/2}_{\ell\pi}(t) - \Delta_{K\pi}\Delta_{\ell\pi} \right) \right) , \\
		u(t,z) &= \frac{1}{2} \left( \Sigma_0 - t - \frac{1}{t} \left( z \, \lambda^{1/2}_{K\pi}(t) \lambda^{1/2}_{\ell\pi}(t) - \Delta_{K\pi}\Delta_{\ell\pi} \right) \right) , \\
		s(u,z) &= \frac{1}{2} \left( \Sigma_0 - u + \frac{1}{u} \left( z \, \lambda^{1/2}_{K\pi}(u) \lambda^{1/2}_{\ell\pi}(u) - \Delta_{K\pi}\Delta_{\ell\pi} \right) \right) , \\
		t(u,z) &= \frac{1}{2} \left( \Sigma_0 - u - \frac{1}{u} \left( z \, \lambda^{1/2}_{K\pi}(u) \lambda^{1/2}_{\ell\pi}(u) - \Delta_{K\pi}\Delta_{\ell\pi} \right) \right) .
	\end{split}
\end{align}
We recall the abbreviations
\begin{align}
	\begin{split}
		\Delta_{K\pi} &= M_K^2 - M_\pi^2, \quad \Delta_{\ell\pi} = s_\ell - M_\pi^2, \quad \Sigma_0 = M_K^2 + 2 M_\pi^2 + s_\ell , \\
		PL &= \frac{1}{2}(M_K^2 - s - s_\ell) , \quad X = \frac{1}{2} \lambda^{1/2}(M_K^2, s, s_\ell), \quad \sigma_\pi = \sqrt{1 - \frac{4M_\pi^2}{s}} .
	\end{split}
\end{align}


\clearpage

\section{Isospin-Breaking Corrected Data Input}

\label{sec:AppendixData}

In this appendix, we list the isospin-corrected data sets on the $K_{\ell4}$ form factors that we use for the fits of the dispersion relation. These are the NA48/2 \cite{Batley2010,Batley2012} and E865 data sets \cite{Pislak2001, Pislak2003}, corrected for isospin-breaking mass effects and (in the case of NA48/2) the additional radiative effects that were calculated in \cite{Stoffer2014}.

More detailed explanations can be found in section~\ref{sec:ExperimentalKl4Data}.

\subsection{One-Dimensional NA48/2 and E865 Data Sets}

\label{sec:AppendixData1Dim}

\begin{table}[H]
	\centering
	\begin{tabular}{c c c c c c}
		\toprule
		$\sqrt{s}/$MeV & $\sqrt{s_\ell}/$MeV & $F_s$ & $F_p$ & $G_p$ & $\tilde G_p$ \\[0.1cm]
		\hline \\[-0.3cm]
		286.06 &	92.61 &	5.6941(85)(185) &		$-0.181(67)(15)$	&	5.035(257)(66) 			&	4.317(74)(20) 	\\
		295.95 &	92.01 &	5.7878(90)(170) &		$-0.324(62)(34)$	&	5.168(142)(84) 			&	4.404(53)(32) 	\\
		304.88 &	91.51 &	5.8410(89)(171) &		$-0.209(60)(33)$	&	4.924(108)(59) 			&	4.532(46)(26) 	\\
		313.48 &	90.65 &	5.8905(91)(171) &		$-0.156(58)(32)$	&	4.879(\hphantom{0}91)(51)	&	4.627(41)(24) 	\\
		322.02 &	88.32 &	5.9275(90)(166) &		$-0.366(55)(41)$	&	5.227(\hphantom{0}80)(58) 	&	4.692(38)(29) 	\\
		330.80 &	85.59 &	5.9557(93)(168) &		$-0.383(54)(40)$	&	5.265(\hphantom{0}73)(56) 	&	4.748(35)(28) 	\\
		340.17 &	81.02 &	5.9915(92)(166) &		$-0.218(55)(46)$	&	5.036(\hphantom{0}68)(59) 	&	4.762(34)(31) 	\\
		350.94 &	76.16 &	6.0161(92)(163) &		$-0.302(54)(35)$	&	5.246(\hphantom{0}62)(37) 	&	4.889(34)(21) 	\\
		364.57 &	69.80 &	6.0351(91)(162) &		$-0.309(54)(33)$	&	5.338(\hphantom{0}57)(31) 	&	5.000(35)(20) 	\\
		389.95 &	58.96 &	6.1155(93)(224) &		$-0.264(59)(35)$	&	5.400(\hphantom{0}55)(34) 	&	5.144(36)(22) 	\\
		\bottomrule
	\end{tabular}
	\caption[NA48/2 data with additional radiative corrections and isospin breaking mass effects.]{NA48/2 data \cite{Batley2010,Batley2012}, corrected by additional radiative and isospin-breaking mass effects \cite{Stoffer2014}. The uncertainties of the isospin corrections (without the higher order estimate) are added in quadrature to the systematic error. The fully correlated error of the normalisation increases from $0.62\%$ to $0.70\%$. The normalisation of $F_s$ is increased by $0.77\%$ to take the dependence on $s_\ell$ into account.}
	\label{tab:NA48DataIsoCorr}
\end{table}

\begin{table}[H]
	\centering
	\begin{tabular}{c c c c}
		\toprule
		$\sqrt{s}/$MeV & $\sqrt{s_\ell}/$MeV & $F_s$ & $G_p$ \\[0.1cm]
		\hline \\[-0.3cm]
		287.6 &		106.8			&	5.781(13)(42) &	4.702(89)(40) 	\\
		299.5 &		105.7			&	5.825(14)(48) &	4.693(62)(37) 	\\
		311.2 &		103.8			&	5.914(14)(56) &	4.771(54)(41) 	\\
		324.0 &		101.1			&	5.974(16)(62) &	4.999(51)(56)	\\
		339.6 &		\hphantom{0}96.3	&	6.097(17)(63) &	5.002(49)(57) 	\\
		370.0 &		\hphantom{0}84.6	&	6.151(20)(41) &	5.104(50)(42) 	\\
		\bottomrule
	\end{tabular}
	\caption[E865 data with isospin breaking mass effects.]{E865 data \cite{Pislak2001, Pislak2003}, corrected by isospin-breaking mass effects \cite{Stoffer2014}. The uncertainties of the isospin corrections (without the higher order estimate) are added in quadrature to the systematic error. The fully correlated error of the normalisation is $1.2\%$.}
	\label{tab:E865DataIsoCorr}
\end{table}

\subsection{Two-Dimensional NA48/2 Data Set}

\label{sec:AppendixData2DimNA48}

For the fits of the dispersion relation to data, we do not use the above NA48/2 data set on $F_s$ consisting of 10 bins, but the two-dimensional data set, which was recently published as an addendum to \cite{Batley2012}. Here, we list the isospin-corrected values of $F_s$ that we use as input in our fits. The values and uncertainties, shown in table~\ref{tab:2dimNA48FsValues}, are constructed as follows.
\begin{itemize}
	\item With the number of data and Monte Carlo events for each 2D bin \cite{Batley2012}, we compute the relative values and statistical uncertainty of the relative values of $F_s$.
	\item We fix the normalisation by requiring $f_s = 5.705$ in a parametric fit of the form (\ref{eq:NA48PhenomenologicalFit}).
	\item Unfortunately, systematic errors are not available for the 2D data set. We guess a systematic error by assuming that the ratio of systematic and statistical error does not depend on $s_\ell$.
	\item We apply isospin corrections due to photonic and mass-difference effects \cite{Stoffer2014}. The uncertainty from the mass effects is added in quadrature to the systematic error.
\end{itemize}

\begin{table}[H]
	\centering
	\begin{tabular}{c | c c c c c c c c c c}
		\toprule
		$F_s$ & 1 & 2 & 3 & 4 & 5  & 6 \\[0.1cm]
		\hline \\[-0.3cm]
		1 &	5.641(51)(55) &	5.628(30)(35) &	5.700(24)(30) &	5.687(22)(28) &	5.744(21)(27) &	5.707(22)(28) \\
		2 &	5.753(51)(30) &	5.716(30)(22) &	5.747(24)(20) &	5.777(22)(20) &	5.807(22)(20) &	5.833(23)(20) \\
		3 &	5.785(52)(33) &	5.775(31)(24) &	5.831(25)(21) &	5.837(22)(21) &	5.841(22)(20) &	5.875(23)(21) \\
		4 &	5.908(52)(33) &	5.809(31)(23) &	5.877(25)(21) &	5.874(22)(20) &	5.910(22)(20) &	5.894(23)(20) \\
		5 &	5.910(52)(24) &	5.903(31)(19) &	5.891(25)(18) &	5.909(22)(18) &	5.924(21)(18) &	5.961(22)(18) \\
		6 &	5.831(51)(29) &	5.925(30)(22) &	5.912(24)(20) &	5.919(22)(19) &	5.971(21)(19) &	6.032(22)(19) \\
		7 &	6.031(50)(25) &	5.927(29)(20) &	5.941(23)(18) &	5.970(21)(18) &	6.024(20)(18) &	6.045(22)(18) \\
		8 &	6.026(47)(21) &	5.976(28)(18) &	5.990(22)(17) &	6.020(20)(17) &	6.024(20)(17) &	6.067(23)(17) \\
		9 &	6.023(44)(22) &	5.987(26)(18) &	6.037(20)(17) &	6.059(19)(17) &	6.044(20)(17) &	6.077(25)(18) \\
		10 &	6.163(38)(67) &	6.128(22)(41) &	6.107(18)(34) &	6.120(18)(34) &	6.139(23)(42)  &	6.130(45)(79) \\[0.1cm]
		\hline \\[-0.3cm]
		 & 7 & 8 & 9 & 10 \\[0.1cm]
		\hline \\[-0.3cm]
		1 &	5.703(25)(30) &	5.721(\hphantom{0}30)(35) &	5.717(43)(47) &	5.709(\hphantom{0}82)(86) \\
		2 &	5.817(26)(21) &	5.828(\hphantom{0}31)(23) &	5.872(43)(27) &	5.929(\hphantom{0}84)(45) \\
		3 &	5.843(26)(22) &	5.934(\hphantom{0}31)(24) &	5.911(43)(29) &	5.923(117)(68) \\
		4 &	5.905(26)(21) &	5.957(\hphantom{0}31)(24) &	6.111(50)(32) &	 \\
		5 &	6.004(25)(18) &	5.942(\hphantom{0}33)(20) &	6.074(70)(29) &	 \\
		6 &	6.025(26)(20) &	6.009(\hphantom{0}38)(24) &	 &	 \\
		7 &	6.042(28)(20) &	6.124(\hphantom{0}54)(26) &	 &	 \\
		8 &	6.086(33)(19) &	6.024(122)(39) &	 &	 \\
		9 &	6.058(54)(25) &	 &	 &	 \\
		\bottomrule
	\end{tabular}
	\caption{Values of $F_s$ for the two-dimensional data set of NA48/2 \cite{Batley2012} including isospin-breaking corrections \cite{Stoffer2014}. The fully correlated error of the normalisation of $0.70\%$ has to be treated separately.}
	\label{tab:2dimNA48FsValues}
\end{table}


\section{Matching Equations}

\subsection{Subtraction Constants at $\O(p^4)$ in \ChPT{}}

\label{sec:AppendixNLOSubtractionConstantsStandardRep}

In the following expressions for the subtraction constants at NLO, we have used the Gell-Mann--Okubo (GMO) formula $M_\eta^2 = (4 M_K^2 - M_\pi^2)/3$ to simplify the analytic expressions considerably. This introduces an error only at NNLO. In practise, we use the physical $\eta$ mass and not the GMO relation. We do not show the analytic expressions for this case because they are much larger.
\begin{align}
	\footnotesize
	\begin{split}
		m_{0,\mathrm{NLO}}^0 &= \frac{M_K}{\sqrt{2} F_\pi} \Bigg( 1 + \frac{1}{F_\pi^2} \begin{aligned}[t]
			&\bigg(-64 L_1^r M_\pi^2 + 16 L_2^r (M_K^2 + M_\pi^2) + 4 L_3^r (M_K^2 - 3 M_\pi^2) + 32 L_4^r M_\pi^2 + 4 L_5^r M_\pi^2 + 2 L_9^r s_\ell \\
   			& - \frac{161 M_K^6 + 42 M_K^4 M_\pi^2 - 27 M_K^2 M_\pi^4 + 4 M_\pi^6}{384 \pi^2 \Delta_{K\pi}^2} - s_\ell \frac{73 M_K^4 - 14 M_K^2 M_\pi^2 + M_\pi^4}{384 \pi^2 \Delta_{K\pi}^2} \\
			& +  \ln\left(\frac{M_\pi^2}{\mu^2}\right) \bigg(\frac{3 M_\pi^2 (3 M_K^6 - 8 M_K^4 M_\pi^2 + 2 M_K^2 M_\pi^4 + M_\pi^6)}{128 \pi^2 \Delta_{K\pi}^3} - s_\ell \frac{M_\pi^4 (3 M_K^2 - M_\pi^2)}{128 \pi^2 \Delta_{K\pi}^3}\bigg)\\
			& - \ln\left(\frac{M_K^2}{\mu^2}\right) \bigg(\frac{M_K^2 (92 M_K^6 - 15 M_K^2 M_\pi^4 + M_\pi^6)}{64 \pi^2 \Delta_{K\pi}^3} + s_\ell \frac{M_K^4 (41 M_K^2 - 15 M_\pi^2)}{64 \pi^2 \Delta_{K\pi}^3}\bigg) \\
			& +  \ln\left(\frac{M_\eta^2}{\mu^2}\right) \begin{aligned}[t]
				& \bigg(\frac{172 M_K^8 + 17 M_K^6 M_\pi^2 - 12 M_K^4 M_\pi^4 - 22 M_K^2 M_\pi^6 + 7 M_\pi^8}{128 \pi^2 \Delta_{K\pi}^3} \\
				& + s_\ell \frac{(4 M_K^2 - M_\pi^2)^2 (5 M_K^2 + M_\pi^2)}{128 \pi^2 \Delta_{K\pi}^3}\bigg) \bigg) \Bigg) , \end{aligned} \end{aligned}
	\end{split}
\end{align}
\begin{align}
	\footnotesize
	\begin{split}
		m_{0,\mathrm{NLO}}^1 &=  \frac{M_K}{\sqrt{2} F_\pi^3} \begin{aligned}[t]
			& \Bigg( 32 L_1^r M_K^2 + 8 L_3^r M_K^2 \\
			& + \frac{M_K^2 (116 M_K^6 + 273 M_K^4 M_\pi^2 - 258 M_K^2 M_\pi^4 + 49 M_\pi^6)}{384 \pi^2 \Delta_{K\pi}^2 (4 M_K^2 - M_\pi^2)} \\
			& - \ln\left(\frac{M_\pi^2}{\mu^2}\right) \frac{M_K^2 (8 M_K^6 - 24 M_K^4 M_\pi^2 + 21 M_K^2 M_\pi^4 - 7 M_\pi^6)}{128 \pi^2 \Delta_{K\pi}^3} \\
			& + \ln\left(\frac{M_K^2}{\mu^2}\right) \frac{M_K^2 (38 M_K^6 - 6 M_K^4 M_\pi^2 - 9 M_K^2 M_\pi^4 + 3 M_\pi^6)}{64 \pi^2 \Delta_{K\pi}^3} \\
			& - \ln\left( \frac{M_\eta^2}{\mu^2}\right) \frac{M_K^2 (4 M_K^2 - M_\pi^2)^2 (5 M_K^2 + M_\pi^2)}{128 \pi^2 \Delta_{K\pi}^3} \Bigg) , \end{aligned} \\
	\end{split}
\end{align}
\begin{align}
	\footnotesize
	\begin{split}
		m_{1,\mathrm{NLO}}^0 &= \frac{M_K}{\sqrt{2} F_\pi^3} \begin{aligned}[t]
			& \Bigg( -8 L_2^r M_K^2 \\
			& + \frac{M_K^2 (79 M_K^4 - 2 M_K^2 M_\pi^2 + 7 M_\pi^4)}{384 \pi^2 \Delta_{K\pi}^2} \\
			& + \ln\left(\frac{M_\pi^2}{\mu^2}\right) \frac{5 M_K^2 M_\pi^4 (3 M_K^2 - M_\pi^2)}{128 \pi^2 \Delta_{K\pi}^3} \\
			& + \ln\left(\frac{M_K^2}{\mu^2}\right) \frac{M_K^6 (43 M_K^2 - 21 M_\pi^2)}{64 \pi^2 \Delta_{K\pi}^3} \\
			& - \ln\left(\frac{M_\eta^2}{\mu^2}\right) \frac{M_K^2 (4 M_K^2 - M_\pi^2)^2 (5 M_K^2 + M_\pi^2)}{128 \pi^2 \Delta_{K\pi}^3} \Bigg) , \end{aligned} \\
	\end{split}
\end{align}
\begin{align}
	\footnotesize
	\begin{split}
		\tilde m_{1,\mathrm{NLO}}^0 &= \frac{M_K}{\sqrt{2} F_\pi} \Bigg( 1 + \frac{1}{F_\pi^2} \begin{aligned}[t]
			& \bigg(  - 8 L_2^r (M_K^2 + 2 M_\pi^2 + s_\ell) - 4 L_3^r (M_K^2 + M_\pi^2) + 4 L_5^r M_\pi^2 + 2 L_9^r s_\ell \\
			& + \frac{16 M_K^6 - 3 M_K^4 M_\pi^2 + 3 M_K^2 M_\pi^4 + 2 M_\pi^6}{96 \pi^2 \Delta_{K\pi}^2} + s_\ell \frac{(M_K^2 + M_\pi^2)^2}{64 \pi^2 \Delta_{K\pi}^2} \\
			& - \ln\left(\frac{M_\pi^2}{\mu^2}\right) \bigg(\frac{M_\pi^2 (3 M_K^2 - M_\pi^2) (M_K^4 - 8 M_K^2 M_\pi^2 - 5 M_\pi^4)}{128 \pi^2 \Delta_{K\pi}^3} -  s_\ell \frac{M_\pi^4 (3 M_K^2 - M_\pi^2)}{32 \pi^2 \Delta_{K\pi}^3}\bigg) \\
			& + \ln\left(\frac{M_K^2}{\mu^2}\right) \bigg(\frac{M_K^2 (37 M_K^6 - 35 M_K^4 M_\pi^2 - 17 M_K^2 M_\pi^4 + 3 M_\pi^6)}{64 \pi^2 \Delta_{K\pi}^3} + s_\ell \frac{M_K^4 (M_K^2 - 3 M_\pi^2)}{32 \pi^2 \Delta_{K\pi}^3}\bigg) \\
			& - \ln\left(\frac{M_\eta^2}{\mu^2}\right) \frac{68 M_K^6 + 7 M_K^4 M_\pi^2 - 2 M_K^2 M_\pi^4 - M_\pi^6}{128 \pi^2 \Delta_{K\pi}^2} \bigg) \Bigg) , \end{aligned}
	\end{split}
\end{align}
\begin{align}
	\footnotesize
	\begin{split}
		\tilde m_{1,\mathrm{NLO}}^1 &= \frac{M_K}{\sqrt{2} F_\pi^3} \begin{aligned}[t]
			& \Bigg(8 L_2^r M_K^2 \\
			& - \frac{M_K^2 (M_K^4 + M_\pi^4)}{32 \pi^2 \Delta_{K\pi}^2} \\
			& - \ln\left(\frac{M_\pi^2}{\mu^2}\right) \frac{M_K^2 (M_K^6 - 3 M_K^4 M_\pi^2 + 12 M_K^2 M_\pi^4 - 4 M_\pi^6)}{96 \pi^2 \Delta_{K\pi}^3} \\
			& - \ln\left(\frac{M_K^2}{\mu^2}\right) \frac{M_K^2 (7 M_K^6 - 21 M_K^4 M_\pi^2 + 3 M_K^2 M_\pi^4 - M_\pi^6)}{192 \pi^2 \Delta_{K\pi}^3} \Bigg) , \end{aligned} \\
	\end{split}
\end{align}
\begin{align}
	\footnotesize
	\begin{split}
		n_{0,\mathrm{NLO}}^1 &= \frac{M_K}{\sqrt{2} F_\pi^3} \begin{aligned}[t]
			& \Bigg(-24 L_2^r M_K^2 - 6 L_3^r M_K^2 \\
			& - \frac{M_K^2 (16613 M_K^6 - 2179 M_K^4 M_\pi^2 + 29 M_K^2 M_\pi^4 + 69 M_\pi^6)}{2048 \pi^2 \Delta_{K\pi}^3} \\
			& + \ln\left(\frac{M_\pi^2}{\mu^2}\right) \frac{3 M_K^2 M_\pi^4 (37 M_K^4 - 80 M_K^2 M_\pi^2 + 20 M_\pi^4)}{512 \pi^2 \Delta_{K\pi}^4} \\
			& + \ln\left(\frac{M_K^2}{\mu^2}\right) \frac{3 M_K^6 (-4840 M_K^4 + 1216 M_K^2 M_\pi^2 + 83 M_\pi^4)}{512 \pi^2 \Delta_{K\pi}^4} \\
			& + \ln\left(\frac{M_\eta^2}{\mu^2}\right) \frac{3 M_K^2 (4 M_K^2 - M_\pi^2) (304 M_K^6 - 6 M_K^4 M_\pi^2 - M_\pi^6)}{128 \pi^2 \Delta_{K\pi}^4} \Bigg) . \end{aligned}
	\end{split}
\end{align}

\subsection{Matching at NNLO}

\subsubsection{Decomposition of the Two-Loop Result}

\label{sec:AppendixTwoLoopDecomposition}

\paragraph{NLO Contribution}

We have already decomposed the NLO contributions. We apply a gauge transformation to convert the expressions to the second gauge and evaluate the result numerically:
\begin{align}
	\begin{split}
		m_{0,L}^{0,\mathrm{NLO}} &= \frac{M_K}{\sqrt{2} F_\pi} \begin{aligned}[t]
			& \Big( -0.1466 \cdot 10^3 L_1^r + 0.4953 \cdot 10^3 L_2^r + 0.0872 \cdot 10^3 L_3^r + 0.0733 \cdot 10^3 L_4^r \\
			& + 0.0092 \cdot 10^3 L_5^r + 0.0573 \cdot 10^3 L_9^r \frac{s_\ell}{M_K^2} \Big) , \end{aligned} \\
		m_{0,L}^{1,\mathrm{NLO}} &= \frac{M_K}{\sqrt{2} F_\pi} \begin{aligned}[t]
			& \Big( 0.9173 \cdot 10^3 L_1^r + 0.2293 \cdot 10^3 L_3^r \Big) , \end{aligned} \\
		m_{0,L}^{2,\mathrm{NLO}} &= 0 , \\
		m_{1,L}^{0,\mathrm{NLO}} &= \frac{M_K}{\sqrt{2} F_\pi} \begin{aligned}[t]
			& \Big( -0.2293 \cdot 10^3 L_2^r \Big) , \end{aligned} \\
		m_{1,L}^{1,\mathrm{NLO}} &= 0 , \\
		\tilde m_{1,L}^{0,\mathrm{NLO}} &= \frac{M_K}{\sqrt{2} F_\pi} \begin{aligned}[t]
			& \Big( -0.2660 \cdot 10^3 L_2^r - 0.1238 \cdot 10^3 L_3^r + 0.0092 \cdot 10^3 L_5^r \\
			&- ( 0.2293 \cdot 10^3 L_2^r - 0.0573 \cdot 10^3 L_9^r) \frac{s_\ell}{M_K^2} \Big) , \end{aligned} \\
		\tilde m_{1,L}^{1,\mathrm{NLO}} &= \frac{M_K}{\sqrt{2} F_\pi} \begin{aligned}[t]
			& \Big( 0.2293 \cdot 10^3 L_2^r \Big) , \end{aligned} \\
		\tilde m_{1,L}^{2,\mathrm{NLO}} &= 0 , \\
		n_{0,L}^{1,\mathrm{NLO}} &= \frac{M_K}{\sqrt{2} F_\pi} \begin{aligned}[t]
			& \Big( -0.6880 \cdot 10^3 L_2^r - 0.1720 \cdot 10^3 L_3^r \Big) , \end{aligned} \\
		n_{0,L}^{2,\mathrm{NLO}} &= n_{1,L}^{0,\mathrm{NLO}} = \tilde n_{1,L}^{1,\mathrm{NLO}} = 0 , 
	\end{split}
\end{align}
\begin{align}
	\begin{split}
		m_{0,R}^{0,\mathrm{NLO}} &= \frac{M_K}{\sqrt{2} F_\pi} \left( 0.1393 + 0.0444 \frac{s_\ell}{M_K^2} + 0.0256 \frac{s_\ell^2}{M_K^4} \right) , \\
		m_{0,R}^{1,\mathrm{NLO}} &= \frac{M_K}{\sqrt{2} F_\pi} \left( 0.3413 - 0.0512 \frac{s_\ell}{M_K^2} \right) , \\
		m_{0,R}^{2,\mathrm{NLO}} &= \frac{M_K}{\sqrt{2} F_\pi} \left( 0.4080 \right) , \\
		m_{1,R}^{0,\mathrm{NLO}} &= \frac{M_K}{\sqrt{2} F_\pi} \left( -0.0916 - 0.0512 \frac{s_\ell}{M_K^2} \right) , \\
		m_{1,R}^{1,\mathrm{NLO}} &= \frac{M_K}{\sqrt{2} F_\pi} \left( 0.0512 \right) , \\
		\tilde m_{1,R}^{0,\mathrm{NLO}} &= \frac{M_K}{\sqrt{2} F_\pi} \left( -0.0902 - 0.0595 \frac{s_\ell}{M_K^2} - 0.0256 \frac{s_\ell^2}{M_K^4} \right) , \\
		\tilde m_{1,R}^{1,\mathrm{NLO}} &= \frac{M_K}{\sqrt{2} F_\pi} \left( 0.1545 + 0.0512 \frac{s_\ell}{M_K^2} \right) , \\
		\tilde m_{1,R}^{2,\mathrm{NLO}} &= \frac{M_K}{\sqrt{2} F_\pi} \left( 0.0137 \right) , \\
		n_{0,R}^{1,\mathrm{NLO}} &= \frac{M_K}{\sqrt{2} F_\pi} \left( -0.1376 - 0.0384 \frac{s_\ell}{M_K^2} \right) , \\
		n_{0,R}^{2,\mathrm{NLO}} &= \frac{M_K}{\sqrt{2} F_\pi} \left( -0.0796 \right) , \\
		n_{1,R}^{0,\mathrm{NLO}} &= \frac{M_K}{\sqrt{2} F_\pi} \left( 0.0384 \right) , \\
		\tilde n_{1,R}^{1,\mathrm{NLO}} &= \frac{M_K}{\sqrt{2} F_\pi} \left( -0.0282 \right) .
	\end{split}
\end{align}

\paragraph{NNLO LECs}

First, we consider the contribution of the NNLO LECs, the $C_i^r$. We decompose this contribution into the form of the polynomial part in (\ref{eq:FunctionsOfOneVariable3Subtr}):
\begin{align}
	\begin{split}
		m_{0,C}^{0,\mathrm{NNLO}} &= \frac{M_K}{\sqrt{2}F_\pi} \frac{1}{F_\pi^4} \begin{aligned}[t]
			&\bigg(4 M_K^4 \begin{aligned}[t]
				& \big(C_{1}^r-2 C_{3}^r-2C_{4}^r+2C_{5}^r +4 C_{6}^r \\
				& +2 C_{10}^r +8 C_{11}^r-4 C_{12}^r-8 C_{13}^r+2C_{22}^r+4 C_{23}^r-2C_{34}^r\big) \end{aligned} \\
			& - M_\pi^2 M_K^2 \begin{aligned}[t]
				&\big(4 C_{1}^r+64 C_{2}^r+56 C_{3}^r+34 C_{4}^r-8 C_{5}^r+40 C_{6}^r \\
				& +64 C_{7}^r+24 C_{8}^r -16 C_{10}^r-48 C_{11}^r-8 C_{12}^r+112 C_{13}^r \\
				& +16 C_{14}^r-80 C_{15}^r-64 C_{17}^r +8 C_{22}^r-16 C_{23}^r+16 C_{25}^r \\
				& +16 C_{26}^r+32 C_{29}^r+64 C_{30}^r+32 C_{36}^r \\
				& +C_{66}^r+2 C_{67}^r-C_{69}^r-C_{88}^r+C_{90}^r\big) \end{aligned} \\
			& + M_\pi^4 \begin{aligned}[t]
				&\big(-24 C_{1}^r-128 C_{2}^r-32 C_{3}^r-18 C_{4}^r-32 C_{5}^r-24 C_{6}^r -64 C_{7}^r \\
				& +8 C_{8}^r +8 C_{10}^r+16 C_{11}^r-80 C_{12}^r-80 C_{13}^r+32 C_{14}^r+8 C_{15}^r \\
				& +128 C_{16}^r-48 C_{17}^r +16 C_{22}^r+32 C_{23}^r+32 C_{26}^r+128 C_{28}^r \\
				&-5 (C_{66}^r+2 C_{67}^r - C_{69}^r -C_{88}^r +C_{90}^r) \big) \end{aligned} \\
			& + s_\ell \begin{aligned}[t]
				&\Big( M_K^2 \begin{aligned}[t]
					&\big( 4 C_{1}^r-8 C_{3}^r-6 C_{4}^r-8 C_{12}^r-32 C_{13}^r \\
					& - 8 C_{63}^r-8 C_{64}^r+C_{66}^r+2 C_{67}^r+3 C_{69}^r-C_{88}^r-3 C_{90}^r\big) \end{aligned} \\
				& - M_\pi^2 \begin{aligned}[t]
					&\big(12 C_{1}^r+64 C_{2}^r+72 C_{3}^r+10 C_{4}^r-48 C_{13}^r-8 C_{22}^r \\
					& -32 C_{23}^r+8 C_{25}^r + 4 C_{64}^r+4 C_{65}^r+9 C_{66}^r+2 C_{67}^r \\
					& +16 C_{68}^r+3 C_{69}^r + 8 C_{83}^r+16 C_{84}^r+3 C_{88}^r+C_{90}^r \big) \Big) \end{aligned} \end{aligned} \\
			& - 2 s_\ell^2 \big(8 C_{3}^r+2 C_{4}^r+C_{66}^r+2 C_{67}^r-C_{69}^r+C_{88}^r-C_{90}^r \big) \bigg) , \end{aligned} \\
		m_{0,C}^{1,\mathrm{NNLO}} &= \frac{M_K}{\sqrt{2} F_\pi} \frac{1}{F_\pi^4} \begin{aligned}[t]
			& \bigg( 8 M_K^4 \begin{aligned}[t]
				&\big(C_{1}^r+4 C_{2}^r+4 C_{3}^r+2 C_{4}^r+4 C_{6}^r+4 C_{7}^r+2 C_{8}^r \\
				& +2 C_{12}^r+8 C_{13}^r-2 C_{23}^r+C_{25}^r\big) \end{aligned} \\
				&+ 2 M_\pi^2 M_K^2 \begin{aligned}[t]
					&\big(24 C_{1}^r+96 C_{2}^r+32 C_{3}^r-6 C_{4}^r+8 C_{5}^r \\
					& +8 C_{6}^r+16 C_{7}^r -16 C_{13}^r-16 C_{23}^r+8 C_{25}^r \\
					& +C_{66}^r+2 C_{67}^r-C_{69}^r-C_{88}^r+C_{90}^r \big) \end{aligned} \\
			&+2 s_\ell M_K^2 \begin{aligned}[t]
				&\big(4 C_{1}^r+16 C_{2}^r+16 C_{3}^r-2 C_{4}^r \\
				& +3 C_{66}^r+2 C_{67}^r+4 C_{68}^r+C_{69}^r+C_{88}^r-C_{90}^r \big) \bigg) , \end{aligned} \end{aligned} \\
		m_{0,C}^{2,\mathrm{NNLO}} &= \frac{M_K}{\sqrt{2}F_\pi} \frac{1}{F_\pi^4} 16 M_K^4 \big(-C_{1}^r-4 C_{2}^r-C_{3}^r+C_{4}^r\big), \\
		m_{1,C}^{0,\mathrm{NNLO}} &= \frac{M_K}{\sqrt{2} F_\pi} \frac{1}{F_\pi^4} 2 M_K^2 \begin{aligned}[t]
			& \bigg( M_K^2  \begin{aligned}[t]
					& \big(4 C_{3}^r+2 C_{4}^r-4 C_{10}^r-16 C_{11}^r+12 C_{12}^r+32 C_{13}^r \\
					& -2 C_{22}^r-4 C_{23}^r+2 C_{63}^r+C_{66}^r+C_{67}^r-C_{69}^r-2 C_{83}^r+C_{90}^r \big) \end{aligned} \\
				& + M_\pi^2 \begin{aligned}[t]
					& \big(16 C_{3}^r+16 C_{4}^r-4 C_{10}^r-8 C_{11}^r+4 C_{12}^r+16 C_{13}^r \\
					& -10 C_{22}^r-8 C_{23}^r+6 C_{25}^r-2 C_{63}^r+4 C_{67}^r+2 C_{83}^r-C_{88}^r \big) \end{aligned} \\
			& + s_\ell \big(12 C_{3}^r+2 C_{4}^r+C_{66}^r+C_{67}^r-C_{69}^r+C_{88}^r-C_{90}^r \big) \bigg), \end{aligned} \\
		m_{1,C}^{1,\mathrm{NNLO}} &= \frac{M_K}{\sqrt{2} F_\pi} \frac{1}{F_\pi^4} 2 M_K^4\big(-16 C_{3}^r-6 C_{4}^r+C_{66}^r-C_{67}^r-C_{69}^r-C_{88}^r+C_{90}^r\big) , \\
	\end{split}
\end{align}
\begin{align}
	\begin{split}
		\tilde m_{1,C}^{0,\mathrm{NNLO}} &= \frac{M_K}{\sqrt{2} F_\pi} \frac{1}{F_\pi^4}  \begin{aligned}[t]
			& \bigg(M_\pi^4  \begin{aligned}[t]
				& \big(-8 C_{1}^r+10 C_{4}^r-8 C_{6}^r-8 C_{8}^r-8 C_{10}^r-16 C_{11}^r+16 C_{12}^r \\
				& +16 C_{13}^r+8 C_{15}^r+16 C_{17}^r -16 C_{22}^r-32 C_{23}^r \\
				& +C_{66}^r+2 C_{67}^r-C_{69}^r-C_{88}^r+C_{90}^r\big) \end{aligned} \\
			& + M_\pi^2 M_K^2 \begin{aligned}[t]
				& \big(-12 C_{1}^r+24 C_{3}^r+34 C_{4}^r-8 C_{5}^r-24 C_{6}^r-8 C_{8}^r \\
				& -8 C_{10}^r-32 C_{11}^r+16 C_{12}^r+16 C_{13}^r+16 C_{14}^r+16 C_{15}^r \\
				& -36 C_{22}^r-32 C_{23}^r+20 C_{25}^r+16 C_{26}^r-32 C_{29}^r \\
				& +4 C_{63}^r+C_{66}^r-6 C_{67}^r-C_{69}^r-4 C_{83}^r+C_{88}^r+C_{90}^r\big) \end{aligned} \\
			& -2 M_K^4 \begin{aligned}[t]
				& \big(2 C_{1}^r-2 C_{4}^r+4 C_{5}^r+8 C_{6}^r+12 C_{12}^r+16 C_{13}^r+2 C_{22}^r \\
				& +4 C_{23}^r+4 C_{34}^r +2 C_{63}^r+C_{66}^r+C_{67}^r-C_{69}^r-2 C_{83}^r+C_{90}^r\big) \end{aligned} \\
			& +s_\ell  \begin{aligned}[t]
				& \Big( M_\pi^2 \begin{aligned}[t]
					& \big(-4 C_{1}^r+40 C_{3}^r+26 C_{4}^r -8 C_{10}^r-16 C_{11}^r+8 C_{12}^r+16 C_{13}^r \\
					& -12 C_{22}^r-16 C_{23}^r+4 C_{25}^r -4 C_{63}^r-4 C_{64}^r-4 C_{65}^r+C_{66}^r \\
					& -6 C_{67}^r-5 C_{69}^r-4 C_{83}^r+C_{88}^r-7 C_{90}^r\big) \end{aligned} \\
				& - M_K^2 \begin{aligned}[t]
					& \big(4 C_{1}^r+8 C_{3}^r-6 C_{4}^r+8 C_{10}^r+32 C_{11}^r-16 C_{12}^r-32 C_{13}^r \\
					& +4 C_{22}^r+8 C_{23}^r +4 C_{63}^r+8 C_{64}^r+C_{66}^r+2 C_{67}^r+3 C_{69}^r \\
					& +4 C_{83}^r+C_{88}^r+C_{90}^r\big) \Big) \end{aligned} \end{aligned} \\
			& + s_\ell^2 \big(8 C_{3}^r-2 C_{67}^r\big) \bigg) , \end{aligned} \\
		\tilde m_{1,C}^{1,\mathrm{NNLO}} &= \frac{M_K}{\sqrt{2} F_\pi} \frac{1}{F_\pi^4} 2 M_K^2 \begin{aligned}[t]
			& \bigg( -M_K^2 \begin{aligned}[t]
				& \big(4 C_{3}^r+8 C_{4}^r-4 C_{10}^r-16 C_{11}^r +4 C_{12}^r+32 C_{13}^r -6 C_{22}^r \\
				& -4 C_{23}^r+4 C_{25}^r -2 C_{63}^r-2 C_{67}^r+2 C_{83}^r-C_{88}^r \big) \end{aligned} \\
			& -M_\pi^2 \begin{aligned}[t]
				& \big(16 C_{3}^r+10 C_{4}^r-4 C_{10}^r-8 C_{11}^r \\
				& +12 C_{12}^r+16 C_{13}^r-6 C_{22}^r-8 C_{23}^r+2 C_{25}^r \\
				& +2 C_{63}^r+C_{66}^r-2 C_{67}^r-C_{69}^r-2 C_{83}^r+C_{90}^r\big) \end{aligned} \\
			& -s_\ell \big(12 C_{3}^r+2 C_{4}^r+C_{66}^r-2 C_{67}^r-C_{69}^r+C_{88}^r-C_{90}^r\big) \bigg), \end{aligned} \\
		\tilde m_{1,C}^{2,\mathrm{NNLO}} &= \frac{M_K}{\sqrt{2} F_\pi} \frac{1}{F_\pi^4} 2 M_K^4 \big(8 C_{3}^r+2 C_{4}^r+C_{66}^r-C_{67}^r-C_{69}^r-C_{88}^r+C_{90}^r\big) , \\
		n_{0,C}^{1,\mathrm{NNLO}} &= \frac{M_K}{\sqrt{2}F_\pi} \frac{1}{F_\pi^4} 3 M_K^2 \begin{aligned}[t]
			& \bigg( - 2M_K^2 \begin{aligned}[t]
				& \big(3 C_{1}^r -4 C_{4}^r + 2C_{5}^r +4 C_{6}^r \\
				& +2C_{10}^r +8 C_{11}^r -2C_{12}^r -8 C_{13}^r +2C_{22}^r +4 C_{23}^r \big) \end{aligned} \\
			& - \frac{1}{2} M_\pi^2 \begin{aligned}[t]
				& \big(16 C_{1}^r +8 C_{3}^r -18 C_{4}^r +8 C_{6}^r +8 C_{8}^r \\
				& +8 C_{10}^r +16 C_{11}^r -16 C_{12}^r -16 C_{13}^r +16 C_{22}^r +32 C_{23}^r \\
				& -C_{66}^r -2 C_{67}^r +C_{69}^r +C_{88}^r -C_{90}^r \big) \end{aligned} \\
			& - \frac{1}{2} s_\ell \begin{aligned}[t]
				&\big(4 C_{1}^r -8 C_{3}^r -6 C_{4}^r +C_{66}^r +2 C_{67}^r +3 C_{69}^r \\
				& -C_{88}^r +C_{90}^r \big) \bigg) , \end{aligned} \end{aligned} \\
		n_{0,C}^{2,\mathrm{NNLO}} &= \frac{M_K}{\sqrt{2}F_\pi} \frac{1}{F_\pi^4} 12 M_K^4 \big(C_{1}^r +C_{3}^r -C_{4}^r \big), \\
		n_{1,C}^{0,\mathrm{NNLO}} &= \frac{M_K}{\sqrt{2}F_\pi} \frac{1}{F_\pi^4} \frac{-3 M_K^4}{2} \big(16 C_{3}^r +6 C_{4}^r -C_{66}^r -2 C_{67}^r +C_{69}^r +C_{88}^r -C_{90}^r \big) , \\
		\tilde n_{1,C}^{1,\mathrm{NNLO}} &= \frac{M_K}{\sqrt{2}F_\pi} \frac{1}{F_\pi^4} 3 M_K^4 \big(8 C_{3}^r +2 C_{4}^r +C_{66}^r +2 C_{67}^r -C_{69}^r -C_{88}^r +C_{90}^r \big) .
	\end{split}
\end{align}
Unfortunately, a lot of NNLO LECs enter the polynomial. In total, there appear 24 linearly independent combinations of the $C_i^r$.

If we use the resonance estimate of \cite{Bijnens2012}, we obtain the following values for the NNLO counterterm contribution:
\begin{align}
	\begin{split}
		m_{0,\mathrm{reso}}^{0,\mathrm{NNLO}} &= \frac{M_K}{\sqrt{2} F_\pi} \left( -0.1546 - 0.1716 \frac{s_\ell}{M_K^2} + 0.0316 \frac{s_\ell^2}{M_K^4} \right) , \\
		m_{0,\mathrm{reso}}^{1,\mathrm{NNLO}} &= \frac{M_K}{\sqrt{2} F_\pi} \left( 0.1747 - 0.0316 \frac{s_\ell}{M_K^2} \right) , \\
		m_{0,\mathrm{reso}}^{2,\mathrm{NNLO}} &= \frac{M_K}{\sqrt{2} F_\pi} \left( 0.0310 \right) , \\
		m_{1,\mathrm{reso}}^{0,\mathrm{NNLO}} &= \frac{M_K}{\sqrt{2} F_\pi} \left( 0.1657 - 0.0316 \frac{s_\ell}{M_K^2} \right) , \\
		m_{1,\mathrm{reso}}^{1,\mathrm{NNLO}} &= \frac{M_K}{\sqrt{2} F_\pi} \left( -0.0104 \right) , \\
		\tilde m_{1,\mathrm{reso}}^{0,\mathrm{NNLO}} &= \frac{M_K}{\sqrt{2} F_\pi} \left( -0.0900 - 0.0135 \frac{s_\ell}{M_K^2} \right) , \\
		\tilde m_{1,\mathrm{reso}}^{1,\mathrm{NNLO}} &= \frac{M_K}{\sqrt{2} F_\pi} \left( -0.1712 - 0.0316 \frac{s_\ell}{M_K^2} \right) , \\
		\tilde m_{1,\mathrm{reso}}^{2,\mathrm{NNLO}} &= \frac{M_K}{\sqrt{2} F_\pi} \left( 0.1805 \right) , \\
		n_{0,\mathrm{reso}}^{1,\mathrm{NNLO}} &= \frac{M_K}{\sqrt{2} F_\pi} \left( 0.1502 - 0.0237 \frac{s_\ell}{M_K^2} \right) , \\
		n_{0,\mathrm{reso}}^{2,\mathrm{NNLO}} &= \frac{M_K}{\sqrt{2} F_\pi} \left( -0.0233 \right) , \\
		n_{1,\mathrm{reso}}^{0,\mathrm{NNLO}} &= \frac{M_K}{\sqrt{2} F_\pi} \left( -0.0078 \right) , \\
		\tilde n_{1,\mathrm{reso}}^{1,\mathrm{NNLO}} &= \frac{M_K}{\sqrt{2} F_\pi} \left( 0.2707 \right) .
	\end{split}
\end{align}

Alternatively, if we use the `preferred values' of the BE14 fit \cite{Bijnens2014} (complemented with $C_{88}^r-C_{90}^r = -55 \cdot 10^{-6}$ \cite{BijnensTalavera2002} and the remaining LECs that appear in the $s_\ell$-dependence set to zero), we obtain the following values for the NNLO counterterm contribution:
\begin{align}
	\begin{split}
		m_{0,\mathrm{BE14}}^{0,\mathrm{NNLO}} &= \frac{M_K}{\sqrt{2} F_\pi} \left( -0.4108 - 0.1823 \frac{s_\ell}{M_K^2} - 0.0033 \frac{s_\ell^2}{M_K^4} \right) , \\
		m_{0,\mathrm{BE14}}^{1,\mathrm{NNLO}} &= \frac{M_K}{\sqrt{2} F_\pi} \left( 0.7959 + 0.0986 \frac{s_\ell}{M_K^2} \right) , \\
		m_{0,\mathrm{BE14}}^{2,\mathrm{NNLO}} &= \frac{M_K}{\sqrt{2} F_\pi} \left( -0.1709 \right) , \\
		m_{1,\mathrm{BE14}}^{0,\mathrm{NNLO}} &= \frac{M_K}{\sqrt{2} F_\pi} \left( 0.2627 + 0.0296 \frac{s_\ell}{M_K^2} \right) , \\
		m_{1,\mathrm{BE14}}^{1,\mathrm{NNLO}} &= \frac{M_K}{\sqrt{2} F_\pi} \left( -0.1709 \right) , \\
		\tilde m_{1,\mathrm{BE14}}^{0,\mathrm{NNLO}} &= \frac{M_K}{\sqrt{2} F_\pi} \left( 0.0356 + 0.1050 \frac{s_\ell}{M_K^2} + 0.0263 \frac{s_\ell^2}{M_K^4} \right) , \\
		\tilde m_{1,\mathrm{BE14}}^{1,\mathrm{NNLO}} &= \frac{M_K}{\sqrt{2} F_\pi} \left( -0.2942 - 0.0296 \frac{s_\ell}{M_K^2} \right) , \\
		\tilde m_{1,\mathrm{BE14}}^{2,\mathrm{NNLO}} &= \frac{M_K}{\sqrt{2} F_\pi} \left( 0.1841 \right) , \\
		n_{0,\mathrm{BE14}}^{1,\mathrm{NNLO}} &= \frac{M_K}{\sqrt{2} F_\pi} \left( 0.3505 + 0.0296 \frac{s_\ell}{M_K^2} \right) , \\
		n_{0,\mathrm{BE14}}^{2,\mathrm{NNLO}} &= \frac{M_K}{\sqrt{2} F_\pi} \left( 0.0099 \right) , \\
		n_{1,\mathrm{BE14}}^{0,\mathrm{NNLO}} &= \frac{M_K}{\sqrt{2} F_\pi} \left( -0.1282 \right) , \\
		\tilde n_{1,\mathrm{BE14}}^{1,\mathrm{NNLO}} &= \frac{M_K}{\sqrt{2} F_\pi} \left( 0.2761 \right) .
	\end{split}
\end{align}

\paragraph{Vertex Integrals}

Let us study in more detail the contribution of the vertex integrals. They can be decomposed into functions of one Mandelstam variable according to
{ \small
\begin{align}
	\begin{split}
		F_V^\mathrm{NNLO}(s,t,u) &= F_{VS,0}^\mathrm{NNLO}(s,s_\ell) + \frac{u-t}{M_K^2} F_{VS,1}^\mathrm{NNLO}(s,s_\ell) + F_{VT,0}^\mathrm{NNLO}(t,s_\ell) + \frac{s-u}{M_K^2} F_{VT,1}^\mathrm{NNLO}(t,s_\ell) + F_{VU}^\mathrm{NNLO}(u,s_\ell) , \\
		G_V^\mathrm{NNLO}(s,t,u) &= G_{VS}^\mathrm{NNLO}(s,s_\ell) + G_{VT,0}^\mathrm{NNLO}(t,s_\ell) + \frac{s-u}{M_K^2} G_{VT,1}^\mathrm{NNLO}(t,s_\ell) + G_{VU}^\mathrm{NNLO}(u,s_\ell) .
	\end{split}
\end{align} }%
The $u$-channel vertex integrals fulfil $F^\mathrm{NNLO}_{VU} = G^\mathrm{NNLO}_{VU}$. In the following, we treat them numerically. The contribution to $R_0$ is obtained by subtracting the constant, linear and quadratic terms:
\begin{align}
	\begin{split}
		R_0^V(u,s_\ell) &= F_{VU}^\mathrm{NNLO}(u,s_\ell) - P_{VU}^\mathrm{NNLO}(u,s_\ell) , \\
		P_{VU}^\mathrm{NNLO}(u,s_\ell) &= F_{VU}^\mathrm{NNLO}(0,s_\ell) + u {F_{VU}^\mathrm{NNLO}}^\prime(0,s_\ell) + \frac{1}{2} u^2 {F_{VU}^\mathrm{NNLO}}^\dprime(0,s_\ell) ,
	\end{split}
\end{align}
where ${}^\prime$ stands for the derivative with respect to the first argument ($u$). The polynomial $P_{VU}^\mathrm{NNLO}$ has to be lumped into the overall polynomial and finally reshuffled into the subtraction constants. Numerically, we find
\begin{align}
	\begin{split}
		P_{VU}^\mathrm{NNLO}(u,s_\ell) &\approx \frac{M_K}{\sqrt{2} F_\pi} \left( 0.4008 + 0.0119 \frac{s_\ell}{M_K^2} + \left( -0.2521 - 0.0130 \frac{s_\ell}{M_K^2} \right) \frac{u}{M_K^2} + 0.0569 \frac{u^2}{M_K^4} \right).
		\raisetag{-0.1cm}
	\end{split}
\end{align}
As we have checked again numerically, the polynomial-subtracted $u$-channel contribution of the vertex integrals fulfils the dispersion relation
\begin{align}
	\begin{split}
		R_0^V(u,s_\ell) &= \frac{u^3}{\pi} \int_{u_0}^\infty \frac{\Im R_0^V(u^\prime,s_\ell)}{(u^\prime-u-i\epsilon){u^\prime}^3} du^\prime .
	\end{split}
\end{align}

Next, we consider the $s$-channel vertex integrals: apart from a polynomial, they belong to either $M_0$, $M_1$ or $\tilde M_1$. Again, we subtract the first few terms of the Taylor expansion:
\begin{align}
	\begin{split}
		M_0^V(s,s_\ell) &= F_{VS,0}^\mathrm{NNLO}(s,s_\ell) - P_{F,VS0}^\mathrm{NNLO}(s,s_\ell) , \\
		M_1^V(s,s_\ell) &= F_{VS,1}^\mathrm{NNLO}(s,s_\ell) - P_{F,VS1}^\mathrm{NNLO}(s,s_\ell) , \\
		\tilde M_1^V(s,s_\ell) &= G_{VS}^\mathrm{NNLO}(s,s_\ell) - P_{G,VS}^\mathrm{NNLO}(s,s_\ell) , \\
		P_{F,VS0}^\mathrm{NNLO}(s,s_\ell) &= F_{VS,0}^\mathrm{NNLO}(0,s_\ell) + s {F_{VS,0}^\mathrm{NNLO}}^\prime(0,s_\ell) + \frac{1}{2} s^2 {F_{VS,0}^\mathrm{NNLO}}^\dprime(0,s_\ell) , \\
		P_{F,VS1}^\mathrm{NNLO}(s,s_\ell) &= F_{VS,1}^\mathrm{NNLO}(0,s_\ell) + s {F_{VS,1}^\mathrm{NNLO}}^\prime(0,s_\ell) , \\
		P_{G,VS}^\mathrm{NNLO}(s,s_\ell) &= G_{VS}^\mathrm{NNLO}(0,s_\ell) + s {G_{VS}^\mathrm{NNLO}}^\prime(0,s_\ell) + \frac{1}{2} s^2 {G_{VS}^\mathrm{NNLO}}^\dprime(0,s_\ell) .
	\end{split}
\end{align}
We find numerically
\begin{align}
	\begin{split}
		P_{F,VS0}^\mathrm{NNLO}(s,s_\ell) &\approx \frac{M_K}{\sqrt{2} F_\pi} \left( 0.2663 + 0.0992 \frac{s_\ell}{M_K^2} + \left( -1.7763 - 0.0450 \frac{s_\ell}{M_K^2} \right) \frac{s}{M_K^2} - 0.5385 \frac{s^2}{M_K^4} \right) , \\
		P_{F,VS1}^\mathrm{NNLO}(s,s_\ell) &\approx \frac{M_K}{\sqrt{2} F_\pi} \left( 0.0029 + 0.0006 \frac{s_\ell}{M_K^2} + 0.0006 \frac{s}{M_K^2} \right) , \\
		P_{G,VS}^\mathrm{NNLO}(s,s_\ell) &\approx \frac{M_K}{\sqrt{2} F_\pi} \left( -0.3197 - 0.0727 \frac{s_\ell}{M_K^2} + \left( 0.1457 + 0.0163 \frac{s_\ell}{M_K^2} \right) \frac{s}{M_K^2} + 0.0003 \frac{s^2}{M_K^4} \right) .
		\raisetag{-0.1cm}
	\end{split}
\end{align}
A numerical check shows that the polynomial-subtracted $s$-channel contributions of the vertex integrals fulfil the dispersion relations
\begin{align}
	\begin{split}
		M_0^V(s,s_\ell) &= \frac{s^3}{\pi} \int_{s_0}^\infty \frac{\Im M_0^V(s^\prime,s_\ell)}{(s^\prime - s - i\epsilon) {s^\prime}^3} ds^\prime , \\
		M_1^V(s,s_\ell) &= \frac{s^2}{\pi} \int_{s_0}^\infty \frac{\Im M_1^V(s^\prime,s_\ell)}{(s^\prime - s - i\epsilon) {s^\prime}^2} ds^\prime , \\
		\tilde M_1^V(s,s_\ell) &= \frac{s^3}{\pi} \int_{s_0}^\infty \frac{\Im \tilde M_1^V(s^\prime,s_\ell)}{(s^\prime - s - i\epsilon) {s^\prime}^3} ds^\prime .
	\end{split}
\end{align}
Finally, we consider the $t$-channel, which is a bit more intricate: the reason is that not all linear and quadratic terms of a simple Taylor expansion in $t$ belong to the subtraction polynomial. The $t$-channel contributions can be written as
{ \small
\begin{align}
	\begin{split}
		F_{VT,0}^\mathrm{NNLO}(t,s_\ell) &= \frac{2}{3} N_0^V(t,s_\ell) + \frac{2}{3} \frac{\Delta_{K\pi}\Delta_{\ell\pi}}{M_K^4} N_1^V(t,s_\ell) - \frac{2}{3} \frac{\Delta_{K\pi} - 3 t}{2M_K^2} \tilde N_1^V(t,s_\ell) + \frac{1}{3} R_0^V(t,s_\ell) + P_{F,VT0}^\mathrm{NNLO}(t,s_\ell) , \\
		F_{VT,1}^\mathrm{NNLO}(t,s_\ell) &= \frac{2t}{3M_K^2} N_1^V(t,s_\ell) + P_{F,VT1}^\mathrm{NNLO}(t,s_\ell) , \\
		G_{VT,0}^\mathrm{NNLO}(t,s_\ell) &= -\frac{2}{3} N_0^V(t,s_\ell) - \frac{2}{3} \frac{\Delta_{K\pi}\Delta_{\ell\pi}}{M_K^4} N_1^V(t,s_\ell) + \frac{2}{3} \frac{\Delta_{K\pi} + t}{2M_K^2} \tilde N_1^V(t,s_\ell) - \frac{1}{3} R_0^V(t,s_\ell) + P_{G,VT0}^\mathrm{NNLO}(t,s_\ell) , \\
		G_{VT,1}^\mathrm{NNLO}(t,s_\ell) &= -\frac{2t}{3M_K^2} N_1^V(t,s_\ell) + P_{G,VT1}^\mathrm{NNLO}(t,s_\ell) , \\
	\end{split}
\end{align} }%
where $P_{F,VT0}^\mathrm{NNLO}$, $P_{G,VT0}^\mathrm{NNLO}$ are second order and $P_{F,VT1}^\mathrm{NNLO}$, $P_{G,VT1}^\mathrm{NNLO}$ are first order polynomials. The Taylor expansion of $N_0^V$ starts with a cubic term, the one of $\tilde N_1^V$ with a quadratic and the one of $N_1^V$ with a linear term. Numerically, we find
\begin{align}
	\begin{split}
		P_{F,VT1}^\mathrm{NNLO}(t,s_\ell) &= -P_{G,VT1}^\mathrm{NNLO}(t,s_\ell) \approx \frac{M_K}{\sqrt{2} F_\pi} \left( 0.0044 + 0.0002 \frac{s_\ell}{M_K^2} + 0.0003 \frac{t}{M_K^2} \right)
	\end{split}
\end{align}
and also identify the linear and the quadratic term of the Taylor expansion of $N_1^V$. In the sum
\begin{align}
	\begin{split}
		F_{VT,0}^\mathrm{NNLO}(t,s_\ell) + G_{VT,0}^\mathrm{NNLO}(t,s_\ell) &= \frac{4 t}{3 M_K^2} \tilde N_1^V(t,s_\ell) + P_{F,VT0}^\mathrm{NNLO}(t,s_\ell) + P_{G,VT0}^\mathrm{NNLO}(t,s_\ell) ,
	\end{split}
\end{align}
we can easily separate $\tilde N_1^V$ from the sum of the polynomials. After having identified $\tilde N_1^V$ (in particular the quadratic term of its Taylor expansion), we can also separate the difference of the polynomials using
\begin{align}
	\begin{split}
		F_{VT,0}^\mathrm{NNLO}(t,s_\ell) - G_{VT,0}^\mathrm{NNLO}(t,s_\ell) &= \frac{4}{3} N_0^V(t,s_\ell) + \frac{4}{3} \frac{\Delta_{K\pi}\Delta_{\ell\pi}}{M_K^4} N_1^V(t,s_\ell) - \frac{2}{3} \frac{\Delta_{K\pi} - t}{M_K^2} \tilde N_1^V(t,s_\ell) \\
			&\quad + \frac{2}{3} R_0^V(t,s_\ell) + P_{F,VT0}^\mathrm{NNLO}(t,s_\ell) - P_{G,VT0}^\mathrm{NNLO}(t,s_\ell) .
	\end{split}
\end{align}
Numerically, we find
{ \small
\begin{align}
	\begin{split}
		P_{F,VT0}^\mathrm{NNLO}(t,s_\ell) &\approx \frac{M_K}{\sqrt{2} F_\pi} \left( -0.6831 - 0.1136 \frac{s_\ell}{M_K^2} - 0.0013 \frac{s_\ell^2}{M_K^4} + \left( 0.2841 - 0.0006 \frac{s_\ell}{M_K^2} \right) \frac{t}{M_K^2} + 0.0190 \frac{t^2}{M_K^4} \right) , \\
		P_{G,VT0}^\mathrm{NNLO}(t,s_\ell) &\approx \frac{M_K}{\sqrt{2} F_\pi} \left( -0.0055 - 0.0146 \frac{s_\ell}{M_K^2} - 0.0006 \frac{s_\ell^2}{M_K^4} + \left( 0.0131 + 0.0095 \frac{s_\ell}{M_K^2} \right) \frac{t}{M_K^2} - 0.0356 \frac{t^2}{M_K^4} \right) .
	\end{split}
\end{align} }%
Again, the following dispersion relations can be checked numerically:
\begin{align}
	\begin{split}
		N_0^V(t,s_\ell) &= \frac{t^3}{\pi} \int_{t_0}^\infty \frac{\Im N_0^V(t^\prime,s_\ell)}{(t^\prime - t - i\epsilon) {t^\prime}^3} dt^\prime , \\
		N_1^V(t,s_\ell) &= \frac{t}{\pi} \int_{t_0}^\infty \frac{\Im N_1^V(t^\prime,s_\ell)}{(t^\prime - t - i\epsilon) {t^\prime}} dt^\prime , \\
		\tilde N_1^V(t,s_\ell) &= \frac{t^2}{\pi} \int_{t_0}^\infty \frac{\Im \tilde N_1^V(t^\prime,s_\ell)}{(t^\prime - t - i\epsilon) {t^\prime}^2} dt^\prime .
	\end{split}
\end{align}
Reshuffling the polynomial contributions into the subtraction constants leads to
\begin{align}
	\begin{split}
		m_{0,V}^{0,\mathrm{NNLO}} &= \frac{M_K}{\sqrt{2} F_\pi} \left( 0.0705 + 0.0667 \frac{s_\ell}{M_K^2} - 0.0539 \frac{s_\ell^2}{M_K^4} \right) , \\
		m_{0,V}^{1,\mathrm{NNLO}} &= \frac{M_K}{\sqrt{2} F_\pi} \left( -1.7841 + 0.0648 \frac{s_\ell}{M_K^2} \right) , \\
		m_{0,V}^{2,\mathrm{NNLO}} &= \frac{M_K}{\sqrt{2} F_\pi} \left( -0.5954 \right) , \\
		m_{1,V}^{0,\mathrm{NNLO}} &= \frac{M_K}{\sqrt{2} F_\pi} \left( -0.2658 + 0.0969 \frac{s_\ell}{M_K^2} \right) , \\
		m_{1,V}^{1,\mathrm{NNLO}} &= \frac{M_K}{\sqrt{2} F_\pi} \left( -0.1132 \right) , \\
		\tilde m_{1,V}^{0,\mathrm{NNLO}} &= \frac{M_K}{\sqrt{2} F_\pi} \left( -0.1310 - 0.2580 \frac{s_\ell}{M_K^2} + 0.0435 \frac{s_\ell^2}{M_K^4} \right) , \\
		\tilde m_{1,V}^{1,\mathrm{NNLO}} &= \frac{M_K}{\sqrt{2} F_\pi} \left( 0.2570 - 0.0849 \frac{s_\ell}{M_K^2} \right) , \\
		\tilde m_{1,V}^{2,\mathrm{NNLO}} &= \frac{M_K}{\sqrt{2} F_\pi} \left( 0.0572 \right) , \\
		n_{0,V}^{1,\mathrm{NNLO}} &= \frac{M_K}{\sqrt{2} F_\pi} \left( -0.2587 + 0.0519 \frac{s_\ell}{M_K^2} \right) , \\
		n_{0,V}^{2,\mathrm{NNLO}} &= \frac{M_K}{\sqrt{2} F_\pi} \left( 0.0898 \right) , \\
		n_{1,V}^{0,\mathrm{NNLO}} &= \frac{M_K}{\sqrt{2} F_\pi} \left( -0.0849 \right) , \\
		\tilde n_{1,V}^{1,\mathrm{NNLO}} &= \frac{M_K}{\sqrt{2} F_\pi} \left( 0.0729 \right) .
	\end{split}
\end{align}

\paragraph{Remaining Two-Loop Integrals}

Next, we consider the remaining two-loop parts, $X_P^\mathrm{NNLO}$. It is easy to decompose them into functions of one Mandelstam variable:
\begin{align}
	\begin{split}
		F_P^\mathrm{NNLO}(s,t,u) &= F_{PS}^\mathrm{NNLO}(s,s_\ell) + F_{PT,0}^\mathrm{NNLO}(t,s_\ell) + \frac{s-u}{M_K^2} F_{PT,1}^\mathrm{NNLO}(t,s_\ell) \\
			&\quad + F_{PU}^\mathrm{NNLO}(u,s_\ell) + P_{F,P}^\mathrm{NNLO}(s,t,u) , \\
		G_P^\mathrm{NNLO}(s,t,u) &= G_{PS}^\mathrm{NNLO}(s,s_\ell) + G_{PT,0}^\mathrm{NNLO}(t,s_\ell) + \frac{s-u}{M_K^2} G_{PT,1}^\mathrm{NNLO}(t,s_\ell) \\
			&\quad + G_{PU}^\mathrm{NNLO}(u,s_\ell) + P_{G,P}^\mathrm{NNLO}(s,t,u) , \\
	\end{split}
\end{align}
where $P_{F,P}^\mathrm{NNLO}$ and $P_{G,P}^\mathrm{NNLO}$ are second order polynomials. The remaining steps are analogous to the case of the vertex integrals. Again, we apply subtractions to the different functions:
\begin{align}
	\begin{split}
		M_0^P(s,s_\ell) &= F_{PS}^\mathrm{NNLO}(s,s_\ell) - P_{F,PS}^\mathrm{NNLO}(s,s_\ell) , \\
		\tilde M_1^P(s,s_\ell) &= G_{PS}^\mathrm{NNLO}(s,s_\ell) - P_{G,PS}^\mathrm{NNLO}(s,s_\ell) , \\
		R_0^P(u,s_\ell) &= F_{PU}^\mathrm{NNLO}(u,s_\ell) - P_{F,PU}^\mathrm{NNLO}(u,s_\ell) \\
			&= G_{PU}^\mathrm{NNLO}(u,s_\ell) - P_{G,PU}^\mathrm{NNLO}(u,s_\ell) , \\
	\end{split}
\end{align}
where
\begin{align}
	\begin{split}
		P_{F,PS}^\mathrm{NNLO}(s,s_\ell) &= F_{PS}^\mathrm{NNLO}(0,s_\ell) + s {F_{PS}^\mathrm{NNLO}}^\prime(0,s_\ell) + \frac{1}{2} s^2 {F_{PS}^\mathrm{NNLO}}^\dprime(0,s_\ell) , \\
		P_{G,PS}^\mathrm{NNLO}(s,s_\ell) &= G_{PS}^\mathrm{NNLO}(0,s_\ell) + s {G_{PS}^\mathrm{NNLO}}^\prime(0,s_\ell) + \frac{1}{2} s^2 {G_{PS}^\mathrm{NNLO}}^\dprime(0,s_\ell) , \\
		P_{F,PU}^\mathrm{NNLO}(u,s_\ell) &= F_{PU}^\mathrm{NNLO}(0,s_\ell) + u {F_{PU}^\mathrm{NNLO}}^\prime(0,s_\ell) + \frac{1}{2} u^2 {F_{PU}^\mathrm{NNLO}}^\dprime(0,s_\ell) , \\
		P_{G,PU}^\mathrm{NNLO}(u,s_\ell) &= G_{PU}^\mathrm{NNLO}(0,s_\ell) + u {G_{PU}^\mathrm{NNLO}}^\prime(0,s_\ell) + \frac{1}{2} u^2 {G_{PU}^\mathrm{NNLO}}^\dprime(0,s_\ell) .
	\end{split}
\end{align}
Numerically, we find
{\small
\begin{align}
	\begin{split}
		P_{F,PS}^\mathrm{NNLO}(s,s_\ell) &\approx \frac{M_K}{\sqrt{2} F_\pi} \left( -0.1660 + 0.0002 \frac{s_\ell}{M_K^2} + 0.0007 \frac{s_\ell^2}{M_K^4} + \left( 1.1629 + 0.0343 \frac{s_\ell}{M_K^2} \right) \frac{s}{M_K^2} + 0.7815 \frac{s^2}{M_K^4} \right) , \\
		P_{G,PS}^\mathrm{NNLO}(s,s_\ell) &\approx \frac{M_K}{\sqrt{2} F_\pi} \left( 0.0609 + 0.0118 \frac{s_\ell}{M_K^2} + \left( -0.0514 - 0.0058 \frac{s_\ell}{M_K^2} \right) \frac{s}{M_K^2} - 0.0007 \frac{s^2}{M_K^4} \right) , \\
		P_{F,PU}^\mathrm{NNLO}(u,s_\ell) &= P_{G,PU}^\mathrm{NNLO}(u,s_\ell) \approx \frac{M_K}{\sqrt{2} F_\pi} \left( 0.0585 - 0.0089 \frac{s_\ell}{M_K^2} + \left( 0.0165 + 0.0069 \frac{s_\ell}{M_K^2} \right) \frac{u}{M_K^2} - 0.0442 \frac{u^2}{M_K^4} \right) .
	\end{split}
\end{align} }%
The $t$-channel contributions can be written as
{ \small
\begin{align}
	\begin{split}
		F_{PT,0}^\mathrm{NNLO}(t,s_\ell) &= \frac{2}{3} N_0^P(t,s_\ell) + \frac{2}{3} \frac{\Delta_{K\pi}\Delta_{\ell\pi}}{M_K^4} N_1^P(t,s_\ell) - \frac{2}{3} \frac{\Delta_{K\pi} - 3 t}{2M_K^2} \tilde N_1^P(t,s_\ell) + \frac{1}{3} R_0^P(t,s_\ell) + P_{F,PT0}^\mathrm{NNLO}(t,s_\ell) , \\
		F_{PT,1}^\mathrm{NNLO}(t,s_\ell) &= \frac{2t}{3M_K^2} N_1^P(t,s_\ell) + P_{F,PT1}^\mathrm{NNLO}(t,s_\ell) , \\
		G_{PT,0}^\mathrm{NNLO}(t,s_\ell) &= -\frac{2}{3} N_0^P(t,s_\ell) - \frac{2}{3} \frac{\Delta_{K\pi}\Delta_{\ell\pi}}{M_K^4} N_1^P(t,s_\ell) + \frac{2}{3} \frac{\Delta_{K\pi} + t}{2M_K^2} \tilde N_1^P(t,s_\ell) - \frac{1}{3} R_0^P(t,s_\ell) + P_{G,PT0}^\mathrm{NNLO}(t,s_\ell) , \\
		G_{PT,1}^\mathrm{NNLO}(t,s_\ell) &= -\frac{2t}{3M_K^2} N_1^P(t,s_\ell) + P_{G,PT1}^\mathrm{NNLO}(t,s_\ell) , \\
	\end{split}
\end{align} }%
where $P_{F,PT0}^\mathrm{NNLO}$, $P_{G,PT0}^\mathrm{NNLO}$ are second order and $P_{F,PT1}^\mathrm{NNLO}$, $P_{G,PT1}^\mathrm{NNLO}$ are first order polynomials. Numerically, we find
\begin{align}
	\begin{split}
		P_{F,PT1}^\mathrm{NNLO}(t,s_\ell) &= -P_{G,PT1}^\mathrm{NNLO}(t,s_\ell) \approx \frac{M_K}{\sqrt{2} F_\pi} \left( -0.0010 - 0.0001 \frac{t}{M_K^2} \right)
	\end{split}
\end{align}
and also identify the linear and the quadratic term of the Taylor expansion of $N_1^P$. In the sum
\begin{align}
	\begin{split}
		F_{PT,0}^\mathrm{NNLO}(t,s_\ell) + G_{PT,0}^\mathrm{NNLO}(t,s_\ell) &= \frac{4 t}{3 M_K^2} \tilde N_1^P(t,s_\ell) + P_{F,PT0}^\mathrm{NNLO}(t,s_\ell) + P_{G,PT0}^\mathrm{NNLO}(t,s_\ell) ,
	\end{split}
\end{align}
we can separate $\tilde N_1^P$ from the polynomials. We obtain the difference of the polynomials with
\begin{align}
	\begin{split}
		F_{PT,0}^\mathrm{NNLO}(t,s_\ell) - G_{PT,0}^\mathrm{NNLO}(t,s_\ell) &= \frac{4}{3} N_0^P(t,s_\ell) + \frac{4}{3} \frac{\Delta_{K\pi}\Delta_{\ell\pi}}{M_K^4} N_1^P(t,s_\ell) - \frac{2}{3} \frac{\Delta_{K\pi} - t}{M_K^2} \tilde N_1^P(t,s_\ell) \\
			&\quad + \frac{2}{3} R_0^P(t,s_\ell) + P_{F,PT0}^\mathrm{NNLO}(t,s_\ell) - P_{G,PT0}^\mathrm{NNLO}(t,s_\ell)
	\end{split}
\end{align}
and find numerically
\begin{align}
	\begin{split}
		P_{F,PT0}^\mathrm{NNLO}(t,s_\ell) &\approx \frac{M_K}{\sqrt{2} F_\pi} \left( 0.2047 + 0.0339 \frac{s_\ell}{M_K^2} + \left( -0.1781 + 0.0019 \frac{s_\ell}{M_K^2} \right) \frac{t}{M_K^2} - 0.0211 \frac{t^2}{M_K^4} \right) , \\
		P_{G,PT0}^\mathrm{NNLO}(t,s_\ell) &\approx \frac{M_K}{\sqrt{2} F_\pi} \left( -0.0662 + 0.0085 \frac{s_\ell}{M_K^2} + \left( 0.0757 - 0.0054 \frac{s_\ell}{M_K^2} \right) \frac{t}{M_K^2} + 0.0251 \frac{t^2}{M_K^4} \right) .
	\end{split}
\end{align}
Finally, the additional polynomials are given by
\begin{align}
	\begin{split}
		P_{F,P}^\mathrm{NNLO}(s,t,u) &\approx \frac{M_K}{\sqrt{2} F_\pi} \begin{aligned}[t]
			&\bigg( 0.2640 - 0.0510 \frac{s_\ell}{M_K^2} - 0.0002 \frac{s_\ell^2}{M_K^4} + 0.0561 \frac{s}{M_K^2} - 0.0700 \frac{t}{M_K^2} \bigg) , \end{aligned} \\
		P_{G,P}^\mathrm{NNLO}(s,t,u) &\approx \frac{M_K}{\sqrt{2} F_\pi} \begin{aligned}[t]
			&\bigg( 0.0686 + 0.0287 \frac{s_\ell}{M_K^2} + 0.0006 \frac{s_\ell^2}{M_K^4} - 0.0396 \frac{s}{M_K^2} + 0.0169 \frac{t}{M_K^2} \bigg) . \end{aligned}
	\end{split}
\end{align}
Reshuffling all polynomial contributions into the subtraction constants leads to
\begin{align}
	\begin{split}
		m_{0,P}^{0,\mathrm{NNLO}} &= \frac{M_K}{\sqrt{2} F_\pi} \left( 0.3349 - 0.0426 \frac{s_\ell}{M_K^2} + 0.0429 \frac{s_\ell^2}{M_K^4} \right) , \\
		m_{0,P}^{1,\mathrm{NNLO}} &= \frac{M_K}{\sqrt{2} F_\pi} \left( 1.1922 - 0.0523 \frac{s_\ell}{M_K^2} \right) , \\
		m_{0,P}^{2,\mathrm{NNLO}} &= \frac{M_K}{\sqrt{2} F_\pi} \left( 0.8257 \right) , \\
		m_{1,P}^{0,\mathrm{NNLO}} &= \frac{M_K}{\sqrt{2} F_\pi} \left( -0.0083 - 0.0797 \frac{s_\ell}{M_K^2} \right) , \\
		m_{1,P}^{1,\mathrm{NNLO}} &= \frac{M_K}{\sqrt{2} F_\pi} \left( 0.0884 \right) , \\
		\tilde m_{1,P}^{0,\mathrm{NNLO}} &= \frac{M_K}{\sqrt{2} F_\pi} \left( 0.0771 + 0.0016 \frac{s_\ell}{M_K^2} - 0.0367 \frac{s_\ell^2}{M_K^4} \right) , \\
		\tilde m_{1,P}^{1,\mathrm{NNLO}} &= \frac{M_K}{\sqrt{2} F_\pi} \left( -0.0030 + 0.0757 \frac{s_\ell}{M_K^2} \right) , \\
		\tilde m_{1,P}^{2,\mathrm{NNLO}} &= \frac{M_K}{\sqrt{2} F_\pi} \left( -0.0449 \right) , \\
		n_{0,P}^{1,\mathrm{NNLO}} &= \frac{M_K}{\sqrt{2} F_\pi} \left( -0.2217 - 0.0478 \frac{s_\ell}{M_K^2} \right) , \\
		n_{0,P}^{2,\mathrm{NNLO}} &= \frac{M_K}{\sqrt{2} F_\pi} \left( -0.0693 \right) , \\
		n_{1,P}^{0,\mathrm{NNLO}} &= \frac{M_K}{\sqrt{2} F_\pi} \left( 0.0662 \right) , \\
		\tilde n_{1,P}^{1,\mathrm{NNLO}} &= \frac{M_K}{\sqrt{2} F_\pi} \left( -0.0633 \right) .
	\end{split}
\end{align}

\paragraph{NNLO One-Loop Integrals}

The last NNLO piece that we have to decompose is the part containing the $L_i^r$. Similar to the two-loop parts, it can be easily decomposed into functions of one variables. Since this contribution contains only one-loop integrals, we can express it in terms of $A_0$ and $B_0$ functions, which can be treated analytically. After decomposing the NNLO one-loop part according to
\begin{align}
	\begin{split}
		F_L^\mathrm{NNLO}(s,t,u) &= F_{LS,0}^\mathrm{NNLO}(s,s_\ell) + \frac{u-t}{M_K^2} F_{LS,1}^\mathrm{NNLO}(s,s_\ell) + F_{LT,0}^\mathrm{NNLO}(t,s_\ell) + \frac{s-u}{M_K^2} F_{LT,1}^\mathrm{NNLO}(t,s_\ell) \\
			&\quad + F_{LU}^\mathrm{NNLO}(u,s_\ell) + P_{F,L}^\mathrm{NNLO}(s,t,u) , \\
		G_L^\mathrm{NNLO}(s,t,u) &= G_{LS}^\mathrm{NNLO}(s,s_\ell) + G_{LT,0}^\mathrm{NNLO}(t,s_\ell) + \frac{s-u}{M_K^2} G_{LT,1}^\mathrm{NNLO}(t,s_\ell) \\
			&\quad + G_{LU}^\mathrm{NNLO}(u,s_\ell) + P_{G,L}^\mathrm{NNLO}(s,t,u) ,
	\end{split}
\end{align}
the polynomial contribution is found in analogy to the two-loop part. Reshuffling the polynomial gives very long expressions for the subtraction constants. We perform a Taylor expansion in $s_\ell$ and evaluate the expressions numerically, using the physical masses and $\mu=770$~MeV:
\begin{align}
	\footnotesize
	\begin{split}
		m_{0,L}^{0,\mathrm{NNLO}} &= \frac{M_K}{\sqrt{2} F_\pi} \begin{aligned}[t]
			&\bigg( (0.0243 + 0.0155 \cdot 10^3 L_5^r) \cdot 10^3 L_1^r + (0.3528 - 0.0523 \cdot 10^3 L_5^r) \cdot 10^3 L_2^r \\
			& + (0.0831 -0.0092 \cdot 10^3 L_5^r) \cdot 10^3 L_3^r + (0.0400 + 0.0350 \cdot 10^3 L_4^r - 0.0020 \cdot 10^3 L_5^r ) \cdot 10^3 L_4^r \\
			& + ( 0.0066 + 0.0048 \cdot 10^3 L_5^r ) \cdot 10^3 L_5^r - (0.0012 + 0.0699 \cdot 10^3 L_4^r + 0.0087 \cdot 10^3 L_5^r) \cdot 10^3 L_6^r \\
			& + 0.0213 \cdot 10^3 L_7^r + ( 0.0100 - 0.0027 \cdot 10^3 L_4^r - 0.0003 \cdot 10^3 L_5^r ) \cdot 10^3 L_8^r \\
			&+ \frac{s_\ell}{M_K^2}  \begin{aligned}[t]
				& \Big( 0.0213 \cdot 10^3 L_1^r - 0.0161 \cdot 10^3 L_2^r + 0.0230 \cdot 10^3 L_3^r + 0.0139 \cdot 10^3 L_4^r + 0.0018 \cdot 10^3 L_5^r \\
				&- 0.0017 \cdot 10^3 L_6^r - 0.0008 \cdot 10^3 L_8^r + (0.0229 - 0.0060 \cdot 10^3 L_5^r ) \cdot 10^3 L_9^r \Big) \end{aligned} \\
			&+ \frac{s_\ell^2}{M_K^4} \begin{aligned}[t]
				& \Big(-0.0053 \cdot 10^3 L_1^r - 0.0029 \cdot 10^3 L_2^r - 0.0029 \cdot 10^3 L_3^r + 0.0065 \cdot 10^3 L_4^r \\
				& + 0.0010 \cdot 10^3 L_5^r - 0.0012 \cdot 10^3 L_6^r - 0.0006 \cdot 10^3 L_8^r + 0.0025 \cdot 10^3 L_9^r \Big)  \bigg) ,\end{aligned} \end{aligned} \\
		m_{0,L}^{1,\mathrm{NNLO}} &= \frac{M_K}{\sqrt{2} F_\pi} \begin{aligned}[t]
			&\bigg(  - ( 0.1644 + 0.0968 \cdot 10^3 L_5^r) \cdot 10^3 L_1^r - 0.2921 \cdot 10^3 L_2^r - ( 0.1665 + 0.0242 \cdot 10^3 L_5^r) \cdot 10^3 L_3^r \\
			& - 0.0353 \cdot 10^3 L_4^r + 0.0049 \cdot 10^3 L_5^r + 0.0185 \cdot 10^3 L_6^r - 0.0033 \cdot 10^3 L_7^r + 0.0076 \cdot 10^3 L_8^r \\
			& + \frac{s_\ell}{M_K^2} \begin{aligned}[t]
				&\Big(0.0138 \cdot 10^3 L_1^r - 0.0575 \cdot 10^3 L_2^r - 0.0087 \cdot 10^3 L_3^r - 0.0130 \cdot 10^3 L_4^r \\
				& - 0.0020 \cdot 10^3 L_5^r + 0.0024 \cdot 10^3 L_6^r + 0.0012 \cdot 10^3 L_8^r + 0.0196 \cdot 10^3 L_9^r \Big) \bigg) , \end{aligned} \end{aligned} \\
		m_{0,L}^{2,\mathrm{NNLO}} &= \frac{M_K}{\sqrt{2} F_\pi} \begin{aligned}[t]
			&\bigg( 0.3345 \cdot 10^3 L_1^r + 0.2734 \cdot 10^3 L_2^r + 0.1618 \cdot 10^3 L_3^r + 0.0863 \cdot 10^3 L_4^r \\
			&+ 0.0096 \cdot 10^3 L_5^r + 0.0067 \cdot 10^3 L_6^r - 0.0003 \cdot 10^3 L_7^r + 0.0032 \cdot 10^3 L_8^r \bigg) ,\end{aligned} \\
		m_{1,L}^{0,\mathrm{NNLO}} &= \frac{M_K}{\sqrt{2} F_\pi} \begin{aligned}[t]
			&\bigg( - 0.1203 \cdot 10^3 L_1^r + ( - 0.2247 + 0.0242 \cdot 10^3 L_5^r) \cdot 10^3 L_2^r - 0.0727 \cdot 10^3 L_3^r \\
			& - 0.0241 \cdot 10^3 L_4^r - 0.0046 \cdot 10^3 L_5^r + 0.0078 \cdot 10^3 L_6^r + 0.0039 \cdot 10^3 L_8^r \\
			& + \frac{s_\ell}{M_K^2}  \begin{aligned}[t] 
				& \Big( 0.0121 \cdot 10^3 L_1^r + 0.0044 \cdot 10^3 L_2^r + 0.0063 \cdot 10^3 L_3^r - 0.0130 \cdot 10^3 L_4^r \\
				& - 0.0020 \cdot 10^3 L_5^r + 0.0024 \cdot 10^3 L_6^r + 0.0012 \cdot 10^3 L_8^r - 0.0053 \cdot 10^3 L_9^r \Big) \bigg) , \end{aligned} \end{aligned} \\
		m_{1,L}^{1,\mathrm{NNLO}} &= \frac{M_K}{\sqrt{2} F_\pi} \begin{aligned}[t]
			&\bigg( -0.0198 \cdot 10^3 L_1^r - 0.0059 \cdot 10^3 L_2^r - 0.0056 \cdot 10^3 L_3^r + 0.0130 \cdot 10^3 L_4^r \\
			& + 0.0020 \cdot 10^3 L_5^r - 0.0024 \cdot 10^3 L_6^r - 0.0012 \cdot 10^3 L_8^r \bigg) ,\end{aligned} \\
	\end{split}
\end{align}
\begin{align}
	\footnotesize
	\begin{split}
		\tilde m_{1,L}^{0,\mathrm{NNLO}} &= \frac{M_K}{\sqrt{2} F_\pi} \begin{aligned}[t]
			&\bigg( 0.0440 \cdot 10^3 L_1^r + ( - 0.1488 + 0.0281 \cdot 10^3 L_5^r) \cdot 10^3 L_2^r + ( - 0.0140 + 0.0131 \cdot 10^3 L_5^r) \cdot 10^3 L_3^r \\
			& + (0.0001 + 0.0044 \cdot 10^3 L_5^r) \cdot 10^3 L_4^r + ( - 0.0033 + 0.0048 \cdot 10^3 L_5^r ) \cdot 10^3 L_5^r \\
			& + ( 0.0186 - 0.0087 \cdot 10^3 L_5^r ) \cdot 10^3 L_6^r - 0.0135 \cdot 10^3 L_7^r + ( 0.0026 - 0.0003 \cdot 10^3 L_5^r ) \cdot 10^3 L_8^r \\
			& + \frac{s_\ell}{M_K^2} \begin{aligned}[t]
				& \Big( -0.0957 \cdot 10^3 L_1^r - ( 0.2423 - 0.0242 \cdot 10^3 L_5^r) \cdot 10^3 L_2^r - 0.0520 \cdot 10^3 L_3^r - 0.0134 \cdot 10^3 L_4^r \\
				& - 0.0033 \cdot 10^3 L_5^r + 0.0067 \cdot 10^3 L_6^r + 0.0033 \cdot 10^3 L_8^r + ( 0.0098 - 0.0060 \cdot 10^3 L_5^r ) \cdot 10^3 L_9^r \Big) \end{aligned} \\
			& + \frac{s_\ell^2}{M_K^4} \begin{aligned}[t]
				& \Big( 0.0057 \cdot 10^3 L_1^r + 0.0010 \cdot 10^3 L_2^r + 0.0029 \cdot 10^3 L_3^r - 0.0065 \cdot 10^3 L_4^r \\
				& - 0.0010 \cdot 10^3 L_5^r + 0.0012 \cdot 10^3 L_6^r + 0.0006 \cdot 10^3 L_8^r - 0.0034 \cdot 10^3 L_9^r \Big) \bigg) , \end{aligned} \end{aligned} \\
		\tilde m_{1,L}^{1,\mathrm{NNLO}} &= \frac{M_K}{\sqrt{2} F_\pi} \begin{aligned}[t]
			&\bigg( 0.0987 \cdot 10^3 L_1^r + (0.2328 - 0.0242 \cdot 10^3 L_5^r) \cdot 10^3 L_2^r + 0.0581 \cdot 10^3 L_3^r + 0.0213 \cdot 10^3 L_4^r \\
			& + 0.0062 \cdot 10^3 L_5^r - 0.0078 \cdot 10^3 L_6^r - 0.0039 \cdot 10^3 L_8^r \\
			& + \frac{s_\ell}{M_K^2} \begin{aligned}[t]
				& \Big(-0.0138 \cdot 10^3 L_1^r - 0.0029 \cdot 10^3 L_2^r - 0.0026 \cdot 10^3 L_3^r + 0.0130 \cdot 10^3 L_4^r \\
				& + 0.0020 \cdot 10^3 L_5^r - 0.0024 \cdot 10^3 L_6^r - 0.0012 \cdot 10^3 L_8^r + 0.0089 \cdot 10^3 L_9^r \Big) \bigg) , \end{aligned} \end{aligned} \\
		\tilde m_{1,L}^{2,\mathrm{NNLO}} &= \frac{M_K}{\sqrt{2} F_\pi} \begin{aligned}[t]
			&\bigg( -0.0070 \cdot 10^3 L_1^r + 0.0114 \cdot 10^3 L_2^r - 0.0097 \cdot 10^3 L_3^r - 0.0044 \cdot 10^3 L_4^r \\
			& + 0.0001 \cdot 10^3 L_5^r + 0.0012 \cdot 10^3 L_6^r + 0.0006 \cdot 10^3 L_8^r \bigg) ,\end{aligned} \\
		n_{0,L}^{1,\mathrm{NNLO}} &= \frac{M_K}{\sqrt{2} F_\pi} \begin{aligned}[t]
			&\bigg( - 0.0796 \cdot 10^3 L_1^r + ( - 0.4712 + 0.0726 \cdot 10^3 L_5^r) \cdot 10^3 L_2^r + ( - 0.1097 + 0.0181 \cdot 10^3 L_5^r) \cdot 10^3 L_3^r \\
			& - 0.0262 \cdot 10^3 L_4^r - 0.0049 \cdot 10^3 L_5^r + 0.0075 \cdot 10^3 L_6^r - 0.0117 \cdot 10^3 L_7^r - 0.0021 \cdot 10^3 L_8^r \\
			& + \frac{s_\ell}{M_K^2} \begin{aligned}[t]
				& \Big( 0.0095 \cdot 10^3 L_1^r + 0.0035 \cdot 10^3 L_2^r + 0.0040 \cdot 10^3 L_3^r - 0.0098 \cdot 10^3 L_4^r \\
				& - 0.0015 \cdot 10^3 L_5^r + 0.0018 \cdot 10^3 L_6^r + 0.0009 \cdot 10^3 L_8^r - 0.0079 \cdot 10^3 L_9^r \Big)  \bigg) , \end{aligned} \end{aligned} \\
		n_{0,L}^{2,\mathrm{NNLO}} &= \frac{M_K}{\sqrt{2} F_\pi} \begin{aligned}[t]
			&\bigg( -0.0003 \cdot 10^3 L_1^r - 0.0010 \cdot 10^3 L_2^r + 0.0101 \cdot 10^3 L_3^r - 0.0007 \cdot 10^3 L_4^r \\
			& - 0.0022 \cdot 10^3 L_5^r + 0.0003 \cdot 10^3 L_6^r - 0.0010 \cdot 10^3 L_7^r - 0.0004 \cdot 10^3 L_8^r \bigg) ,\end{aligned} \\
		n_{1,L}^{0,\mathrm{NNLO}} &= \frac{M_K}{\sqrt{2} F_\pi} \begin{aligned}[t]
			&\bigg( -0.0125 \cdot 10^3 L_1^r - 0.0051 \cdot 10^3 L_2^r - 0.0069 \cdot 10^3 L_3^r + 0.0098 \cdot 10^3 L_4^r \\
			& + 0.0015 \cdot 10^3 L_5^r - 0.0018 \cdot 10^3 L_6^r - 0.0009 \cdot 10^3 L_8^r \bigg) ,\end{aligned} \\
		\tilde n_{1,L}^{1,\mathrm{NNLO}} &= \frac{M_K}{\sqrt{2} F_\pi} \begin{aligned}[t]
			&\bigg( 0.0059 \cdot 10^3 L_1^r + 0.0096 \cdot 10^3 L_2^r + 0.0051 \cdot 10^3 L_3^r - 0.0072 \cdot 10^3 L_4^r \\
			& - 0.0012 \cdot 10^3 L_5^r + 0.0018 \cdot 10^3 L_6^r + 0.0009 \cdot 10^3 L_8^r \bigg) . \end{aligned}
	\end{split}
\end{align}
Note that there are no quadratic terms in $L_1^r$, $L_2^r$ or $L_3^r$.

\subsubsection{Chiral Expansion of the Omnès Representation}

\label{sec:AppendixNNLOChiralExpansionOmnes}

In order to derive the NNLO chiral expansion of the Omnès representation (\ref{eq:FunctionsOfOneVariableOmnes3Subtr}), we first expand the Omnès function chirally:
\begin{align}
	\begin{split}
		\Omega^\mathrm{NNLO}(s) &= 1 + \omega \frac{s}{M_K^2} +  \bar\omega \frac{s^2}{M_K^4} +  \frac{s^3}{\pi} \int_{s_0}^\infty \frac{\delta_\mathrm{NLO}(s^\prime)}{(s^\prime - s - i \epsilon) {s^\prime}^3} ds^\prime \\
			&\quad + \frac{1}{2} \left( \omega \frac{s}{M_K^2} + \frac{s^2}{\pi} \int_{s_0}^\infty \frac{\delta_\mathrm{LO}(s^\prime)}{(s^\prime - s - i \epsilon) {s^\prime}^2} ds^\prime \right)^2 ,
	\end{split}
\end{align}
where the subtraction terms $\omega$ and $\bar \omega$ are defined in (\ref{eqn:ThriceSubtractedOmnesFunction}).

In the quadratic term of the expansion, only the LO phase enters and therefore only two subtractions are needed. The NLO expansion of the modulus of the inverse Omnès function is given by
\begin{align}
	\begin{split}
		\frac{1}{| \Omega^\mathrm{NLO}(s)|} &= 1 - \omega \frac{s}{M_K^2} - \frac{s^2}{\pi} \pvint_{s_0}^\infty \frac{\delta_\mathrm{NLO}(s^\prime)}{(s^\prime - s - i \epsilon) {s^\prime}^2} ds^\prime .
	\end{split}
\end{align}
Therefore, the NNLO chiral expansion of the argument of the dispersive integrals reads
\begin{align}
	\begin{split}
		\frac{\hat M(s) \sin\delta(s)}{|\Omega(s)|}\Bigg|_\mathrm{NNLO} &= \hat M^\mathrm{LO}(s) \delta_\mathrm{NLO}(s) + \hat M^\mathrm{NLO}(s) \delta_\mathrm{LO}(s) \\
			&\quad - \hat M^\mathrm{LO}(s) \delta_\mathrm{LO}(s) \left( 1 + \omega \frac{s}{M_K^2} + \frac{s^2}{\pi} \pvint_{s_0}^\infty \frac{\delta_\mathrm{LO}(s^\prime)}{(s^\prime - s - i\epsilon) {s^\prime}^2} ds^\prime \right) .
	\end{split}
\end{align}

This leads to
\begin{align}
	\scalebox{0.8}{
	\begin{minipage}{1.1\textwidth}
	$
	\begin{split}
		M_0^\mathrm{NNLO}(s) &= a_\mathrm{LO}^{M_0} \Bigg( \omega_0^0 \frac{s}{M_K^2} +  \bar\omega_0^0 \frac{s^2}{M_K^4} +  \frac{s^3}{\pi} \int_{s_0}^\infty \frac{\delta_{0,\mathrm{NLO}}^0(s^\prime)}{(s^\prime - s - i \epsilon) {s^\prime}^3} ds^\prime + \frac{1}{2} \bigg( \omega_0^0 \frac{s}{M_K^2} + \frac{s^2}{\pi} \int_{s_0}^\infty \frac{\delta_{0,\mathrm{LO}}^0(s^\prime)}{(s^\prime - s - i \epsilon) {s^\prime}^2} ds^\prime \bigg)^2  \Bigg) \\
			&\quad + \bigg( \Delta a_\mathrm{NLO}^{M_0} + b_\mathrm{NLO}^{M_0} \frac{s}{M_K^2} + c_\mathrm{NLO}^{M_0} \frac{s^2}{M_K^4} \bigg) \Bigg( \omega_0^0 \frac{s}{M_K^2} + \frac{s^2}{\pi} \int_{s_0}^\infty \frac{\delta_{0,\mathrm{LO}}^0(s^\prime)}{(s^\prime - s - i \epsilon) {s^\prime}^2} ds^\prime \Bigg) \\
			&\quad + a_\mathrm{NNLO}^{M_0} + b_\mathrm{NNLO}^{M_0} \frac{s}{M_K^2} + c_\mathrm{NNLO}^{M_0} \frac{s^2}{M_K^4} + d_\mathrm{NNLO}^{M_0} \frac{s^3}{M_K^6} + \frac{s^4}{\pi} \int_{s_0}^\infty \frac{\hat M_0^\mathrm{NLO}(s^\prime) \delta_{0,\mathrm{LO}}^0(s^\prime)}{(s^\prime - s - i\epsilon) {s^\prime}^4} ds^\prime , \\
		M_1^\mathrm{NNLO}(s) &= \bigg( a_\mathrm{NLO}^{M_1} + b_\mathrm{NLO}^{M_1} \frac{s}{M_K^2} \bigg) \Bigg( \omega_1^1 \frac{s}{M_K^2} + \frac{s^2}{\pi} \int_{s_0}^\infty \frac{\delta_{1,\mathrm{LO}}^1(s^\prime)}{(s^\prime - s - i \epsilon) {s^\prime}^2} ds^\prime \Bigg) \\
			& \quad + a_\mathrm{NNLO}^{M_1} + b_\mathrm{NNLO}^{M_1}  \frac{s}{M_K^2} + c_\mathrm{NNLO}^{M_1} \frac{s^2}{M_K^4} + \frac{s^3}{\pi} \int_{s_0}^\infty \frac{\hat M_1^\mathrm{NLO}(s^\prime) \delta_{1,\mathrm{LO}}^1(s^\prime)}{(s^\prime - s - i\epsilon) {s^\prime}^3} ds^\prime , \\
		 \tilde M_1^\mathrm{NNLO}(s) &= a_\mathrm{LO}^{\tilde M_1} \Bigg( \omega_1^1 \frac{s}{M_K^2} +  \bar\omega_1^1 \frac{s^2}{M_K^4} +  \frac{s^3}{\pi} \int_{s_0}^\infty \frac{\delta_{1,\mathrm{NLO}}^1(s^\prime)}{(s^\prime - s - i \epsilon) {s^\prime}^3} ds^\prime + \frac{1}{2} \bigg( \omega_1^1 \frac{s}{M_K^2} + \frac{s^2}{\pi} \int_{s_0}^\infty \frac{\delta_{1,\mathrm{LO}}^1(s^\prime)}{(s^\prime - s - i \epsilon) {s^\prime}^2} ds^\prime \bigg)^2 \Bigg) \\
			& \quad + \bigg( \Delta a_\mathrm{NLO}^{\tilde M_1} + b_\mathrm{NLO}^{\tilde M_1}  \frac{s}{M_K^2} + c_\mathrm{NLO}^{\tilde M_1}  \frac{s^2}{M_K^4} \bigg) \Bigg( \omega_1^1 \frac{s}{M_K^2} + \frac{s^2}{\pi} \int_{s_0}^\infty \frac{\delta_{1,\mathrm{LO}}^1(s^\prime)}{(s^\prime - s - i \epsilon) {s^\prime}^2} ds^\prime \Bigg) \\
			& \quad + a_\mathrm{NNLO}^{\tilde M_1} + b_\mathrm{NNLO}^{\tilde M_1}  \frac{s}{M_K^2} + c_\mathrm{NNLO}^{\tilde M_1}  \frac{s^2}{M_K^4} + d_\mathrm{NNLO}^{\tilde M_1} \frac{s^3}{M_K^6} + \frac{s^4}{\pi} \int_{s_0}^\infty \frac{\hat{\tilde M}_1^\mathrm{NLO}(s^\prime) \delta_{1,\mathrm{LO}}^1(s^\prime)}{(s^\prime - s - i\epsilon) {s^\prime}^4} ds^\prime , \\
		N_0^\mathrm{NNLO}(t) &= \Bigg( b_\mathrm{NLO}^{N_0} \frac{t}{M_K^2} + c_\mathrm{NLO}^{N_0} \frac{t^2}{M_K^4} + \frac{t^3}{\pi} \int_{t_0}^\infty \frac{\hat N_0^\mathrm{LO}(t^\prime) \delta_{0,\mathrm{LO}}^{1/2}(t^\prime)}{(t^\prime - t - i\epsilon) {t^\prime}^3} dt^\prime \Bigg) \Bigg( \omega_0^{1/2} \frac{t}{M_K^2} + \frac{t^2}{\pi} \int_{t_0}^\infty \frac{\delta_{0,\mathrm{LO}}^{1/2}(t^\prime)}{(t^\prime - t - i \epsilon) {t^\prime}^2} dt^\prime \Bigg) \\
			&\quad + b_\mathrm{NNLO}^{N_0} \frac{t}{M_K^2} + c_\mathrm{NNLO}^{N_0} \frac{t^2}{M_K^4} + \frac{t^3}{\pi} \int_{t_0}^\infty \frac{\hat N_0^\mathrm{LO}(t^\prime) \delta_{0,\mathrm{NLO}}^{1/2}(t^\prime)}{(t^\prime - t - i\epsilon) {t^\prime}^3} dt^\prime + \frac{t^3}{\pi} \int_{t_0}^\infty \frac{\hat N_0^\mathrm{NLO}(t^\prime) \delta_{0,\mathrm{LO}}^{1/2}(t^\prime)}{(t^\prime - t - i\epsilon) {t^\prime}^3} dt^\prime \\
			&\quad - \frac{t^3}{\pi} \int_{t_0}^\infty \frac{\hat N_0^\mathrm{LO}(t^\prime) \delta_{0,\mathrm{LO}}^{1/2}(t^\prime)}{(t^\prime - t - i\epsilon) {t^\prime}^3} \bigg( 1 + \omega_0^{1/2} \frac{t^\prime}{M_K^2} + \frac{{t^\prime}^2}{\pi} \pvint_{t_0}^\infty \frac{\delta_{0,\mathrm{LO}}^{1/2}(t^\dprime)}{(t^\dprime - t^\prime - i \epsilon){t^\dprime}^2} dt^\dprime \bigg) dt^\prime , \\
		N_1^\mathrm{NNLO}(t) &= a_\mathrm{NLO}^{N_1} \Bigg( \omega_1^{1/2} \frac{t}{M_K^2} + \frac{t^2}{\pi} \int_{t_0}^\infty \frac{\delta_{1,\mathrm{LO}}^{1/2}(t^\prime)}{(t^\prime - t - i \epsilon) {t^\prime}^2} dt^\prime \Bigg) + a_\mathrm{NNLO}^{N_1} + \frac{t}{\pi} \int_{t_0}^\infty \frac{\hat N_1^\mathrm{NLO}(t^\prime) \delta_{1,\mathrm{LO}}^{1/2}(t^\prime)}{(t^\prime - t - i\epsilon){t^\prime}} dt^\prime, \\
		\tilde N_1^\mathrm{NNLO}(t) &= \Bigg( b_\mathrm{NLO}^{\tilde N_1} \frac{t}{M_K^2} + \frac{t^2}{\pi} \int_{t_0}^\infty \frac{\hat{\tilde N}_1^\mathrm{LO}(t^\prime) \delta_{1,\mathrm{LO}}^{1/2}(t^\prime)}{(t^\prime - t - i\epsilon) {t^\prime}^2} dt^\prime \Bigg) \Bigg( \omega_1^{1/2} \frac{t}{M_K^2} + \frac{t^2}{\pi} \int_{t_0}^\infty \frac{\delta_{1,\mathrm{LO}}^{1/2}(t^\prime)}{(t^\prime - t - i \epsilon) {t^\prime}^2} dt^\prime \Bigg) \\
			&\quad +  b_\mathrm{NNLO}^{\tilde N_1} \frac{t}{M_K^2} + \frac{t^2}{\pi} \int_{t_0}^\infty \frac{\hat{\tilde N}_1^\mathrm{LO}(t^\prime) \delta_{1,\mathrm{NLO}}^{1/2}(t^\prime)}{(t^\prime - t - i\epsilon) {t^\prime}^2} dt^\prime + \frac{t^2}{\pi} \int_{t_0}^\infty \frac{\hat{\tilde N}_1^\mathrm{NLO}(t^\prime) \delta_{1,\mathrm{LO}}^{1/2}(t^\prime)}{(t^\prime - t - i\epsilon) {t^\prime}^2} dt^\prime \\
			&\quad - \frac{t^2}{\pi} \int_{t_0}^\infty \frac{\hat{\tilde N}_1^\mathrm{LO}(t^\prime) \delta_{1,\mathrm{LO}}^{1/2}(t^\prime)}{(t^\prime - t - i\epsilon) {t^\prime}^2} \bigg( 1 + \omega_1^{1/2} \frac{t^\prime}{M_K^2} + \frac{{t^\prime}^2}{\pi} \pvint_{t_0}^\infty \frac{\delta_{1,\mathrm{LO}}^{1/2}(t^\dprime)}{(t^\dprime-t^\prime-i\epsilon){t^\dprime}^2} dt^\dprime \bigg) dt^\prime , \\
		R_0^\mathrm{NNLO}(t) &= \Bigg( \frac{t^3}{\pi} \int_{t_0}^\infty \frac{\hat R_0^\mathrm{LO}(t^\prime) \delta_{0,\mathrm{LO}}^{3/2}(t^\prime)}{(t^\prime - t - i\epsilon) {t^\prime}^3} dt^\prime \Bigg) \Bigg( \omega_0^{3/2} \frac{t}{M_K^2} + \frac{t^2}{\pi} \int_{t_0}^\infty \frac{\delta_{0,\mathrm{LO}}^{3/2}(t^\prime)}{(t^\prime - t - i \epsilon) {t^\prime}^2} dt^\prime \Bigg) \\
			&\quad +  \frac{t^3}{\pi} \int_{t_0}^\infty \frac{\hat R_0^\mathrm{LO}(t^\prime) \delta_{0,\mathrm{NLO}}^{3/2}(t^\prime)}{(t^\prime - t - i\epsilon) {t^\prime}^3} dt^\prime +  \frac{t^3}{\pi} \int_{t_0}^\infty \frac{\hat R_0^\mathrm{NLO}(t^\prime) \delta_{0,\mathrm{LO}}^{3/2}(t^\prime)}{(t^\prime - t - i\epsilon) {t^\prime}^3} dt^\prime \\
			& \quad - \frac{t^3}{\pi} \int_{t_0}^\infty \frac{\hat R_0^\mathrm{LO}(t^\prime) \delta_{0,\mathrm{LO}}^{3/2}(t^\prime)}{(t^\prime - t - i\epsilon) {t^\prime}^3} \bigg( 1 + \omega_0^{3/2} \frac{t^\prime}{M_K^2} + \frac{{t^\prime}^2}{\pi} \pvint_{t_0}^\infty \frac{\delta_{0,\mathrm{LO}}^{3/2}(t^\dprime)}{(t^\dprime-t^\prime-i\epsilon){t^\dprime}^2} dt^\dprime \bigg) dt^\prime  , \\
		R_1^\mathrm{NNLO}(t) &= 0 , \\
		\tilde R_1^\mathrm{NNLO}(t) &= 0 ,
	\end{split}
	$
	\end{minipage}
	}
\end{align}
where we use the following notation for the contributions to the subtraction constants:
\begin{align}
	\begin{split}
		a_\mathrm{NLO} &= a_\mathrm{LO} + \Delta a_\mathrm{NLO} , \\
		a_\mathrm{NNLO} &= a_\mathrm{LO} + \Delta a_\mathrm{NLO} + \Delta a_\mathrm{NNLO} .
	\end{split}
\end{align}
Remember that $b_\mathrm{NLO}^{M_1}$ and $a_\mathrm{NLO}^{N_1}$ are non-zero after the gauge transformation.

We further define
\begin{align}
	\begin{split}
		b^{M_0}_\mathrm{NLO} &=: -\omega_0^0 \frac{M_K}{\sqrt{2} F_\pi} + \bar b^{M_0}_\mathrm{NLO} , \\
		b^{M_0}_\mathrm{NNLO} &=: -\omega_0^0 \frac{M_K}{\sqrt{2} F_\pi} + \bar b^{M_0}_\mathrm{NNLO} , \\
		b^{\tilde M_1}_\mathrm{NLO} &=: -\omega_1^1 \frac{M_K}{\sqrt{2} F_\pi} + \bar b^{\tilde M_1}_\mathrm{NLO} , \\
		b^{\tilde M_1}_\mathrm{NNLO} &=: -\omega_1^1 \frac{M_K}{\sqrt{2} F_\pi} + \bar b^{\tilde M_1}_\mathrm{NNLO} ,
	\end{split}
\end{align}
which allows the simplifications
\begin{align}
	\scalebox{0.75}{
	\begin{minipage}{1.2\textwidth}
	$
	\begin{split}
		\label{eq:OmnesRepresentationChirallyExpanded3Subtr}
		M_0^\mathrm{NNLO}(s) &= \frac{M_K}{\sqrt{2} F_\pi} \begin{aligned}[t]
				& \Bigg( \left( \bar\omega_0^0 - \frac{1}{2}{\omega_0^0}^2 \right) \frac{s^2}{M_K^4} +  \frac{s^3}{\pi} \int_{s_0}^\infty \frac{\delta_{0,\mathrm{NLO}}^0(s^\prime)}{(s^\prime - s - i \epsilon) {s^\prime}^3} ds^\prime + \frac{1}{2}\left(\frac{s^2}{\pi} \int_{s_0}^\infty \frac{\delta_{0,\mathrm{LO}}^0(s^\prime)}{(s^\prime - s - i \epsilon) {s^\prime}^2} ds^\prime \right)^2  \Bigg) \end{aligned} \\
			&\quad + \bigg( \Delta a_\mathrm{NLO}^{M_0} + \bar b^{M_0}_\mathrm{NLO} \frac{s}{M_K^2} + c_\mathrm{NLO}^{M_0} \frac{s^2}{M_K^4} \bigg) \Bigg( \omega_0^0 \frac{s}{M_K^2} + \frac{s^2}{\pi} \int_{s_0}^\infty \frac{\delta_{0,\mathrm{LO}}^0(s^\prime)}{(s^\prime - s - i \epsilon) {s^\prime}^2} ds^\prime \Bigg) \\
			&\quad + a_\mathrm{NNLO}^{M_0} + \bar b_\mathrm{NNLO}^{M_0} \frac{s}{M_K^2} + c_\mathrm{NNLO}^{M_0} \frac{s^2}{M_K^4} + d_\mathrm{NNLO}^{M_0} \frac{s^3}{M_K^6} + \frac{s^4}{\pi} \int_{s_0}^\infty \frac{\hat M_0^\mathrm{NLO}(s^\prime) \delta_{0,\mathrm{LO}}^0(s^\prime)}{(s^\prime - s - i\epsilon) {s^\prime}^4} ds^\prime , \\
		M_1^\mathrm{NNLO}(s) &= \bigg( a_\mathrm{NLO}^{M_1} + b_\mathrm{NLO}^{M_1} \frac{s}{M_K^2} \bigg) \Bigg( \omega_1^1 \frac{s}{M_K^2} + \frac{s^2}{\pi} \int_{s_0}^\infty \frac{\delta_{1,\mathrm{LO}}^1(s^\prime)}{(s^\prime - s - i \epsilon) {s^\prime}^2} ds^\prime \Bigg) \\
			& \quad + a_\mathrm{NNLO}^{M_1} + b_\mathrm{NNLO}^{M_1}  \frac{s}{M_K^2} + c_\mathrm{NNLO}^{M_1}  \frac{s^2}{M_K^4} + \frac{s^3}{\pi} \int_{s_0}^\infty \frac{\hat M_1^\mathrm{NLO}(s^\prime) \delta_{1,\mathrm{LO}}^1(s^\prime)}{(s^\prime - s - i\epsilon) {s^\prime}^3} ds^\prime , \\
		\tilde M_1^\mathrm{NNLO}(s) &= \frac{M_K}{\sqrt{2} F_\pi} \begin{aligned}[t]
		 		& \Bigg( \left( \bar\omega_1^1 - \frac{1}{2} {\omega_1^1}^2 \right) \frac{s^2}{M_K^4} +  \frac{s^3}{\pi} \int_{s_0}^\infty \frac{\delta_{1,\mathrm{NLO}}^1(s^\prime)}{(s^\prime - s - i \epsilon) {s^\prime}^3} ds^\prime + \frac{1}{2} \bigg( \frac{s^2}{\pi} \int_{s_0}^\infty \frac{\delta_{1,\mathrm{LO}}^1(s^\prime)}{(s^\prime - s - i \epsilon) {s^\prime}^2} ds^\prime \bigg)^2 \Bigg) \end{aligned} \\
			& \quad + \bigg( \Delta a_\mathrm{NLO}^{\tilde M_1} + \bar b^{\tilde M_1}_\mathrm{NLO} \frac{s}{M_K^2} + c_\mathrm{NLO}^{\tilde M_1}  \frac{s^2}{M_K^4} \bigg) \Bigg( \omega_1^1 \frac{s}{M_K^2} + \frac{s^2}{\pi} \int_{s_0}^\infty \frac{\delta_{1,\mathrm{LO}}^1(s^\prime)}{(s^\prime - s - i \epsilon) {s^\prime}^2} ds^\prime \Bigg) \\
			& \quad + a_\mathrm{NNLO}^{\tilde M_1} + \bar b_\mathrm{NNLO}^{\tilde M_1}  \frac{s}{M_K^2} + c_\mathrm{NNLO}^{\tilde M_1}  \frac{s^2}{M_K^4} + d_\mathrm{NNLO}^{\tilde M_1}  \frac{s^3}{M_K^6} + \frac{s^4}{\pi} \int_{s_0}^\infty \frac{\hat{\tilde M}_1^\mathrm{NLO}(s^\prime) \delta_{1,\mathrm{LO}}^1(s^\prime)}{(s^\prime - s - i\epsilon) {s^\prime}^4} ds^\prime , \\
		N_0^\mathrm{NNLO}(t) &= b_\mathrm{NNLO}^{N_0} \frac{t}{M_K^2} + c_\mathrm{NNLO}^{N_0} \frac{t^2}{M_K^4} + \omega_0^{1/2} \frac{t}{M_K^2} \Bigg( b_\mathrm{NLO}^{N_0} \frac{t}{M_K^2} + \delta c_\mathrm{NLO}^{N_0} \frac{t^2}{M_K^4} \Bigg) \\
			&\quad +  \Bigg(\frac{t^2}{\pi} \int_{t_0}^\infty \frac{\delta_{0,\mathrm{LO}}^{1/2}(t^\prime)}{(t^\prime - t - i \epsilon) {t^\prime}^2} dt^\prime \Bigg) \Bigg( b_\mathrm{NLO}^{N_0} \frac{t}{M_K^2} + c_\mathrm{NLO}^{N_0} \frac{t^2}{M_K^4} + \frac{t^3}{\pi} \int_{t_0}^\infty \frac{\hat N_0^\mathrm{LO}(t^\prime) \delta_{0,\mathrm{LO}}^{1/2}(t^\prime)}{(t^\prime - t - i\epsilon) {t^\prime}^3} dt^\prime \Bigg) \\
			&\quad + \frac{t^3}{\pi} \int_{t_0}^\infty \frac{\hat N_0^\mathrm{LO}(t^\prime) \delta_{0,\mathrm{NLO}}^{1/2}(t^\prime)}{(t^\prime - t - i\epsilon) {t^\prime}^3} dt^\prime + \frac{t^3}{\pi} \int_{t_0}^\infty \frac{\hat N_0^\mathrm{NLO}(t^\prime) \delta_{0,\mathrm{LO}}^{1/2}(t^\prime)}{(t^\prime - t - i\epsilon) {t^\prime}^3} dt^\prime \\
			&\quad - \frac{t^3}{\pi} \int_{t_0}^\infty \frac{\hat N_0^\mathrm{LO}(t^\prime) \delta_{0,\mathrm{LO}}^{1/2}(t^\prime)}{(t^\prime - t - i\epsilon) {t^\prime}^3} \bigg( 1 + \frac{{t^\prime}^2}{\pi} \pvint_{t_0}^\infty \frac{\delta_{0,\mathrm{LO}}^{1/2}(t^\dprime)}{(t^\dprime - t^\prime - i \epsilon){t^\dprime}^2} dt^\dprime \bigg) dt^\prime , \\
		N_1^\mathrm{NNLO}(t) &= a_\mathrm{NNLO}^{N_1} + a_\mathrm{NLO}^{N_1} \omega_1^{1/2} \frac{t}{M_K^2} \\
			&\quad + a_\mathrm{NLO}^{N_1} \frac{t^2}{\pi} \int_{t_0}^\infty \frac{\delta_{1,\mathrm{LO}}^{1/2}(t^\prime)}{(t^\prime - t - i \epsilon) {t^\prime}^2} dt^\prime + \frac{t}{\pi} \int_{t_0}^\infty \frac{\hat N_1^\mathrm{NLO}(t^\prime) \delta_{1,\mathrm{LO}}^{1/2}(t^\prime)}{(t^\prime - t - i\epsilon){t^\prime}} dt^\prime, \\
		\tilde N_1^\mathrm{NNLO}(t) &= b_\mathrm{NNLO}^{\tilde N_1} \frac{t}{M_K^2} + \omega_1^{1/2} \delta b_\mathrm{NLO}^{\tilde N_1} \frac{t^2}{M_K^4} \\
			&\quad + \Bigg( \frac{t^2}{\pi} \int_{t_0}^\infty \frac{\delta_{1,\mathrm{LO}}^{1/2}(t^\prime)}{(t^\prime - t - i \epsilon) {t^\prime}^2} dt^\prime \Bigg) \bigg( b_\mathrm{NLO}^{\tilde N_1} \frac{t}{M_K^2} + \frac{t^2}{\pi} \int_{t_0}^\infty \frac{\hat{\tilde N}_1^\mathrm{LO}(t^\prime) \delta_{1,\mathrm{LO}}^{1/2}(t^\prime)}{(t^\prime - t - i\epsilon) {t^\prime}^2} dt^\prime \bigg) + \frac{t^2}{\pi} \int_{t_0}^\infty \frac{\hat{\tilde N}_1^\mathrm{LO}(t^\prime) \delta_{1,\mathrm{NLO}}^{1/2}(t^\prime)}{(t^\prime - t - i\epsilon) {t^\prime}^2} dt^\prime \\
			&\quad + \frac{t^2}{\pi} \int_{t_0}^\infty \frac{\hat{\tilde N}_1^\mathrm{NLO}(t^\prime) \delta_{1,\mathrm{LO}}^{1/2}(t^\prime)}{(t^\prime - t - i\epsilon) {t^\prime}^2} dt^\prime - \frac{t^2}{\pi} \int_{t_0}^\infty \frac{\hat{\tilde N}_1^\mathrm{LO}(t^\prime) \delta_{1,\mathrm{LO}}^{1/2}(t^\prime)}{(t^\prime - t - i\epsilon) {t^\prime}^2} \bigg( 1 + \frac{{t^\prime}^2}{\pi} \pvint_{t_0}^\infty \frac{\delta_{1,\mathrm{LO}}^{1/2}(t^\dprime)}{(t^\dprime-t^\prime-i\epsilon){t^\dprime}^2} dt^\dprime \bigg) dt^\prime , \\
		R_0^\mathrm{NNLO}(t) &= - \omega_0^{3/2} \frac{t}{M_K^2} \frac{t^2}{\pi} \int_{t_0}^\infty \frac{\hat R_0^\mathrm{LO}(t^\prime) \delta_{0,\mathrm{LO}}^{3/2}(t^\prime)}{{t^\prime}^3} dt^\prime \\
			&\quad + \Bigg( \frac{t^2}{\pi} \int_{t_0}^\infty \frac{\delta_{0,\mathrm{LO}}^{3/2}(t^\prime)}{(t^\prime - t - i \epsilon) {t^\prime}^2} dt^\prime \Bigg) \bigg( \frac{t^3}{\pi} \int_{t_0}^\infty \frac{\hat R_0^\mathrm{LO}(t^\prime) \delta_{0,\mathrm{LO}}^{3/2}(t^\prime)}{(t^\prime - t - i\epsilon) {t^\prime}^3} dt^\prime \bigg) + \frac{t^3}{\pi} \int_{t_0}^\infty \frac{\hat R_0^\mathrm{LO}(t^\prime) \delta_{0,\mathrm{NLO}}^{3/2}(t^\prime)}{(t^\prime - t - i\epsilon) {t^\prime}^3} dt^\prime \\
			&\quad +  \frac{t^3}{\pi} \int_{t_0}^\infty \frac{\hat R_0^\mathrm{NLO}(t^\prime) \delta_{0,\mathrm{LO}}^{3/2}(t^\prime)}{(t^\prime - t - i\epsilon) {t^\prime}^3} dt^\prime - \frac{t^3}{\pi} \int_{t_0}^\infty \frac{\hat R_0^\mathrm{LO}(t^\prime) \delta_{0,\mathrm{LO}}^{3/2}(t^\prime)}{(t^\prime - t - i\epsilon) {t^\prime}^3} \bigg( 1 + \frac{{t^\prime}^2}{\pi} \pvint_{t_0}^\infty \frac{\delta_{0,\mathrm{LO}}^{3/2}(t^\dprime)}{(t^\dprime-t^\prime-i\epsilon){t^\dprime}^2} dt^\dprime \bigg) dt^\prime , \\
		R_1^\mathrm{NNLO}(t) &= 0 , \\
		\tilde R_1^\mathrm{NNLO}(t) &= 0 ,
	\end{split}
	$
	\end{minipage}
	}
\end{align}
where $\delta c_\mathrm{NLO}^{N_0}$ and $\delta b_\mathrm{NLO}^{\tilde N_1}$ are given by (\ref{eq:NLOGaugeTransformationSubtractionConstants}) and the remaining subtraction constants denote the quantities after the gauge transformation. Note that $\omega$ and $\bar\omega$ appear only in polynomial terms. In $M_0$, $M_1$ and $\tilde M_1$, they can be reabsorbed into the NNLO subtraction constants. However, this is not the case for $N_0$, $N_1$, $\tilde N_1$ and $R_0$. Here, we are required to fix the $\omega$-terms by imposing that the chirally expanded Omnès representation agrees with the standard dispersive representation (or finally the two-loop representation). This somewhat awkward situation is just another manifestation of the fact that we identify the chiral representation with the Omnès dispersion relation although the phase shifts of the former have a wrong asymptotic behaviour.

The comparison of the Taylor expansions of (\ref{eq:OmnesRepresentationChirallyExpanded3Subtr}) and (\ref{eq:FunctionsOfOneVariable3Subtr}) leads to the relation (\ref{eqn:NNLORelationOmnesStandardSubtrConst}) for the subtraction constants.

\end{appendices}


\bibliographystyle{my-physrev}
\phantomsection
\addcontentsline{toc}{section}{References}
\bibliography{Literature}

\begin{thebibliography}{10}

\bibitem{Weinberg1968}
S.~Weinberg,
\newblock Phys.~Rev. {\bf 166}, 1568 (1968).

\bibitem{GasserLeutwyler1984}
J.~Gasser and H.~Leutwyler,
\newblock Annals Phys. {\bf 158}, 142 (1984).

\bibitem{GasserLeutwyler1985}
J.~Gasser and H.~Leutwyler,
\newblock Nucl.~Phys. {\bf B250}, 465 (1985).

\bibitem{Shabalin1963}
E.~P. Shabalin,
\newblock J.~Exp.~Theor.~Phys.~(USSR) {\bf 44}, 765 (1963),
\newblock [Sov.~Phys.~JETP {\bf 17}, 517 (1963)].

\bibitem{Cabibbo1965}
N.~Cabibbo and A.~Maksymowicz,
\newblock Phys.~Rev. {\bf 137}, B438 (1965),
\newblock \newline [Erratum-ibid. {\bf 168}, 1926 (1968)].

\bibitem{Batley2010}
J.~Batley {\em et~al.} (NA48-2 Collaboration),
\newblock Eur.~Phys.~J. {\bf C70}, 635 (2010).

\bibitem{Pislak2001}
S.~Pislak {\em et~al.} (BNL-E865 Collaboration),
\newblock Phys.~Rev.~Lett. {\bf 87}, 221801 (2001),
\newblock \newline [Erratum-ibid. {\bf 105}, 019901 (2010)],
  \href{http://arXiv.org/abs/hep-ex/0106071}{[arXiv:hep-ex/0106071]}.

\bibitem{Pislak2003}
S.~Pislak {\em et~al.} (BNL-E865 Collaboration),
\newblock Phys.~Rev. {\bf D67}, 072004 (2003),
\newblock \newline [Erratum-ibid. {\bf D81}, 119903 (2010)],
  \href{http://arxiv.org/abs/hep-ex/0301040}{[arXiv:hep-ex/0301040]}.

\bibitem{Batley2012}
J.~Batley {\em et~al.} (NA48/2 Collaboration),
\newblock Phys.~Lett. {\bf B715}, 105 (2012),
\newblock \href{http://arxiv.org/abs/1206.7065}{[arXiv:1206.7065 [hep-ex]]},
  [Addendum-ibid.~{\bf B740}, 364 (2015)].

\bibitem{Bijnens1994}
J.~Bijnens, G.~Colangelo and J.~Gasser,
\newblock Nucl.~Phys. {\bf B427} (1994),
\newblock {\href{http://arxiv.org/abs/hep-ph/9403390}{[arXiv:hep-ph/9403390]}}.

\bibitem{Amoros2000}
G.~Amoros, J.~Bijnens and P.~Talavera,
\newblock Nucl.~Phys. {\bf B585}, 293 (2000),
\newblock \\
  \href{http://arXiv.org/abs/hep-ph/0003258}{[arXiv:hep-ph/0003258]}.

\bibitem{Stoffer2010}
P.~Stoffer,
\newblock {\em {A Dispersive Treatment of $K_{\ell4}$ Decays}},
\newblock Master's thesis, University of Bern, 2010.

\bibitem{Colangelo2012}
G.~Colangelo, E.~Passemar and P.~Stoffer,
\newblock EPJ Web Conf. {\bf 37}, 05006 (2012),
\newblock \\ \href{http://arxiv.org/abs/1209.0755}{[arXiv:1209.0755 [hep-ph]]}.

\bibitem{Stoffer2013}
P.~Stoffer, G.~Colangelo and E.~Passemar,
\newblock PoS {\bf CD12}, 058 (2013).

\bibitem{Stoffer2014a}
P.~Stoffer,
\newblock {\em Dispersive Treatments of $K_{\ell4}$ Decays and Hadronic
  Light-by-Light Scattering},
\newblock PhD thesis, University of Bern, 2014,
\newblock \href{http://arxiv.org/abs/1412.5171}{arXiv:1412.5171 [hep-ph]}.

\bibitem{Riggenbach1992}
C.~Riggenbach,
\newblock {\em Formfaktoren und Zerfallsbreite von $K_{l4}$-Zerf\"allen in der
  chiralen St\"orungstheorie},
\newblock PhD thesis, University of Bern, 1992.

\bibitem{Jacob1959}
M.~Jacob and G.~Wick,
\newblock Annals Phys. {\bf 7}, 404 (1959).

\bibitem{Martin1970}
A.~D. Martin and T.~D. Spearman,
\newblock {\em Elementary Particle Theory} (North-Holland Publishing Company,
  Amsterdam, 1970).

\bibitem{Kacser1963}
C.~Kacser,
\newblock Phys.~Rev. {\bf 132}, 2712 (1963).

\bibitem{Stern1993}
J.~Stern, H.~Sazdjian and N.~H. Fuchs,
\newblock Phys.~Rev. {\bf D47}, 3814 (1993),
  \href{http://arxiv.org/abs/hep-ph/9301244}{[arXiv:hep-ph/9301244]}.

\bibitem{Ananthanarayan2001}
B.~Ananthanarayan and P.~Buettiker,
\newblock Eur.~Phys.~J. {\bf C19}, 517 (2001),
  {\href{http://arxiv.org/abs/hep-ph/0012023}{[arXiv:hep-ph/0012023]}}.

\bibitem{Froissart1961}
M.~Froissart,
\newblock Phys.~Rev. {\bf 123}, 1053 (1961).

\bibitem{Ananthanarayan2001a}
B.~Ananthanarayan, G.~Colangelo, J.~Gasser and H.~Leutwyler,
\newblock Phys.~Rept. {\bf 353}, 207 (2001), \newline
  \href{http://arxiv.org/abs/hep-ph/0005297}{[arXiv:hep-ph/0005297]}.

\bibitem{Caprini2012}
I.~Caprini, G.~Colangelo and H.~Leutwyler,
\newblock Eur.~Phys.~J. {\bf C72}, 1860 (2012), \\
  \href{http://arxiv.org/abs/1111.7160}{[arXiv:1111.7160 [hep-ph]]}.

\bibitem{Oller2007}
J.~A. Oller and L.~Roca,
\newblock Phys.~Lett. {\bf B651}, 139 (2007),
  \href{http://arxiv.org/abs/0704.0039}{[arXiv:0704.0039 [hep-ph]]}.

\bibitem{Buettiker2004}
P.~Buettiker, S.~Descotes-Genon and B.~Moussallam,
\newblock Eur.~Phys.~J. {\bf C33}, 409 (2004), \\
  \href{http://arxiv.org/abs/hep-ph/0310283}{[arXiv:hep-ph/0310283]}.

\bibitem{Boito2010}
D.~Boito, R.~Escribano and M.~Jamin,
\newblock JHEP {\bf 1009}, 031 (2010),
  \href{http://arxiv.org/abs/1007.1858}{[arXiv:1007.1858 [hep-ph]]}.

\bibitem{Stoffer2014}
P.~Stoffer,
\newblock Eur.~Phys.~J. {\bf C74}, 2749 (2014),
  \href{http://arxiv.org/abs/1312.2066}{[arXiv:1312.2066 [hep-ph]]}.

\bibitem{Treiman1972}
S.~B. Treiman, E.~Witten, R.~Jackiw and B.~Zumino,
\newblock {\em {Current Algebra and Anomalies}} (World Scientific Publishing,
  1985).

\bibitem{DeAlfaro1973}
V.~de~Alfaro, S.~Fubini, G.~Furlan and C.~Rossetti,
\newblock {\em Currents in Hadron Physics} (North-Holland Publishing Company,
  1973).

\bibitem{Bijnens2003}
J.~Bijnens and P.~Talavera,
\newblock Nucl.~Phys. {\bf B669}, 341 (2003),
  \href{http://arxiv.org/abs/hep-ph/0303103}{[arXiv:hep-ph/0303103]}.

\bibitem{DAgostini1994}
G.~D'Agostini,
\newblock Nucl.~Instrum.~Meth. {\bf A346}, 306 (1994).

\bibitem{Ball2010}
R.~D. Ball {\em et~al.} (NNPDF Collaboration),
\newblock JHEP {\bf 1005}, 075 (2010),
  \href{http://arxiv.org/abs/0912.2276}{[arXiv:0912.2276 [hep-ph]]}.

\bibitem{Bernard1991}
V.~Bernard, N.~Kaiser and U.-G. Mei{\ss}ner,
\newblock Nucl.~Phys. {\bf B357}, 129 (1991).

\bibitem{Bijnens1990}
J.~Bijnens,
\newblock Nucl.~Phys. {\bf B337}, 635 (1990).

\bibitem{Riggenbach1991}
C.~Riggenbach, J.~Gasser, J.~F. Donoghue and B.~R. Holstein,
\newblock Phys.~Rev. {\bf D43}, 127 (1991).

\bibitem{Bijnens2014}
J.~Bijnens and G.~Ecker,
\newblock Ann.~Rev.~Nucl.~Part.~Sci. {\bf 64}, 149 (2014),
  \href{http://arxiv.org/abs/1405.6488}{[arXiv:1405.6488 [hep-ph]]}.

\bibitem{MILC2009}
A.~Bazavov {\em et~al.} (MILC Collaboration),
\newblock PoS {\bf CD09}, 007 (2009),
\newblock \href{http://arxiv.org/abs/0910.2966}{[arXiv:0910.2966 [hep-ph]]}.

\bibitem{Aoki2013}
S.~Aoki {\em et~al.},
\newblock {\em {Review of lattice results concerning low energy particle
  physics}}, 2013,
\newblock \\ \href{http://arxiv.org/abs/1310.8555}{arXiv:1310.8555 [hep-lat]}.

\bibitem{BijnensTalavera2002}
J.~Bijnens and P.~Talavera,
\newblock JHEP {\bf 0203}, 046 (2002),
\newblock \href{http://arxiv.org/abs/hep-ph/0203049}{[arXiv:hep-ph/0203049]}.

\bibitem{Jiang2010}
S.-Z. Jiang, Y.~Zhang, C.~Li and Q.~Wang,
\newblock Phys.~Rev. {\bf D81}, 014001 (2010), \\
  \href{http://arxiv.org/abs/0907.5229}{[arXiv:0907.5229 [hep-ph]]}.

\bibitem{Bijnens2012}
J.~Bijnens and I.~Jemos,
\newblock Nucl.~Phys. {\bf B854}, 631 (2012),
\newblock \href{http://arxiv.org/abs/1103.5945}{[arXiv:1103.5945 [hep-ph]]}.

\end{thebibliography}

\end{document}